\documentclass[12pt,preprint]{aastex}

\slugcomment{Accepted for publication in ApJS}

\shorttitle{Spitzer Gould Belt Survey of Corona Australis}
\shortauthors{Peterson et al.}

\begin{document}


\title{The \textit{Spitzer} Survey of Interstellar Clouds in the Gould Belt. III. A Multi-Wavelength View of Corona Australis}


\author{Dawn E. Peterson\altaffilmark{1}, Alessio Caratti o Garatti\altaffilmark{2}, Tyler L. Bourke\altaffilmark{1}, Jan Forbrich\altaffilmark{1}, Robert A. Gutermuth\altaffilmark{3}, Jes K. J{\o}rgensen\altaffilmark{4}, Lori E. Allen\altaffilmark{5}, Brian M. Patten\altaffilmark{1}, Michael M. Dunham\altaffilmark{6}, Paul M. Harvey\altaffilmark{6}, Bruno Mer\'{\i}n\altaffilmark{7}, Nicholas L. Chapman\altaffilmark{8}, Lucas A. Cieza\altaffilmark{9}, Tracy L. Huard\altaffilmark{10}, Claudia Knez\altaffilmark{10}, Brian Prager\altaffilmark{10}, and Neal J. Evans II\altaffilmark{6}}

\email{dpeterson@cfa.harvard.edu}

\altaffiltext{1}{Harvard-Smithsonian Center for Astrophysics, 60 Garden Street, Cambridge, MA 02138, USA}
\altaffiltext{2}{Dublin Institute for Advanced Studies, 31 Fitzwilliam Place, Dublin 2, Ireland}
\altaffiltext{3}{Five College Astronomy Dept., Smith College, Northampton, MA 01063, USA; Department of Astronomy, University of Massachusetts, Amherst, MA 01002, USA}
\altaffiltext{4}{Centre for Star and Planet Formation, Natural History Museum of Denmark, University of Copenhagen, {\O}ster Voldgade 5-7, DK-1350 Copenhagen K., Denmark}
\altaffiltext{5}{National Optical Astronomy Observatory, 950 N. Cherry Avenue, Tucson, AZ 85719, USA}
\altaffiltext{6}{Department of Astronomy, University of Texas at Austin, 1 University Station, C1400, Austin, TX 78712-0259, USA}
\altaffiltext{7}{Herschel Science Centre, European Space Astronomy Centre (ESA), P.O. Box 78, 28691, Villanueva de la Ca\~nada, Madrid, Spain}
\altaffiltext{8}{Jet Propulsion Laboratory, California Institute of Technology, 4800 Oak Grove Drive, MS 301-429, Pasadena, CA 91109, USA}
\altaffiltext{9}{Institute for Astronomy, University of Hawaii at Manoa, Honolulu, HI 96822, USA}
\altaffiltext{10}{Department of Astronomy, University of Maryland, College Park, MD 20742, USA}



\begin{abstract}

We present \textit{Spitzer Space Telescope} IRAC and MIPS observations of a 0.85 deg$^2$ field including the Corona Australis (CrA) star-forming region.  At a distance of 130~pc, CrA is one of the closest regions known to be actively forming stars, particularly within its embedded association, the Coronet.  Using the \textit{Spitzer} data, we identify 51 young stellar objects (YSOs) in CrA which include sources in the well-studied Coronet cluster as well as distributed throughout the molecular cloud.  Twelve of the YSOs discussed are new candidates, one of which is located in the Coronet.  Known YSOs retrieved from the literature are also added to the list, and a total of 116 candidate YSOs in CrA are compiled.  Based on these YSO candidates, the star formation rate is computed to be 12 M$_{\odot}$Myr$^{-1}$, similar to that of the Lupus clouds.  A clustering analysis was also performed, finding that the main cluster core, consisting of 68 members, is elongated (having an aspect ratio of 2.36), with a circular radius of 0.59 pc and mean surface density of 150 pc$^{-2}$.

In addition, we analyze outflows and jets in CrA by means of new CO and H$_2$ data.  We present 1.3~mm interferometric continuum observations made with the Submillimeter Array (SMA) covering R\,CrA, IRS\,5, IRS\,7, and IRAS\,18595-3712 (IRAS\,32).  We also present multi-epoch H$_2$ maps and detect jets and outflows, study their proper motions, and identify exciting sources.  The \textit{Spitzer} and \textit{ISAAC}/VLT observations of IRAS\,32 show a bipolar precessing jet, which drives a CO (2-1) outflow detected in the SMA observations.  There is also clear evidence for a parsec-scale precessing outflow, E-W oriented, and originating in the SMA\,2 region, likely driven by SMA\,2 or IRS\,7A.

\end{abstract}


\keywords{infrared: general --- open clusters and associations: individual (Corona Australis, CrA) --- stars: formation}



\section{Introduction \label{intro}}

The Gould Belt {\it Spitzer} Legacy program is a GO-4 program designed to 
extend the earlier {\it Spitzer} Cores to Disks 
\citep[c2d;][]{evans03,evans09} program, thereby completing a census of 
star-forming regions within 500 pc.  The Gould Belt is a band of stars and 
molecular clouds located within $\sim$20$^{\circ}$ of the Galactic Plane 
\citep{herschel,gould}.  Although, at a declination of approximately 
$-$40$^{\circ}$, Corona Australis (CrA) is not technically within the Gould 
Belt, it is discussed in this paper as part of the survey.  We present an 
extensive study of the entire Corona Australis star-forming region, 
investigating the overall young population, its spatial distribution along the 
molecular cloud, and the young stellar object (YSO) outflows.

\citet{rossano78} used star counts to create an extinction map, and identified 
five clouds in CrA, named clouds A-E, noting that CrA is highly elongated, 
oriented nearly east to west in the sky.  Cloud A is in the west, and it is 
this cloud which corresponds to the R\,CrA/Coronet region.  Large-scale CO 
mapping by \citet{loren79} also shows this elongation, but their high 
spatial resolution observations of the velocity field indicated no velocity 
gradient.  \citet{loren79} interpreted this to mean that the elongation is not 
due to contraction along the rotational axis.  However, \citet{harju93} argued 
that the \citet{loren79} data were not sufficient, and suggest from their own 
C$^{18}$O observations that the R\,CrA core is, in fact, a fragmented disk, 
explaining the observed elongation along the major axis.  It is somewhat 
surprising that CrA is not associated with the Gould Belt, considering its 
close distance of 130~pc\footnote{The distance to CrA is based on measurements 
to several stars, ranging from 85 to 190 pc.  A comprehensive summary of the 
various measurements can be found in \citet{nf08}; throughout this paper we 
use their suggested distance of 130 pc.}; \citet{mf01} argue that instead it 
formed as part of expanding Sco-Cen superbubbles, specifically Loop I, citing 
evidence from the \citet{harju93} millimeter observations, and radio 
observations from \citet{cappa91}.  

Studies from the literature have mainly focused on R\,CrA, the brightest star 
in the cluster, the Coronet region, and its population.  The variability of 
the nebula surrounding the Herbig Ae star, R\,CrA, has been known since the 
early 1900s \citep{ks16,reynolds16}.  Many years later, the two variable stars 
R\,CrA and T\,CrA were identified by \citet{h60} to be young, and he then 
concluded that the associated stars should also be young.  This prompted an 
interest in studying the R\,CrA region, and in 1973, the first major optical 
and infrared study of the main stars near R\,CrA was conducted, finding a 
total of 11 stars in the young stellar group: TY\,CrA, S\,CrA, T\,CrA, R\,CrA, 
DG\,CrA, VV\,CrA, KS-15, HR\,7169, HR\,7170, Anon\,1, and Anon\,2 
\citep{knacke73}.  The following year, IRS\,1 was suggested by \citet{ssg74} 
to be the driving source for the Herbig-Haro (HH) object, HH\,100.  Subsequent 
infrared observations were made by many groups \citep{gp75,vrba76,ts84,wts86} 
as well as H$\alpha$ observations \citep{mr81} and emission-line observations 
\citep{graham93}.  These were followed by early X-ray 
\citep{walter86,koyama96,np97,walter97,patten98}, radio 
\citep{brown87,cappa91}, millimeter \citep{harju93}, and far-infrared studies 
\citep[][who first mentioned IRAS\,32]{wilking92}.

\citet{nf08} have recently reviewed the literature on the entire Corona 
Australis star-forming region, although many of the studies cover only a 
subset of the region which we include here as part of our {\it Spitzer} IRAC 
and MIPS study.  The most relevant recent studies include: deep infrared 
observations \citep{wilking97,haas08}, millimeter and submillimeter 
observations \citep{chini03,groppi04,nutter05}, including Submillimeter Array 
(SMA) observations \citep{groppi07}, mid-infrared observations with the 
{\it Infrared Space Observatory (ISO)} \citep{olof99}, and a series of papers 
focusing on {\it Chandra} X-ray studies of the region \citep{fpm06,f07,fp07}.  
Finally, some spectroscopic work has been done to determine association 
memberships \citep{patten98,nisini05,s-a08,mw09}.

The most massive stars in CrA are the Herbig Ae/Be stars, R\,CrA and TY\,CrA.  
R\,CrA has a spectral type of A5 \citep{knacke73,mr81} and is located at the 
tip of a cometary shaped reflection nebula, NGC\,6729.  TY\,CrA, located 
$\sim$5$^{\prime}$ to the northwest of R\,CrA, is at least a quadruple 
system \citep{casey95,chauvin03} where the primary has a spectral type of 
B8-B9 \citep{hr72,knacke73,mr81}, and is associated with the reflection nebula 
NGC 6726/7.  Located 1$^{\prime}$ south of TY\,CrA, is HD 176386, which is 
also surrounded by reflection nebulosity from NGC 6727 \citep{knacke73,mr81}.  
HD 176386 is a binary system, first recognized as such by \citet{wilking97}; 
HD 176386A/B is a visual pair separated by 3.$^{\prime\prime}$7 with spectral 
types of A0V and K7, respectively \citep{mw09}.  Additionally, a little 
further from the Coronet ($\sim$12$^{\prime}$ to the southwest of R\,CrA) lie 
two B8V stars, HR\,7169 and HR\,7170, which are discussed in more detail in 
Appendix~\ref{sec:bstars}.  Evidence for heating of the molecular cloud by 
these two B8 stars has been seen \citep{loren79}, making it likely that there 
is a physical association and that they are therefore located at the same 
distance as R\,CrA \citep[see further discussion in][]{neu00}.

CrA is also well-known to harbor many active YSOs with outflows.  To date, 
twenty HH objects, including 48 different knots, have been discovered in the 
CrA star-forming region.  Eight objects (HH 82, 96-101, 104) were detected in 
the 1970s and 1980s \citep{ssg74,schwartz,hartigan87,reipurth}, while the 
remaining twelve objects were observed by \citet{wang} in a large optical 
survey.  The majority of the detected objects are located close to the Coronet 
and seem to be driven by YSOs inside it or in its outskirts.  A few more HH 
objects are positioned close to HH100-IR (IRS\,1), S\,CrA, VV\,CrA and 
IRAS\,18595-3712 (IRAS 32), which seem to drive outflows as well.  On the 
other hand, there are very few and sparse studies on H$_2$ jets and outflows 
in CrA \citep[see e.\,g.][]{wilking90,gredel,davis99,caratti06}.  In these 
papers, the H$_2$ counterparts of HH\,99, 100, 101 and 104 were identified and 
studied, and a few new jets in the Coronet were detected~\citep{caratti06}.  
\citet{davis09b} cataloged five molecular hydrogen objects (MHOs, MHO 
2000-2004), but so far, an extensive H$_2$ map of the region has not been 
made.  Thus, a complete census of outflows and their driving sources is 
lacking.

We present {\it Spitzer} observations of a 0.85 deg$^2$ region in the Corona 
Australis molecular cloud, identifying the {\it Spitzer}-selected YSOs 
distributed throughout the cloud, as well as the outflows and their driving 
sources.  This study includes infrared imaging of a much larger portion of the 
molecular cloud than many previous studies have included.  In 
\S~\ref{sec:observations}, we discuss the {\it Spitzer} observations and data 
reduction, including basic statistics for the sources detected, as well as the 
ancillary SMA observations, and H$_2$ observations of outflows.  In 
\S~\ref{sec:ysoselect}, we discuss methods used to select YSO candidates from 
color-magnitude diagrams constructed from the {\it Spitzer} data, as well as 
other methods of selection.  We also discuss the addition of known YSOs and 
YSO candidates from the literature because they were not classified from 
\textit{Spitzer} data due to saturation or other observational issues.  The 
distribution of YSOs is discussed in \S~\ref{sec:yso_distribution}.  There we 
present an extinction map created from near-infrared and {\it Spitzer} data, 
an analysis of the spatial distribution of the YSOs in the cloud, and the 
clustering analysis performed using the YSO candidates in CrA.  In 
\S~\ref{sec:sma_analysis}, we present an analysis of the SMA observations.  In 
\S~\ref{sec:outflows_jets}, we present an analysis of the jets and outflows, 
the proper motions of the H$_2$ knots, and the driving sources for the 
outflows and jets seen in CrA.  Finally, in \S~\ref{sec:cloudproperties}, we 
discuss the overall cloud properties and how they compare with the other, c2d 
and Gould Belt regions, and summarize the results in \S~\ref{sec:summary}.

\section{Observations and Data Reduction \label{sec:observations}}

\subsection{\textit{Spitzer} IRAC and MIPS \label{sec:spitzerobs}}

%% In a manner similar to \objectname authors can provide links to dataset
%% hosted at participating data centers via the \dataset{} command.  The
%% second curly bracket argument is printed in the text while the first
%% parentheses argument serves as the valid data set identifier.  Large
%% lists of data set are best provided in a table (see Table 3 for an example).
%% Valid data set identifiers should be obtained from the data center that
%% is currently hosting the data.
%%
%% Note that AASTeX interprets everything between the curly braces in the 
%% macro as regular text, so any special characters, e.g. "#" or "_," must be 
%% preceded by a backslash. Otherwise, you will get a LaTeX error when you 
%% compile your manuscript.  Special characters do not 
%% need to be escaped in the optional, square-bracket argument.

\begin{figure}
\epsscale{1}
\plotone{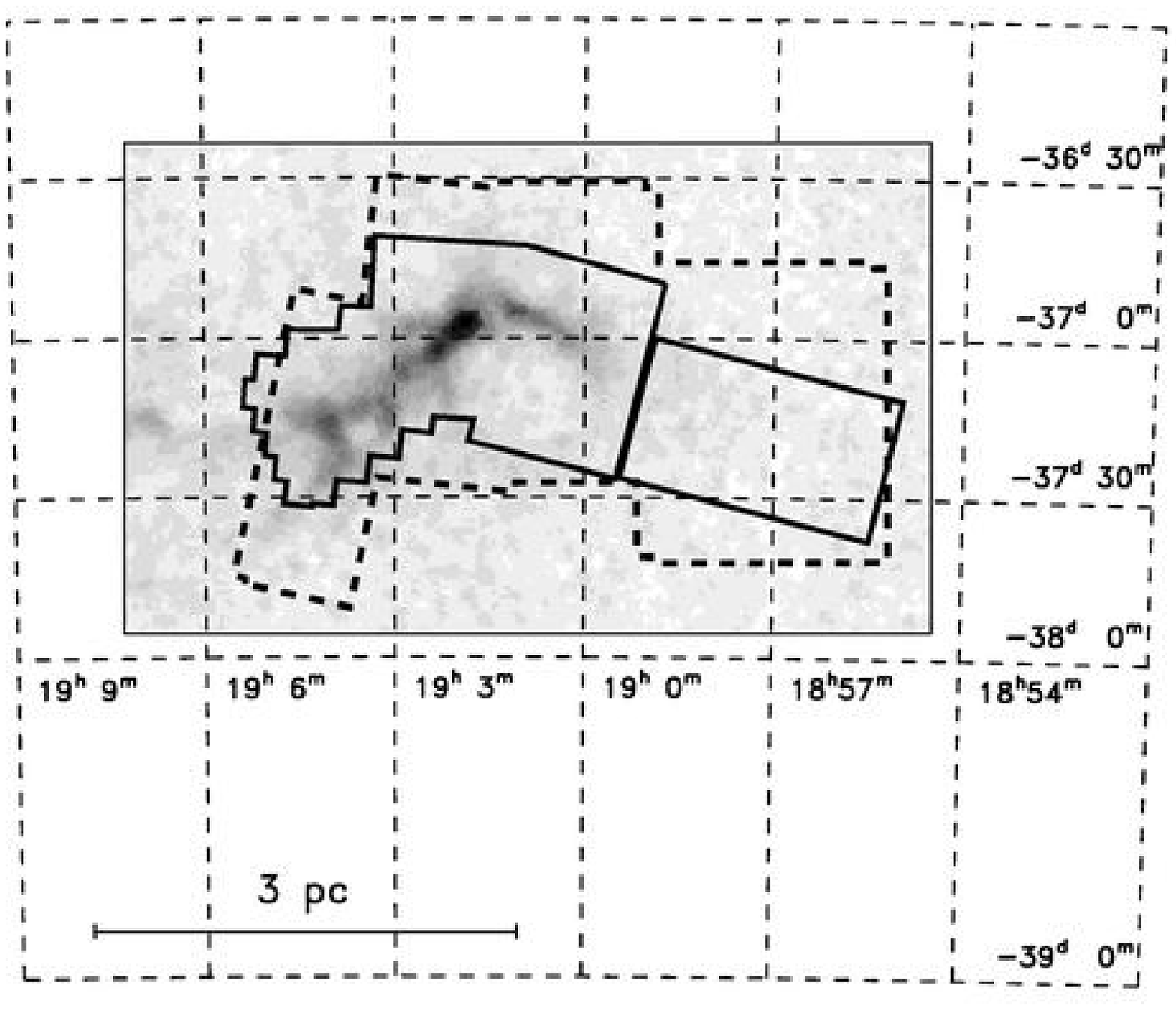}
\caption{Coverage map for the {\it Spitzer} data taken in CrA.  The thick black line shows the IRAC coverage in bands 1 and 3 (3.6/5.8 $\mu$m) and the thick black dotted line shows the MIPS 24 $\mu$m coverage.  The underlying greyscale is a near-infrared extinction map created using 2MASS point sources.  \label{fig:coverage}}
\end{figure}

Corona Australis was observed with the {\it Spitzer} Infrared Array Camera 
\citep[IRAC;][]{fazio04} at 3.6, 4.5, 5.8, and 8.0 $\mu$m and the Multiband 
Imaging Photometer for {\it Spitzer} \citep[MIPS;][]{rieke04} at 24, 70, and 
160 $\mu$m as part of two guaranteed time observation (GTO) programs (PID 6, 
30784; PI Fazio) as well as the Gould Belt Legacy Survey (PID 30574; 
PI Allen).  Table \ref{tab:spitzerobs} summarizes the program identification 
numbers, AOR identification numbers and observation dates for all of the CrA 
data included in this paper.  The IRAC mapping includes one epoch, two 
dithers, with 12 second integration times per exposure, in high-dynamic range 
mode.  The MIPS mapping was done with medium scan rate, one epoch, with 
148$^{\prime \prime}$ return and forward leg cross scan steps.  The size of 
the overlap area from IRAC band 1 - MIPS band 1 is 0.85 deg$^2$; Figure 
\ref{fig:coverage} shows the IRAC and MIPS coverage in CrA.  

The basic calibrated data (BCD) are downloaded from the {\it Spitzer} archive 
and processed by the team using the same, custom pipeline processing programs 
and techniques as used by c2d 
\citep[see the c2d Final Delivery Document,][]{evans07}.  Briefly, the data 
were inspected, custom masks were created to identify bad pixels, and 
corrected for instrumental effects.  Mosaics were created from the improved 
data using the MOPEX package \citep{makovoz06}, and source extraction was 
performed using the c2dphot tool \citep{evans07}, which is a derivative of 
DoPHOT \citep{schechter93}.  Source lists for detections at each wavelength 
were band-merged together with the 2MASS catalog \citep{skruts06} and cross 
identifications are accurate within 2$^{\prime \prime}$ \citep[see Section 2.4 
of the c2d Final Delivery Document,][]{evans07}.  Also note that the 
final catalog is ``band-filled'' (see e.\,g., the c2d Final Delivery Document) 
to produce flux estimates of objects that were not found in the original 
source extraction processing, but were detected in the other bands (see 
further discussion in \S~\ref{sec:ysoselect}).  This procedure is described in 
detail in the delivery documentation for the c2d Final Delivery Document 
\citep{evans07}.  Briefly, c2dphot fits the PSF profile at the position of the 
known source.  Note, however, that all these flux densities, flagged as 
band-filled in our tables, should be considered as ``bad photometry,'' in the 
same way as the 5$\sigma$ upper limits in this catalog.

Statistics for the sources detected in CrA with S/N 
of at least 3 can be found in Table \ref{tab:spitzerdetect}.  This corresponds 
to selecting all sources with detection quality ``A,'' ``B,'' or ``C'' 
(S/N $\geq$ 7, 5, and 3, respectively) in any of the IRAC bands from the 
final delivered catalogs (cf. the c2d Final Delivery Document).

\begin{figure}
\epsscale{1}
\plotone{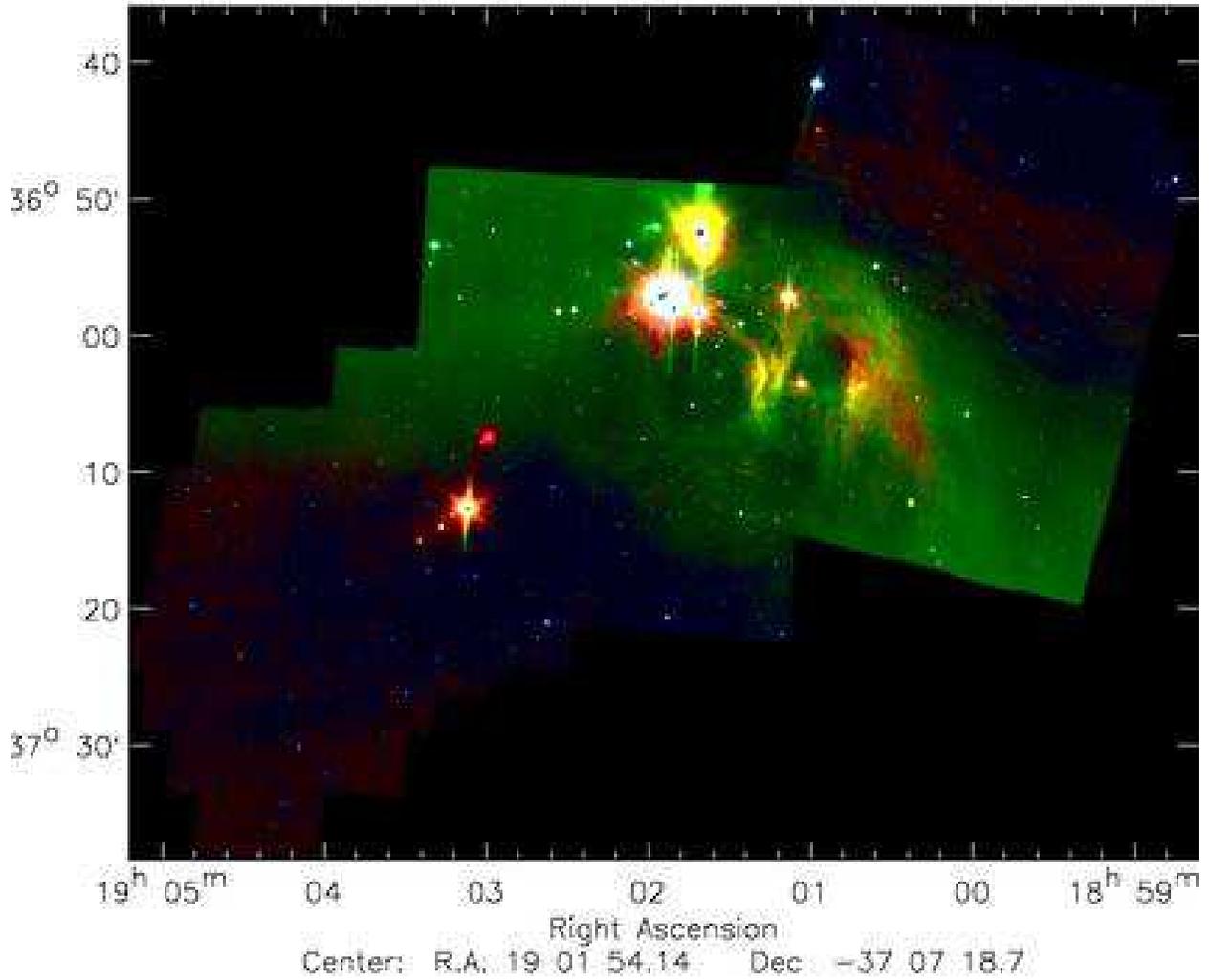}
\caption{Three-color image of the main region in CrA, including the Coronet (bright white saturated region toward the center), made from the {\it Spitzer} IRAC 4.5 $\mu$m (blue), 8.0 $\mu$m (green), and MIPS 24 $\mu$m (red) bands. \label{fig:rgbmain}}
\end{figure}

\begin{figure}
\epsscale{1}
\plotone{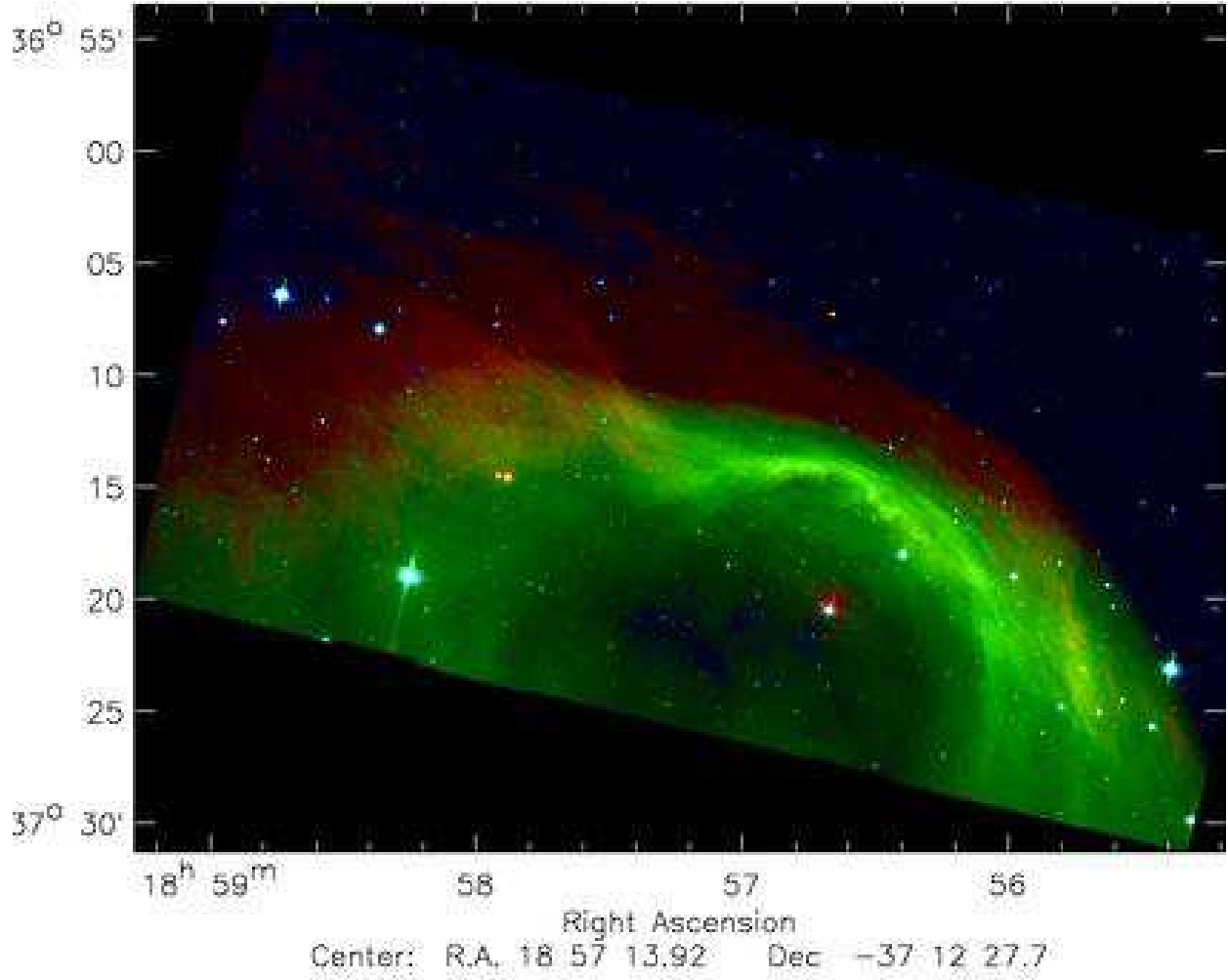}
\caption{Three-color image of the streamer to the west of the main CrA region shown in Figure \ref{fig:rgbmain}, made from the {\it Spitzer} IRAC 4.5 $\mu$m (blue), 8.0 $\mu$m (green), and MIPS 24 $\mu$m (red) bands. \label{fig:rgbstreamer}}
\end{figure}

Figures \ref{fig:rgbmain} and \ref{fig:rgbstreamer} show color mosaics of the 
entire CrA region (mapped with IRAC and MIPS), using IRAC bands 2 (4.5 $\mu$m) 
and 4 (8.0 $\mu$m), and MIPS 24 $\mu$m (blue, green, and red, respectively).  
Figure \ref{fig:rgbmain} includes the Coronet region, the SE and SW filaments, 
and Figure \ref{fig:rgbstreamer} shows the less known region called the 
``streamer,'' positioned further to the southwest.

\subsection{Ancillary Submillimeter Array Observations \label{sec:smaobs}}

The regions around the previously identified YSOs IRS\,5 
\citep{wilking97,fpm06}, IRS\,7 \citep{wilking97,groppi07}, and IRAS\,32 
\citep[IRAS 18595-3712;][]{wilking92,con07,vankempen} were observed with the 
SMA\footnote{The Submillimeter Array is a joint project between the Smithsonian
Astrophysical Observatory and the Academia Sinica Institute of Astronomy
and Astrophysics and is funded by the Smithsonian Institution and the
Academia Sinica.} \citep{ho04} in the dust continuum near 225 GHz
($\sim1.3$ mm).  Observations of the IRS\,7 region were made on 2006 August
20 (program 2006-03-S046), while IRAS\,32 and IRS\,5 were observed on 2008
June 10 and 14 (program 2008A-S074).  In this paper, we present only the 
continuum data for IRS\,7, which were obtained as part of a line survey 
program; the results from that program will be reported elsewhere (Lindberg et 
al., in preparation).  The data were edited and calibrated in the IDL-based 
software package 
MIR.\footnote{http://cfa-www.harvard.edu/$\sim$cqi/mircook.html} 

For the observations of IRS\,5 and IRAS\,32, the SMA was in its compact-north
configuration, with eight antennas, covering baselines of 16--139 m.  The
receiver's two sidebands were centered near 221 and 231 GHz.  Chunks 
containing obvious spectral lines were removed before forming the continuum, 
resulting in a total continuum coverage of $\sim3.6$ GHz centered at 226 GHz.  
Observations of the quasar 3C~279 were used for bandpass calibration, and the 
quasars J1924-292 and J1937-399 were used for complex gain calibration.  
3C~279 was used for absolute flux calibration, assuming a flux of 10.3 Jy 
based on observations at dates near in time to these that were flux calibrated 
using planets and moons.\footnote{See
http://sma1.sma.hawaii.edu/callist/callist.html?data=1256-057} 

For the observations of the IRS\,7 region, the SMA was in its compact
configuration with six antennas, covering baselines of 16--69 m.  A
two-point mosaic pattern was used to cover the region.  The receiver's two
sidebands were centered near 219 and 229 GHz, with each sideband consisting
of 24 overlapping chunks of 109 MHz each, or about 2 GHz for each sideband.
Spectral lines were removed from the data, resulting in a total continuum
coverage of $\sim3.3$ GHz centered at 224 GHz.  Observations of the quasar
3C~454.3 were used for bandpass calibration, and the quasars J1924-292 and
J1957-387 were used for complex gain calibration.  Uranus was used for
absolute flux calibration.  

The MIRIAD software package was used for Fourier inversion of the
visibilities, CLEAN deconvolution, and restoration with a synthesized beam.
A Briggs robust weighting of 0 was used when mapping the continuum
emission.  The data were scaled by the inverse response of the primary beam
(``primary-beam corrected"), to account for the loss of sensitivity away
from the phase center.  Synthesised beam sizes and 1\,$\sigma$ rms 
sensitivities are: $5\farcs5 \times 2\farcs$3 and 7.4 mJy for IRS\,7, 
$4\farcs6 \times 2\farcs6$ and 3.7 mJy for IRS\,5, and 
$3\farcs7 \times 2\farcs2$ and 4.4 mJy for IRAS\,32, respectively. 

\subsection{Ancillary H$_2$ Data \label{sec:ancillary}}

As a complement to the {\it Spitzer} and SMA data, we make use of multi-epoch 
H$_2$ (2.12\,$\mu$m) maps to detect jets and outflows in the CrA star-forming 
region to study their proper motions (P.M.s) and identify the exciting 
sources.  Most of the data collected also have K$_s$ or narrow-band filter 
images to remove the continuum from the H$_2$ maps.  The relevant information 
on these ancillary data is reported in Table~\ref{obs_ancillary:tab}.

Our narrow-band H$_2$ image archive is composed of data collected at ESO-NTT 
with \textit{SofI}~\citep{moor1} in 
1999~\citep[already published in][]{caratti06}, and in 2007.  Additional 
images were retrieved from the ESO science archive 
facility\footnote{http://archive.eso.org/} taken at the ESO-VLT with 
\textit{ISAAC}~\citep{moor2} in 2000, 2003, and 2005.

\begin{figure}
\includegraphics[width=16.0 cm]{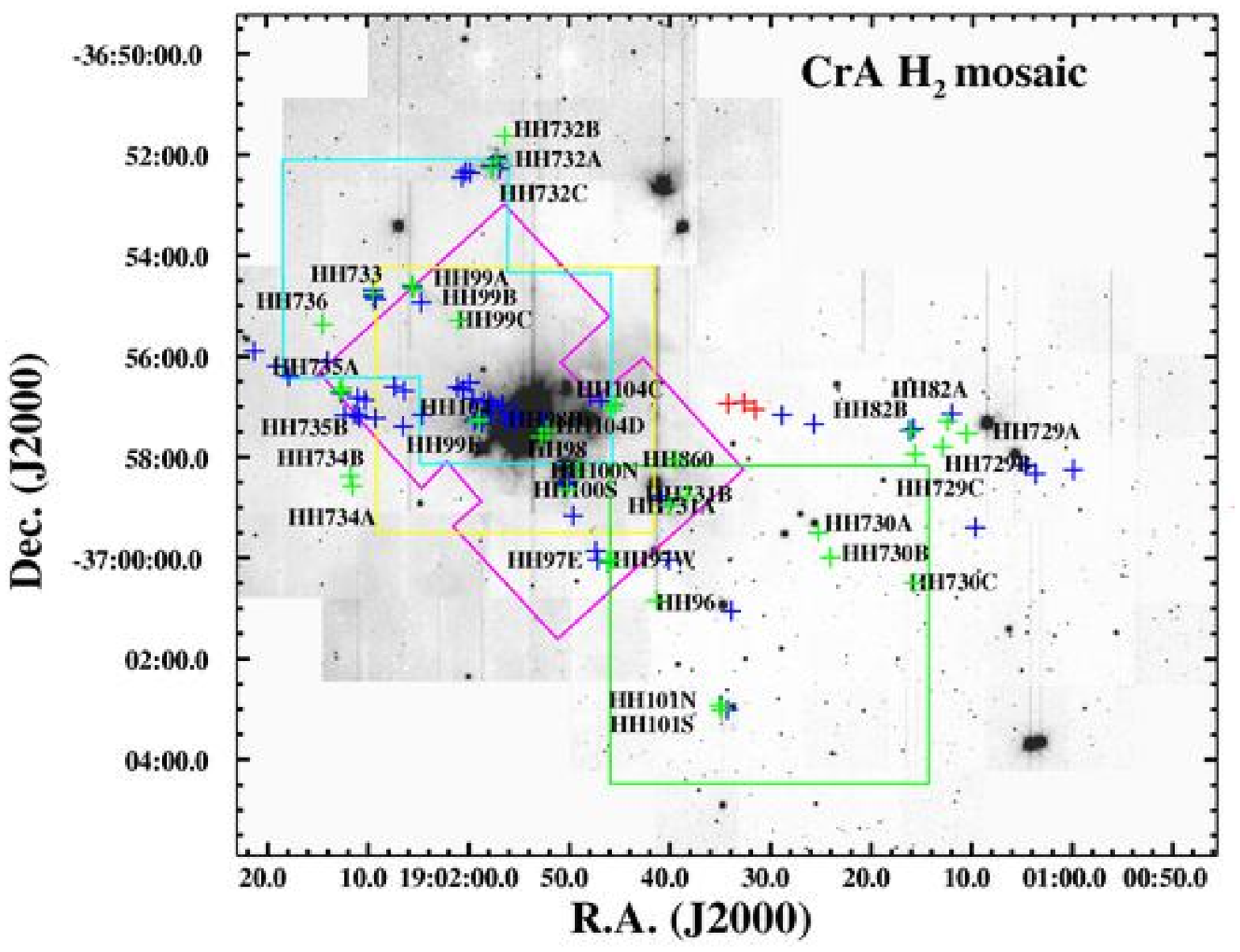}
\caption{\textit{SofI} (2007) H$_2$ mosaic.  Different polygons show regions mapped in different epochs (Yellow: 1999, Cyan: 2000, Green: 2003, Magenta: 2005).  Known HH objects are indicated with green crosses.  Newly detected H$_2$ and {\it Spitzer} knots are also displayed as blue and red crosses, respectively. \label{SofI07map:fig}}
\end{figure}

\textit{SofI} 2007 images were observed in service mode between June and 
August 2007, and cover four regions of CrA ($\sim$10$\arcmin\times$10$\arcmin$ 
each), mapping in total an area of $\sim$20$\arcmin$$\times$17$\arcmin$ 
(about one tenth of the area mapped by {\it Spitzer}), for a total integration 
time of 5760\,s.  Figure\,~\ref{SofI07map:fig} shows the area covered by the 
H$_2$ mosaic.

Earlier epoch maps enclose much smaller regions, centered on R\,CrA 
(\textit{ISAAC} 2005 and \textit{SofI} 1999, 
$\sim$5$\arcmin$$\times$7$\arcmin$ and 5.2$\arcmin$$\times$5.2$\arcmin$ FoV, 
respectively), on HH\,99 (\textit{ISAAC} 2000, 
$\sim$6.8$\arcmin$$\times$6$\arcmin$) or HH\,101 (\textit{ISAAC} 2003, 
6.3$\arcmin$$\times$6.3$\arcmin$), the details of which can be found in 
Table~\ref{obs_ancillary:tab}.  These areas are indicated in 
Figure\,~\ref{SofI07map:fig}, where different polygons are superimposed on the 
\textit{SofI} 2007 mosaic, showing regions mapped in different epochs (yellow: 
1999, cyan: 2000, green: 2003, magenta: 2005).  As a consequence, the proper 
motion (P.M.) analysis was mostly performed on knots inside and close to the 
Coronet region.  

Finally, additional \textit{ISAAC} images (H$_2$ and K$_s$ filters) around 
IRAS\,18595-3712 (IRAS\,32; located outside the \textit{SofI} 2007 map), 
taken in 2005, were acquired to detect a possible H$_2$ jet from that source.

All the raw data were reduced using \emph{IRAF} packages applying standard 
procedures for sky subtraction, dome flat-fielding, bad pixel and cosmic ray 
removal, and making image mosaics.  Continuum-subtracted images, H$_2$-$K_s$ 
and H$_2$-NB(2.09$\mu$m), were obtained by subtracting $K_s$ and 
NB (2.09\,$\mu$m) images, appropriately scaled and registered.  Each pair of 
images was matched by means of tens to hundreds of field stars (depending on 
the mosaic size and crowding), selected excluding the YSO candidates.  The 
scaling has been done by performing relative photometry on the selected 
stars.  The images were not flux calibrated.

\section{YSO Selection and Classification \label{sec:ysoselect}}

\begin{figure}
\plotone{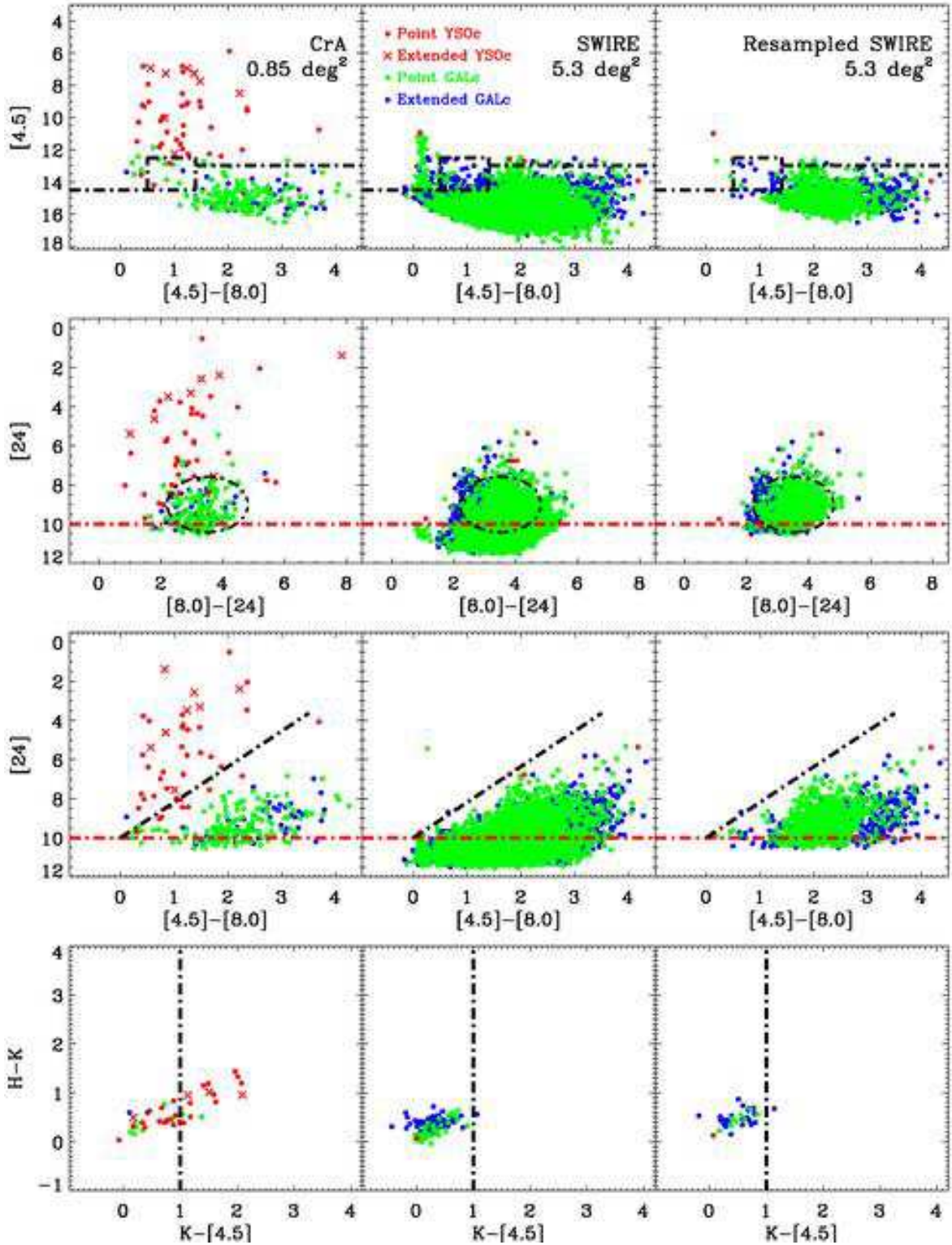}
\caption{Color-magnitude diagrams showing the color space where YSOs are selected from {\it Spitzer} data along with the full SWIRE and re-sampled SWIRE catalogs.  Based on the c2d YSO identification scheme by \citet{harv07}, where red and black dot-dashed lines are hard and fuzzy identification limits, respectively.\label{fig:ysosel}}
\end{figure}

There are 45 sources in the observed Corona Australis region which were 
initially classified as YSO candidates from their {\it Spitzer} plus 
near-infrared colors using the technique outlined in \citet[][and references 
therein]{harv07,harv08}.  This technique has been successfully adopted for 
the c2d and Gould Belt \textit{Spitzer} surveys.  Indeed, 33 out of the 45 
sources in CrA ($\sim$73\%) have already been confirmed as YSOs in other 
studies (see \S~\ref{sec:spitzer_ysos} and Appendix \ref{sec:appA}).  Briefly, 
the selection method uses criteria based on a combination of infrared excess 
with a brightness limit, in order to minimize extragalactic contamination.  
The selection method requires a S/N of 3 or higher detection in all four IRAC 
bands as well as in the MIPS 24~$\mu$m band.  Figure~\ref{fig:ysosel} shows 
the color-magnitude space used to separate the YSO population in CrA from 
contaminants, which are mainly background giant stars and galaxies.  To 
illustrate where the contaminating galaxy population is located in the same 
color-magnitude space, Figure~\ref{fig:ysosel} also shows color-magnitude 
diagrams for the full and re-sampled {\it Spitzer} Wide-area InfraRed 
Extragalactic Survey (SWIRE) catalogs \citep{surace04}.  The SWIRE 
observations were processed in exactly the same way as our data were 
processed in order to be used for a direct comparison (except that 
band-filling was not performed).  Note that the SWIRE data are considerably 
deeper than ours, so the SWIRE catalog was trimmed down assuming the SWIRE 
observations had been obtained with sensitivities similar to those of our 
observations (this results in the ``re-sampled SWIRE'' catalog).  A full 
discussion of the specifics of this process can be found in the c2d Final 
Delivery Document \citep{evans07}.

Table~\ref{tab:ysos_wfluxes} lists all YSO candidates selected through this 
method in CrA along with their {\it Spitzer} fluxes, known names, and classes 
\citep{lada87,greene94}.  Class is determined from the spectral slope, 
$\alpha$, which is calculated over the widest range possible where data are 
available between 2.2 and 24~$\mu$m, and is defined as:

\begin{equation}
\alpha = \frac{d \: \log (\lambda F_{\lambda})}{d \: \log (\lambda)}.
\label{eqn:sed}
\end{equation}

\noindent
The value of $\alpha$ is calculated from a linear fit to the logarithms, 
taking into account the uncertainties in the flux measurements.  We classify 
sources with $\alpha$~$\geq$~0.3 as Class~I, $-$0.3~$\leq$~$\alpha$~$<$~0.3 as 
Flat spectrum, $-$1.6~$\leq$~$\alpha$~$<$~$-$0.3 as Class~II, and 
$\alpha$~$<$~$-$1.6 as Class~III.  Note that deeply embedded protostars, 
Class~0 sources, which require submillimeter data for identification, cannot 
be distinguished from Class~I objects in this analysis.

It is important to note that the $\alpha$ quoted in the tables and in our 
catalogs is computed by selecting all valid wavelengths (having a quality of 
detection not ``N'' or ``U'', meaning a non-detection or an upper limit, 
respectively) in the combined epochs between K$_s$ and 24 $\mu$m (up to 6 
possible bands).  Any fluxes or their uncertainties that are NaN, zero, or 
negative are excluded.  In some cases, for the calculation of $\alpha$, a 
band-filled flux has been used (in the c2d catalogs this is indicated as 
having an image type of ``-2'').  Band-filling typically happens for sources 
that have clear shorter wavelength IRAC fluxes, but are undetected at longer 
wavelengths \citep[see the c2d Final Delivery Document,][]{evans07}.  If a 
source in Table \ref{tab:ysos_wfluxes} has a band-filled flux, it will be 
indicated as such with a footnote.

The use of the terms ``YSO'' and ``YSO candidate'' will be somewhat 
interchanged throughout this paper.  To be clear, not all the sources 
discussed as YSOs are spectroscopically identified YSOs or confirmed as 
members of the CrA cloud, although some of them have, through other studies.  
This is why the term ``YSO candidate'' is used most frequently when describing 
specific sources.  Ample references to the literature are provided in the 
text, appendices, and the tables in order to facilitate further studies of 
these YSOs and YSO candidates.

\subsection{\textit{Spitzer} Classified YSO Candidates \label{sec:spitzer_ysos}}

There are 45 YSO candidates classified using the standard technique, as 
discussed above, and are listed in Table \ref{tab:ysos_wfluxes}.  In two cases 
(CrA-2 and CrA-39), the candidates turned out to be known galaxies; they are 
discussed in Appendix~\ref{sec:appA} and appear in the table, although are not 
counted in our final list of YSO candidates.  There are also three YSO 
candidates which, upon visual inspection, were determined not to be YSOs: 
CrA-17, CrA-25 (which is a likely HH object), and CrA-32 (see descriptions for 
all in Appendix~\ref{sec:appA}).  This gives a final total of 40 YSO 
candidates selected by our method with {\it Spitzer}.  Many of the YSO 
candidates are already known YSOs from the literature, but there are 7 new YSO 
candidates (CrA-1, CrA-7, CrA-9, CrA-22, CrA-24, CrA-36, CrA-37) that have not 
been selected by any other survey, in most cases because the {\it Spitzer} 
observations extend beyond the region included in other surveys. CrA-24 is a 
new YSO candidate which is located within the Coronet, and further discussion 
of it can be found in \ref{cra24}.  For a more detailed description of each of 
the 45 sources selected using this method, see Appendix \ref{sec:appA}.

\begin{figure}
\plotone{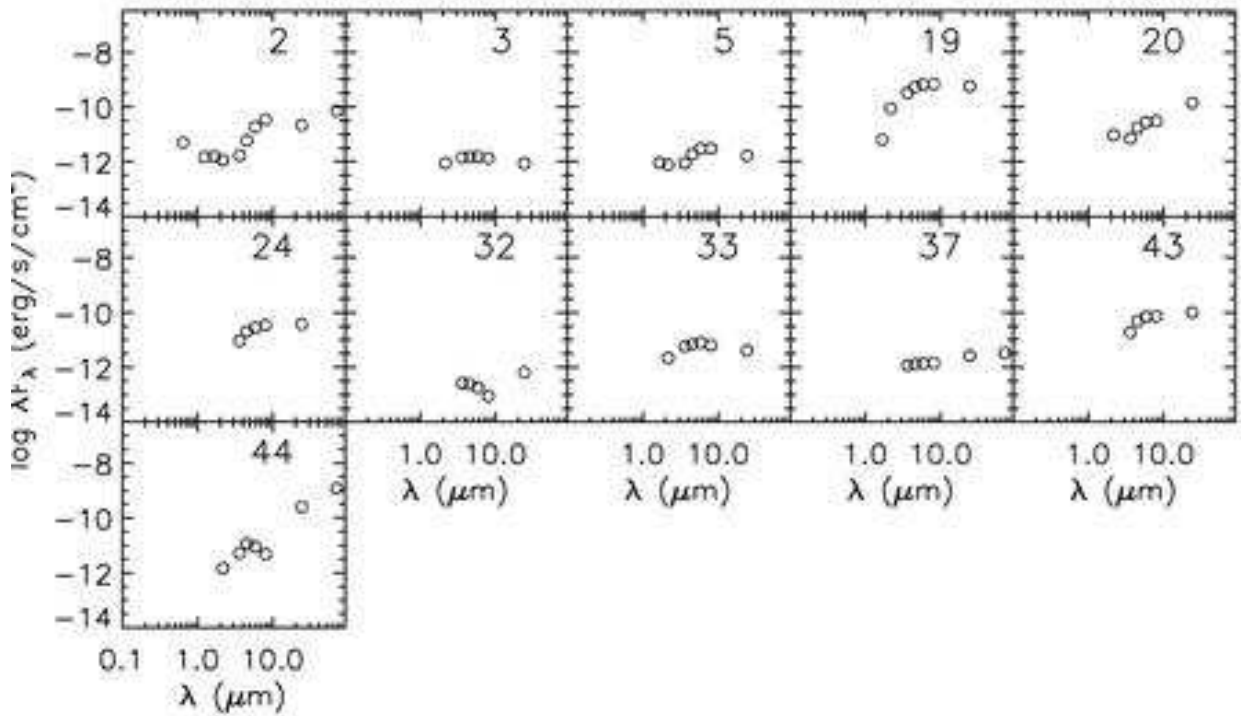}
\caption{SEDs are shown for the Class I and Flat spectrum sources in CrA selected from the {\it Spitzer} data as discussed in Section \ref{sec:spitzer_ysos}.  Open circles represent the {\it Spitzer} and ancillary data.  The number that appears in each panel corresponds to the object number in Table \ref{tab:ysos_wfluxes}.  Note that sources CrA-2 (a known galaxy, see discussion in \S \ref{sec:leda90315}) and CrA-32 (see discussion in \S \ref{sec:cra32}) have been removed from our sample of YSO candidates. \label{fig:classI}}
\end{figure}

\begin{figure}
\plotone{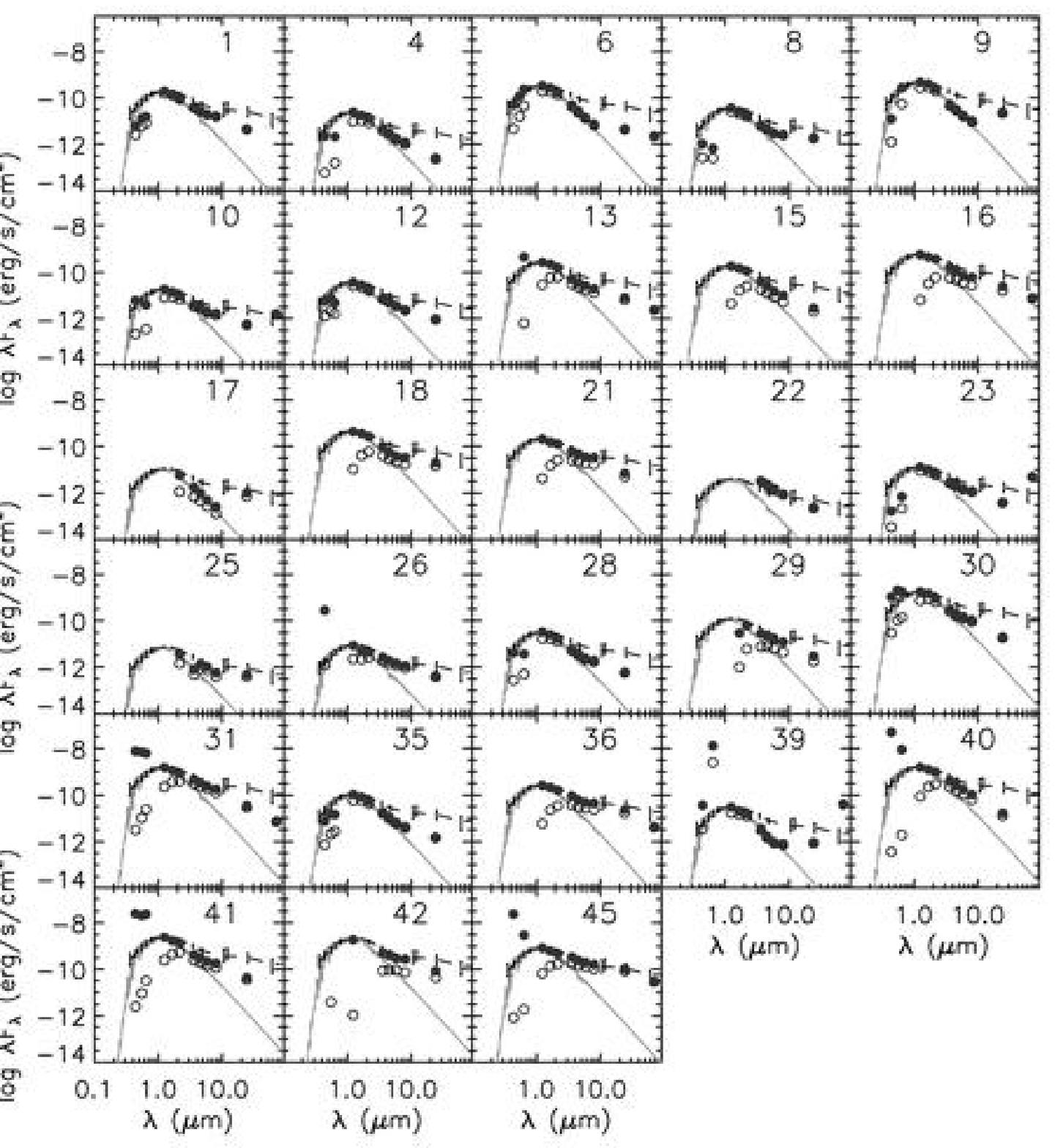}
\caption{SEDs are shown for the Class II sources in CrA selected from the {\it Spitzer} data as discussed in Section \ref{sec:spitzer_ysos}. Open circles represent the {\it Spitzer} and ancillary data, filled circles represent the dereddened fluxes, the grey line represents the stellar photospheric emission expected from the best-fit NextGen model, and the dashed black line represents the average SED for T Tauri stars in Taurus \citep{dalessio99}.  The number that appears in each panel corresponds to the object number in Table \ref{tab:ysos_wfluxes}.  Note that sources CrA-17 (see discussion in \S \ref{sec:cra17}), CrA-25 (see discussion in \S \ref{sec:cra25}), and CrA-39 (a known galaxy, see discussion in \S \ref{sec:cra39}) have all been removed from our sample of YSO candidates. \label{fig:classII}}
\end{figure}

\begin{figure}
\plotone{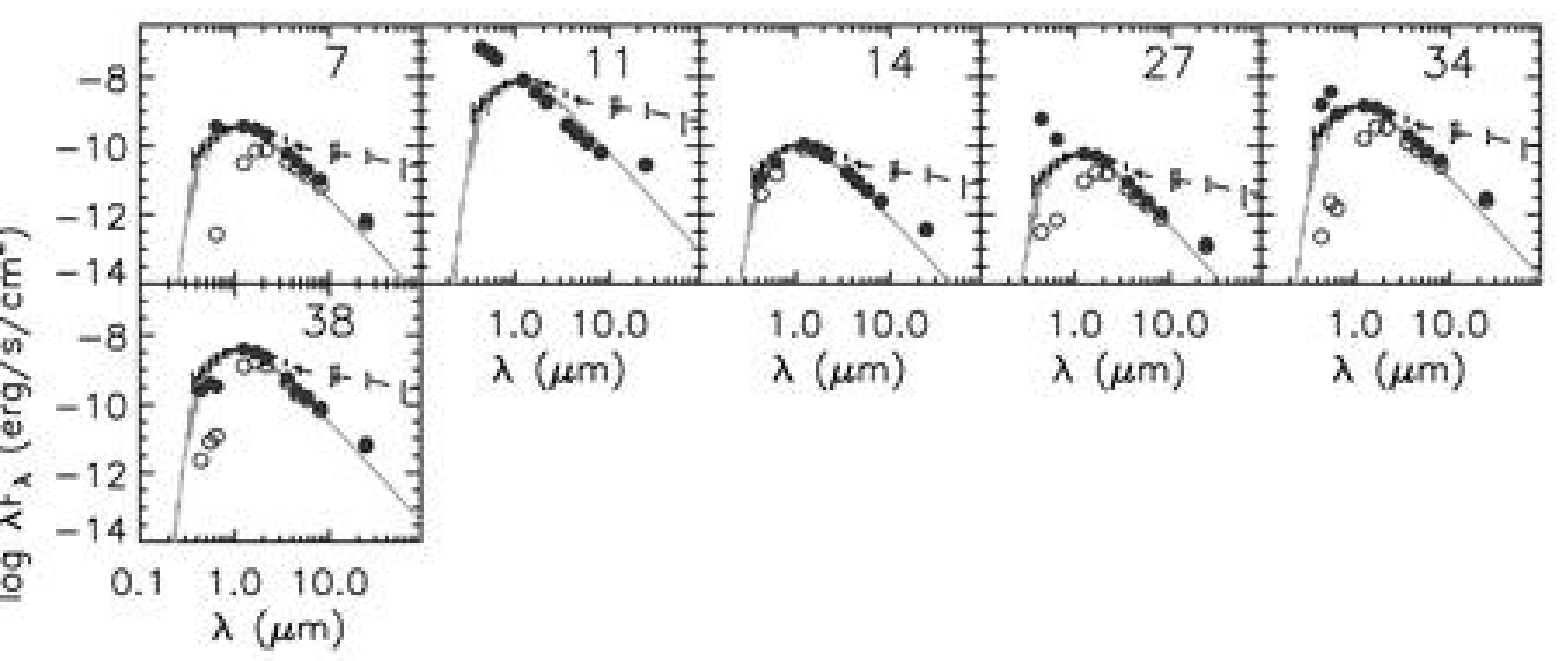}
\caption{SEDs are shown for the Class III sources in CrA selected from the {\it Spitzer} data as discussed in Section \ref{sec:spitzer_ysos}.  Open circles represent the {\it Spitzer} and ancillary data, filled circles represent the dereddened fluxes, the grey line represents the stellar photospheric emission expected from the best-fit NextGen model, and the dashed black line represents the average SED for T Tauri stars in Taurus \citep{dalessio99}.  The number that appears in each panel corresponds to the object number in Table \ref{tab:ysos_wfluxes}.  The fit to CrA-11 is not very good in the visible wavelength range because it has a known spectral type of B8V (see \S \ref{sec:bstars}), whereas the model assumes a spectral type of K7 (which is a reasonable approximation for most of the sources).\label{fig:classIII}}
\end{figure}

The spectral energy distributions (SEDs) for the 45 sources are shown in 
Figure~\ref{fig:classI} (Class~I and Flat spectrum candidates), 
Figure~\ref{fig:classII} (Class~II candidates), and Figure~\ref{fig:classIII} 
(Class~III candidates).  In these figures, the open circles represent the 
{\it Spitzer} and ancillary data, and the filled circles represent the 
dereddened fluxes \citep[using the extinction law of][with R$_V$ =5.5]{wd01}.  
In Figures~\ref{fig:classII} and \ref{fig:classIII}, a grey line represents 
the photosphere of a K7 main sequence star, and the dashed black line 
represents the average SED for T Tauri stars in Taurus \citep{dalessio99}.  

\subsection{YSO Candidates Classified from 2MASS and MIPS \label{sec:2mass_mips_ysos}}

\begin{figure}
\plotone{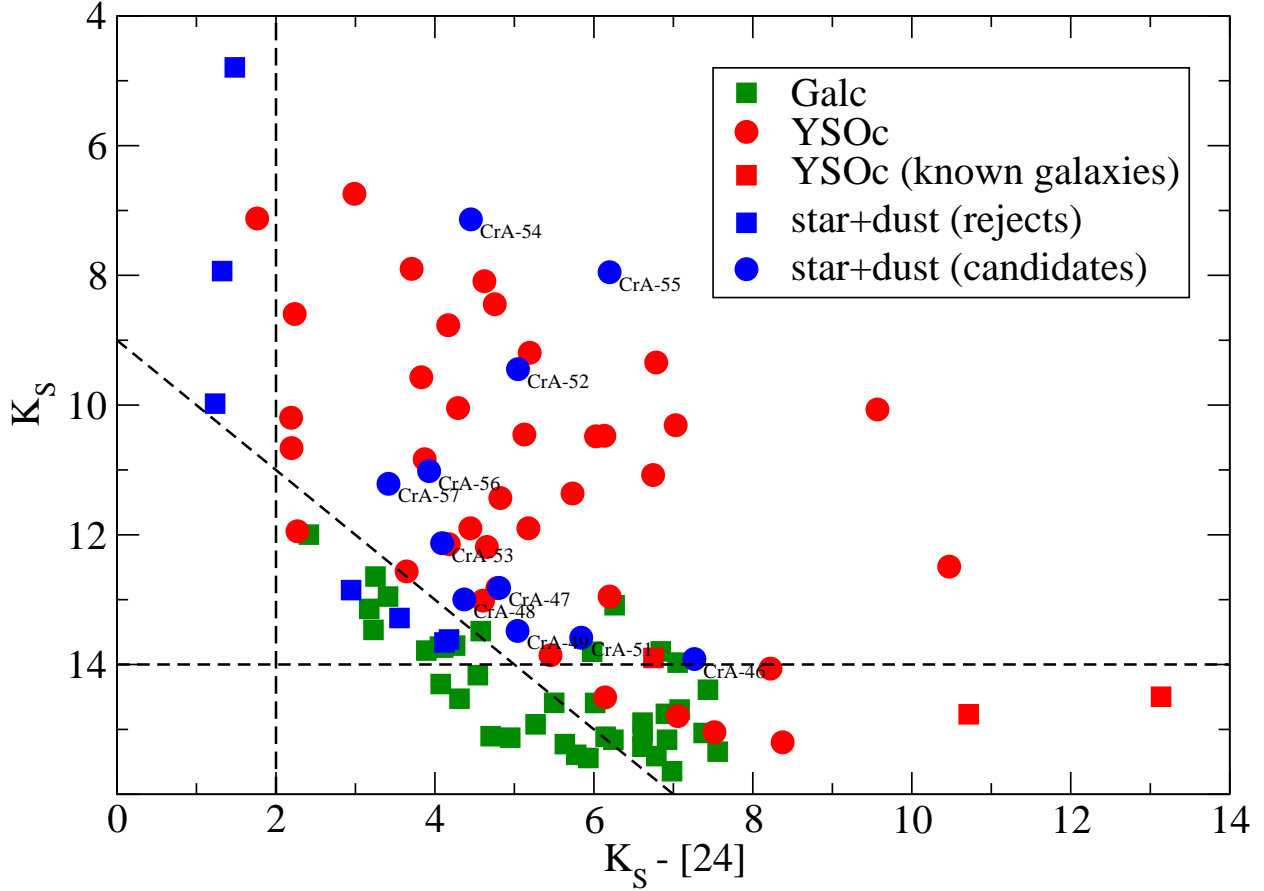}
\caption{A K$_s$ vs. K$_s$ - [24 $\mu$m] color-magnitude diagram for CrA.  Symbols represent the following classifications: ``Galc'', or galaxies (green squares), YSO candidates that appear in Table \ref{tab:ysos_wfluxes} (filled red circles), and YSO candidates that turned out to be known galaxies (filled red squares).  The phase space used to select star+dust sources as candidate YSOs, as discussed in \S \ref{sec:2mass_mips_ysos}, is shown by the dashed lines.  Those star+dust sources that were rejected as candidate YSOs are represented by filled blue squares.  The star+dust sources that met the criteria to be considered YSO candidates, and which appear in Table \ref{tab:mips_ysos}, are annotated and represented by filled blue circles. \label{fig:2mass_mips_cmd}}
\end{figure}

In some cases, there are regions covered by MIPS that are not covered in IRAC 
(or else are only covered in IRAC bands 1/3 and not in 2/4, and vice versa).  
To have as complete a sample as possible, we use 2MASS + the MIPS data in 
order to select additional YSO candidates.  A K$_s$ vs. K$_s$ - [24] 
color-magnitude diagram for CrA is shown in Figure~\ref{fig:2mass_mips_cmd}.  
Following the analysis of \citet{rebull07} for Perseus and \citet{padgett08} 
for Ophiuchus, we find several new YSO candidates based on this diagram.  
The specific criteria used to select YSO candidates in this way are as 
follows: K$_s$ $<$ 14 magnitudes, 24 $\mu$m $<$ 9 magnitudes, and 
K$_s$ - [24] $>$ 2.  The region of the color-magnitude diagram which meets 
these criteria is outlined in Figure~\ref{fig:2mass_mips_cmd}.  There are 12 
YSO candidates which meet these criteria, and they are listed in 
Table~\ref{tab:mips_ysos}.  However, as was for the case of CrA-2, source 
CrA-50 is part of the extended galaxy Leda 90315, and so it has been rejected 
from the list of candidates (see also \S~\ref{sec:leda90315} for a discussion 
of CrA-2).  Thus we yield a total of 11 new YSO candidates from this 
technique, 5 of which have never been identified as candidate YSOs before.

For the sources discussed in this section and in 
Section~\ref{sec:spitzer_ysos}, the final list of {\it Spitzer}-selected 
sources includes 51 YSO candidates, 12 of which are new.  Using the values of 
$\alpha$ defined at the beginning of Section~\ref{sec:ysoselect}, we find 10 
Class~I or Flat spectrum, 35 Class~II, and 6~Class III candidate YSOs.  This 
high fraction of Class~II sources compared with other classes, is consistent 
with that seen in the previously studied c2d star-forming regions 
\citep{evans09}.

\subsection{Class~III Population in CrA \label{sec:classIII}}

While infrared surveys are excellent for identifying embedded YSOs still 
surrounded by circumstellar dust, they are less successful for classifying the 
Class~III YSO population.  Class~III objects do not have the significant 
infrared excess associated with optically thick disks, and thus cannot be 
clearly distinguished from main sequence stars.  That is apparent from the 
analysis discussed in \S~\ref{sec:spitzer_ysos} (see 
Table~\ref{tab:ysos_wfluxes}), where only 6 Class~III YSO candidates are 
identified.  On the other hand, diskless Class~III objects can be readily 
identified using X-ray observations in addition to the infrared 
\citep[e.g.][]{grosso2000}.  Therefore in order to obtain a census of the 
young population in CrA, we combine our {\it Spitzer} catalog with X-ray 
ROSAT data (Patten et al., in preparation) to select additional Class~III YSO 
candidates.  The completeness of our selected sample of Class~III YSOs will be 
discussed in \S~\ref{sec:completeness}.

There are 34 ROSAT detections within our {\it Spitzer} CrA field.  Of those, 3 
do not have a {\it Spitzer} counterpart; two sources are off the main cloud 
and have \textit{Spitzer} detections on the border of the positional 
uncertainty circle, so we do not make the association.  The third (which is 
located within the high nebulosity region of TY\,CrA) is likely to be 
detected, but there are two \textit{Spitzer} detections within the positional 
uncertainty, making the association ambiguous.  Three ROSAT detections are the 
well-known YSOs in CrA known as IRS\,2, S\,CrA, and TY\,CrA.  Six are YSO 
candidates which we classified as Class~II or III using {\it Spitzer} and 
2MASS:  CrA-6, 11, 14, 16, 30, and 40 (see Table \ref{tab:ysos_wfluxes}).  Two 
Class~II sources, CrA-52 and CrA-54 (listed in Table~\ref{tab:mips_ysos}), 
were selected using K$_s$ and 24 $\mu$m, as described in 
\S~\ref{sec:2mass_mips_ysos}.  The remaining 20 sources are listed in 
Table~\ref{tab:classIII} as YSO candidates and are included in the clustering 
analysis later in this paper (see \S~\ref{sec:cluster_analysis}).  All 20 of 
these sources are classified as ``star'' in our {\it Spitzer} catalogs 
(meaning they have SEDs that are well-fitted by a reddened stellar 
photosphere) and have $\alpha$ values within the range of a Class~III YSO, as 
calculated from their {\it Spitzer} + 2MASS SEDs.

\subsection{Known YSOs Not Classified with \textit{Spitzer} \label{sec:known_ysos}}

The Coronet is a well-known young stellar cluster, with several YSOs 
previously studied by several authors at visible and near-infrared 
wavelengths.  Unfortunately this region is very bright in the longer 
wavelength {\it Spitzer} bands and many sources are saturated in our images, 
which may result in a band-filled flux.  Therefore we do not have enough SED 
coverage to classify them as YSO candidates (remember from 
\S~\ref{sec:ysoselect} that we require a S/N of 3 or higher detection in all 
four IRAC bands as well as in the MIPS 24~$\mu$m band before even considering 
a source as a YSO candidate).  Nevertheless, since we want to analyze the 
entire CrA population, these sources should be included in a comprehensive 
list of YSO candidates.  We reviewed the literature to recover those YSOs (and 
YSO candidates) which are not already included in 
Tables~\ref{tab:ysos_wfluxes}, \ref{tab:mips_ysos}, and \ref{tab:classIII}.  
Our findings are listed in Table~\ref{tab:knownysos_spitzercounterpart}.  

In Table~\ref{tab:knownysos_spitzercounterpart}, we summarize the known YSOs 
along with the coordinates and fluxes of their {\it Spitzer} counterparts as 
well as information on how they are classified in the delivered catalog.  Many 
of the most massive stars in CrA were added in this way, including R\,CrA.  In 
the case of R\,CrA, because many of the {\it Spitzer} bands are saturated, 
$\alpha$ is computed from the K- and 5.8~$\mu$m bands alone, resulting in a 
classification of Class~III.  However, it should be noted that R\,CrA has long 
been known to be a pre-main sequence (PMS) star with an accretion disk 
\citep[e.g.][]{knacke73}, and an SED slope of a Class~II 
\citep{hillenbrand92}.  In addition, there are three known YSOs in CrA which 
do not have a counterpart in our {\it Spitzer} catalog: IRS\,9, SMA\,2, and 
Anon\,2 (also known as VSSt\,14 and HBC\,675).  In the case of IRS\,9 and 
SMA\,2, the nebulosity from R\,CrA is too bright to distinguish the sources, 
and in the case of Anon\,2, it is a diffraction spike from S\,CrA that causes 
a problem.  IRS\,9 is classified as a Class~I \citep[and detected with 
{\it Chandra};][]{fp07}, and SMA\,2 is classified as 
a Class~0 \citep{groppi07}.  Whether or not Anon\,2 is a YSO is debated: both 
\citet{patten98} and \citet{neu00} claim it is not a young member 
\citep[lithium is not detected in its spectrum,][]{neu00}.  Unfortunately 
without the {\it Spitzer} fluxes available, we cannot determine a 
classification for Anon\,2.

Not all sources that have ever been identified as a candidate YSO are 
included in Table~\ref{tab:knownysos_spitzercounterpart}.  For example, IRS\,3 
and IRS\,4 are not included because they are spectroscopically recognized as 
background giants \citep{mw09}.  Additionally, any candidate YSOs in the 
literature which our {\it Spitzer} observations classify as a galaxy (noted as 
``Galc'' in the catalog) are not included.  There is an exception to this, and 
that is when additional observations show that a source classified as ``Galc'' 
in our catalog should be a member, as in the case of LS-RCrA 1 (discussed 
at the end of this section).

Moreover, we compare our candidates to the \citet{haas08} list (their 
Table~3), which was selected using a millimeter excess technique.  Of those 38 
candidates, 8 are already included in our tables.  There are 5 sources in 
their list which are classified in our {\it Spitzer} GB catalogs with SED 
slopes described as ``star'' or ``cup-down,'' and their fluxes are reasonable 
enough (i.e. are not band-filled and have good quality detections) to include 
them as candidate YSOs in Table~\ref{tab:knownysos_spitzercounterpart}.  Note 
that a ``cup-down'' classification means that the SED has a shape that is 
convex and can not be classified by any other category \citep[see the c2d 
Final Delivery Document,][]{evans07}.  However, many of the remaining sources 
are listed in our catalog as ``one'' or ``two,'' meaning they are only 
detected in one or two of the IRAC bands (generally in the 3.6~$\mu$m and/or 
4.5~$\mu$m bands).  Because there is such little information in the 
{\it Spitzer} bands for a proper classification, none of these sources has 
been included in our Table~\ref{tab:knownysos_spitzercounterpart}.

Finally, Table~\ref{tab:knownysos_spitzercounterpart} also reports 2 out of 
35 sources (namely G09 CrA-9, 11; from their Table~4) classified as candidate 
YSOs in \citet{rguter09} that are not included in our previous tables.  
\citet{rguter09} used {\it Spitzer} data of CrA available from the GTO PID 6 
program (see Table \ref{tab:spitzerobs}), which is included in the area 
studied in this paper, but covers a much smaller field, and adopted a slightly 
different selection technique.  They found 35 YSO candidates in the region 
covered by PID 6, and of those, 25 are also selected by our method to be YSO 
candidates.  Two of their sources have been added to our 
Table~\ref{tab:knownysos_spitzercounterpart}: G09 CrA-11, which we classify as 
``star'' and G09 CrA-9, which we classify as ``cup-down.''  Of the remaining 
8 candidates, 4 are known YSOs which already appear in 
Table~\ref{tab:knownysos_spitzercounterpart} (IRS\,7A, IRS\,7B, S\,CrA, and 
T\,CrA), and 4 are classified as ``Galc'' in our catalog.  However, one of the 
``Galc'' sources is a star of late spectral type, LS-RCrA 1 \citep{fc01}, and 
thus has also been included in Table~\ref{tab:knownysos_spitzercounterpart}.

Table~\ref{tab:knownysos_spitzercounterpart}, along with 
Tables~\ref{tab:ysos_wfluxes}, \ref{tab:mips_ysos}, and \ref{tab:classIII}, 
are intended to make up the most comprehensive list of YSOs and YSO candidates 
in CrA to date.  The sources included in these tables (minus those sources 
identified as likely galaxies) will be used to analyze clustering in CrA 
(\S~\ref{sec:cluster_analysis}) and to determine the driving sources for the 
outflows/jets (\S~\ref{outflows_and_sources:sec}).

The total number of candidate YSOs in CrA is 116, where 14 are classified as 
Class~I, 5 are Flat spectrum, 43 are Class~II, and 54 are Class~III.  All 116 
sources are shown overlaid on a 24~$\mu$m image in Figure~\ref{fig:allysos}.

\subsection{Completeness \label{sec:completeness}}

\citet{evans07,evans09} estimated that the c2d (and similarly, GB) selected 
sample of YSO candidates is 90\% complete down to a luminosity of 
0.05L$_{\odot}$, integrated from 1 to 30 $\mu$m at a distance of the Serpens 
star-forming region (260~pc).  However, the observing strategy for CrA is 
slightly different from the rest of the GB regions because it was originally 
observed as part of a GTO program.  There is only one epoch, as opposed to 
the usual two, and therefore the photometric depth is $\sim$30\% less.  
Accounting for the difference in photometric depth and scaling for the 
distance of CrA (130~pc), we estimate that our 2MASS/\textit{Spitzer} sample, 
selected using the c2d/GB method (discussed in \S~\ref{sec:spitzer_ysos} and 
reported in Table~\ref{tab:ysos_wfluxes}), is $\sim$90\% complete down to 
$\sim$0.015L$_{\odot}$.

To illustrate the completeness of the CrA sample, we constructed bolometric 
luminosity functions for the Class II and III candidates 
\citep[Figure~\ref{fig:lumi_histo}; see also][]{bmerin08,spezzi11}.  First, we 
estimate the bolometric luminosity of the YSO candidates by integrating over 
all the observed SED data points, assuming a distance of 130 pc to CrA.  The 
solid line in Figure~\ref{fig:lumi_histo} indicates the bolometric luminosity 
distribution of Class II and III YSO candidates discussed in 
\S~\ref{sec:spitzer_ysos} (and reported in Table~\ref{tab:ysos_wfluxes}) 
rebinned to a 0.03 size bin.  Then, following the \citet{harv07} method, we 
apply the c2d completeness correction to the sample of sources selected using 
the c2d/GB method discussed in \S~\ref{sec:spitzer_ysos}.  The \citet{harv07} 
completeness correction compares, for each luminosity bin, the number of 
counts from a trimmed version of the deep SWIRE catalog of extragalactic 
sources with the number of counts for the c2d catalogs in Serpens.  The 
completeness correction estimate is then applied to the five molecular clouds 
observed by c2d.  

For all the regions in the GB survey, this completeness correction translates, 
since they are homogeneous in terms of photometric depth.  However, as stated 
previously, this is not precisely the case for CrA.  If we assume that the 
c2d/GB survey samples are homogeneous with CrA down to $\sim$0.015L$_{\odot}$, 
the completeness correction, indicated by dashed lines in 
Figure~\ref{fig:lumi_histo}, provides a completeness estimate of $\sim$90\%.  
We would like to emphasize here that CrA is peculiar when compared with the 
c2d and GB star-forming regions previously studied, in terms of the initial 
observing strategy, and mostly because of various observational limitations 
(discussed in detail in the following paragraphs), which led to the inclusion 
of many additional YSO candidates, selected using different techniques (see 
\S~\ref{sec:2mass_mips_ysos}, \S~\ref{sec:classIII}, 
\S~\ref{sec:known_ysos}).  To address the inclusion of these sources 
quantitatively, the bolometric luminosity function for {\it all} sources is 
overplotted in Figure \ref{fig:lumi_histo} (dotted lines).  However, because 
of the addition of these sources, the final sample of YSO candidates was not 
uniformly selected, and therefore we cannot speak to the completeness of the 
sample as a whole.  It is clear from Figure \ref{fig:lumi_histo} that many YSO 
candidates in CrA are not selected using solely the c2d/GB method, and we 
enumerate the reasons why in the following paragraphs.  

The greatest sources of incompleteness in the standard c2d/GB method 
(\S~\ref{sec:spitzer_ysos}) manifest in four ways.  First, very low luminosity 
objects may be lost in the group of extragalactic sources.  An additional bias 
originates from the Coronet region, with its bright, diffuse nebula.  Second, 
in \S~\ref{sec:2mass_mips_ysos} we discussed the addition of YSO candidates 
based on their 2MASS + MIPS 24 $\mu$m photometry because they were not covered 
in IRAC (or else are only covered in IRAC bands 1/3 and not in 2/4, and vice 
versa).  These sources (see Table~\ref{tab:mips_ysos}) were not selected by 
the standard technique because the method requires detections in all four IRAC 
bands.  Third, as discussed briefly in \S~\ref{sec:known_ysos}, the brightest 
sources (e.g. R\,CrA), which saturate the camera arrays, were not selected as 
YSO candidates by the c2d/GB method due to poor photometry and subsequent flux 
measurements.  To remedy such incompleteness, we added known YSOs and YSO 
candidates from the literature (see \S~\ref{sec:known_ysos} and 
Table~\ref{tab:knownysos_spitzercounterpart}).  Finally, in order to select a 
sample of YSO candidates with a minimum of contamination from extragalactic 
sources and background stars, the c2d/GB method required at least a small 
infrared excess component.  Therefore, as mentioned in \S~\ref{sec:classIII}, 
our method fails to detect PMS objects that do not have infrared excess, but 
still have some accretion activity, indicated by X-ray emission, H$\alpha$ 
emission, etc.  Consequently, the sample of Class~III candidates in 
Table~\ref{tab:ysos_wfluxes} is heavily biased, missing many Class~III 
candidates.  To correct this bias, we used the ROSAT data to include 
additional Class~III candidates (see Table~\ref{tab:classIII}).

As discussed in \S~\ref{sec:classIII}, there are 34 ROSAT detections within 
our {\it Spitzer} field for CrA, of which 31 have {\it Spitzer} counterparts.  
To assess the completeness of the ROSAT All-Sky Survey at the distance of 
CrA, we determine the limiting unabsorbed X-ray luminosities for different 
levels of foreground extinction.  Sources listed in the Faint Source Catalog 
have a minimum of six counts, corresponding to a limiting count rate of about 
0.03~s$^{-1}$ for a typical exposure time of 218~s.  The typical exposure time 
is the median exposure time of Faint Source Catalog sources within 3$\degr$ 
of R\,CrA. To find the limiting unabsorbed X-ray fluxes and luminosities, we 
simulate X-ray sources with assumed spectra using the Portable, Interactive 
Multi-Mission Simulator \citep[PIMMS;][]{mukai93}.  The spectra are assumed 
to be absorbed APEC thermal plasma spectra.  The visual extinction of newly 
identified T Tauri stars in \citet{neu00} is A$_V<$ 1~mag, which is the value 
that we are assuming for this estimate, after conversion into absorbing 
hydrogen column densities using the empirical relation N$_H({\rm cm}^{-2}) 
\approx  2\times10^{21}\times A_V$ (mag) \citep[e.g.,][]{ryter96,vuong03}.  
Conservatively assuming a plasma temperature of 1~keV 
\citep[$\sim10$~MK; e.g.,][]{feigelson05}, the limiting luminosity is log 
(L$_X$) = 30.2 erg s$^{-1}$ for A$_V$ = 1~mag, assuming a distance of 130~pc. 

Subsequently, we use the cumulative X-ray luminosity distribution functions 
derived for subsets of the PMS population of the Orion Nebula Cluster 
\citep{pf05} to estimate the completeness of the ROSAT data for G, K, and M 
PMS stars.  The limiting unabsorbed X-ray luminosity determined above 
corresponds to a completeness of 60-70\% for G and K stars and 20\% for M 
stars.  For the deep \textit{Chandra} data, \citet{fp07} quote a considerably 
better limiting luminosity of 4.3$\times$10$^{26}$ erg s$^{-1}$ for no 
intervening extinction which would be complete for all G, K, and M PMS stars.  
Assuming a foreground extinction of A$_V $= 10~mag yields a limiting 
luminosity of 6.1$\times$10$^{27}$~erg~s$^{-1}$ for 5-count detections, still 
complete for all G, K, and M PMS stars.  For our purposes, the 
\textit{Chandra} data are complete for the PMS population of the inner 
Coronet, while the much less sensitive ROSAT data still yield a good estimate 
of the outlying YSO population.

\section{Distribution of YSOs \label{sec:yso_distribution}}

\subsection{Extinction Map and YSO Distribution \label{Av_ysodist}}

With a census of the YSOs in CrA, we can examine the spatial distribution of 
the sources with respect to each other and the dense gas.  Usually such a 
distribution reflects the underlying structure of the cloud, providing us 
with clues about the physical processes responsible for its creation.  In 
particular, it can be of help in discriminating between two well-known basic 
structures: hierarchical-type clusters and centrally-condensed embedded 
clusters \citep[see e.g.,][and references therein]{ll03}.  In the first case, 
clouds exhibit surface density distributions with multiple peaks and more 
complex shapes, whereas the centrally-condensed clusters have a highly 
concentrated surface density distribution, with a relatively smooth radial 
profile ($\rho_* \sim r^{-a}$).  Moreover, the latter usually correlates with 
mass segregation, with massive stars found near the cluster center 
\citep[see e.g.,][and references therein]{ll03}.

\begin{figure}
\plotone{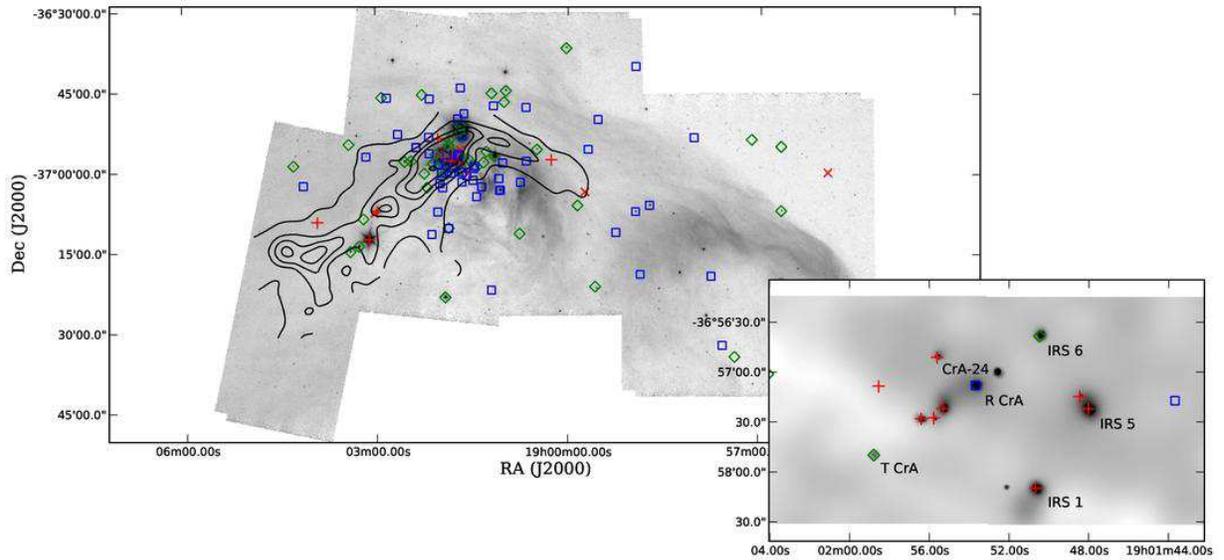}
\caption{All 116 YSO candidates are overlaid on a 24 $\mu$m image; the image 
inset is the {\it Chandra} X-ray image of the Coronet (the blue square near 
the center is R\,CrA).  The Class~I (+) and Flat spectrum (x) sources are 
shown in red, Class~II (diamond) sources are shown in green, and Class~III 
(square) sources are shown in blue.  The contour shown is the extinction map 
made from 2MASS + {\it Spitzer} data with levels A$_V$=[5, 10, 15, 20].   This 
extinction map has a resolution of 180$^{\prime\prime}$.  Most Class~I/Flat 
sources are located in the high density cluster, while Class~II and III 
sources are found spread further out.\label{fig:allysos}}
\end{figure}

\begin{figure}
\plotone{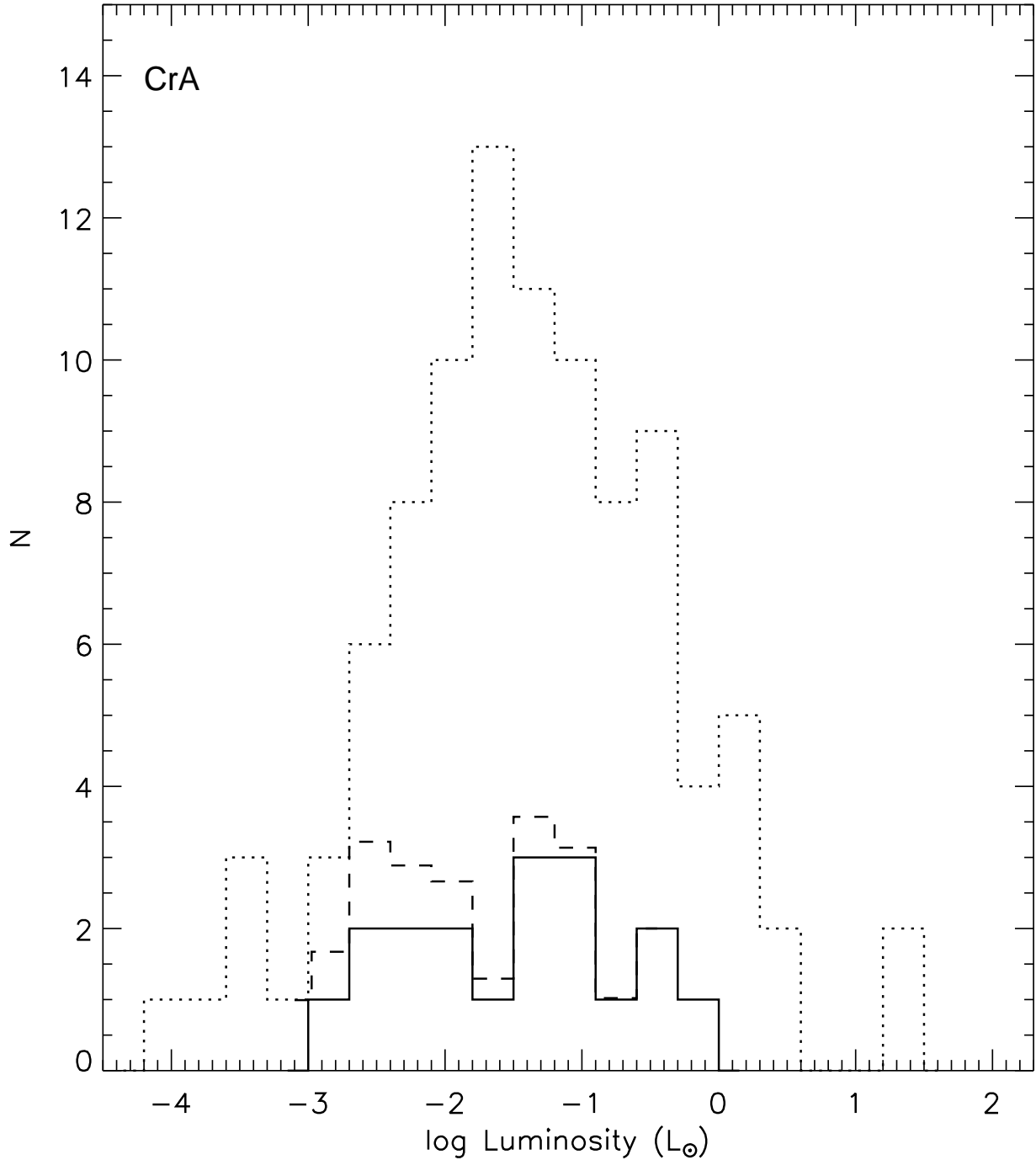}
\caption{Bolometric luminosity function for the YSO candidates in CrA selected using the c2d/GB method (discussed in \S~\ref{sec:spitzer_ysos}, solid line), with the completeness correction included for comparison (dashed line), as discussed in \S~\ref{sec:completeness}.  In addition, the bolometric luminosity function for {\it all} YSO candidates is shown (dotted line). \label{fig:lumi_histo}}
\end{figure}

In Figure~\ref{fig:allysos}, all 116 YSO candidates are overlaid on the 
24~$\mu$m mosaic, along with contours showing the extinction map we have 
created as part of the data pipeline using 2MASS and {\it Spitzer} data.  To 
make this map, the line-of-sight visual extinction (A$_V$) is estimated for 
each source classified in the catalog as a ``star'' (e.g., sources classified 
as YSO candidates are excluded) based on its SED from 1.25 to 24~$\mu$m, 
adopting the \citet{wd01} R$_V$ = 5.5 extinction law, which has been shown to 
be consistent with data from the c2d/GB studies \citep{chap09,cm09}.  These 
line-of-sight extinctions are then convolved with uniformly spaced Gaussian 
beams (with FWHM = 180$^{\prime\prime}$) to construct the extinction map 
\citep[refer to][for details]{evans07}.  The mean A$_V$ in our CrA map is 
$\sim$5 mag, and the maximum A$_V$ is $\sim$30 mag, which is found in the 
Coronet region.  The A$_V$ values in our map have not been corrected for any 
extinction offsets because a nearby off-cloud field was not imaged for 
comparison, and may be systematically high by as much as 1--2 mag 
\citep[see Table~27 in][]{evans07}.  For comparison, the extinction map 
presented by \citet[][and shown in our Figure~\ref{fig:coverage}]{rguter09}, 
which is based only on 2MASS observations, exhibits a mean A$_V$ of $\sim$0.3 
mag and maximum A$_V$ of $\sim$20 mag over the same region covered by our 
{\it Spitzer} observations.  The lower values can perhaps be explained by the 
use of a different extinction law, R$_V$ = 3.1 \citep{rl85}, for the 
near-infrared observations.  Correcting for the different extinction laws, the 
near-infrared extinction map would have A$_V$ values that are $\sim$ 25\% 
higher. 

As can be seen in Figure~\ref{fig:allysos}, the morphology of the entire 
region is mainly a clustered one, extending toward the southeast, although the 
stars are mainly clustered near the Coronet, which indeed harbors a Herbig Ae 
star (R\,CrA) in its very center.  It is striking how the Class~I sources 
(red +) are found in the highest extinction regions, highly clustered in the 
Coronet, the Class~II sources (green diamonds) are spread a little further 
out, and the Class~III sources (blue squares), spread even further out into 
the molecular cloud, forming a sort of halo around the Class~I and Class~II 
sources.  A distribution where the Class~I sources are more densely clustered 
than the Class~II sources is consistent with observations of other nearby 
star-forming regions such as Lupus, Serpens, and NGC 1333 
\citep[e.g.][]{bmerin08,rguter08,rguter09,winston09,bressert10}.  A 
quantitative analysis of the clustering in CrA will be discussed in the next 
section, \S~\ref{sec:cluster_analysis}.

\subsection{Clustering Analysis \label{sec:cluster_analysis}}

Using all 116 YSO candidates selected and/or compiled in this paper (see 
Figure \ref{fig:allysos}) and listed in Tables~\ref{tab:ysos_wfluxes}, 
\ref{tab:mips_ysos}, \ref{tab:classIII}, and 
\ref{tab:knownysos_spitzercounterpart}, we perform the clustering analysis 
identical in method to that presented in \citet{rguter09} for comparison.  The 
sample sizes are different because \citet{rguter09} used only the 
{\it Spitzer} data of CrA available from the GTO PID 6 program (see 
Table~\ref{tab:spitzerobs}), which covers a much smaller field of view than 
the region we include in our study.  For this smaller field, \citet{rguter09} 
classified a total of 35 YSO candidates: 25 Class~II and 10 Class~I, for a 
Class~II/Class~I ratio of 2.5.  They found one core of more than 10 stars in 
CrA, consisting of 24 members.  For the core of 24 members, there is a 
Class~II/Class~I ratio of 2.0.

In their analysis, which we also employ, \citet{rguter09} isolated the dense 
``cores'' of clustered star-forming regions via analysis of the Minimal 
Spanning Trees (MST) of the {\it Spitzer}-identified YSO positions in each 
region.  The MST is defined as the network of lines, or branches, that connect 
a set of points together such that the total length of the branches is 
minimized, containing no closed loops.  Once the MST is constructed for all 
YSO positions, a critical branch length, L$_{crit}$, is adopted whereby all 
branches in the network that are longer than L$_{crit}$ are excluded.  In the 
case of the \citet{rguter09} analysis, L$_{crit}$ is computed based on a 
two-line fit to the cumulative distribution function of the lengths of the MST 
branches.  After the branches longer than L$_{crit}$ are excluded, smaller 
sub-networks generally remain connected.  Any of these sub-networks that are 
composed of 10 or more members 
\citep[a common, but arbitrary choice, e.g.][]{allen07,rguter09} are 
defined as ``cluster cores.''  The size of a core is quantified in two 
ways, by its circular radial size, R$_{circ}$, and by its effective radial 
size, R$_{hull}$.  The circular radial size, R$_{circ}$, is defined as the 
maximal radial distance to a group member from the cluster center point, where 
the cluster center point is the geometric center of the convex hull polygon 
enclosing each grouping \citep[the geometric center is the position of the 
center of mass of the convex hull polygon;][]{rguter09}.  The effective radial 
size, R$_{hull}$, is defined as the square root of the area (divided by $\pi$) 
of the convex hull polygon that contains each grouping.  The ratio 
R$_{circ}^2$/R$_{hull}^2$ is the aspect ratio of elongated distributions, i.e. 
it is used quantify the degree to which a core is elongated.  

When we analyze our list of {\it all} 116 YSO candidates (which includes 
Class~I, Flat spectrum, Class~II, and Class~III candidates), we find one 
significantly sized core consisting of 68 members, which is located in the 
Coronet region.  The ratio of Class~II to Class~I candidates of the entire 
sample is: 2.26 (43 Class~II, 19 Class~I and Flat spectrum sources); for the 
core of 68 members, Class~II/Class~I = 1.77.  If we include cores that consist 
of fewer than 10 members, the next two largest groups consist of 6 members 
(the region near IRAS\,32) and 5 members (a region to the northwest of 
TY\,CrA).  The CrA cluster is clearly dominated by the cluster core around the 
Coronet.

\begin{figure}
\begin{center}$
\begin{array}{c}
\includegraphics[width=6.0 in]{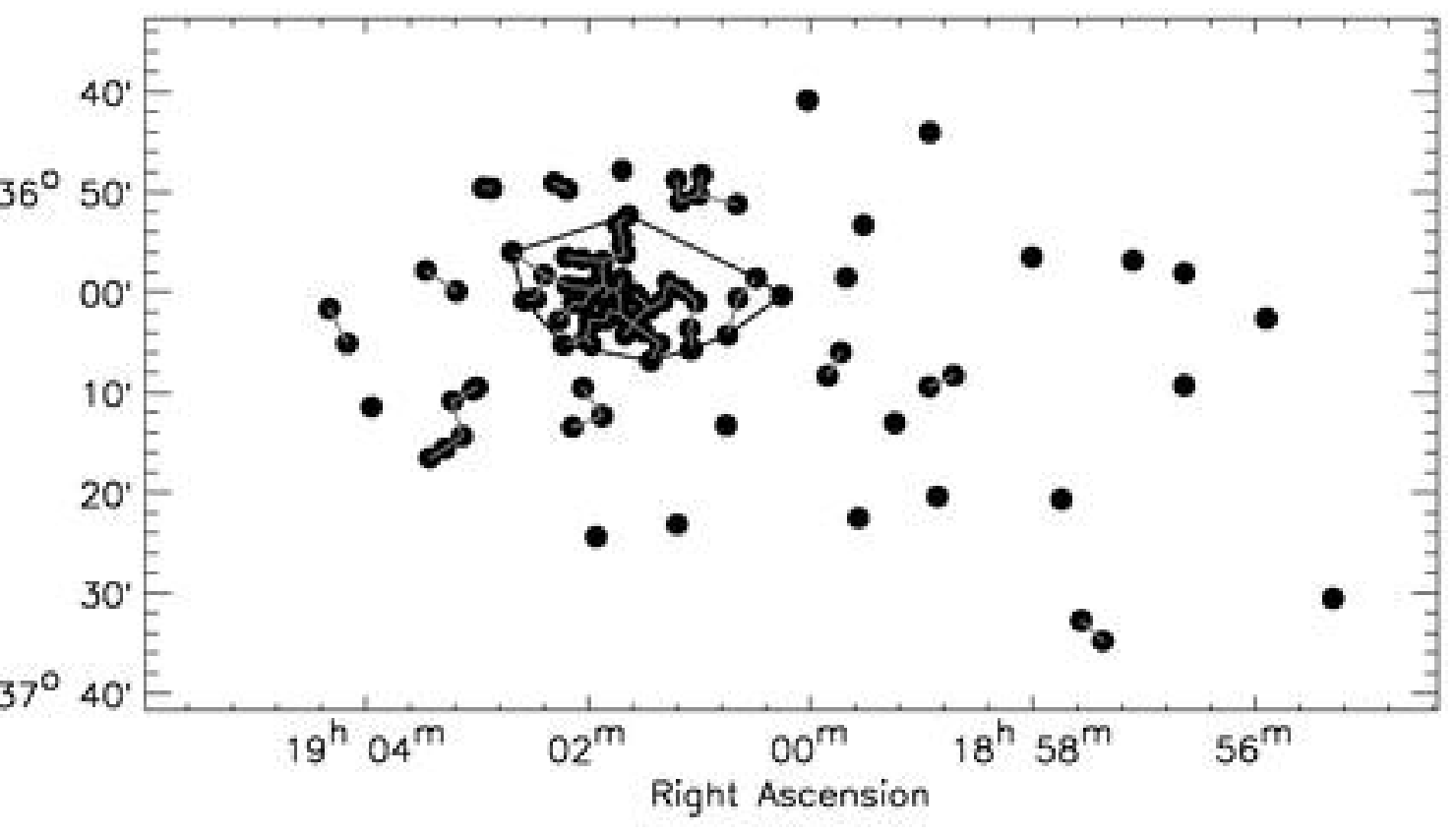}\\
\includegraphics[width=6.0 in]{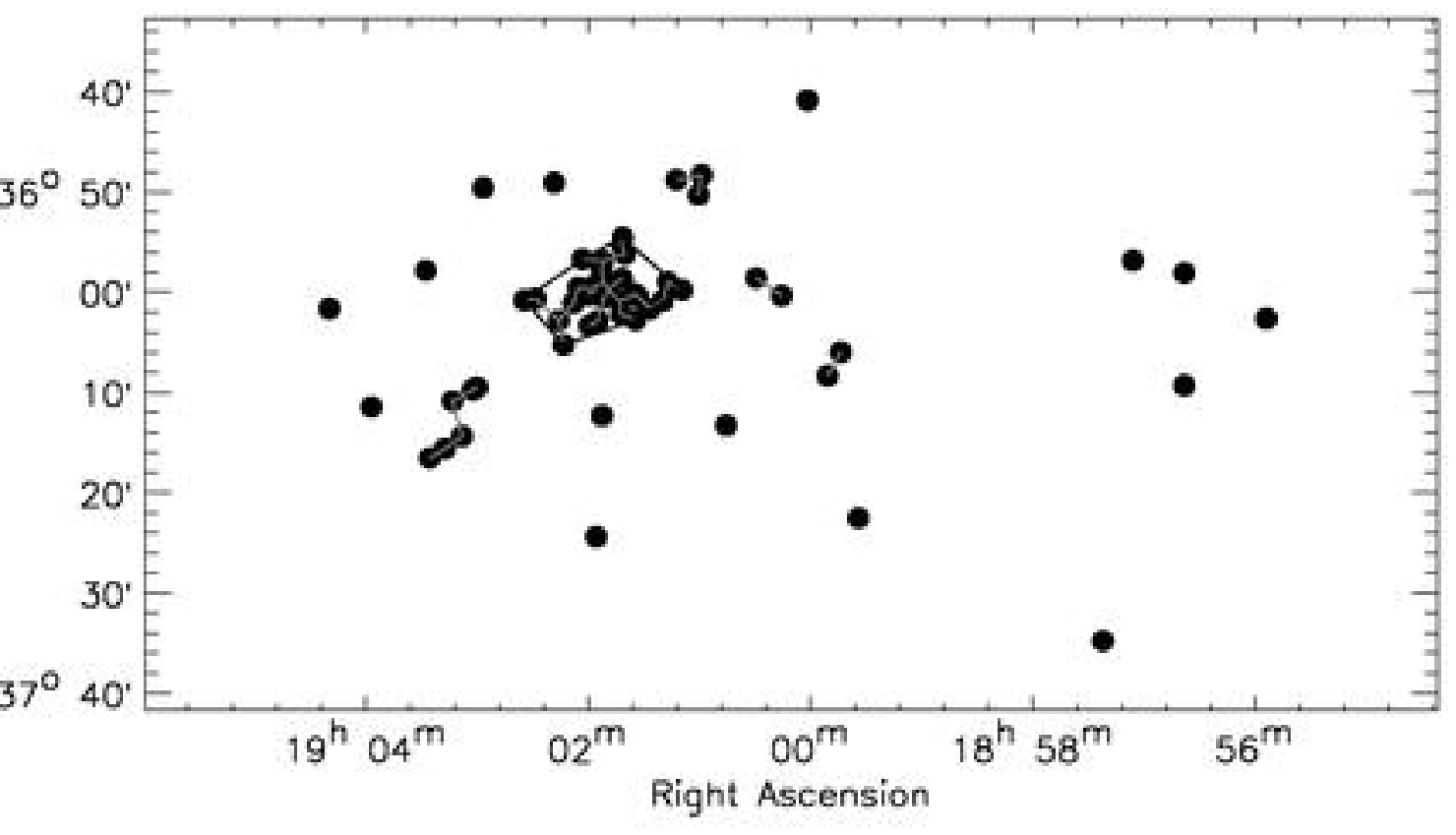}
\end{array}$
\end{center}
\caption{The residual MST sub-networks for the YSO positions after those 
longer than L$_{crit}$ have been excluded for the case of all 116 YSO 
candidates {\it (upper panel)} and for the 62 Class~I, Flat spectrum, and 
Class~II sources {\it (lower panel)}.  The branch lines connecting sources 
that are closer than the critical branch length (see 
Figure~\ref{fig:mst_crit}) are shown as grey lines.  The convex hull 
(solid black line), which is discussed in \S~\ref{sec:cluster_analysis}, is 
overlaid on both figures.\label{fig:mst}}
\end{figure}

\begin{figure}
\begin{center}$
\begin{array}{c}
\includegraphics[width=6.0 in]{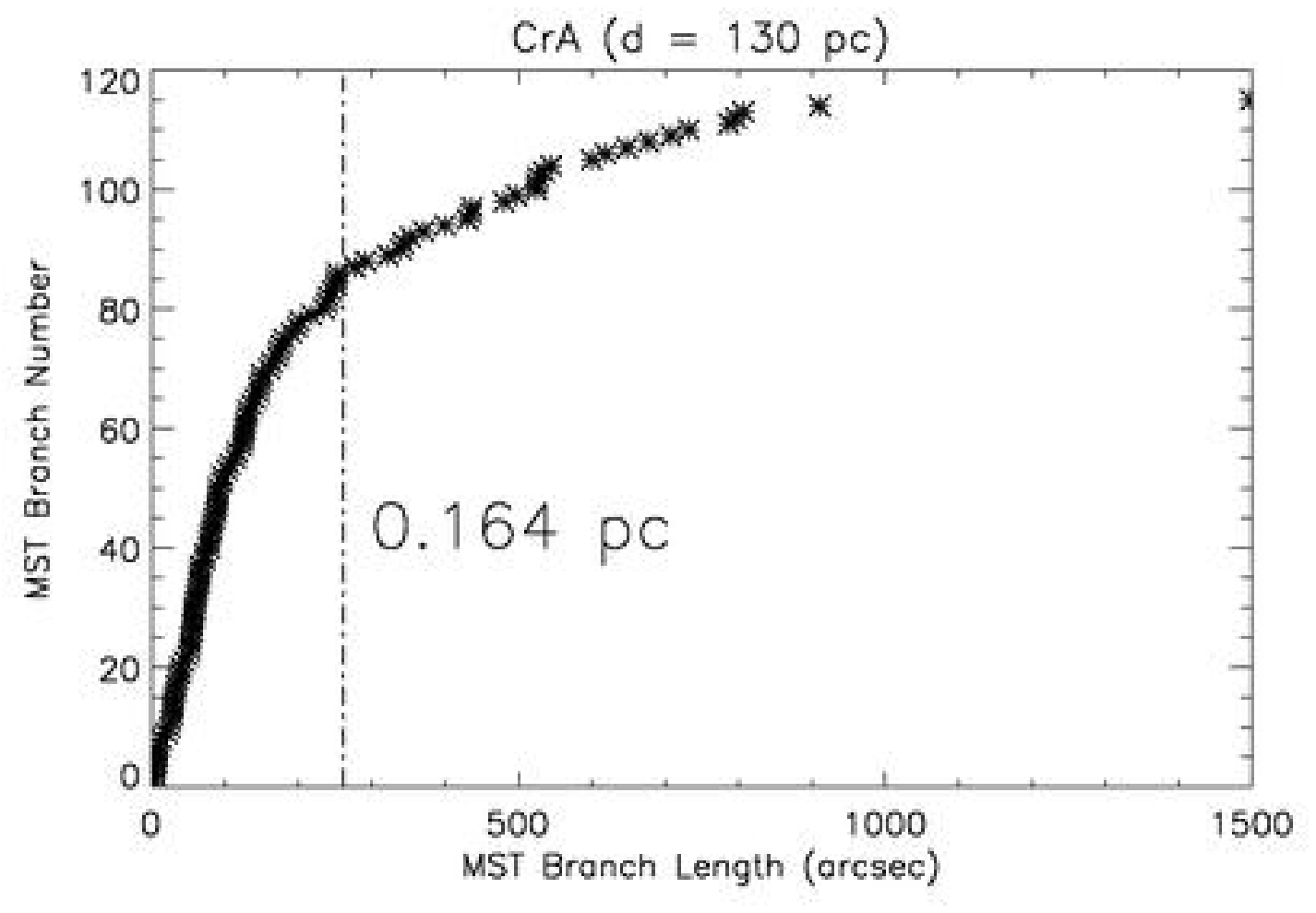} \\
\includegraphics[width=6.0 in]{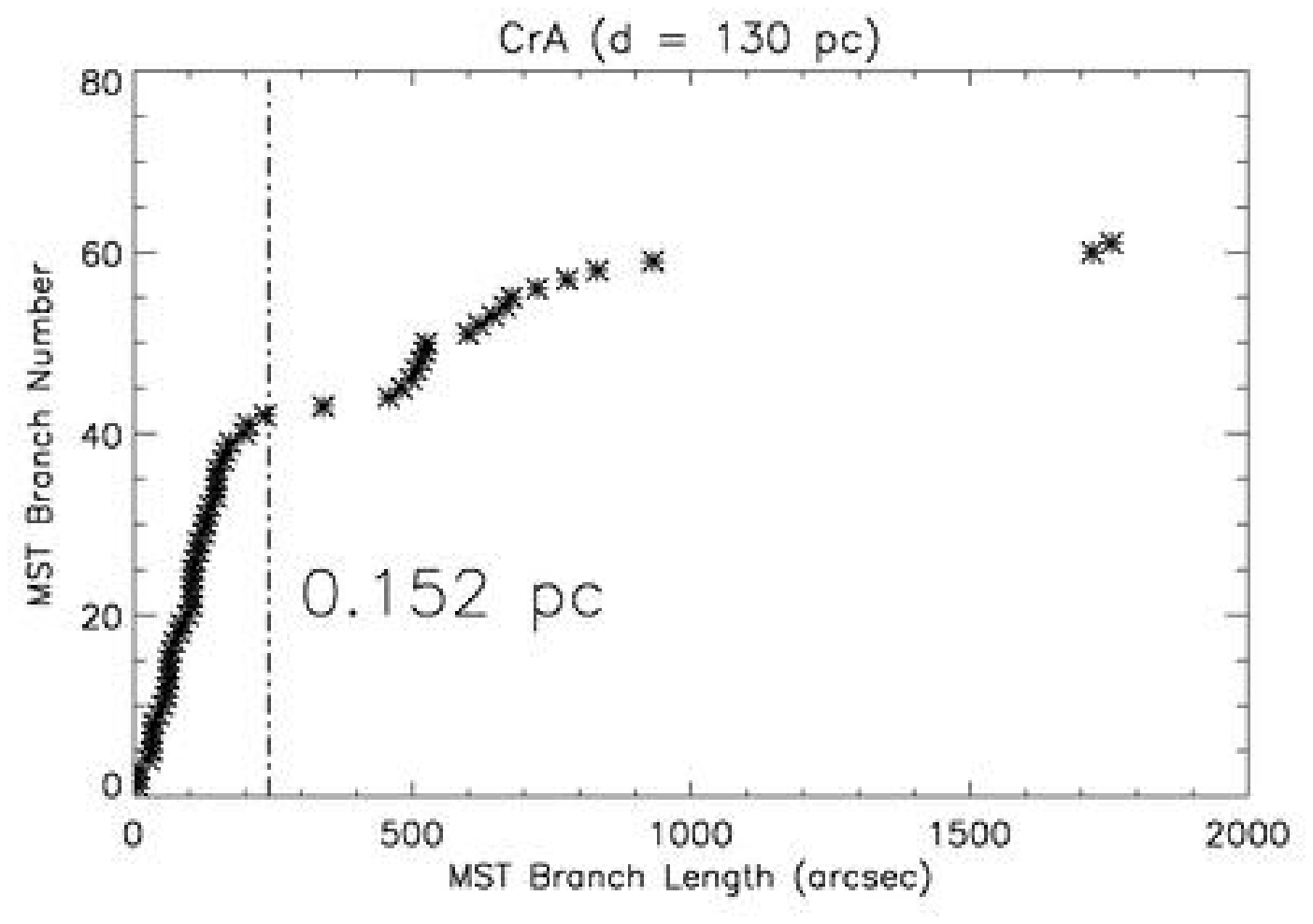}
\end{array}$
\end{center}
\caption{{\it (Upper panel)} MST critical branch length for all 116 YSO 
candidates. {\it (Lower panel)} MST critical branch length for the 62 Class~I, 
Flat spectrum, and Class~II sources (Class~III sources are excluded). 
\label{fig:mst_crit}}
\end{figure}

Figure~\ref{fig:mst} shows the residual MST sub-networks for the YSO positions 
after those longer than L$_{crit}$ have been excluded.  Black points represent 
each source, and the branches (grey lines) connect those sources that are more 
closely spaced than the critical length measured in the cumulative 
distribution of MST branch lengths, which is shown in 
Figure~\ref{fig:mst_crit}.  The critical branch length found in 
\citet{rguter09} for CrA, corrected for a distance of 130~pc (a distance of 
170~pc was used in their paper) was 0.116~pc (see their Figure~1 for CrA).  
Using all 116 YSO candidates, we obtained a L$_{crit}$ = 0.164~pc (see the top 
panel of Figure~\ref{fig:mst_crit}).

We performed the same analysis for the 62 YSO candidates that had either 
Class~I, Flat spectrum, or Class~II SEDs (i.e. the Class~III sources were 
excluded).  Again, only one significant core was found, consisting of 34 
members.  The bottom panel of Figure~\ref{fig:mst} shows the MST of the YSO 
positions for these sources, and in this case, we obtained a L$_{crit}$ = 
0.152~pc (see the bottom panel of Figure~\ref{fig:mst_crit}).  
Table~\ref{tab:clustering_analysis} summarizes the results of this comparison, 
as well as the results of the \citet{rguter09} study, including values of 
R$_{circ}$, R$_{hull}$, the aspect ratio R$_{circ}^2$/R$_{hull}^2$, and the 
density for each core.  For CrA, when looking at the core of 68 members, the 
value for R$_{circ}$ = 0.59~pc, with an aspect ratio of 2.36, indicating that 
the cluster is elongated.  To illustrate this elongation, the convex hull for 
each grouping is shown (solid black line) in Figure~\ref{fig:mst}.

The addition of new members from this paper changes the membership of the main 
CrA core from 24 members in the \citet{rguter09} study, to 68 members in our 
study of all classes of YSOs.  However, the mean densities are quite similar 
since our analysis covers a larger area and includes known YSOs from the 
literature (see Table~\ref{tab:clustering_analysis}).  In addition, the 
elongation of the core changes only slightly, from an aspect ratio of 2.51 to 
2.36.  A careful look at the differences in the top and bottom panels of 
Figure~\ref{fig:mst} shows that the Class~III population extends further to 
the west, indicating an evolutionary stage gradient to the west in CrA.  This 
confirms our qualitative analysis of \S~\ref{Av_ysodist}, obtained from 
looking at the spatial distribution of YSOs (see also 
Figure~\ref{fig:allysos}), and the noted elongation of the CrA cloud from 
previous studies, as discussed in \S~\ref{intro} \citep{neu00,mf01}.

\section{Submillimeter Analysis \label{sec:sma_analysis}}

\begin{figure}
\centering
\includegraphics[angle=0,width=6.6in]{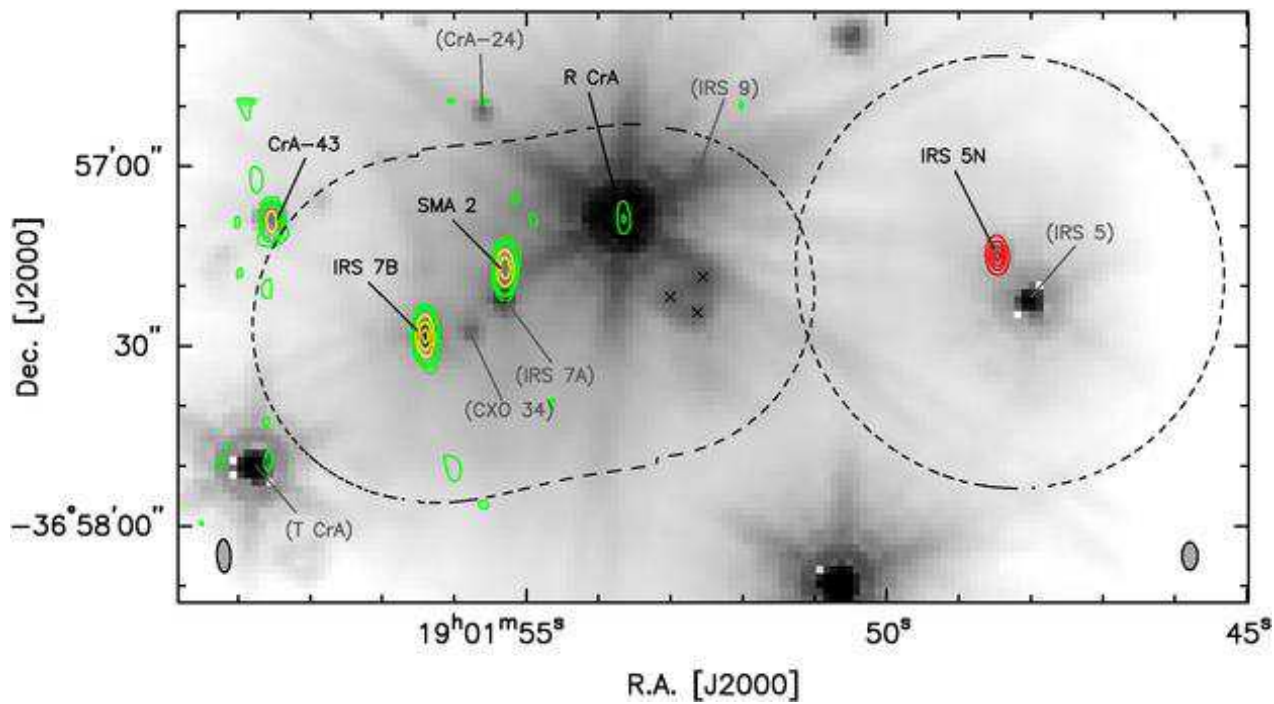}
\caption{SMA observations of the R\,CrA region. The image is the {\it Spitzer} 
4.5~$\micron$ image.  Two SMA mosaics are shown. On the right, indicated by the
large dashed black circle, are the SMA observations of IRS\,5 at 226~GHz.  On 
the left, indicated by the large dashed oval-like shape, are SMA observations 
of the IRS\,7 region at 224~GHz. The circle and oval, respectively, indicate 
the 50\% sensitivity levels of the observations.  In the IRS\,5 region, the 
red contours indicate emission at the 6$\sigma$, 9$\sigma$, ..., 18$\sigma$ 
levels, where $\sigma$ = 3.7~mJy/beam ($4\farcs6~\times~2\farcs6$).  In the 
IRS 7 region, the green contours indicate emission at the 3$\sigma$, 
5$\sigma$, and 7$\sigma$ levels, , while the yellow contours indicate emission 
at the 15$\sigma$, 27$\sigma$, and 39$\sigma$ levels, where $\sigma$ = 
7.4~mJy/beam ($5\farcs5~\times~2\farcs3$).  Sources with black labels were 
detected by the SMA, while those with grey labels (in parentheses) were not. 
R CrA is detected at the 5$\sigma$ level.  The sources IRS\,7B and SMA\,2 
\citep[Radio Source 9;][]{brown87} are equivalent to SMA\,1 and SMA\,2 
\citep[][271 GHz]{groppi07}.  The synthesised beams are shown as grey ovals at 
lower right (IRS\,5) and lower left (IRS\,7).  Crosses indicate artifacts in 
the {\it Spitzer} data. \label{fig:sma_irs5}}
\end{figure}

\begin{figure}
\centering
\includegraphics[angle=0,width=6.6in]{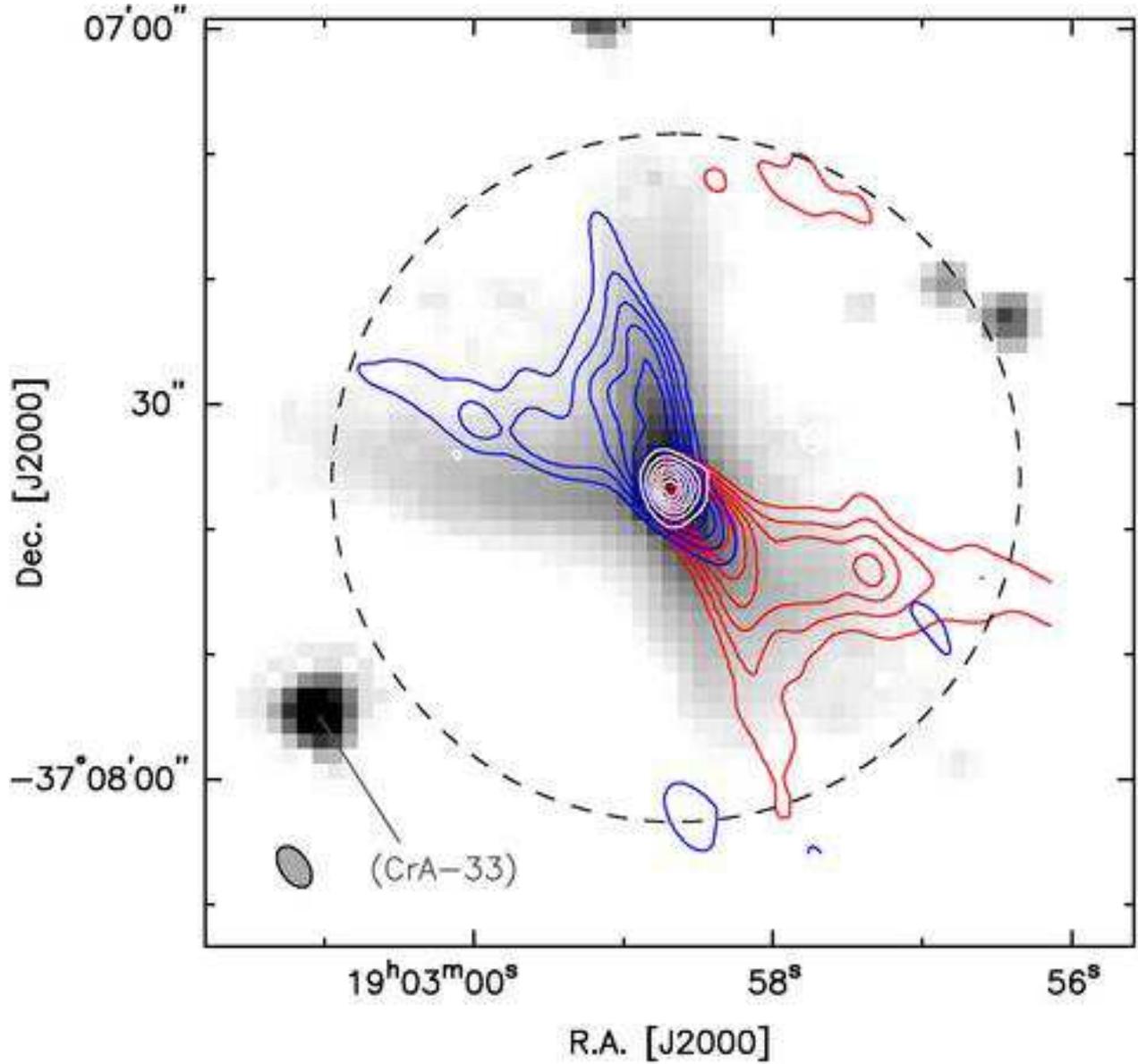}
\caption{SMA observations of the CrA IRAS\,32 region.  The location of YSO 
candidate CrA-33 is also shown, although it was not detected with the SMA.  
The image is the {\it Spitzer} 4.5~$\micron$ image.  The blue and red contours 
represent blue- and red-shifted integrated $^{12}$CO~2-1 emission, 
respectively.  The white contours represent 226 GHz continuum emission.  
Contour levels for the blue- and red-shifted emission are 3$\sigma$, 
6$\sigma$, etc, where 1$\sigma$ = 1.0~Jy km/s and 0.7~Jy km/s, respectively.  
The CO is integrated over LSR velocities of [0,4.2] km/s and [6.7,11.6] km/s 
for blue- and red-shifted emission, respectively.  The contours of continuum 
emission are 3$\sigma$, 6$\sigma$, etc, where 1$\sigma$ is 4.4~mJy/beam.  The 
large dashed circle shows the primary beam size of the SMA at 226~GHz, while 
the small grey oval at lower left shows the synthesised beam of 
$3\farcs2~\times~2\farcs2$. \label{fig:sma_iras32}}
\end{figure}

Six YSOs were detected in the dust continuum at 225 GHz by the SMA, and are 
listed with their flux densities in Table~\ref{tab-sma}, and shown in 
Figures~\ref{fig:sma_irs5} and \ref{fig:sma_iras32}.  The YSOs IRS\,5, IRS\,9, 
CXO\,34, and CrA-24 were not detected, above a 3$\sigma$ rms level of 11 mJy 
(IRS\,5) or 22 mJy (IRS\,9, CXO\,34, and CrA-24).  The detected emission in 
all cases is compact and centrally concentrated, and is thus assumed to be 
primarily due to compact disk emission.  Point source and 2-D Gaussian models 
were fitted to the detected YSOs, and for all sources but IRAS\,32, the 
results were consistent with a point source, indicating unresolved emission on 
size scales of about 400 AU, consistent with disk sizes measured for other 
embedded YSOs \citep{jorg07,jorg09}, and for T Tauri stars in nearby regions 
\citep{aw07}.  For IRAS\,32, an extended component is also present (more flux 
is recovered from fitting a Gaussian than from a point source alone), and is 
assumed to be due to inner envelope emission; this extended component was 
also seen in infrared imaging by \citet{wilking92}.  To remove this component, 
IRAS\,32 was re-imaged using only baselines $>$ 20 k$\lambda$, with the result 
being consistent with a point source.  More detailed model fitting of 
IRAS\,32, including 2D radiative transfer, will be presented in a later 
paper.  The flux densities reported in Table~\ref{tab-sma} are consistent with 
the values for SMA\,2 ($\equiv$ RS\,9) and SMA\,1 ($\equiv$ IRS\,7B) reported 
by \citet{groppi07}, who observed with the SMA at a slightly higher frequency 
of $\sim$271 GHz.

If the disks are optically thin, the continuum fluxes provide a direct
measure of the masses.  For marginally optically thick emission, possible
for such disks \citep{jorg07,aw07}, the mass becomes a lower limit to the true 
mass.  The corrections are usually small, however.  Previous studies of 
T~Tauri stars \citep{aw07,lommen07} indicate that the ratio of optically 
thick-to-thin emission is typically of order 0.2-0.3.  This translates 
into an underestimation of the mass by up to a factor of two.  Given the other 
assumptions (constant disk temperature, no radiative transfer modeling), this 
is a small correction.  To compute disk masses, we use equations (3) and (4) 
from \citet{jorg07} with their assumptions for the dust temperature (30 K) and 
1.3 mm opacity (assuming a gas-to-dust ratio of 100).  For the assumed 
distance of 130 pc for CrA, the disk mass at $\sim$ 225 GHz (1.3 mm) is

\begin{equation}
M = 0.236\,\, M_{\sun}\,\, \left(\frac{F}{\rm 1 Jy}\right)
\label{eqn:sma}
\end{equation}

\noindent
where $F$ is the flux density measured in Jy.  The results for the detected 
sources are listed in Table~\ref{tab-sma}.  The uncertainties listed in that 
table are those returned by the fitting procedure in MIRIAD, and are similar 
to the 3$\sigma$ rms values reported earlier.  So although the flux density of 
R\,CrA reported in this table is only 1.5 times the reported uncertainty, it 
is a well detected source in the maps (about 5$\sigma$).  

For the non-detected sources, the upper limits to the disk mass are 
0.003~M$_{\sun}$ for IRS\,5 and 0.005~M$_{\sun}$ for IRS\,9, CrA-24, and 
CXO\,34, that is, only a few Jupiter masses (M$_{J}$), and significantly less 
than the minimum mass solar nebula (MMSN), of 10 - 70~M$_{J}$ 
\citep[$\sim$0.01 - 0.07~M$_\odot$;][]{weidenschilling77}.  In these cases, 
the masses are either very low, the dust has been processed into bodies 
significantly larger than a millimeter, and are thus mostly invisible to our 
observations \citep{lommen07}, or the dust is absent, perhaps due to clearing 
in a close binary \citep{jensen96}.  It is known that IRS\,5 is a close binary 
system ($\sim$100 AU), and so this last scenario is a real possibility 
\citep[see further discussion of IRS\,5 in 
\S~\ref{sec:irs5};][]{cg93,fpm06,chlt08}.  The disk masses reported for the 
detected sources, aside from R\,CrA, are consistent with the MMSN.  

The masses reported here are similar to masses of Class~0 and Class~I 
protostars derived in an identical manner \citep[0.017 - 0.089~M$_{\odot}$, 
and 0.009 - 0.076~M$_{\odot}$, respectively;][]{jorg09}.  For Class~II YSOs 
(T Tauri stars) also observed with the SMA, but analyzed with more detailed 
models, the masses are similar or slightly higher 
\citep[0.003 - 0.18~M$_{\odot}$;][]{aw07}.  The fluxes reported for the 
Class~II YSOs are similar to those reported here \citep{aw07}, and lie at a 
similar distance (Taurus and Ophiuchus), so calculating masses using equation 
(\ref{eqn:sma}) would produce masses similar to our results.

Observations at centimeter wavelengths show that the emission we have observed 
is most likely due to thermal dust emission, with essentially no contamination 
from free-free emission, except perhaps for R\,CrA 
\citep{chlt08,miettinen08}.  All sources, other than R\,CrA, are either weakly 
detected at cm wavelengths ($<$ 1~mJy), or not at all, with spectral indices 
between 4.9 and 8.5~GHz that are flat or negative.  Although R\,CrA is also 
only weakly detected at cm wavelengths, it has a positive spectral index in 
the range 0.4-0.9, which implies that up to 5~mJy of the emission observed at 
1.3~mm could be due to free-free emission, and so the disk mass should be 
slightly reduced to 0.008~M$_{\sun}$, which is consistent with the upper limit 
of 0.012~M$_{\sun}$ derived by \citet{groppi07}.  

\section{Outflows and Jets \label{sec:outflows_jets}}

\subsection{Outflows and Jets Detected in H$_2$ and \textit{Spitzer} maps \label{sec:outflows}}

\begin{figure}
\includegraphics[width=16.0 cm]{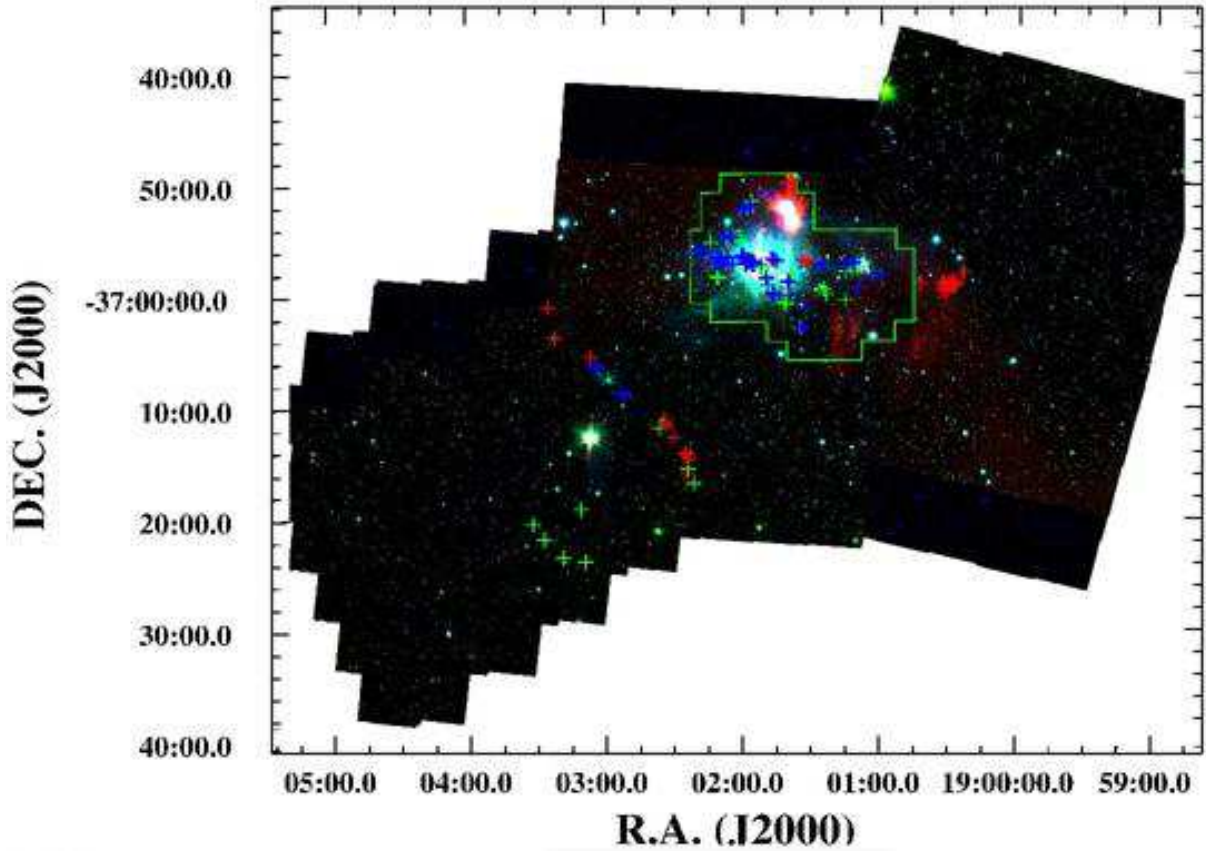}
\caption{{\it Spitzer} three-color image of the CrA star-forming region, made from the IRAC 3.6\,$\mu$m (blue), 4.5\,$\mu$m (green), and 8\,$\mu$m (red) images.  The green polygon shows the area covered by our \textit{SofI} H$_2$ map.  Green, blue, and red crosses indicate the position of HH objects, H$_2$, and {\it Spitzer} knots, respectively.\label{irac124:fig}}
\end{figure}

Our H$_2$ survey maps an area of about 20$\arcmin$$\times$17$\arcmin$, which 
spans from $\sim$5$\arcmin$ east of the Coronet region to $\sim$4$\arcmin$ 
west of S\,CrA, and from $\sim$3$\arcmin$ north of TY\,CrA to $\sim$3$\arcmin$ 
south of HH\,101 (see Figure~\ref{SofI07map:fig}).  The mapped region is about 
one tenth of the \textit{Spitzer} area, and is shown in 
Figure~\ref{irac124:fig} as a green polygon on an IRAC/\textit{Spitzer} 
three-color map of CrA.

Our H$_2$ continuum-subtracted mosaic reveals nearly 100 knots, some of them 
coincident with, or close to previously known HH objects.  These knots are 
shocked regions along the outflows, usually traced by H$_2$ and/or \ion{Fe}{2} 
emission lines in the near-infrared \citep[see e.\,g.,][]{caratti06}, and 
ionic species in the optical (as \ion{O}{1}, \ion{S}{2}, and H${\alpha}$), 
also known as HH objects, which trace dissociative shocked regions with lower 
extinction and higher temperatures \citep[see e.\,g.,][]{eisloffel94}.  Once 
the continuum is removed from the H$_2$ narrow-band images, what is left is 
the emission line alone, and thus knots and jets can be unambiguously 
identified.  We detect 8 H$_2$ counterparts out of 14 HH objects located in 
the mosaic FoV, while no evidence of molecular emission is observed from the 
remaining 6, namely HH\,96, HH\,729, HH\,730, HH\,734, HH\,736, HH\,860.  
Following \cite{davis09b}, we name the detected H$_2$ flows MHO\,2000--2014, 
followed by letters indicating knots and sub-structures inside each flow.  It 
is worth noting that such a grouping is based on the flow morphology alone, 
and does not necessarily imply a real association to a single outflow.  
\citet{davis09b} designated five MHOs (2000-2004) in the Coronet region, on 
the basis of morphology alone, because no PM analysis was available.  Usually 
such a grouping is effective for isolated outflows and sources, where the 
outflows do not overlap and it is straightforward to recognize the outflow 
structure (chains of knots more or less aligned, bow-shocks indicating the 
outflow direction, etc.), which also points to the exciting source.  On the 
other hand, it is often particularly difficult to disentangle different flows 
in crowded regions such as the Coronet, where several flows overlap, and the 
exciting sources are enclosed in a relatively small area.  Here, we also use 
our P.M. analysis (see \S~\ref{PMs:sec} and Appendix \ref{sec:appB}) to 
recognize and group each flow.  However, our analysis and thus our groupings 
are not resolute for several cases (see e.\,g., 
\S~\ref{outflows_and_sources:sec}, and Appendix \ref{sec:appB}) due to the 
crowding in the Coronet and the uncertainties in our PM measurements.  As a 
result, we keep the \citet{davis09b} nomenclature, adding ten more newly 
detected MHOs, i.\,e. MHO2005-2014, and include letters to pinpoint recognized 
knots and sub-structures within each MHO.  We stress that such a grouping does 
not necessarily imply a real association to a single outflow.

The majority of the flows originate from the Coronet region and are radially 
distributed around it, with a larger concentration in the east and north-east 
regions (namely MHO\,2000--2002, 2009, 2011--2014), and west and south-west 
regions (namely MHO\,2003--2008), with MHO\,2005 positioned around S\,CrA.  
Finally, MHO\,2010 is positioned around IRAS\,18595-3712 (IRAS\,32).  We 
identify about 14 H$_2$ flows in CrA.

Most of the known HH objects, and all the detected H$_2$ objects, are visible 
in our IRAC maps.  Indeed, the IRAC maps contain molecular hydrogen lines that 
may be shock-excited in protostellar outflows \citep[see e.\,g.][and 
references within]{dv10}.  Generally, the H$_2$ lines in band 2 (4.5\,$\mu$m) 
and, partially, in band 1 (3.6\,$\mu$m) have excitation conditions quite 
similar to the 1-0\,S(1) line (2.12\,$\mu$m), tracing shocks at T$\sim$2000\,K
\citep[see e.\,g.][]{caratti06,takami2010}.  In particular, IRAC band 2 
contains bright molecular hydrogen lines and can be used to detect knots 
\citep[see e.\,g.][]{cyganowski08,cyganowski09,ybarra}, often called 
``extended green objects"~\citep[EGOs, ][]{cyganowski08}, for the common 
coding of the 4.5\,$\mu$m band as green in the IRAC three-color images.  To 
better identify these features in the IRAC/\textit{Spitzer} mosaics, we thus 
constructed IRAC three-color images, using 3.6\,$\mu$m, 4.5\,$\mu$m, and 
8\,$\mu$m (blue, green, and red, respectively, see Figure \ref{irac124:fig}), 
and identified the new knots/EGOs by means of their colors and morphologies.  
Usually the outflow morphologies in band 2 and, partially, in band 1 resemble 
those observed in our H$_2$ images, mostly delineating relatively 
high-excitation regions as bow shock heads or clumpy structures/knots along 
the flow axes.  On the other hand, band 4 and 3, which contain relatively 
low-excitation H$_2$ lines, often delineate wings and wakes of bow shocks, and 
in general, milder shock regions \citep[for a detailed analysis 
see][]{takami2010}.

In addition to the H$_2$ detected features, our \textit{Spitzer} maps reveal 
three more flows positioned outside the H$_2$ mosaic, and named 
\textit{Spitzer} outflow 1, 2, and IRAS\,32 outflow, which include 30 
additional new knots.  Because we lack spectral confirmation, these flows 
could not be named MHO.

H$_2$ and \textit{Spitzer} newly detected knots and all previously known HH 
objects (including those not detected in H$_2$) are labeled in our H$_2$ map 
(see Figure~\ref{SofI07map:fig}) and in our three-color IRAC mosaic 
(Figure~\ref{irac124:fig}) as blue, red, and green crosses, respectively.  
Additionally, a list of known HH objects of the CrA star-forming region is 
reported in Table~\ref{HHs:tab}, and a complete list of knots detected in our 
\textit{Spitzer} and H$_2$ images is given in Table~\ref{identified_knots:tab}.

In Appendix~\ref{sec:appB}, Sections~\ref{NE_MHOs:sec}, \ref{SW_MHOs:sec}, and 
\ref{Spitzer_flows:sec}, a description of individual MHOs and \textit{Spitzer} 
flows is given.

\subsection{Proper motions of H$_2$ knots \label{PMs:sec}}

To derive proper motions from our multi-epoch set of images, we followed the 
technique described in detail by \citet{caratti09}. As shown in 
Figure \ref{SofI07map:fig}, the latest epoch map (\textit{SofI} 2007) encloses 
all the remaining four, thus we identify the \textit{SofI} 2007 map as the 
common reference frame for our PM analysis.  All earlier epoch maps were 
scaled and aligned to the reference frame with sub-pixel accuracy.  The match 
was done using several field stars observed in each map (see 
\S~\ref{sec:ancillary}), and optical distortions were corrected by means of 
the \emph{geomap} (with a polynomial fit of 3rd order) and \emph{geotran} 
routines in \emph{IRAF}.  The resulting errors are represented by the 
residuals of the fits, and range between $\sim$0.2 and $\sim$0.4 pixels 
(i.\,e. $\sim$0\farcs06-0\farcs12), depending on the seeing and the number of 
stars in each map.

Knot shifts were determined between image pairs (i.\,e. last epoch and 
previous epoch maps) using a cross-correlation method. Knots were identified 
in the continuum-subtracted maps, and a rectangle was defined around each 
knot, enclosing its 3\,$\sigma$ contour.  The first epoch image was then 
shifted with respect to the last epoch image, with a sub-pixel accuracy of 0.1 
pix and then multiplied by it. For each shift (x,y), the total integrated flux 
(f) in the rectangle around each knot was measured in the product image.  For 
the final shift for each knot, we used the position of the maximum of f(x,y), 
determined via a Gaussian fit. 

Systematic errors for shift measurements were obtained by modifying the size 
and shape of the rectangle that defines each object.  The resulting range of 
values gives a systematic error, which depends on the S/N, shape, and flux 
variability of the knot.  This error is usually comparable to the alignment 
error.  Thus, each single shift error was derived combining the alignment 
error of the two epoch images and the uncertainties in the centering routine.  
Usually, for each pair, the final error was around 20\% of the P.M. value.  It 
is worth noting that the number of P.M. measurements differs from knot to 
knot, depending on the number of matching maps.  Thus knots located in the 
Coronet, with three or four P.M. measurements, usually have smaller errors.

\begin{figure}
\includegraphics[width=14.0 cm]{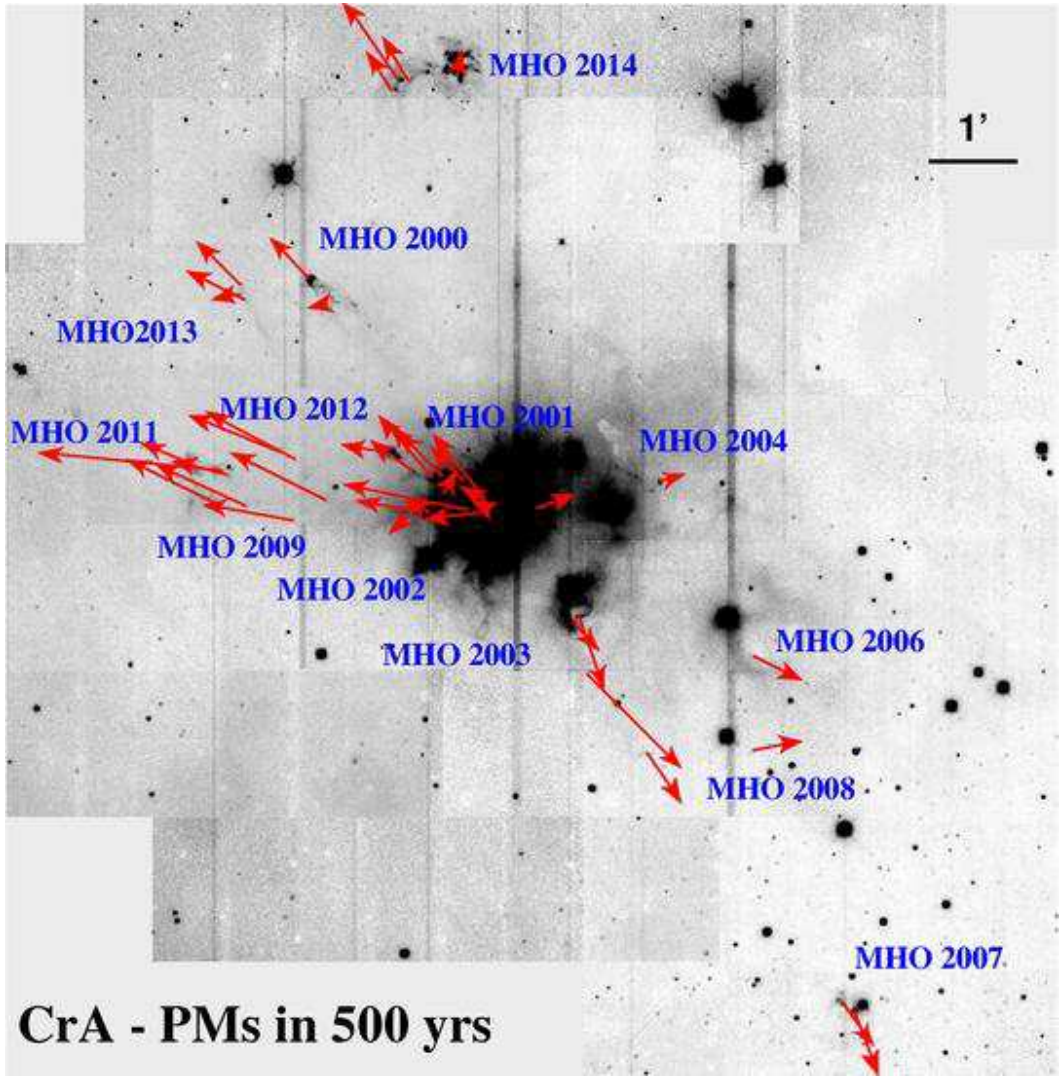}
\caption{H$_2$ flow chart of the CrA star-forming region. Proper motions in 500\,yrs are indicated by arrows. \label{PM_all:fig}}
\end{figure}

In Figure~\ref{PM_all:fig}, we show the flow charts of the CrA star-forming 
region.  To display MHO motions for the entire region, proper motions, 
indicated by arrows, are reported in units of 500\,yrs.  The results of our 
P.M. measurements are given in Table~\ref{PMs:tab}, which lists the knot ID, 
the measured P.M. in arcsec\,yr$^{-1}$, the derived tangential velocity 
($v_\mathrm{tan}$) in km\,s$^{-1}$ at a distance of 130\,pc, and the position 
angle (P.A.) in degrees of the P.M. vector.  P.M.s were measured for knots 
observed in at least two different epochs.  Moreover Table~\ref{PMs:tab} 
reports only P.M.s with S/N ratios larger than 2$\sigma$.  As a result, 44 
measurements are reported, namely those from MHO\,2000, 2002, 2003, 2007, 
2013, the majority of knots in MHO\,2001, 2009, 2011, 2012, 2014, and some 
knots in MHO\,2004, 2006, and 2008.

The derived P.M. values range between 0.024 and 0.256\,arcsec\,yr$^{-1}$, 
corresponding to a $v_\mathrm{tan}$ between 15 and 158\,km\,s$^{-1}$ (at a 
distance of 130 pc).  The uncertainties vary from 0.006 to 
0.045\,arcsec\,yr$^{-1}$ (i.\,e. $\sim$4 to $\sim$28\,km\,s$^{-1}$), 
depending on the knot S/N ratio, the number of epochs, and the accuracy in the 
alignment.  Such uncertainties, in turn, reflect on P.A. errors, which span 
from 2$\degr$ to 27$\degr$.

A quick inspection of Figure~\ref{PM_all:fig} confirms that all the studied 
knots, except MHO\,2008\,D, are launched by YSOs inside or close to the 
Coronet.  Several flows originate in the eastern part of the Coronet, in 
particular in a region a few arcseconds around SMA\,2, which encloses 
several early stage YSOs \citep[e.\,g.][]{nutter05,groppi07}.

A detailed analysis of P.M.s in single outflows is reported in 
Appendix~\ref{sec:appB}.  There we show in more detail the flow charts of each 
region, where P.M.s (in 100\,yr) and their error bars are indicated by arrows 
and ellipses, respectively 
(Figures~\ref{PM_MHO2000_13:fig}--\ref{PM_MHO2007:fig}).

Finally, detailed discussion on outflow occurrence in CrA, and possible 
driving sources can be found in \S~\ref{outflows_and_sources:sec}.

\subsection{Outflows and possible driving sources \label{outflows_and_sources:sec}}

Our H$_2$ survey along with the \textit{Spitzer} IRAC maps and proper motion 
analysis shows at least 14 different H$_2$ flows in CrA.  This number 
increases to 17 if we also include HH objects that do not have H$_2$ 
counterparts but likely originate from one of the detected YSOs.  However, the 
number of flows could be larger than this because we are not considering some 
of the HH objects which do not have a viable candidate driving source.  
Outflows and their possible driving sources were selected on the basis of the 
outflow morphology and proper motion analysis (see \S~\ref{sec:outflows} and 
Appendix~\ref{sec:appB}).  Candidate driving sources, their possible flows and 
position angles, are reported in Table~\ref{YSOs_flows:tab}.  Also blue- and 
red-shifted lobes (hereafter blue and red lobes, respectively) have been 
tentatively indicated considering radial velocity measurements from this work 
(see \S~\ref{sec:sma_analysis} and Figure~\ref{fig:sma_iras32}) and from the 
literature~\citep{hartigan87,leverault,anderson,davis99,groppi04,groppi07,knee,
vankempen}.  Several question marks indicate the uncertainty in matching flows 
and sources.  A clear match was obtained for those flows close to the driving 
source (IRS\,1, IRS\,2, IRS\,6, SMM\,2, SMA\,2, IRS\,7B, CrA-41), or driven by 
isolated YSOs (IRAS\,32).

\begin{figure}
\includegraphics[width=16.0 cm]{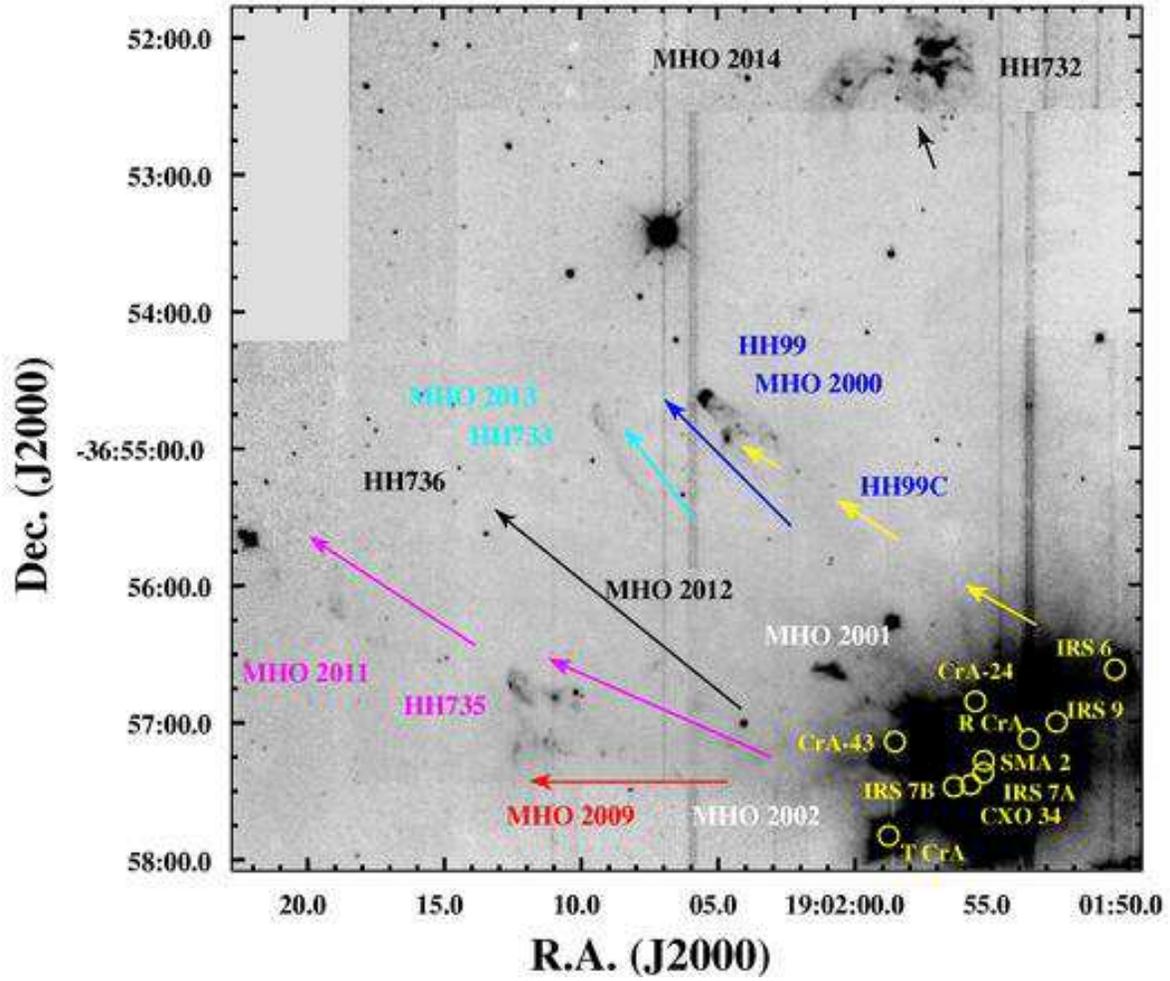}
\caption{\textit{SofI} H$_2$ image of the north-eastern flows detected outside 
the Coronet.\label{NE_flows:fig}}
\end{figure}

The uncertainties in Table~\ref{YSOs_flows:tab} are mainly due to the 
particular morphology of the Coronet region, as well as the uncertainties in 
the P.M. measurements.  The majority of the flows originate from sources 
inside, or close to, the Coronet, in particular from the regions around SMA\,2 
and IRS\,7A, where four known YSOs are located (namely SMA\,2, IRS\,7A, B and 
CXO\,34).  However, we note that at least six different overlapping flows are 
launched from this region of the Coronet, likely indicating the presence of 
other embedded YSOs~\citep[see e.\,g.,][see also Figures~\ref{NE_flows:fig} 
and \ref{MHO2001-2:fig}]{groppi07,chlt08}.  The overlap of the flows and 
the errors on the P.M.s do not allow us to pinpoint the driving sources, with 
the possible exceptions of SMA\,2, IRS\,7A, and IRS\,7B, where the detected 
jets appear to be launched from or close to the positions of the sources (see 
Figures~\ref{MHO2001-2:fig} and \ref{PM_MHO2001_2:fig}, and 
Appendix~\ref{NE_MHOs:sec}).  The SMA\,2 flow should include  MHO\,2001\,J, A, 
and MHO\,2009 to the east, and MHO\,2001\,Q to the west.  The IRS\,7B flow 
should include MHO\,2002\,C (see Figure~\ref{MHO2001-2:fig}), MHO\,2012, and 
possibly HH\,736 in one lobe (see Figure~\ref{NE_flows:fig}), and, possibly, 
HH\,731\,B in the other (see Figure~\ref{SW_flows:fig} and 
Table~\ref{YSOs_flows:tab}). 

There is also strong evidence of a very extended ($\sim$1\,pc) east-west wide 
precessing flow ($\sim$15$\degr$), with blue and red lobes positioned east and 
west of the Coronet, respectively.  This claim is also supported by CO 
observations from literature 
\citep[see e.\,g.,][]{knee,leverault,groppi04,groppi07}.  MHO\,2011 matches 
well the CO blue lobe, whereas MHO\,2004\,D, \textit{Spitzer} outflows 1 and 
2, and MHO\,2005 (see Figures~\ref{spitzer1:fig}, \ref{spitzer2:fig}, and 
\ref{westflow:fig}) match the more extended CO red lobe.  The driving source 
is likely to be one of the youngest sources in the region around SMA\,2 and 
IRS\,7A.  SMA\,2 would be the obvious candidate, but it already drives another 
jet, which is also east-west oriented, but with smaller size and precession 
angle ($\sim$4$\degr$) (MHO\,2002 and MHO\,2009).  However we note that SMA\,2 
could harbor a binary or multiple system, as e.\,g. reported by 
\citet{chlt08}.  In this work, they detect three different sources, namely 
B9a/b in the centimeter continuum and CT 2 in the 7\,mm continuum, positioned 
a few arcseconds north of IRS\,7A, and at least two outflows (FPM 10 and 
FPM 13).  We observe at least three different precessing flows escaping from 
the SMA\,2 - IRS\,7A region (namely knots MHO\,2001 M, N, and O).  Another 
possibility is that this large outflow is driven by IRS 7A. 

Another large CO flow extends roughly SSW to 
NNE~\citep{hartigan87,leverault,anderson,knee}, with HH\,101 (MHO\,2007) 
coincident with the blue lobe.  HH\,99 (MHO\,2000) was previously believed to 
be the counterflow, but both geometric and kinematic considerations ruled 
out this hypothesis~\citep[see e.\,g.,][]{hartigan87,davis99}.  More recent 
CO observations~\citep{knee} place the peak of the red lobe far to the 
north-west, i.\,e. close to the position of the newly discovered MHO\,2014, 
which is likely the counterflow, as both morphology and P.M. analysis also 
suggest.  IRS\,5 and 5N are positioned between the two lobes, thus one of the 
two is likely the driving source.  Based on our SMA results, IRS\,5N seems to 
be the most likely candidate, since no continuum emission from IRS\,5 is 
detected (see also \S~\ref{sec:sma_analysis}, \ref{sec:irs5}, and 
\ref{sec:irs5n}).     

There is no clear evidence of the HH\,99\,B (MHO\,2000\,A) counter-flow in our 
analysis.  The P.M. analysis indicates that MHO\,2000\,A and B (HH\,99\,B and 
A) are likely part of two distinct flows.  MHO\,2000\,A could be driven by 
CrA-24, one of the newly discovered Class I YSOs, or by IRS\,9 (see also 
Appendix~\ref{sec:appB} and Figure~\ref{NE_flows:fig}).  MHO\,2000\,B 
seems to be part of a larger flow driven by IRS\,6~\citep[see also, Appendix 
\ref{sec:appA}, \ref{sec:appB}, Figure \ref{NE_flows:fig}, and][]{wang}.  
Another Class I YSO, namely CrA-43 (or SMM\,2) in the Coronet outskirts (see 
e.g. Figures \ref{fig:sma_irs5} and \ref{MHO2001-2:fig}), is also driving an 
H$_2$ outflow, SW-NE oriented (P.A. $\sim$125$\degr$), which includes 
MHO\,2002\,B, G, and possibly H, and HH\,734~\citep[see also, 
Appendix~\ref{sec:appA}, \ref{sec:appB}, and][]{wang}. 

On the basis of our morphological and kinematical analysis (see 
Appendix~\ref{sec:appB}), in Table~\ref{YSOs_flows:tab} we tentatively 
associate the remaining observed H$_2$ and HH objects with nearby detected 
YSOs.  These possible matches are labeled in the table with a question mark.  
It is worth noting that we clearly detect only one lobe in several flows.  
Indeed this has already been observed in relatively crowded and confused 
regions~\citep[see e.\,g.,][]{davis09a}, but our sample has an extremely high 
rate of single-lobe flows (9 out of 17), which is particularly disturbing.  
Although this can be ascribed both to the morphology of the region, the high 
extinction (see Section~\ref{Av_ysodist}), and the 
limited sensitivity of our survey, it could also indicate mismatches between 
flows and sources.

\subsection{Incidence of outflows vs number \& class of YSOs}

A more detailed analysis of Table~\ref{YSOs_flows:tab} confirms that H$_2$ 
outflows are mostly driven by early stage 
YSOs~\citep[see also,][]{davis09a}.  Indeed, Class~0 source SMA\,2 drives 
an H$_2$ flow, as well as 10 out of the 14 YSO candidates classified as 
Class~I.  All of the Class~I YSOs located near or in the Coronet seem 
to drive outflows.  Three out of 43 Class~II YSOs have H$_2$ flows, which 
becomes 6 out of 43 Class II YSOs if we also consider flows with HH objects.  
None of the Class~III sources shows any sign of outflow activity, indicating 
that accretion activity has reached completion in this stage, as one would 
expect.  On the other hand, the percentage of H$_2$ and atomic jets detected 
in Class~II sources by this, and previous surveys \citep[e.\,g.][]{wang} seem 
to be too small ($\sim$14\%) for this class of objects, because accretion 
should still be ongoing in Class~II sources, and thus outflows should be 
expected.  This would then suggest that these surveys are not sensitive enough 
to detect such outflows, or that there are mechanisms, for example the cloud 
magnetic field, that inhibit the outflow activity in more evolved YSOs.  
Nevertheless our data confirm that the accretion/ejection activity decreases 
with protostellar age.  The H$_2$ outflow sources are protostars (Class~0/I 
sources) rather than disk-excess sources (mostly Class~II T Tauri stars), 
though a few of the latter source sample do seem to power H$_2$ flows, and HH 
objects.  However, if we exclude the IRS\,6 flow, the H$_2$ emission in these 
Class~II flows is extremely reduced, whereas the HH and ionic emission are 
still well detected.  

\section{Overall Cloud Properties \label{sec:cloudproperties}} 

\subsection{Disk Fraction \label{sec:disk_fraction}}

We have compiled a fairly complete estimate of the Class~III population in 
CrA (see \S~\ref{sec:completeness}), and as a result, can examine the disk 
fraction of YSO candidates.  There are 90 YSO candidates of all classes within 
the A$_V$ = 2\,mag contours; of those, 42 are Class~III YSO candidates.  
Therefore the disk fraction in CrA is 53\%.  Recent disk fraction estimates 
for CrA include 53\% by \citet{lopezmarti10}, who also included a 
population of diskless sources, and 67\% by \citet{mw09}, who obtained the 
disk fraction from an extinction-limited sample of sources and used 
spectroscopic information to estimate an age range of 0.3$-$3~Myr from their 
H-R diagrams.  Disk fraction values similar to ours have been inferred for 
young clusters NGC~7129 \citep[54\%;][]{gutermuth04} and NGC~2264 
\citep[52\%;][]{haisch01}.  Based on comparison with theoretical PMS tracks, 
\citet{haisch01} estimate an age of $\sim$2.5 Myr for NGC~2264, which is 
consistent with the age range of 0.3$-$3~Myr estimated for CrA \citep{mw09}.  

\subsection{Star Formation Rate and Efficiency \label{sec:sfr_sfe}}

The mass of the CrA cloud can be estimated, using the {\it Spitzer} + 2MASS 
extinction map discussed in \S~\ref{Av_ysodist}.  To compare our results to 
those for the c2d and GB clouds, we employ the same method used by 
\citet{heiderman10}.  They use the lowest resolution extinction map 
(FWHM = 270$^{\prime\prime}$) to calculate M$_{\rm gas,cloud}$ by summing up 
all the pixels ($\Sigma$A$_V$) above an A$_V$ = 2\,mag threshold, convert to 
column density using N$_H$/A$_V$ = 1.37$\times$10$^{21}$ cm$^{-2}$mag$^{-1}$ 
\citep{draine03} for a \citet{wd01} R$_V$ = 5.5 extinction law, and obtain 
M$_{\rm gas,cloud}$ = 279~M$_{\odot}$ \citep{heiderman10}.  By following this 
procedure we obtain the same value for the cloud mass and the area of the 
cloud within the A$_V$ = 2\,mag contour (i.e. 0.6~deg$^2$), which translates 
to A$_{\rm cloud}$ = 3.03~pc$^2$ at a distance of 130~pc.  By dividing 
M$_{\rm gas,cloud}$ by the area of the cloud, we obtain the average gas 
surface density of the cloud, $\Sigma_{\rm gas,cloud}$ = 
92.1~M$_{\odot}$pc$^{-2}$.  The range of gas surface densities for the c2d and 
GB clouds discussed in \citet{heiderman10} is ${\Sigma}_{\rm gas,cloud} 
\sim$~45$-$140~M$_{\odot}$pc$^{-2}$, and CrA falls in the middle of that 
range, similarly to Chamaeleon I, Auriga, and Perseus.
  
Because we include additional YSO candidates not included in the 
\citet{heiderman10} study (i.e. N$_{\rm YSO,tot}$ = 48, versus 41), the 
resulting star formation rate (SFR) for CrA is a factor of $\sim$1.2 higher 
than theirs.  Note that only Class~I, Flat spectrum, and Class~II sources 
located in the area of the extinction map above the A$_V$ = 2\,mag contours 
are considered.  Following the analysis of \citet{heiderman10}, we assume a 
mean YSO mass of 0.5$\pm$0.1~M$_{\odot}$ over a period of star formation of 
2$\pm$1~Myr, based on estimates of the time span of the Class~II phase 
\citep{evans09}.  Using equation (8) in \citet{heiderman10}, we compute the 
SFR surface density to be ${\Sigma}_{\rm SFR}$ = 
3.96~M$_{\odot}$yr$^{-1}$kpc$^{-2}$, which translates to a SFR of 
12.0~M$_{\odot}$Myr$^{-1}$.  CrA has a relatively low SFR, compared with 
Perseus, Serpens and Ophiuchus \citep[96, 56, and 73~M$_{\odot}$Myr$^{-1}$, 
respectively;][]{evans09,heiderman10}, and is similar to the Lupus clouds, in 
particular Lupus III, Lupus V, and Lupus VI 
\citep[17, 11, and 11~M$_{\odot}$Myr$^{-1}$, respectively;][]{heiderman10}.  
The SFR surface density, however, is the highest of all the clouds, with only 
Serpens showing similar values \citep[${\Sigma}_{\rm SFR}$ = 
3.29~M$_{\odot}$yr$^{-1}$kpc$^{-2}$;][]{heiderman10}.  Similar comparisons 
were also made by \citet{lada10}, where a useful graphical comparison of CrA 
with the c2d and GB clouds can be found \citep[][particularly, their Figures 
2 and 4]{lada10}.

For comparison with the other c2d/GB regions, we also compute the star 
formation efficiency (SFE) for CrA using the following relation:

\begin{equation}
SFE = \frac{M_*}{M_* + M(cloud)}.
\label{eqn:sfe}
\end{equation}

\noindent
The total mass of the stars, M$_*$ is estimated by multiplying the total 
number of stars (N$_{\rm YSO,tot}$ = 48) by an assumed average stellar mass of 
0.5M$_{\odot}$.  Using the entire cloud mass computed above 
(M$_{\rm gas,cloud}$ = 279~M$_{\odot}$), we obtain a SFE for CrA of 
$\sim$0.08.  \citet{evans09} compiled SFE for several of the c2d clouds (see 
their Table 4), which range from 0.03 (Chamaeleon I) to 0.063 (Ophiuchus).  
The overall value of SFE for the GB cloud IC~5146 is $\sim$0.05 \citep{harv08}.
In general, CrA has the highest SFE of all the other clouds observed in c2d/GB 
surveys.  This result is still valid even if we use the smaller, 
N$_{\rm YSO,tot}$ = 41 value used in the \citet{heiderman10} study, so the 
effect is not merely a result of the addition of YSO candidates (although this 
may represent a useful measure of the uncertainties that the studies for the 
other c2d/GB regions may face due to incomplete YSO lists).  \citet{kirk09} 
found an SFE of 8\% for one of the GB regions, the Cepheus Flare region, when 
looking only at cores where groups of YSOs were located (and an SFE of 1\% for 
the quiescent protostellar cores).  In the case of CrA, perhaps we are seeing 
a small cloud of gas efficiently producing a cluster of stars, and no 
additional regions of diffuse gas diluting the SFE.

The high SFE in CrA has been noted in the past \citep[e.g. 45\%, 40\% and 
14\%:][respectively]{wts86,harju93,tachihara02}.  The very high estimate by 
both the \citet{wts86} and \citet{harju93} studies is due to the fact that 
they used only the main Coronet core surrounding R\,CrA to estimate SFE 
\citep[specifically, the area corresponding to the 2 $\mu$m survey 
by][]{ts84}.  \citet{wilking92} compare the SFE of CrA with that of the $\rho$ 
Ophiuchi cloud, which is similar in terms of distance and angular extent; 
also, neither cloud has a star more massive than B0V.  They note that CrA 
has fewer YSOs than $\rho$~Oph, and attribute that to the overall lower mass 
of both the low- and high-density molecular gas in the cloud 
\citep{dame87,loren79}.  

\citet{tachihara02} go a step further and compare the SFEs for a large sample 
of 179 cores in Taurus, the $\rho$~Oph cloud, the Ophiuchus north region, the 
Lupus clouds, L1333, Southern Coalsack, the Pipe nebula, and CrA star-forming 
regions. They find an average SFE of roughly $\sim$~10\% for all the cores, 
with an SFE of 14\% for all of CrA, and 40\% for the cluster-forming core of 
CrA.  Using C$^{18}$O observations, \citet{tachihara02} note that the regions 
with cluster-forming cores (in addition to CrA, this includes $\rho$~Oph, 
Lupus III and Chamaeleon I) have similar cloud structures, i.e., a head, where 
the active star formation occurs, and a tail that extends from it, suggesting 
that star formation has been triggered by an external shock.  In order to 
explore this observation, they measure the average line widths, $\Delta$V, for 
the starless cores which have not been disturbed by outflows from YSOs, and 
therefore represent the initial turbulent velocity in the cloud.  The SFE, 
estimated using the total cloud mass from the $^{13}$CO clouds, is then 
plotted versus $\Delta$V \citep[see Figure 11 in][]{tachihara02}.  For CrA, 
they estimate a cloud mass by applying the ratio of the total C$^{18}$O core 
mass to the $^{13}$CO cloud mass to be 30\%, and estimate SFE to be $\sim$4\%.

In general, \citet{tachihara02} find that the regions actively forming stars, 
and whose $^{13}$CO-based SFE is greater than 2\% (which includes CrA), all 
have $\Delta$V $<$ 0.7 km~s$^{-1}$ (conversely, the less active regions have 
$\Delta$V $>$ 0.7 km~s$^{-1}$).  Based on these results, they suggest that 
a star-forming region with an initial turbulent velocity that is low will form 
stars spontaneously; if hit with a shock wave, the shock wave will then 
trigger the formation of a cluster in the head, forming a head-tail 
structure.  Our observations and clustering analysis of CrA (see 
\S~\ref{sec:cluster_analysis}) are consistent with this picture: we see 
an elongated cluster core which is dominated by star formation in the 
``head,'' the Coronet, with a more extended population of PMS stars, and a 
quiescent clump in the southeastern ``tail'' of the cloud \citep{harju93}.  As 
mentioned in \S~\ref{intro}, it has been shown that CrA is moving away from 
the Upper Centaurus Lupus (UCL) association with an expansion velocity similar 
to that of the Sco-Cen superbubbles \citep{mf01}, specifically the \ion{H}{1} 
shell of Loop I, suggesting that the expanding \ion{H}{1} shell collided with 
the CrA cloud a few million years ago, triggering star formation in the 
Coronet, which is the part of the cloud facing the UCL \citep{harju93}.  The 
``tail'' of CrA is located downstream from the UCL association, which is also 
consistent with the triggering scenario.

\section{Summary \label{sec:summary}}

In this paper, we present the {\it Spitzer Space Telescope}, Submillimeter 
Array, and H$_2$ observations, combined with ROSAT and 2MASS, of the Corona 
Australis YSO population, its distribution, and outflows.

\begin{itemize}

\item A total of 116 YSO candidates are identified, where 14 are classified as 
Class~I, 5 are Flat spectrum, 43 are Class~II, and 54 are Class~III.  These 
candidates were selected based on {\it Spitzer}, 2MASS, ROSAT, and 
{\it Chandra} observations, as well as an extensive search of the literature.  
Of the 116 YSO candidates, there are 12 which have not been selected as 
candidates by any previous study, where 3 are classified as Class~I/Flat, 8 
are Class~II, and 1 is a Class~III.

\item Six of these YSOs: R\,CrA, IRS\,5N, IRS\,7B, SMA\,2 (RS\,9), CrA-43 
(SMM\,2), and IRAS\,32, were detected in the dust continuum at 225 GHz by the 
SMA.  All sources except IRAS\,32 have emission that is compact and centrally 
concentrated, assumed to be primarily due to compact disk emission.  Disk 
masses for these sources were computed and range from 
0.009-0.076~M$_{\odot}$.  SMA data of IRAS\,32 show an extended component, 
assumed to be due to inner envelope emission.  The computed disk mass for 
IRAS\,32 is 0.024~M$_{\odot}$.

\item An extinction map created using 2MASS and {\it Spitzer} data is 
presented and compared with the distribution of YSOs.  The measured A$_V$ over 
the entire field has a mean of $\sim$5 mag, with a maximum of $\sim$30 mag, 
peaking on the Coronet.  Using this extinction map, we calculated the cloud 
mass in the area of the extinction map above the A$_V$ = 2 contours to be 
279~M$_{\odot}$, confirming the result found in \citet{heiderman10}.  There is 
a clear radial spread of sources from the Coronet, with Class~I sources 
clustered most tightly in the center, surrounded by Class~II sources.  
Class~III sources can be found spread throughout the entire field.

\item We perform a clustering analysis mirroring that of \citet{rguter09}, 
using the 116 YSO candidates identified in this paper, and confirm the 
\citet{rguter09} result that the CrA cluster is elongated (having an aspect 
ratio of 2.36) with a circular radius of 0.59~pc and mean surface density of 
150 pc$^{-2}$.  By looking at the MST branches in the case where all YSOs are 
included compared with the case where the Class III sources have been 
excluded, it is clear that there is an evolutionary stage gradient in CrA, 
where the older Class III population extends to the west of the Coronet.

\item The star formation rate is calculated to be 12~M$_{\odot}$Myr$^{-1}$, 
similar to that of the Lupus clouds.  The SFR surface density, however, is 
quite high: ${\Sigma}_{\rm SFR}$ = 3.96~M$_{\odot}$yr$^{-1}$kpc$^{-2}$, 
similar to that of Serpens.  A disk fraction is also calculated, and found 
to be 53\% for CrA.  The star formation efficiency for CrA is $\sim$8\%, which 
is the highest of any of the c2d/GB regions surveyed, and is perhaps due to 
the triggering of star formation by an expanding \ion{H}{1} shell associated 
with the Sco-Cen superbubbles.

\item Our H$_2$ multi-epoch survey maps an area around CrA of about 
20$\arcmin$$\times$17$\arcmin$ (i.\,e. $\sim$10\% of the {\it Spitzer} map), 
and the continuum-subtracted mosaic reveals nearly 100 H$_2$ knots, and we 
study P.M.s in 44 of them.  The derived P.M. values range between 0.024 and 
0.256~$\arcsec$\,yr$^{-1}$, corresponding to a v$_{tan}$ between 15 and 
158~km\,s$^{-1}$ (at a distance of 130~pc). 

\item We identify at least 17 outflows with their candidate driving sources.  
All the Class~0/I sources inside the Coronet, and the majority of Class~I 
sources in the CrA region drive H$_2$ flows.  Most of the known HH objects, 
and all the detected H$_2$ objects, are visible in our IRAC maps.  Outflow 
morphologies in the 3.6 and 4.5~$\mu$m bands resemble those observed in our 
H$_2$ images, mostly delineating relatively high-excitation regions, whereas 
the 5.8 and 8.0~$\mu$m bands mostly delineate wings and wakes of the bow 
shocks and, in general, milder shock regions.

\item There is clear evidence for a parsec-scale precessing outflow, E-W 
oriented, and originating in the SMA\,2 region: IRS\,7A or SMA\,2 are likely 
the driving sources.  Furthermore we likely identify the HH\,101 lobe 
counterpart (MHO\,2014), which is likely driven by IRS\,5 or IRS\,5N.  By 
means of {\it Spitzer} mapping, we identify new flows in those regions not 
covered by our H$_2$ maps: {\it Spitzer} outflows 1, and 2, which are likely 
part of the parsec-scale precessing outflow, E-W oriented, and the IRAS\,32 
outflow, which extends $\sim$0.8 pc and appears to precess as well.

\end{itemize}


%% Included in this acknowledgments section are examples of the
%% AASTeX hypertext markup commands. Use \url without the optional [HREF]
%% argument when you want to print the url directly in the text. Otherwise,
%% use either \url or \anchor, with the HREF as the first argument and the
%% text to be printed in the second.

\acknowledgments

The authors would like to thank Karl Stapelfeldt, Eli Bressert, and the 
anonymous referee for their helpful comments.  In addition, we thank Amanda 
Heiderman for very helpful discussions of the overall cloud properties and 
comparison with her work.  Support for this work, part of the \textit{Spitzer} 
Legacy Science Program, was provided by NASA through contract numbers 1288820 
and 1298236 issued by the Jet Propulsion Laboratory, California Institute of 
Technology, under NASA contract 1407.  ACG was supported by the Science 
Foundation of Ireland, grant 07/RFP/PHYF790.  The research at Centre for Star 
and Planet Formation (co-author JKJ) is funded by the Danish National Research 
Foundation and the University of Copenhagen's programme of excellence.

Partially based on observations collected at the European Southern Observatory 
(Paranal and La Silla, Chile, 079.C-00020(A), 075.C-0561(A), 071.B-0274(A), 
065.L-0637(A),63.I-0031(A)).  This research has also made use of NASA's 
Astrophysics Data System Bibliographic Services and the SIMBAD database, 
operated at the CDS, Strasbourg, France, and the 2MASS data, obtained as part 
of the Two Micron All Sky Survey, a joint project of the University of 
Massachusetts and the Infrared Processing and Analysis Center/California 
Institute of Technology, funded by the National Aeronautics and Space 
Administration and the National Science Foundation.


%% in parentheses, after the keyword, but they will not be verified.

{\it Facilities:} \facility{Spitzer (IRAC)}, \facility{Spitzer (MIPS)}.



\appendix

\section{Appendix A \label{sec:appA}}

\subsection{Notes on individual \textit{Spitzer}-identified YSO candidates}

In many cases, e.g. CrA-7, the ISOCAM source listed in 
\citet[][hereafter O99]{olof99} corresponds to one of our {\it Spitzer} YSO 
candidates, but is not exactly coincident with the position of the 
{\it Spitzer} detection.  However, each case was checked and it was found that 
the ISOCAM sources listed as a match in this paper are within the pointing 
errors for both telescopes (and in no case is there another infrared source 
close enough to deem it questionable).

\subsubsection{CrA-1}

CrA-1 is too far afield to have been observed before; it is not included in 
the \citet{fp07} Chandra study.  CrA-1 is located in the western part of the 
CrA molecular cloud called ``the streamer.''  It is not detected at 70 
$\mu$m, but is bright at 24 $\mu$m; we classify it as a Class II, and it is a 
new YSO candidate.

\subsubsection{CrA-2 (Leda Galaxy 90315) \label{sec:leda90315}}

CrA-2 was originally classified as a Class I YSO candidate from the 
{\it Spitzer} plus 2MASS photometry.  However, by means of visual inspection, 
it proved to be a galaxy, and a quick search in SIMBAD found it to be Leda 
galaxy 90315.  Interestingly, as was described in Section 
\ref{sec:2mass_mips_ysos}, another candidate YSO, CrA-50, is also included 
within the area which constitutes this extended source.  Therefore, both CrA-2 
and CrA-50 are clearly misclassified as YSO candidates and have been removed 
from our sample. 

\subsubsection{CrA-3 (ISO-CrA\,55)}

CrA-3 was detected with ISOCAM by O99, and is referred to as ISO-CrA\,55 in 
their survey.  From the ISOCAM photometry, it was found to have a mid-infrared 
excess, and therefore is listed as one of their YSO candidates.

Like CrA-1, it is located in ``the streamer'' so it has not been observed with 
{\it Chandra}.  It is not detected at 70 $\mu$m, but is quite bright at 24 
$\mu$m, and, according to {\it Spitzer} colors, has a Flat spectrum SED.

\subsubsection{CrA-4 (DENIS-P\,J185950.9-370632, ISO-CrA\,63)}

CrA-4 is an association member, a young binary brown dwarf in CrA 
\citep{bouy04}.  Known as DENIS-P\,J185950.9-370632, it was discovered to be a 
binary by \citet{bouy04} using high resolution Hubble Space Telescope (HST) 
observations.  The source was elongated in the WFPC2, ACS and STIS images and 
the companion comes up clearly on the WFPC2 and ACS images after the primary 
PSF was subtracted.  From the images, they found it to be a common proper 
motion pair with a separation of 0$^{\prime\prime}$.060.  The magnitude 
difference between the two components indicates a mass ratio of $\sim$75\%.

High and low resolution optical spectra of CrA-4 were also obtained, and the 
spectral features show that it is young, and likely an association member 
\citep{bouy04}.  The high resolution spectrum shows H$\alpha$ in emission, and 
with a strength that makes it a likely accretor; there is also weak 
\ion{Li}{1} absorption, which indicates a young age.

CrA-4 was detected with ISOCAM by O99, and is referred to as ISO-CrA\,63 in 
their survey.  It is not listed as a candidate YSO by O99 (i.e. it does not 
exhibit a mid-infrared color excess), however \citet{bouy04} analyzed the 
archival images and found there to be a mid-infrared excess in the 
5.0$-$8.5~$\mu$m band.  This is consistent with our {\it Spitzer} 
observations, which detect this source in all IRAC bands and at 24~$\mu$m, 
classifying it as a Class~II YSO candidate based on the shape of its SED. 

\subsubsection{CrA-5 (ISO-CrA\,76)}

This source was detected with ISOCAM by O99, and is referred to as ISO-CrA\,76 
in their survey.  From the ISOCAM photometry, it was found to have a 
mid-infrared excess, and therefore is listed as one of their YSO candidates.

CrA-5 is classified as a Class~I YSO candidate by our method and is outside 
the {\it Chandra} field of view.  Like CrA-1 and CrA-3, it is not detected at 
70~$\mu$m, but is quite bright at 24~$\mu$m.  Along with CrA-6 (Class~II) and 
CrA-7 (Class~III), these three YSO candidates surround ``Spitzer Outflow 1'' 
which is discussed in more detail in \S~\ref{Spitzer_flows:sec} (see 
Figure~\ref{spitzer1:fig}).

\subsubsection{CrA-6 (CrAPMS\,8, GP\,g2, ISO-CrA\,88)}

First observed in the infrared by \citet{gp75}, CrA-6 is listed as source 
[GP]\,g2 in their Table 1.  It was also detected in X-rays by EINSTEIN 
\citep[CrAPMS\,8;][]{walter97} who also obtained spectra and classified it as 
M3V.  \citet{patten98} classifies it as a likely association member: he finds 
a spectral type of M5, and detects H$\alpha$ in its spectrum, a ROSAT X-ray 
counterpart, and a $VRI$ color excess.

Additionally, CrA-6 was detected with ISOCAM by O99, and is referred to as 
ISO-CrA\,88 in their survey.  From the ISOCAM photometry, it was found to have 
a mid-infrared excess, and therefore is listed as one of their YSO 
candidates.  Most recently, it was one of the sources included in a kinematic 
study of the extended R\,CrA association \citep[referred to as (GP75) 
R\,CrA\,g2 in their Table 13;][]{fern08}, and found to be a member.  All of 
this is consistent with our classification of CrA-6 as a Class~II YSO 
candidate from its {\it Spitzer} photometry; we detect this source in all IRAC 
and MIPS bands out to and including 70 $\mu$m.  Finally, CrA-6 was recently 
found to have a companion \citep{kohler08}.

\subsubsection{CrA-7 (ISO-CrA 93)}

CrA-7 was detected with ISOCAM by O99, and is referred to as ISO-CrA 93 in 
their survey.  From the ISOCAM photometry, it was not found to have a 
mid-infrared excess.  However, we classify CrA-7 as a Class~III YSO candidate 
based on its {\it Spitzer} colors, and so it is a new YSO candidate.  Bright 
24 $\mu$m emission is detected, but nothing is detected at 70 $\mu$m.

\subsubsection{CrA-8 (CrA-444)}

YSO candidate CrA-8 was previously identified as a candidate low-mass member 
of CrA \citep[][their source CrA-444]{lopezmarti05}  as part of their optical 
study of the very low mass population of CrA.  Using a method outlined in 
\citet{lopezmarti04}, CrA-8 was classified from its optical spectrum as a 
brown dwarf candidate with spectral type of M8.5.  We classify CrA-8 as a 
Class~II YSO candidate from the shape of its SED, detecting it out to, and 
including, 24 $\mu$m.

\citet{lopezmarti05} postulate that CrA-444 could have a very close companion, 
which they call CrA-444b.  This source, approximately 2$^{\prime\prime}$.3 
away, is only visible in their deep $I$-band exposure.  We do not see any 
conclusive evidence for CrA-444b in any of our {\it Spitzer} images, due to 
the lower spatial resolution. 

\subsubsection{CrA-9 \label{sec:cra9}}

CrA-9, located to the northwest of the Coronet region and TY\,CrA, has not 
been previously observed, thus it is a new YSO candidate, classified as a 
Class~II.  The source is located very close to the edge of the 70~$\mu$m map, 
so there is no catalog entry in this band for it.  However, the MIPS 70~$\mu$m 
image clearly shows emission, and we estimate the flux using aperture 
photometry, finding a 70 $\mu$m flux of 154.2 $\pm$ 19.8 mJy (using a 2 pixel 
aperture radius).

\subsubsection{CrA-10 (CrA-432)}

\citet{lopezmarti05} observed this source (referred to as CrA-432 in their 
Table~3) and found a photometric spectral type classification of M7 (1 
subclass error on the spectral type).  Located near CrA-9, we classify CrA-10 
as a Class II YSO candidate from the shape of its SED; it is detected in all 
bands out to, and including, 70~$\mu$m.

CrA-10 was also observed by \citet[][also called CrA-432]{s-a08} with the 
{\it Spitzer} Infrared Spectrograph (IRS); the silicate feature at 8$-$13 
$\mu$m is not detected in this source and there is only a marginal detection 
of the gas lines in the 13$-$14 $\mu$m IRS band.

Another identifier for this source is DENIS-P\,J190059.7-364711.

\subsubsection{CrA-11 (HD\,176269, HR\,7169) \label{sec:bstars}}

CrA-11, more commonly known as HD\,176269, is an optically detected pre-main 
sequence star with a spectral type of B8V that was first observed by 
\citet{knacke73} as HR\,7169 to be a possible member of the young stellar 
association near R\,CrA ($\sim$12$^{\prime}$ to the SW).  Another B8V star 
$\sim$12$^{\prime\prime}$ away, HR\,7170 (HD\,176270), was also listed as a 
possible member and we include it in 
Table~\ref{tab:knownysos_spitzercounterpart}.  These two sources were 
detected, but not spatially resolved, by the IRAS 
\citep[IRAS\,18;][]{wilking92} and EINSTEIN satellites 
\citep[CrAPMS\,10;][]{walter97}. 

HD\,176269 was observed in the near-infrared by \citet{gp75} and is referred 
to in their Table 1 as source $l$; HD\,176270 was also observed, and listed as 
source $k$.  Later, \citet{patten98} does not classify the HD 176269/HD 
176270 pair (which he calls source R05) as a likely association member: he 
finds a spectral type of B9V for the source and detects a ROSAT X-ray 
counterpart, however the source does not fulfill enough criteria for him to 
consider it an association member.  Note that these stars are far enough 
afield so that they were not included as part of the \citet{fp07} 
{\it Chandra} study.

HD\,176269 was also detected with ISOCAM by O99, and is referred to as 
ISO-CrA\,110 in their survey.  ISO-CrA\,110 was found to have a mid-infrared 
excess based on those observations so it is considered one of their YSO 
candidates.  HD\,176270 is referred to as ISO-CrA\,111 in the same study, 
and it was not found to have a mid-infrared excess.

These studies are consistent with the {\it Spitzer} observations; we 
classify HD\,176269 as a Class~III based on the shape of its SED.  Although we 
do not classify HD\,176270 as a YSO candidate, it very likely is also a 
Class~III based on its colors and SED shape (see 
Table~\ref{tab:knownysos_spitzercounterpart}).  The SEDs of the two stars look 
similar out to 8~$\mu$m.  However, at 24~$\mu$m, HD\,176269 is much brighter 
than HD\,176270.  Neither source is detected at 70~$\mu$m.

\subsubsection{CrA-12 (V667\,CrA, CrA-4110)}

CrA-12 \citep[or V667\,CrA;][]{kkkp72} is a known variable star, detected with 
{\it Chandra} \citep[J190116.26-365628.4;][]{fp07} and with ISOCAM 
(ISO-CrA\,123; O99).  CrA-12 was not found to have a mid-infrared excess based 
on the ISOCAM observations.  However, based on our {\it Spitzer} 
observations, it is classified as a Class~II YSO candidate.  It does not have 
a 70~$\mu$m detection, but is quite bright at 24~$\mu$m.

\citet{lopezmarti05} observed this source (referred to as CrA-4110 in their 
Table~3) and found a photometric spectral type classification of M5 (error of 
2 subclasses on the spectral type).  The same spectral classification was also 
obtained by \citet{s-a08}, who classify it as a weak-line T Tauri star (wTTs), 
due to its small H$\alpha$ equivalent width (EW) and the fact that it has no 
other accretion indicators (see their Table~5).  A {\it Spitzer} IRS spectrum 
was obtained for CrA-12 but only a short-low (5$-$14 $\mu$m) spectrum was 
extracted; the silicate feature from 8$-$13 $\mu$m was not detected in this 
source \citep{s-a08}.

Another identifier for this source is DENIS-P\,J190116.3-365628.

\subsubsection{CrA-13 (CrA-466) \label{sec:cra13}}

CrA-13 is likely ISO-CrA\,127 (O99), which is slightly offset 
($\sim$11$^{\prime\prime}$) from the CrA-13 {\it Spitzer} coordinates.   
ISO-CrA\,127 does fall within the extent of the very bright 24~$\mu$m source, 
though, so it is highly likely this is the same source.  We detect CrA-13 in 
all IRAC bands and in both the 24 and 70 $\mu$m MIPS bands, and classify it as 
a Class~II from its SED.  This source was also detected with {\it Chandra} 
\citep[J190118.90-365828.4;][]{fp07}.

\citet{lopezmarti05} observed this source (referred to as CrA-466 in their 
Table~3) and found a photometric spectral type classification of M4.5 (error 
of 2 subclasses on the spectral type).  However, \citet{s-a08}, who also 
obtained an optical spectrum, classified it as an M2 based on the strength of 
its spectral features.  Moreover they identify it as a classical T Tauri star 
(cTTs) because it has strong H$\alpha$ emission (see their Table~5; it is also 
referred to as CrA-466 in their paper).  They also classify CrA-13 as a 
transition object (TO) due to its lack of infrared excess at wavelengths 
shorter than 6 $\mu$m.  However, \citet{ecr09} run several models to fit the 
SED of the source using the spectral type of M2 given by \citet{s-a08}, and 
find that its SED is well fit by an untruncated disk model, and is likely not 
a TO.  A {\it Spitzer} IRS spectrum was obtained for CrA-13 and shows silicate 
emission in both the 8$-$13 $\mu$m and 20$-$30 $\mu$m regions \citep{s-a08}.

Another identifier for this source is DENIS-P\,J190118.9-365828.

\subsubsection{CrA-14 (HH~101\,IRS 1, G-94)}

CrA-14 is an infrared source first observed by \citet{rw83} to be associated 
with HH~101, and they call it IRS 1.  HH~101 was observed again, in the 
optical, by \citet{hl85} who called it HH~101\,7.

This source was also detected with ISOCAM by O99, and is referred 
to as ISO-CrA\,137 in their survey.  ISO-CrA\,137 was not found to have a 
mid-infrared excess based on those observations.  This is consistent with our 
{\it Spitzer} observations which classify this source as a Class~III YSO 
candidate based on its SED; it is faint at 24~$\mu$m and not detected at 
70~$\mu$m.  This source was also detected with {\it Chandra} 
\citep[J190129.01-370148.8;][]{fp07}, making it a likely YSO candidate.  It 
was observed in multiple epochs with {\it Chandra} and was found to have some 
variable X-ray emission as well as enhanced activity in one of the epochs 
\citep{f07}.

CrA-14 was also observed by \citet{s-a08}, which they call G-94, who 
obtained an optical spectrum and classified it as M3.5 based on its spectral 
features.  They classify it as a wTTs due to its small H$\alpha$ EW and the 
fact that it has no other accretion indicators (see their Table~5).  A 
{\it Spitzer} IRS spectrum was obtained for G-94, but no silicate emission 
was seen \citep{s-a08}.

Another identifier for this source is DENIS-P\,J190129.0-370148.

\subsubsection{CrA-15 (IRS\,14) \label{sec:cra15}}

CrA-15 was first observed by \citet[][hereafter TS84]{ts84}, and called 
IRS\,14 (TS\,2.9), with the 3.9-m Anglo-Australian Telescope, using its 
infrared photometer-spectrometer system.  This same source was observed with 
increasingly better arrays in the near-infrared by \citet{wts86}, and later 
\citet[source 185809.8-370224;][]{wilking97}.  However, each time, IRS\,14 was 
classified as a field star due to its lack of a near-infrared excess.  Based 
on {\it Spitzer} photometry, we classify IRS\,14 as a Class~II YSO candidate; 
it is not detected at 70~$\mu$m but its detection at 24~$\mu$m gives it a 
mid-infrared excess consistent with a Class~II SED.  O99 saw this 
mid-infrared excess as well (their source ISO-CrA\,139, see their Table~1) and 
therefore also considered it a YSO candidate.  IRS\,14 was detected with 
{\it Chandra} \citep[J190132.34-365803.1;][]{fp07}.

IRS\,14 was also observed by \citet{s-a08}, which they call G-87, who obtained 
an optical spectrum and classified it as M3$-$M4 based on the strength of 
its spectral features compared with other, known M stars.  They classify it 
as a wTTs due to its small H$\alpha$ EW and the fact that it has no other 
accretion indicators (see their Table~5).  They also classify IRS\,14 as a 
TO due to its lack of infrared excess at wavelengths shorter than 6~$\mu$m.  
However, as was discussed in Section~\ref{sec:cra13} for CrA-13, \citet{ecr09} 
found that the SED for IRS\,14 was also well fit by an untruncated disk model, 
and is likely not a TO.  A {\it Spitzer} IRS spectrum was obtained for IRS\,14 
and shows silicate emission in the 8$-$13~$\mu$m region as well as marginal 
detections of gas lines in the 13$-$14~$\mu$m region \citep{s-a08}.

Another identifier for this source is DENIS-P\,J190132.2-365803.

\subsubsection{CrA-16 (IRS\,13)}

CrA-16 was also observed by TS84, and is known as IRS\,13 (TS\,2.8).  Like 
IRS\,14, this source was observed with increasingly better arrays in the 
near-infrared by \citet{wts86}, and later 
\citet[source 185811.4-370206;][]{wilking97}.  Unlike IRS\,14, though, a 
near-infrared excess was seen for IRS\,13 and so it was considered a cTTs and 
association member.  IRS\,13 was also detected with {\it Chandra} 
\citep[J190133.84-365745.0;][]{fp07}.

IRS\,13 and IRS\,14 are about 30$^{\prime\prime}$ apart, and CrA-17 (discussed 
in the next section) is just a little bit further away.  Based on 
{\it Spitzer} photometry, we classify IRS\,13 as a Class~II YSO candidate; it 
is not detected at 70~$\mu$m but its detection at 24~$\mu$m gives it a 
mid-infrared excess consistent with a Class~II SED.  An optical spectrum was 
obtained by \citet{s-a08}, who refer to IRS\,13 as G-85, and classify it as 
M2$-$M3 based on the strength of its spectral features compared with other, 
known M stars; it also has strong H$\alpha$ emission, confirming it as a cTTs 
(see their Table~5).  A {\it Spitzer} IRS spectrum was obtained, and as for 
CrA-13, shows silicate emission in both the 8$-$13~$\mu$m and 20$-$30~$\mu$m 
regions \citep{s-a08}.

\subsubsection{CrA-17 \label{sec:cra17}}

CrA-17 was first observed in the near-infrared by 
\citet[185813.8-370225;][]{wilking97}.  Based on the {\it Spitzer} catalog, we 
classify CrA-17 as a Class~II YSO candidate.  The catalog lists a 24~$\mu$m 
flux for CrA-17, but that flux was obtained by band-filling, and after 
examining the images, it appears that the detection may not be real.  In 
addition, although CrA-17 falls within the region covered by the \citet{fp07} 
{\it Chandra} study, it is not detected in X-rays.  Due to these reasons, we 
have removed CrA-17 from our list of candidate YSOs.  

\subsubsection{CrA-18 (G-65)}

CrA-18 is located just north of TY\,CrA (see Figure \ref{MHO2014:fig}), and 
was first observed in the near-infrared by 
\citet[source 185818.2-365603;][]{wilking97}.  Based on {\it Spitzer} 
photometry we classify CrA-18 as a Class~II YSO candidate; it is not detected 
at 70~$\mu$m but its detection at 24~$\mu$m gives it a mid-infrared excess 
consistent with a Class~II SED.  This source was also detected with 
{\it Chandra} \citep[J190140.40-365142.4;][]{fp07}.

An optical spectrum was obtained by \citet{s-a08}, who refer to CrA-18 as 
G-65, and classify the source as M1$-$M2 based on the strength of its spectral 
features compared with other, known M stars; it also has strong H$\alpha$ 
emission, confirming it as a cTTs (see their Table~5).  Also, as was discussed 
in Sections~\ref{sec:cra13} and \ref{sec:cra15} for CrA-13 and IRS\,14, 
\citet{ecr09} found that the SED for this source was also well fit by an 
untruncated disk model, and is likely not a TO.  A {\it Spitzer} IRS spectrum 
was obtained for CrA-18, but no silicate emission was seen \citep{s-a08}.

\subsubsection{CrA-19 (IRS\,5, MMS\,12,\,SMM 4) \label{sec:irs5}}

Best known as IRS\,5, this was one of the first infrared sources to be 
associated with CrA \citep[TS\,2.4;][]{ts84}.  IRS\,5 was first seen to be a 
binary by \citet{cg93}, and later confirmed by \citet{nisini05}, who found a 
separation of $\sim$78 AU between the two components (at a distance of 130 
pc).  \citet{nisini05} also found a spectral type for IRS\,5A of K5$-$K7V 
(IRS\,5B was too faint to classify).  Additionally, IRS\,5 was detected with 
the VLA at 6~cm \citep[called VLA\,7;][]{brown87}, and shows significant 
short- and long-term variability in the radio \citep{suters96}.  
\citet{feigelson98} detected circularly polarized continuum emission at 
centimeter wavelengths in IRS\,5, demonstrating that radio emission from 
protostars can sometimes arise from non-thermal processes.  IRS\,5 is also a 
millimeter source, MMS\,12 \citep{chini03}, and was observed at 450 and 
850~$\mu$m by \citet[][referred to in their survey as SMM\,4]{nutter05}, seen 
as an extended source in their maps.  Using photometry available in the 
literature at the time, \citet{chen97} use a distance of 130~pc to estimate 
T$_{bol}$ = 403~K and L$_{bol}$ = 0.9~L$_{\odot}$ for IRS\,5 (which 
they refer to as TS\,2.4 in their Table~2).

The {\it Chandra} X-ray emission for the binary is only marginally resolved, 
so IRS\,5A/B is listed as a single source in the catalog of 
\citet[J190148.02-365722.4;][]{fp07}.  However, a later study by 
\citet{hamaguchi08} used a subpixel repositioning technique in order to 
resolve the X-ray emission from multiple {\it Chandra} observations.  They 
found that IRS\,5A is flaring, while IRS\,5B is quiescent.  IRS\,5A/B is also 
resolved by \citet{chlt08}, called CHLT\,3 (IRS\,5A) and CHLT\,4 (IRS\,5B), in 
their centimeter imaging survey of CrA.  

With {\it Spitzer}, we classify IRS\,5 as a Class~I YSO candidate; it is 
detected in all bands out to, and including, 24~$\mu$m.  We do not resolve the 
binary.  In addition, IRS\,5 is one of the sources which we observed with the 
SMA (see Section~\ref{sec:sma_analysis}), and surprisingly, there is no clear 
detection, although IRS\,5N does show strong detection (see 
Figure~\ref{fig:sma_irs5} and the following, \S~\ref{sec:irs5n}).

\subsubsection{CrA-20 (IRS\,5N) \label{sec:irs5n}}

IRS\,5N was first seen in the radio \citep[][listed as source 5]{fpm06}, and 
then later in X-rays with {\it Chandra}, and referred to as IRS\,5N 
\citep[J190148.46-365714.5;][]{fp07,f07}.  Based on {\it Spitzer} photometry, 
we classify IRS\,5N as a Class~I YSO candidate; it is detected in all bands 
out to, and including, 24~$\mu$m.  IRS\,5N was one of the sources we observed 
with the SMA, and it has a strong detection (see \S~\ref{sec:sma_analysis} and 
Figure~\ref{fig:sma_irs5}).  We compute a disk mass for IRS\,5N of 
0.023M$_{\odot}$ (see Table~\ref{tab-sma}).

\subsubsection{CrA-21 (IRS\,8)}

CrA-21 was also observed by TS84, and is known as IRS\,8 (TS\,2.2).  Like 
IRS\,13 and IRS\,14, this source was observed with increasingly better arrays 
in the near-infrared by \citet{wts86}, and later 
\citet[source 185828.8-365834;][]{wilking97}.  Unlike IRS\,14, though, a 
near-infrared excess was seen for IRS\,8 and so it was considered an 
association member.  We classify CrA-21 as a Class~II YSO candidate; it is not 
detected at 70~$\mu$m but its detection at 24~$\mu$m gives it a mid-infrared 
excess consistent with a Class~II SED.

\citet{mw09} also classify this source as a likely association member from 
H- and K-band spectra, finding it to be consistent with a G5 star.  
Additionally, it was detected with {\it Chandra} 
\citep[J190151.11-365412.5;][]{fp07}.  

\subsubsection{CrA-22}

CrA-22, which is located directly south of the Coronet, is too far afield to 
have been observed before, nor is it included in the \citet{fp07} 
{\it Chandra} study.  Another source falls very close to CrA-22 in the IRAC 
bands, however the catalog flux detections of CrA-22 are good, and do not 
appear to contaminate the photometry.  CrA-22 has a MIPS detection at 
24~$\mu$m, but is not detected at 70~$\mu$m.  We classify CrA-22 as a 
Class~II, and it is a new YSO candidate.

\subsubsection{CrA-23 (Star\,A)}

First detected by \citet{graham93}, this source, referred to as ``Star\,A'' 
(which was not detected by IRAS), was found to be very bright in H$\alpha$, 
weaker in $R-$band, and very weak, only just visible in the original 
\ion{S}{2} image.  They recognized it as a likely a strong emission-line star, 
but it was too faint to obtain a spectrum.

Later, Star\,A was also detected in the infrared $J, H, K^{\prime}$-bands by 
\citet[][source 185831.1-370456]{wilking97}, who classified it as a brown 
dwarf candidate.  \citet{lopezmarti05} found a photometric spectral 
type classification of M8.5 (1 subclass error on the spectral type) for 
Star\,A (they call it CrA-465).  \citet{s-a08} obtained an optical spectrum 
(also referred to in their paper as CrA-465), classifying it as M7.5 based on 
the strength of its spectral features; it also has strong H$\alpha$ emission 
which is double peaked, confirming it as a cTTs (see their Table~5). With 
{\it Spitzer}, we detect Star\,A in all IRAC bands and at 24~$\mu$m, but not 
at 70~$\mu$m.  We classify it as a Class~II YSO candidate based on its SED.

Another identifier for Star A is DENIS-P J190153.7-370033.

\subsubsection{CrA-24 \label{cra24}}

We classify CrA-24 as a Class~I YSO candidate based on its {\it Spitzer} SED.  
It is definitely detected up through 8~$\mu$m with a rising SED, however it is 
difficult to tell whether or not the 24~$\mu$m MIPS detection is an actual 
detection due to contamination from the nearby R\,CrA nebula.  It is not 
detected in our SMA data (see Figure \ref{fig:sma_irs5}), but it is 
observed by {\it Chandra} \citep[J190155.61-365651.1;][]{fp07}.  Due to its 
position, CrA-24 could be a good candidate driving source for HH\,99 
(MHO\,2000~A; see Figure~\ref{NE_flows:fig}).  Thus we include it as a new YSO 
candidate. 

\subsubsection{CrA-25 \label{sec:cra25}}

CrA-25 is classified as a Class~I YSO candidate based on its {\it Spitzer} 
SED, but that SED is based on almost all band-filled fluxes.  Looking at 
CrA-25, it is clear that it is an extended source, a knot that is associated 
with an HH object, located within 3$^{\prime\prime}$ of HH732A \citep{wang}.  
Therefore CrA-25 has been removed from our list of candidate YSOs.

\subsubsection{CrA-26 (WMB\,185844.2-370304)}

CrA-26 was detected by \citet{wilking97} in the infrared $J, H, 
K^{\prime}$-bands (named 185844.2-370304).  It is classified with 
{\it Spitzer} as a Class~II YSO candidate and is well detected through 
8~$\mu$m.  There is a flux listed at 24~$\mu$m, but it is suspect, due to the 
fact that it falls within a diffraction spike; the 24~$\mu$m flux has been 
band-filled.  After inspecting the SED (see Figure~\ref{fig:classII}), we 
determine that it looks like a YSO candidate out to, and including 8~$\mu$m.  
Therefore we retain CrA-26 as a candidate Class~II YSO.

\subsubsection{CrA-27 (WMB\,185848.1-365808)}

CrA-27 was detected by \citet{wilking97} in the infrared $J, H, 
K^{\prime}$-bands (named 185848.1-365808).  Also detected by O99 
(ISO-CrA\,142), it did not exhibit a mid-infrared color excess and was not 
listed as a candidate YSO.  However, \citet{mw09} obtained an infrared 
spectrum and determined it to be a likely association member, 
``late-type'' $>$ K4.  With {\it Spitzer}, we classify CrA-27 as a Class~III 
YSO candidate, which is consistent with the other studies.

\subsubsection{CrA-28 (ISO-CrA\,143, G-14)}

CrA-28 was detected by \citet{wilking97} in the infrared $J, H, 
K^{\prime}$-bands (named 185849.3-370733).  It was also detected with ISOCAM 
by O99 (ISO-CrA\,143), was found to have a mid-infrared excess, and 
therefore is listed as one of their YSO candidates.  CrA-28 is similar to 
CrA-30 (see \S~\ref{sec:cra30}) because they both have small $H-K$ colors and 
similar SEDs (O99).  On the basis of our {\it Spitzer} data, we classify this 
source as a Class~II YSO candidate; it is detected out to, and including, 
24~$\mu$m.  CrA-28 also has a {\it Chandra} detection 
\citep[J190211.99-370309.4;][]{fp07}.

This source was also observed by \citet{s-a08}, called G-14, who obtained an 
optical spectrum and classified it as an M4.5 based on the strength of its 
spectral features.  They classify it as a wTTs because it has weak/no 
H$\alpha$ emission (see their Table~5).  They also classify it as a TO due 
to its lack of infrared excess at wavelengths shorter than 6~$\mu$m.  However, 
\citet{ecr09} run several models to fit the SED of G-14 using the spectral 
type of M4.5 given by \citet{s-a08}, and find that its SED is well fit by an 
untruncated disk model, and is likely not a TO.  A {\it Spitzer} IRS spectrum 
was obtained and it shows silicate emission in the 8$-$13~$\mu$m region 
\citep{s-a08}.

\subsubsection{CrA-29 (ISO-CrA\,145)}

CrA-29 was detected by \citet{wilking97} in the infrared $J, H, 
K^{\prime}$-bands (named 185852.0-370456).  It was also detected with ISOCAM 
by O99 (ISO-CrA\,145), was found to have a mid-infrared excess, and 
therefore is listed as one of their YSO candidates.  With {\it Spitzer}, we 
classify this source as a Class~II YSO candidate, and it is detected out 
to, and including, 24~$\mu$m.

\subsubsection{CrA-30 (H$\alpha$\,14; HBC\,680) \label{sec:cra30}}

CrA-30 was detected as an emission-line star by \citet[][H$\alpha$\,14]{mr81} 
and is also listed in the Herbig \& Bell Catalog as HBC\,680 \citep{hbc88}.  
It has also been noted as an IRAS detection, source IRAS\,18591-3702, 
\citep[e.g.][]{weintraub90,wilking92}.  Using photometry available in the 
literature at the time, \citet{chen97} use a distance of 130~pc to estimate 
T$_{bol}$ = 2604~K and L$_{bol}$ = 1.2~L$_{\odot}$ for H$\alpha$\,14.

This source was detected by \citet{wilking97} in the infrared $J, H, 
K^{\prime}$-bands (named 185904.3-370238), and by O99 (ISO-CrA\,155).  From 
the ISOCAM photometry, it was found to have a mid-infrared excess, and 
therefore is listed as one of their YSO candidates.  H$\alpha$\,14 has a 
{\it Chandra} detection \citep[J190227.05-365813.2;][]{fp07}, and in addition, 
it was found to have a companion \citep{kohler08}.  A {\it Spitzer} IRS 
spectrum was obtained, and the source, referred to as G-1, shows silicate 
emission at 8$-$13~$\mu$m \citep{s-a08}.

Most recently, H$\alpha$\,14 was one of the sources included in a kinematic 
study of the extended R\,CrA association \citep[referred to as HBC\,680 in 
their Table~13;][]{fern08}, and found to be a member; it was also observed by 
\citet{mw09} and determined from its infrared spectrum to be a M0 member of 
the association.  All of these observations are consistent with our 
{\it Spitzer} observations which show it to be well-detected out to, and 
including, 24~$\mu$m, classifying it as a Class~II YSO candidate from its 
SED.

\subsubsection{CrA-31 (H$\alpha$\,3; ISO-CrA\,159)}

CrA-31 is an emission-line star of M type \citep[][H$\alpha$\,3]{mr81}.  
\citet{chen97} use a distance of 130~pc to estimate T$_{bol}$ = 1875~K and 
L$_{bol}$ = 0.6~L$_{\odot}$ for H$\alpha$\,3.  This source was also detected 
with ISOCAM by O99 (ISO-CrA\,159), was found to have a mid-infrared excess, 
and therefore is listed as one of their YSO candidates.  In addition, it was 
detected with {\it Chandra} 
\citep[][catalog number J190233.07-365821.1]{fp07}.  Our {\it Spitzer} 
observations classify this source as a Class~II YSO candidate and it is 
detected all the way out to, and including, 70~$\mu$m.

\subsubsection{CrA-32 \label{sec:cra32}}

We originally classified CrA-32 as a Flat spectrum YSO candidate based on its 
{\it Spitzer} SED.  However, its MIPS 24~$\mu$m flux has been band-filled and 
is likely not very accurate due to the fact that CrA-32 is close to the very 
bright 24~$\mu$m source, CrA-44 (also known as IRAS\,32, see 
\S~\ref{sec:iras32}).  In the IRAC bands, there is also another source in very 
close proximity making CrA-32 look quite extended.  Due to the complicated 
nature of this source, CrA-32 has been removed from our list of candidate 
YSOs.  

\subsubsection{CrA-33 (IRAS\,32d)}

CrA-33 is classified as a Flat spectrum YSO candidate based on its 
{\it Spitzer} SED, and is well detected out to, and including, 24~$\mu$m.  It 
was first observed by \citet{wilking92} as part of the IRAS\,32 outflow, and 
is referred by to them as IRAS\,32d.  It was also detected with ISOCAM by O99, 
and is referred to as ISO-CrA\,185 in their survey.  From the ISOCAM 
photometry, it was found to have a mid-infrared excess, and therefore is 
listed as one of their YSO candidates.  Located to the southeast of the main 
IRAS\,32 outflow (see Figures~\ref{fig:sma_iras32} and \ref{irasoutflow:fig}), 
CrA-33 is too far afield to have been observed by many other surveys, e.g., it 
is not included in the \citet{fp07} {\it Chandra} study.

\subsubsection{CrA-34 (ISO-CrA\,198)}

This source was detected with ISOCAM by O99, and is referred to as 
ISO-CrA\,198 in their survey.  From the ISOCAM photometry, it was found to 
have a mid-infrared excess, and therefore is listed as one of their YSO 
candidates.

Located to the north of IRAS\,32, CrA-34 is too far afield to have been 
observed by many other surveys, e.g., it is not included in the \citet{fp07} 
{\it Chandra} study.  Based on the shape of its {\it Spitzer} + 2MASS SED, we 
classify it as a Class~III YSO candidate and it is detected in all bands up 
to, and including, 24~$\mu$m.

\subsubsection{CrA-35 (ISO-CrA\,201)}

This source was detected with ISOCAM by O99, and is referred to as
ISO-CrA\,201 in their survey.  From the ISOCAM photometry, it was found to 
have a mid-infrared excess, and therefore is listed as one of their YSO 
candidates.

Located very close to IRAS\,32, just to the southeast, CrA-35 is too far 
afield to have been observed by many other surveys, e.g., it is not included 
in the \citet{fp07} {\it Chandra} study.  Based on the shape of its 
{\it Spitzer} + 2MASS SED, we classify it as a Class~II YSO candidate and it 
is detected in all bands up to, and including, 24~$\mu$m.

\subsubsection{CrA-36}

Also located to the southeast of IRAS 32, CrA-36 is too far afield to have 
been observed by many other surveys, e.g., it is not included in the 
\citet{fp07} {\it Chandra} study.  Based on the shape of its {\it Spitzer} + 
2MASS SED, we classify it as a Class~II YSO candidate and it is detected in 
all bands up to, and including, 70~$\mu$m.  CrA-36 has not been classified as 
a YSO candidate before, therefore it is a new YSO candidate.

\subsubsection{CrA-37}

CrA-37 was possibly detected with ISOCAM by O99, referred to as
ISO-CrA\,232 in their survey.  ISO-CrA\,232 is $\sim$8$^{\prime\prime}$ away 
from CrA-37 which is close to the pointing uncertainties for both telescopes; 
there are no other infrared sources nearby so it is likely that this is the 
same source.  From the ISOCAM photometry, ISO-CrA\,232 was not found to have a 
mid-infrared excess, and therefore is not listed as one of their YSO 
candidates.

CrA-37 is too far afield to have been observed by many other surveys, e.g., it 
is not included in the \citet{fp07} {\it Chandra} study.  It is located to the 
southeast of IRAS\,32.  Based on the shape of its SED, we classify it as 
a Class~I YSO candidate and it is detected in all bands up to, and including, 
70~$\mu$m.  CrA-37 has not been classified as a YSO candidate before, 
therefore it is a new YSO candidate.

\subsubsection{CrA-38 (ISO-CrA\,13)}

This source was detected with ISOCAM by O99, and is referred to as
ISO-CrA\,13 in their survey.  From the ISOCAM photometry, it was found to have 
a mid-infrared excess, and therefore is listed as one of their YSO candidates.

CrA-38 is also too far afield to have been observed by many other surveys, 
e.g., it is not included in the \citet{fp07} {\it Chandra} study.  Like CrA-1 
and CrA-3, it is located in the western part of the CrA molecular cloud, 
``the streamer.''  Based on the shape of its {\it Spitzer} + 2MASS SED, we 
classify it as a Class~III YSO candidate and it is detected in all bands up 
to, and including, 24~$\mu$m.

\subsubsection{CrA-39 (6dFGS gJ190023.5-371224) \label{sec:cra39}}

CrA-39 was originally classified as a Class~II YSO candidate from the 
{\it Spitzer} + 2MASS photometry.  However, a closer look at the images 
reveals that it is a galaxy (indicated in SIMBAD to be an emission-line 
galaxy, 6dFGS gJ190023.5-371224).  Therefore, this source is clearly 
misclassified as a YSO candidate and has been removed from our sample.

\subsubsection{CrA-40 (VSSt\,18, ISO-CrA\,134)}

CrA-40 was first observed at 2.2~$\mu$m by \citet[][known as VSSt\,18]{vrba76} 
and again later, by \citet{wilking97} in the infrared $J, H, K^{\prime}$-bands 
(named 185802.9-370344).  CrA-40 was also detected with ISOCAM by O99, and is 
referred to as ISO-CrA\,134 in their survey.  From the ISOCAM photometry, it 
was found to have a mid-infrared excess, and therefore is listed as one of 
their YSO candidates.  Additionally, it was detected with {\it Chandra} 
\citep[J190125.75-365919.3;][]{fp07}.

Most recently CrA-40 was classified by \citet{mw09} from its infrared 
spectrum as a K3 member of the association.  All of this is consistent with 
our {\it Spitzer} observations, which detect this source in all IRAC bands and 
at 24~$\mu$m, classifying it as a Class~II YSO candidate.

\subsubsection{CrA-41 (H$\alpha$\,2; HBC\,677)}

CrA-41 is an emission-line star, first observed by 
\citet[][H$\alpha$\,2]{mr81}, and listed in the Herbig \& Bell Catalog 
as HBC\,677 \citep{hbc88}.  This infrared source is known to be a part of 
HH~100, which was observed in the optical by \citet{hl85} who called it 
HH~100\,1, and was detected by \citet{wilking97} in the infrared $J, H, 
K^{\prime}$-bands (named 185819.1-370418).  CrA-41 also has a {\it Chandra} 
detection \citep[J190141.62-365953.1;][]{fp07}.

Most recently, CrA-41 was one of the sources included in a kinematic 
study of the extended R\,CrA association \citep[referred to as HBC\,677 in 
their Table~13;][]{fern08}, and found to be a member; it was also classified 
by \citet{mw09} from its infrared spectrum as a M2 member of the association.  
All of these observations are consistent with our {\it Spitzer} observations, 
which detect this source in all IRAC bands and at 24~$\mu$m, classifying it as 
a Class~II YSO candidate based on the shape of its SED.

\subsubsection{CrA-42 (IRS\,6)}

CrA-42 is a binary source, known as IRS\,6A and 6B, first noted as a binary by 
\citet{nisini05}; we do not resolve it in our {\it Spitzer} images.  
\citet{nisini05} found a separation of 97 AU ($<$ 1$^{\prime\prime}$) between 
the two components (at a distance of 130 pc), and a spectral type for IRS\,6A 
of M2V (IRS\,6B was too faint to classify).  Positionally, our YSO candidate 
CrA-42 is closer to the coordinates given by \citet{nisini05} for IRS\,6A, but 
we will discuss both IRS\,6A and 6B here (referring to it as ``IRS\,6,'' since 
it is unresolved in most studies).  

IRS\,6 was first observed by TS84 (TS\,2.3), and then by \citet{wts86}, and 
later \citet[source 185828.2-370058;][]{wilking97}.  Using photometry 
available in the literature at the time, \citet{chen97} use a distance of 
130~pc to estimate T$_{bol}$ = 1055~K and L$_{bol}$ = 0.1~L$_{\odot}$ for 
IRS\,6 (which they refer to as TS\,2.3 in their Table~5).

Along with several Class~I sources in the Coronet cluster (including IRS\,1, 
IRS\,2, IRS\,5, IRS\,7, IRS\,9, and R\,CrA), IRS\,6 was first seen in X-rays 
by \citet{koyama96} with the ASCA satellite.  Interestingly, it was not 
detected in X-rays with ROSAT \citep{np97}, although the ROSAT band has a 
softer energy range than ASCA.  \citet{fpm06} find that IRS\,6 is detected as 
a weak X-ray source with {\it Chandra} and XMM-{\it Newton}, and it is 
referred to as IRS\,6A in Table~2 of 
\citet[source J190150.45-365638.1;][]{fp07}

\citet{wilking97} noted that IRS\,6 is associated with a reflection nebula, 
and \citet{wang} infer that it could be the driving source for a giant outflow 
including HH objects HH\,99, 730, 860, and the HH~104\,D jet (see 
Figure~\ref{MHO2004:fig}).  However, \citet{caratti06} discount IRS\,6 as the 
driving source for HH\,99 based on the tangential velocity and position angle 
of HH\,99 (see their Section~7.11 and their Figure~15).  On the basis of our 
analysis, we suggest that IRS\,6 is driving a large outflow which includes 
MHO\,2000~B, HH\,99~C, and MH0\,2004~B-D (see also Appendix~\ref{sec:appB} and 
Table~\ref{YSOs_flows:tab}).

In the radio, IRS\,6 has been seen as a very weak source at 3.6~cm and 6~cm 
\citep{fpm06,miettinen08}.  Unlike IRS\,5A/B, IRS\,6A/B was not resolved by 
\citet{chlt08} in their centimeter imaging survey of CrA, in which they refer 
to IRS\,6 as CHLT\,5.  In fact, like the previous studies, the single source 
is only weakly detected in their 3.5~cm continuum map, and flux values for 
densities at both 6.2 and 3.5~cm are not listed in their Table~3.

Most recently, IRS\,6 was classified by \citet{mw09} from its infrared 
spectrum as a M1 member of the association.  All of these observations are 
consistent with our {\it Spitzer} observations, which detect this source in 
all bands up to, and including, 24~$\mu$m, and classify it as a Class~II YSO 
candidate.

\subsubsection{CrA-43 (SMM\,2, CHLT\,15, WMB\,55)}

CrA-43 was first observed by \citet{wilking97} in the infrared $J, H, 
K^{\prime}$-bands (named 185836.1-370131).  This source corresponds to SMM\,2 
from the \citet{nutter05} SCUBA 450 and 850~$\mu$m survey of CrA.  
\citet{nutter05} also refer to the near-infrared counterpart of SMM\,2 as 
WMB\,55 because it is source number 55 in Table~1 of \citet{wilking97}.  
\citet{groppi07} mention this source within their study of SMA observations of 
the IRS\,7 region (of which SMM\,2 is nearby), however their field did not 
cover the SMM\,2 region.  Although CrA-43 is part of the field covered by 
{\it Chandra}, it is not detected \citep{fp07}.  

In \citet{haas08}, CrA-43 is identified as a candidate YSO using a millimeter 
excess technique, called A3~51, with a 1.2~mm flux of 101.8$\pm$18.4 MJy/sr.  
It is also weakly detected in 3.5~cm continuum emission by \citet[referred to 
as CHLT\,15;][]{chlt08}, but not in the 6.2~cm continuum.  A close-up of 
CrA-43 and the surrounding flows can be seen in Figures~\ref{MHO2001-2:fig} 
and \ref{PM_MHO2001_2:fig}, as well as in Figure~5 ([SII] narrow-band images) 
of \citet{wang}, clearly indicating that it is driving an outflow (see also 
Appendix~\ref{sec:appB} and Table~\ref{YSOs_flows:tab}).  

In our {\it Spitzer} images, CrA-43 is well-detected out to, and including 
24~$\mu$m, and is classified as a Class~I YSO candidate.  It is also detected 
with the SMA (see Figure \ref{fig:sma_irs5}), and we compute a disk mass for 
CrA-43 of 0.032M$_{\odot}$ (see Table~\ref{tab-sma}).  

\subsubsection{CrA-44 (IRAS\,32c) \label{sec:iras32}}

CrA-44 is classified as a Class~I YSO candidate based on its {\it Spitzer} 
SED, and is well detected out to, and including, 160~$\mu$m.  It was first 
observed by \citet{wilking92} as part of the IRAS\,32 outflow and is referred 
to by them as IRAS\,32c (IRAS\,18595-3712).  Using photometry available in the 
literature at the time, \citet{chen97} use a distance of 130~pc to estimate 
T$_{bol}$ = 148~K and L$_{bol}$ = 1.3~L$_{\odot}$ for IRAS\,32 (see 
their Table~2).  It was also detected with ISOCAM by O99, and is referred to 
as ISO-CrA\,182 in their survey.  From the ISOCAM photometry, it was found to 
have a mid-infrared excess, and therefore is listed as one of their YSO 
candidates.  IRAS\,32 has a near-infrared reflection nebula first reported by 
\citet{con07}, and named the ``Isabelle-Nebula'' by \citet{haas08}, which they 
observed in the H- and K$_s$-bands.  This reflection nebula can be easily 
seen in the infrared {\it Spitzer} bands and ISAAC/VLT narrow-band 2.12~$\mu$m 
image, and it likely delineates the outflow cavity (see 
Figure~\ref{irasoutflow:fig}).

There have been other {\it Spitzer} studies of IRAS\,32 as well.  
\citet{seale08} show that the outflow cavity has an opening angle averaging 
66$\degr$ $\pm$ 4.  Because IRAS\,32 is nearly edge-on, when correcting for 
inclination angle, \citet{seale08} find a similar average opening angle of 
65$\degr$ $\pm$ 4.  In addition, a {\it Spitzer} IRS spectrum of IRAS\,32 
shows a solid CH$_4$ feature at 7.7~$\mu$m \citep{oberg08}.

IRAS\,32 is also a millimeter source, MMS\,23 \citep{chini03}, and was 
observed at 450 and 850~$\mu$m by \citet[][referred to in their survey as 
SMM\,8]{nutter05}.   Using the Atacama Pathfinder EXperiment (APEX), 
\citet{vankempen} observed IRAS\,32 in the submillimeter lines of CO, HCO$^+$ 
and their isotopologues, finding it to be a truly embedded source with an 
exceptionally strong outflow.  This is confirmed by our SMA, H$_2$, 
and {\it Spitzer} observations (see \S~\ref{sec:sma_analysis}, 
Appendix~\ref{sec:appB}, and Figure~\ref{fig:sma_iras32}), which show a 
parsec-scale precessing outflow.  Using the SMA data, we compute a disk mass 
for IRAS\,32 of 0.024M$_{\odot}$ (see Table \ref{tab-sma}).

IRAS\,32 is too far afield to have been observed by many other surveys, e.g., 
it is not included in the \citet{fp07} {\it Chandra} study.

\subsubsection{CrA-45 (VSSt\,10, IRAS\,34)}

CrA-45 was first observed at 2.2~$\mu$m by 
\citet[][known as VSSt\,10]{vrba76}, and is an IRAS source, named IRAS\,34 by 
\citet{wilking92}.  Using photometry available in the literature at the time, 
\citet{chen97} use a distance of 130~pc to estimate T$_{bol}$ = 712~K and 
L$_{bol}$ = 0.5~L$_{\odot}$ for VSSt\,10 (see their Table~2).  

Located to the southeast of IRAS\,32 (CrA-44), CrA-45 is on the outskirts of 
the main CrA cloud core, and is the closest YSO candidate to the quiescent 
``Core C'' discussed in \citet{harju93} at about $\sim$8$^{\prime}$ to the 
northwest.  It is too far afield to have been observed by many other surveys, 
e.g., it is not included in the \citet{fp07} {\it Chandra} study.  Our 
{\it Spitzer} observations classify CrA-45 as a Class~II YSO candidate, and 
it is well observed out to, and including, 70~$\mu$m.  

\section{Appendix B \label{sec:appB}}

\subsection{Notes on individual MHOs - North and east regions, Coronet. \label{NE_MHOs:sec}}

\subsubsection{Morphology}

Figure~\ref{NE_flows:fig} shows a \textit{SofI} H$_2$ image of north-eastern 
flows detected outside and inside the Coronet.  We observe at least seven 
H$_2$ flows radially launched from the Coronet.  The well-known MHO\,2000 A
\citep[i.\,e. HH\,99 B, see e.\,g.][]{davis99,caratti06}, MHO\,2000 B (i.\,e. 
HH\,99 A), MHO\,2013 \citep[i.\,e. HH\,733, see][]{wilking97,davis99,wang}, 
and four newly detected outflows, namely MHO\,2009, 2011--2014. MHO\,2011, and 
2014 are partially coincident with HH\,733, and 732 \citep[see][]{wang}, 
respectively.  Finally, MHO\,2001 and 2002~\citep{caratti06,davis09b}, 
positioned inside the Coronet are labeled with blue crosses in 
Figure~\ref{NE_flows:fig}.

\begin{figure}
\includegraphics[width=16.0 cm]{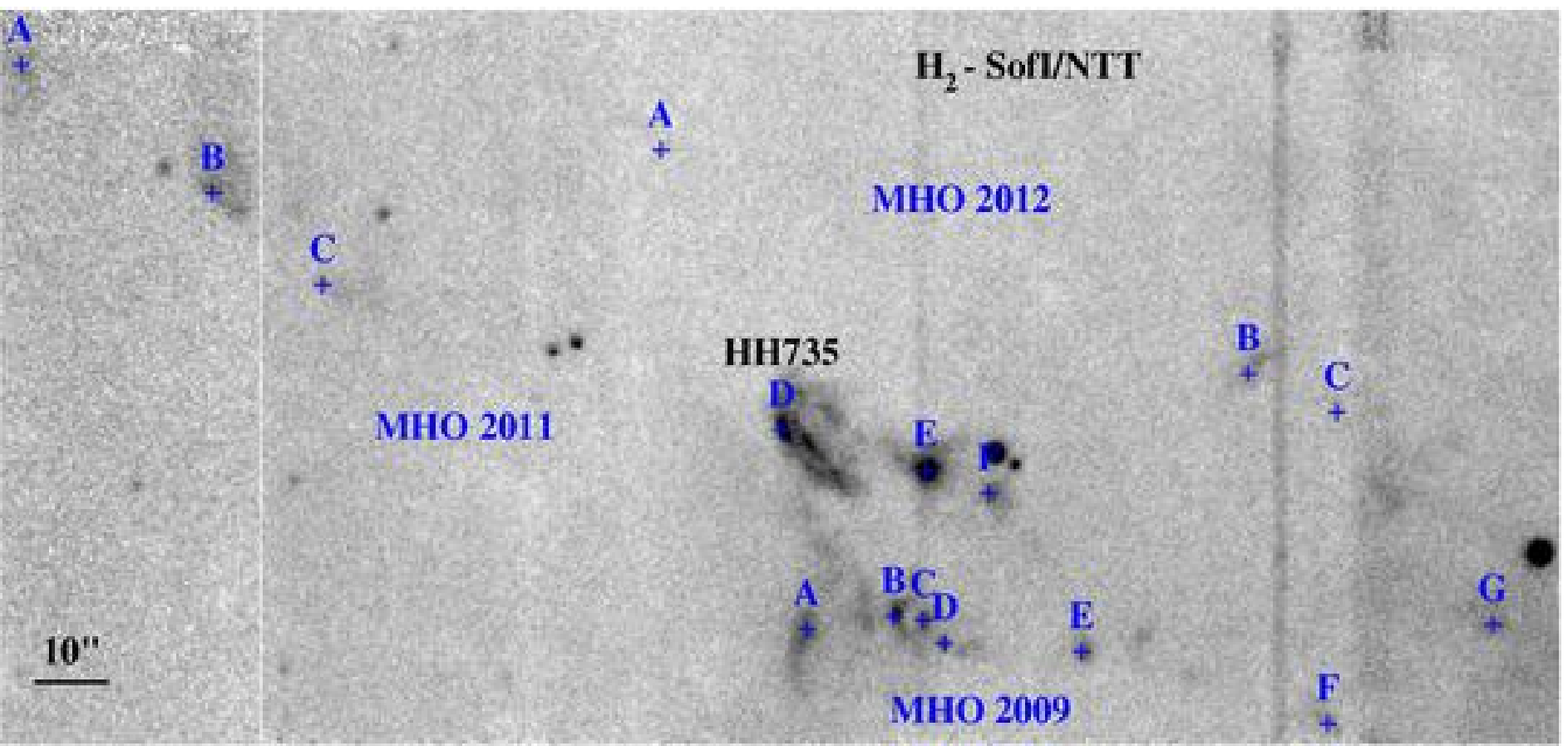}
\caption{Close up view of MHOs 2009, 2011, and 2012 at 2.12\,$\mu$m, from a 
\textit{SofI/NTT} image (see Figure~\ref{NE_flows:fig}).  Labels and crosses 
indicate names and positions of detected knots and sub-structures along the 
flows.\label{MHO2009-12:fig}}
\end{figure}

A close up view of MHO\,2009, 2011, and 2012 is given in 
Figure~\ref{MHO2009-12:fig}.  MHO\,2009 is composed of a chain of precessing 
knots (A-F), roughly elongated west-east for $\sim$1.2$\arcmin$.  About 
20$\arcsec$ northward, MHO\,2011 (knots A-G) shows a curved shape, elongated 
towards ENE, and is more extended ($\sim$ 3.6$\arcmin$). Knots D and F are 
the brightest structures.  The first is a large bow-shock, coincident with 
HH\,735, while the second has a point-like shape. 

\begin{figure}
\includegraphics[width=16.0 cm]{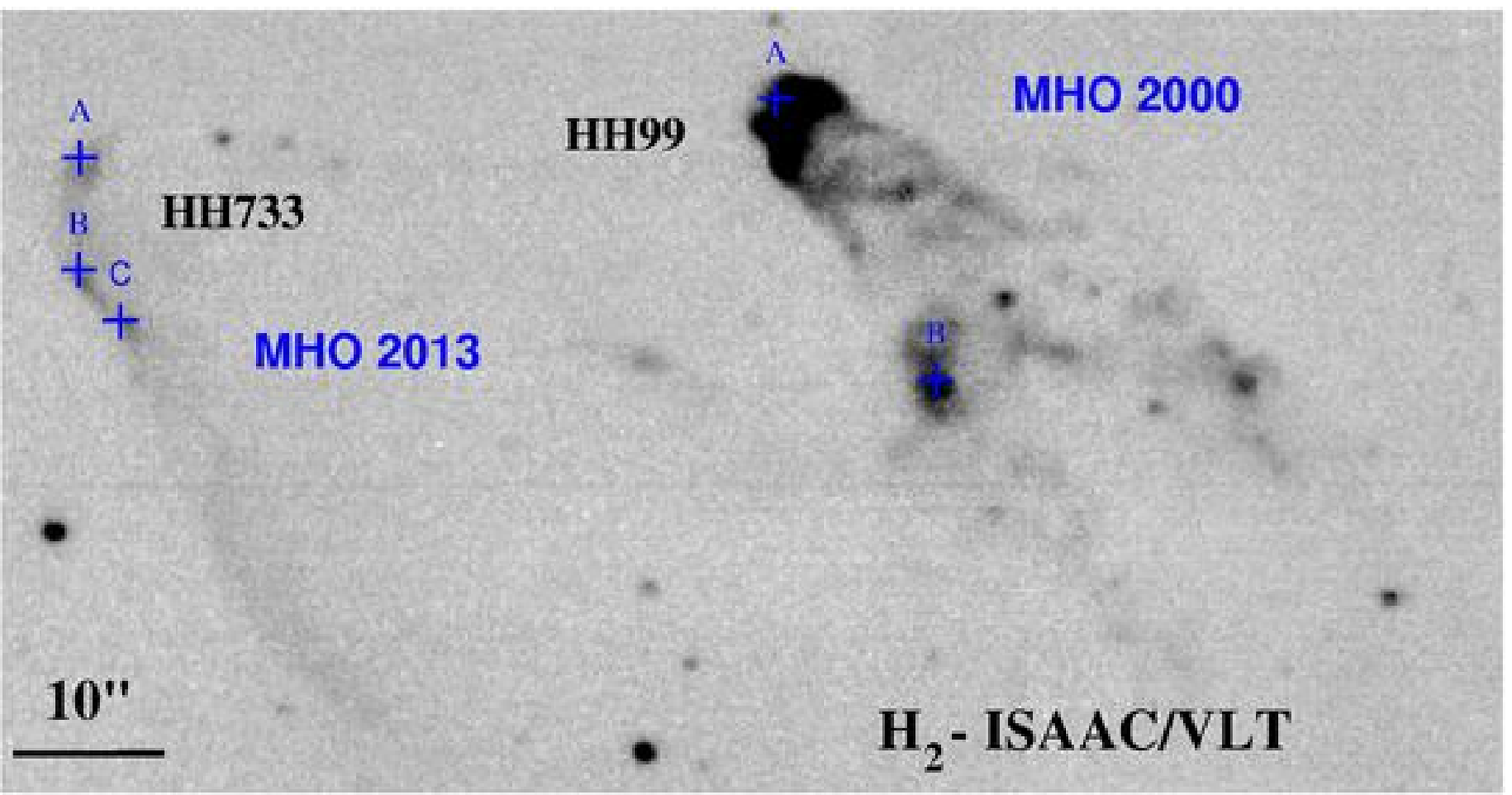}
\caption{Close up view of MHOs 2000 and 2013 at 2.12\,$\mu$m, from a 
\textit{ISAAC/VLT} image (see Figure~\ref{NE_flows:fig}).  Labels and crosses 
indicate names and positions of detected knots and sub-structures along the 
flows.\label{MHO2000-13:fig}}
\end{figure}

\begin{figure}
\includegraphics[width=16.0 cm]{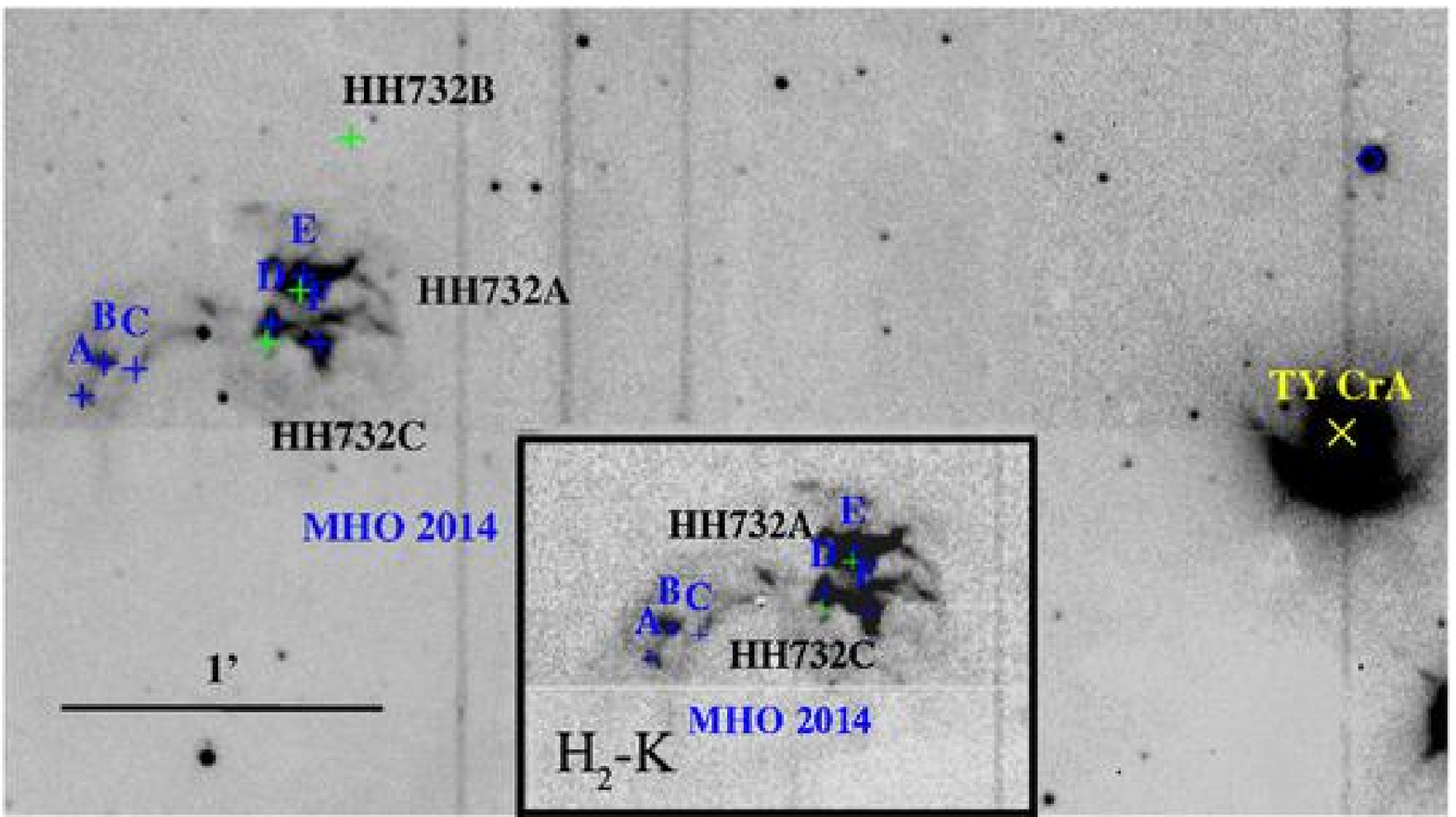}
\caption{Close up view of MHOs 2014 at 2.12\,$\mu$m, from a \textit{SofI/NTT} 
image (see Figure~\ref{NE_flows:fig}).  Labels and crosses indicate names and 
positions of detected knots and sub-structures along the flows.
\label{MHO2014:fig}}
\end{figure}

\begin{figure}
\includegraphics[width=14.0 cm]{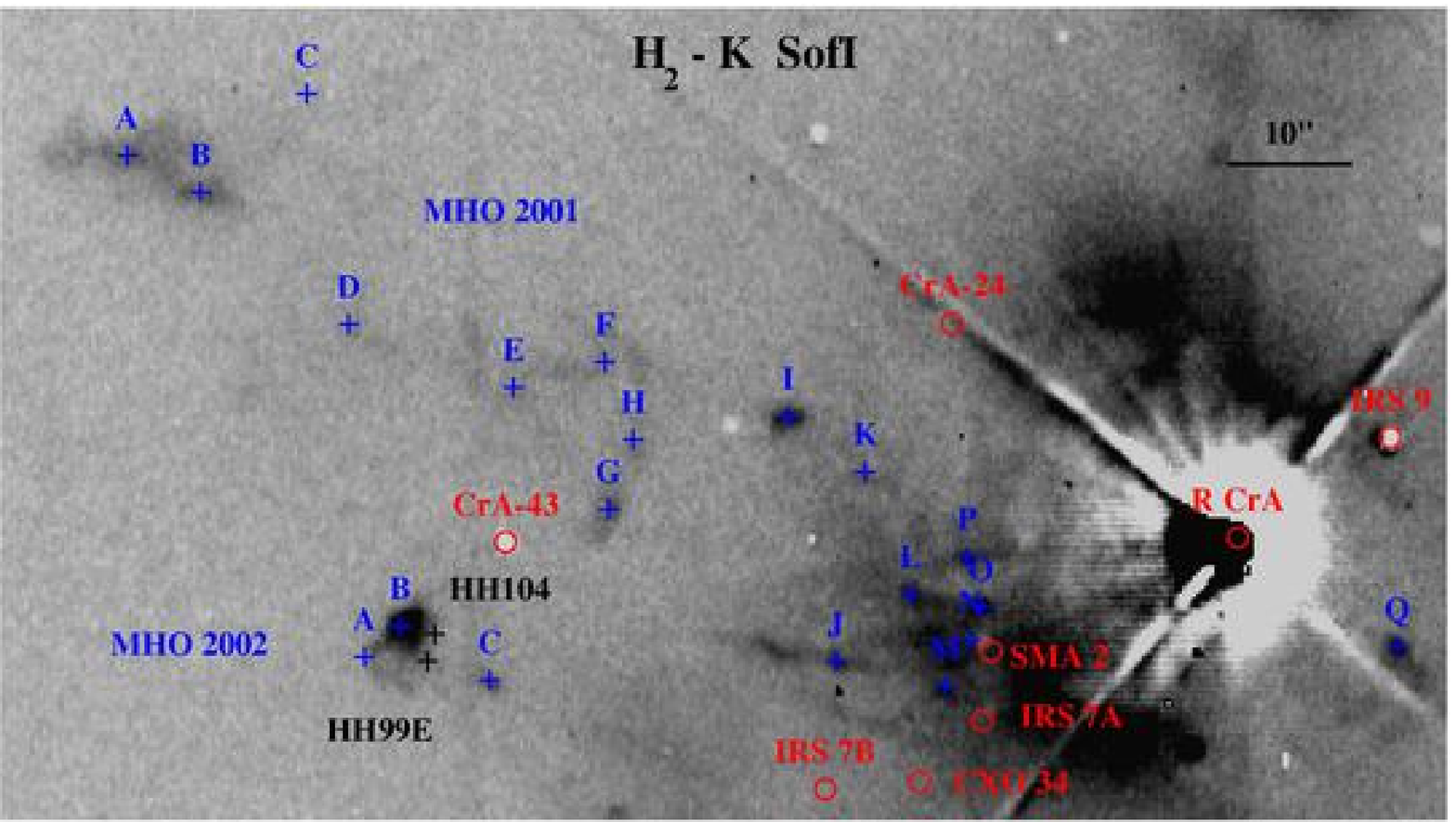}\\
\includegraphics[width=14.0 cm]{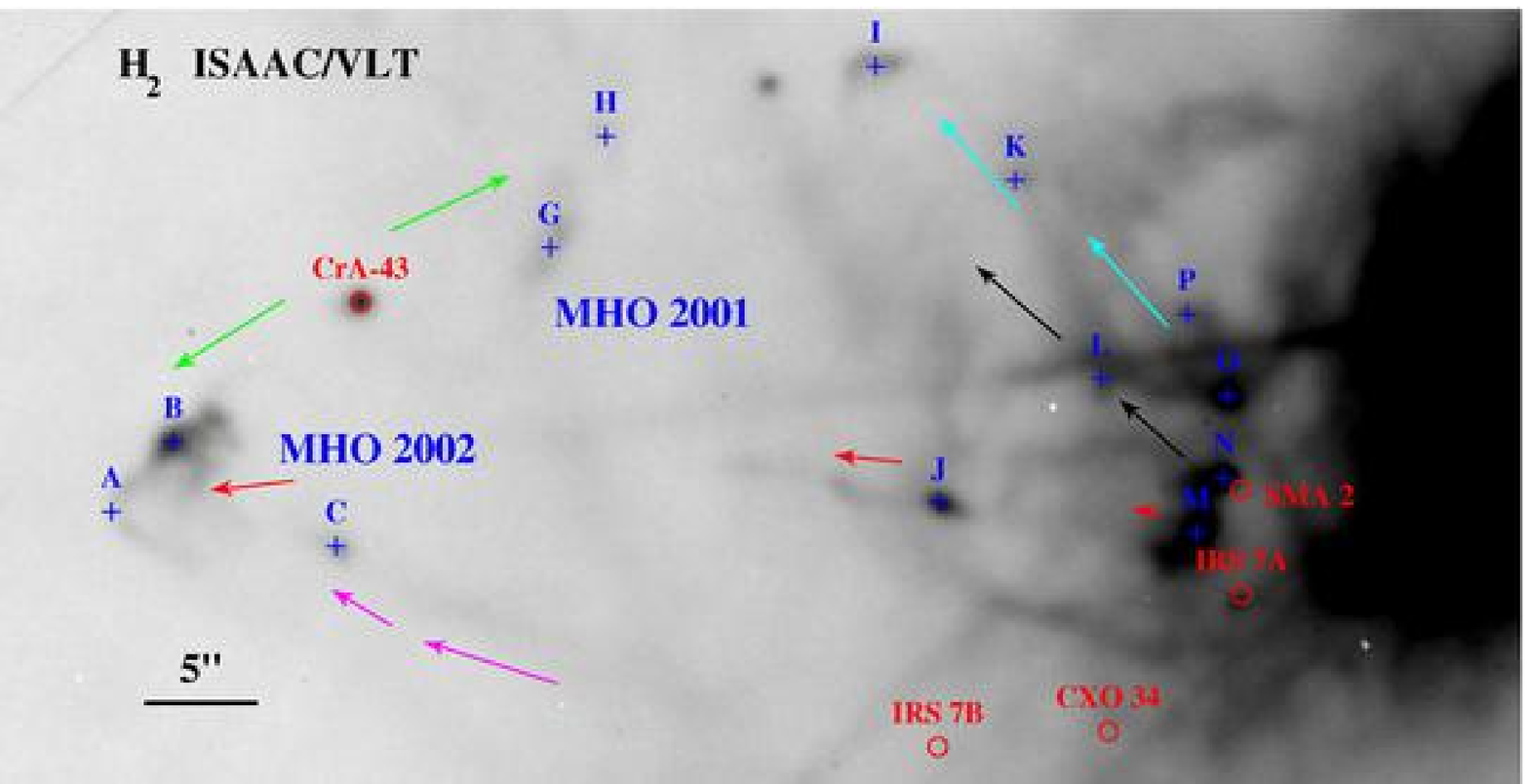}
\caption{\textit{(Upper panel)} \textit{SofI} H$_2$ continuum-subtracted image 
of MHO\,2001 and 2002 in the Coronet region.  Labels and crosses indicate 
names and positions of detected knots and sub-structures along the flows.  
\textit{(Lower panel)} Close up view of the region at 2.12\,$\mu$m, from a 
\textit{ISAAC/VLT} image in logarithmic scale. \label{MHO2001-2:fig}}
\end{figure}

Roughly 25$\arcsec$ northward, we detect three faint aligned knots 
(MHO\,2011 A-C) oriented SW-NE (length $\sim$1.7$\arcmin$).  HH\,736, not 
detected in our H$_2$ images, is positioned farther north-eastward, and 
although not well aligned with the knots, could be part of the flow.  Further 
north (see Figure~\ref{NE_flows:fig}) we detect MHO\,2013 and 2000, molecular 
counterparts to HH\,733 and HH\,99.  A close up view of the two flows is 
given in Figure~\ref{MHO2000-13:fig}.  Both flows are roughly parallel and 
elongated towards the NE.  MHO\,2013 shows a morphology identical to the 
[SII] emission observed by \citet{wang}, with three brighter condensations 
located at the curved head of the jet.  MHO\,2000 is composed of two bright 
structures, a heading bow-shock (A - HH\,99\,B) and a knot (B - HH\,99\,A), 
with a rough bow-shape, oriented towards MHO\,2013.  It is not clear whether 
or not this last is associated with the main flow or with the MHO\,2013 flow, 
although our P.M. analysis marginally indicates that the latter is the case 
(see Appendix~\ref{sec:pm_appendix} and Figure~\ref{PM_MHO2000_13:fig}).

About 5$\arcmin$ northward of the Coronet and $\sim$3.5$\arcmin$ to the east 
of TY\,CrA, we discover MHO\,2014, the brightest MHO in CrA 
(see Figure~\ref{NE_flows:fig}).  The MHO extends for roughly 1$\arcmin$, and 
displays an extremely complex morphology, as shown in 
Figure~\ref{MHO2014:fig}, where H$_2$ and H$_2$ continuum-subtracted images 
are reported.  Three arc-shaped structures are visible in the eastern part of 
the object (knots A-C), while the brightest knots (D-F) are located in the 
western part, partially coincident with HH\,732\,A and C.  Such a puzzling 
structure could be caused by the interaction of several bow-shocks from a 
precessing flow, or the overlapping of different flows.  Notably, HH\,732\,B, 
not detected in H$_2$, is located $\sim$30$\arcsec$ to the north, suggesting 
that more than one flow could be present.

Flow detection and P.M. study inside the Coronet region is particularly 
difficult, due to the presence of R\,CrA and the extremely bright and diffuse 
nebulosity around it, which cannot be completely removed in our H$_2$ 
continuum-subtracted images (see e.\,g. Figure~\ref{MHO2001-2:fig}, upper 
panel).  Several H$_2$ knots E and NE of R\,CrA and around CrA-43 (SMM\,2)
\citep[see e.\,g.,][]{nutter05,groppi07} were previously discovered by 
\citet{caratti06}, and named MHO\,2001 and 2002 by ~\citet{davis09b}.  
Figure~\ref{MHO2001-2:fig} (upper panel) shows a \textit{SofI} H$_2$ 
continuum-subtracted image of the region, where we detect more knots 
positioned a few arcseconds NE and E of the SMA\,2 and IRS\,7A 
region~\citep[see e.\,g.,][]{groppi07}.  Following \cite{davis09b}, 
the new knots are named MHO\,2001 from K to Q, although it is clear that 
MHO\,2001 and 2002 consist of several overlapping flows.  Knot Q is the only 
source of H$_2$ emission detected westward of R\,CrA.  MHO\,2001 is an 
elongated chain of scattered knots (roughly directed SW-NE), with several 
knots grouped around the SMA\,2 and IRS\,7A regions, which are the focal 
point of several flows.  A closer look with \textit{ISAAC/VLT} is given 
in Figure~\ref{MHO2001-2:fig} (lower panel).  The H$_2$ image is shown in 
logarithmic scale to display simultaneously the bright regions close to 
SMA and the fainter knots of MHO\,2002.  Unfortunately, an \textit{ISAAC} 
H$_2$ continuum-subtracted image is not available for this region.  The 
\textit{ISAAC/VLT} image reveals the presence of at least five different 
flows in this region, possibly more.  Knots C, I, K, P, O are possibly 
part of a first flow directed NNE (displayed with white arrows).  A 
second flow could consist of MHO\,2001 A--E, L, N , directed NE (cyan 
arrows).  MHO\,2002\,A is the terminal bow-shock of a third outflow 
(roughly west-east oriented, P.A.$\sim$90$\degr$) delineated by green 
arrows in the figure.  It is not clear whether MHO\,2002\,C could be 
part of this flow, or, as a faint H$_2$ trail seems to suggest could be 
driven by IRS\,7B (yellow arrows). MHO\,2001\,J is the jet of MHO\,2002\,A, 
and knot M could be part of the flow as well (green arrows).  This flow 
matches well the centimeter maps by \cite{chlt08}, who detected a 
west-east jet launched by SMA\,2.  This is also observed in the CO\,3-2 
maps by \citep{vankempen}, who detected a west-east flow 
(P.A.$\sim$ 90$\degr$).  The blue lobe is positioned to the east of 
SMA\,2 region, while the red lobe is to the west.  Another flow (red 
arrows), SE-NW oriented (P.A.$\sim$125$\degr$), and composed of 
MHO\,2002\,B, MHO\,2001\,G, and possibly H, seems to originate from CrA-43 
(SMM\,2).

\subsubsection{P.M.s \label{sec:pm_appendix}}

\begin{figure}
\includegraphics[width=14.0 cm]{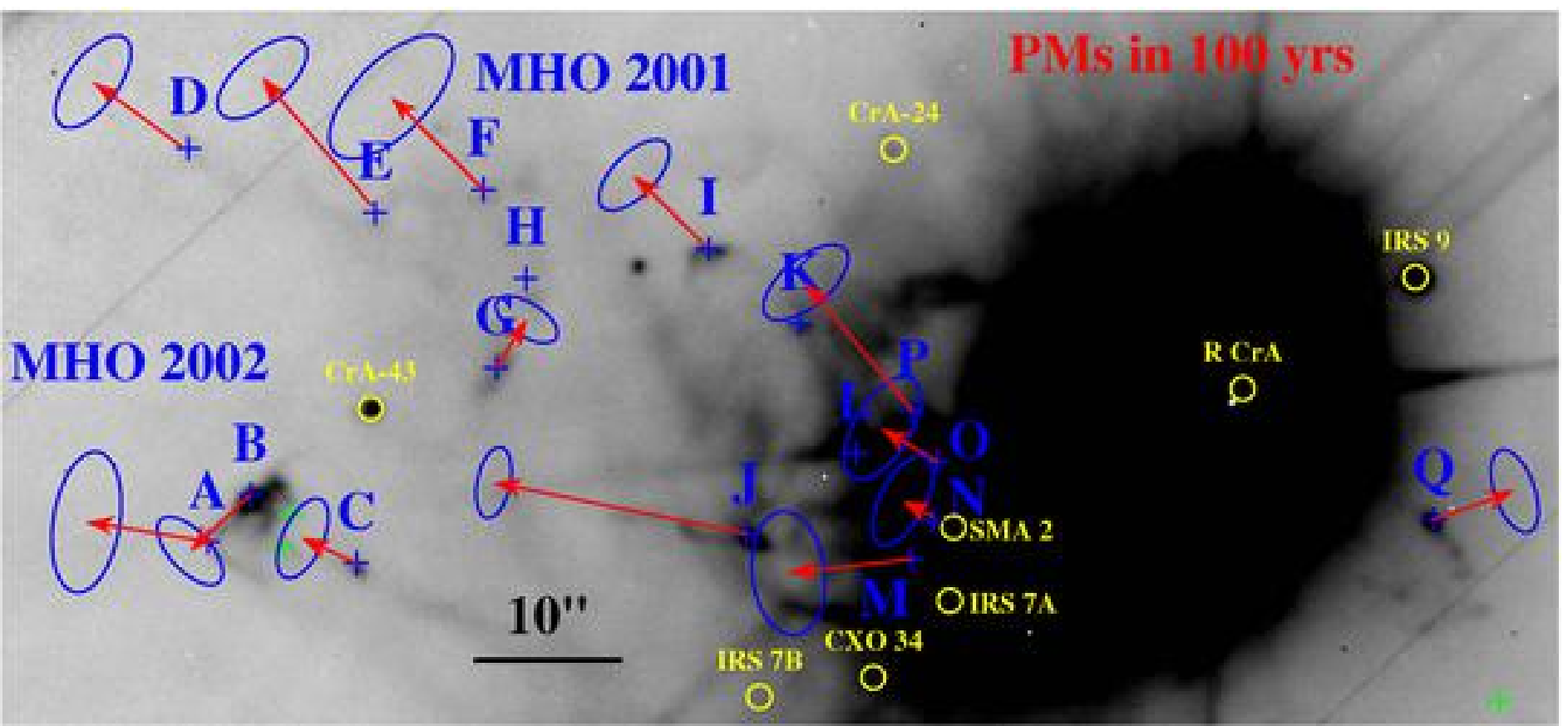}
\caption{H$_2$ flow chart of MHO\,2001 and 2002. Proper motions in 100\,yrs and their error bars are indicated by arrows and ellipses, respectively. \label{PM_MHO2001_2:fig}}
\end{figure}

A quick inspection of Figure~\ref{PM_all:fig} confirms that all the studied 
knots, except MHO\,2008\,D, are launched by YSOs inside or close to the 
Coronet.  As already noted in \S~\ref{sec:outflows}, several flows 
originate in the eastern part of the Coronet, in particular in a region a few 
arcseconds around SMA\,2, which encloses several early stage YSOs 
\citep[see Appendix \ref{sec:appA} and e.\,g.,][]{nutter05,groppi07}.  This 
can be seen in more detail in Figure~\ref{PM_MHO2001_2:fig}.  MHO\,2002\,A is 
the terminal bow-shock of a west-east flow (MHO\,2001\,J, M), which has 
MHO\,2001\,Q as a flow counterpart.  MHO\,2002\,C is likely part of another 
flow from IRS7\,B.  MHO\,2002\,B and MHO\,2001\,G are the south-eastern and 
north-western lobes of the CrA-43 (SMM\,2) outflow.  Then MHO\,2001 D--N are 
part of one or more outflows roughly moving towards the north-east and emanate 
from the SMA\,2 region. 

\begin{figure}
\includegraphics[width=14.0 cm]{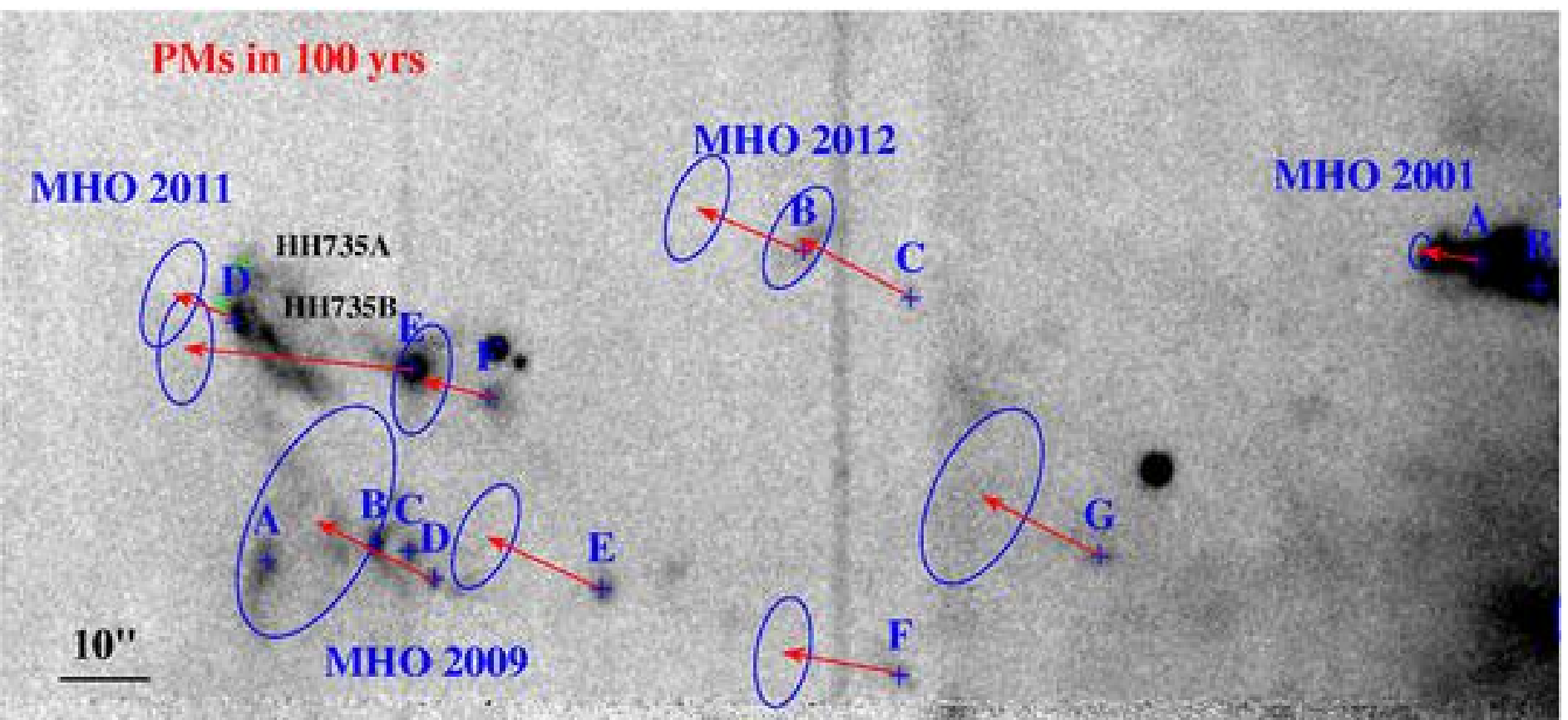}
\caption{H$_2$ flow chart of MHO\,2009, 2011, 2012 and part of MHO\,2001. 
Proper motions in 100\,yrs and their error bars are indicated by arrows and ellipses, respectively. \label{PM_MHO2009_12_2:fig}}
\end{figure}

\begin{figure}
\includegraphics[width=14.0 cm]{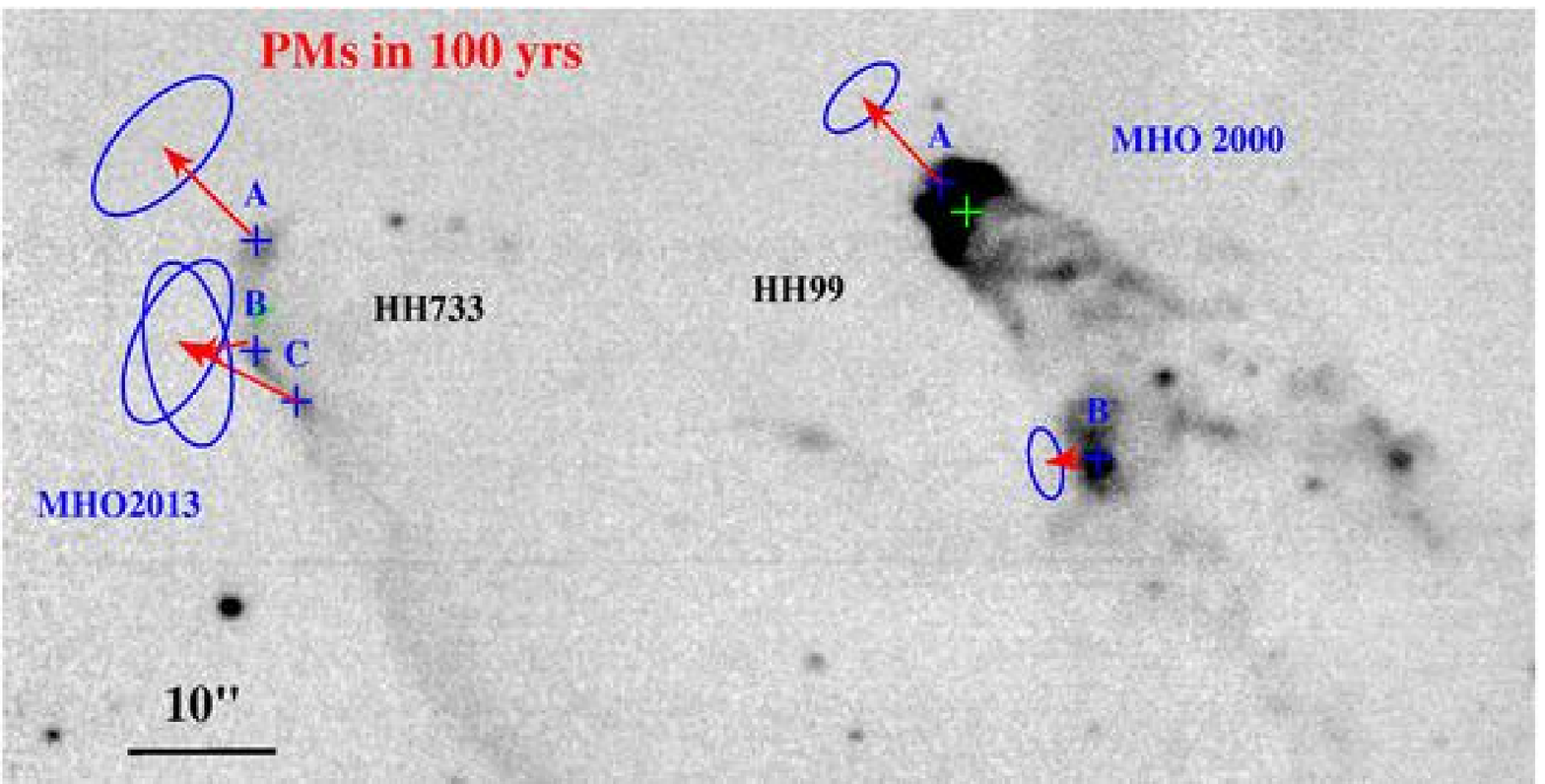}
\caption{H$_2$ flow chart of MHO\,2000 and 2013. Proper motions in 100\,yrs and their error bars are indicated by arrows and ellipses, respectively. \label{PM_MHO2000_13:fig}}
\end{figure}

\begin{figure}
\includegraphics[width=10.0 cm]{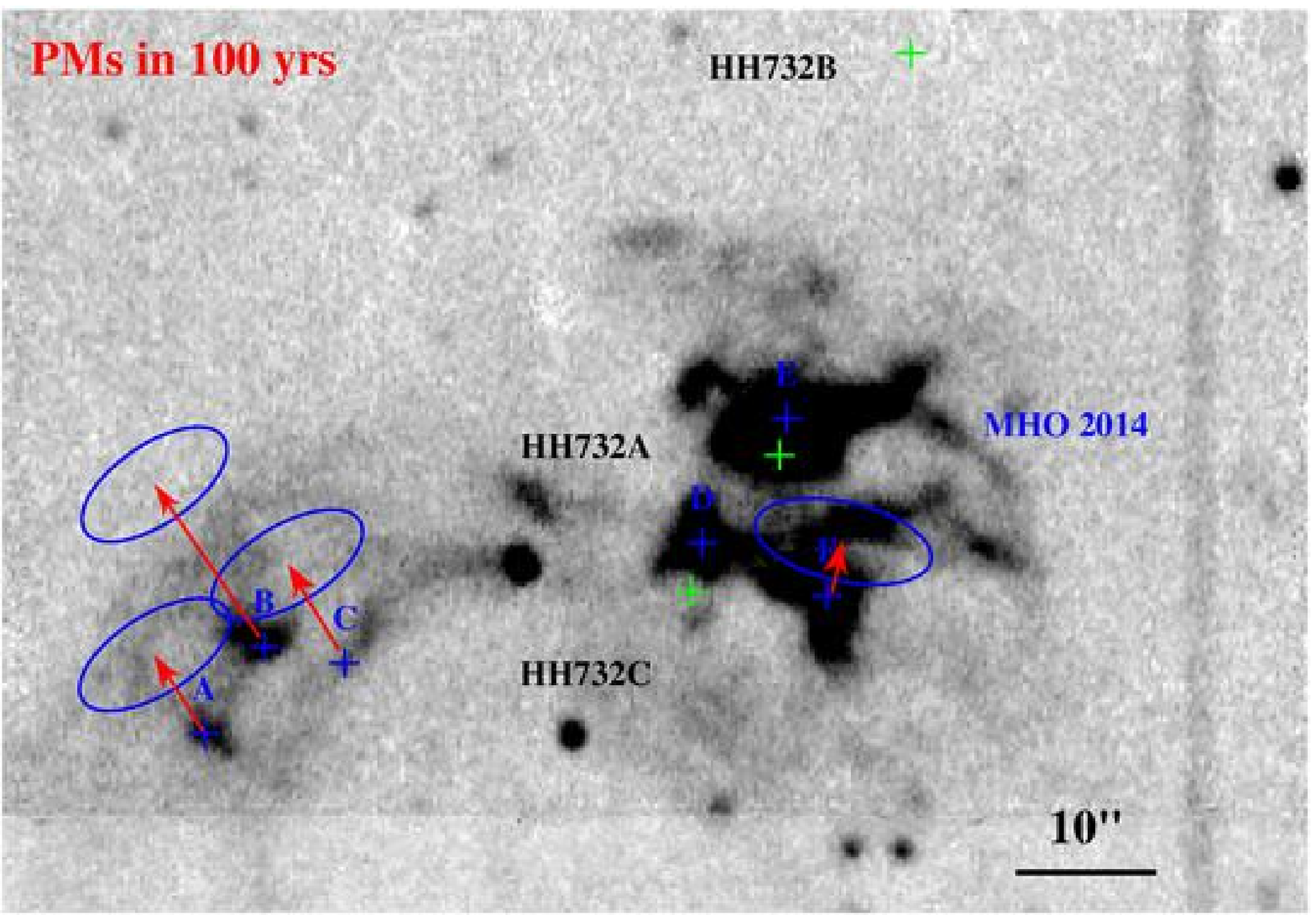}
\caption{H$_2$ flow chart of MHO\,2014. Proper motions in 100\,yrs and their error bars are indicated by arrows and ellipses, respectively. \label{PM_MHO2014:fig}}
\end{figure}

Figure~\ref{PM_MHO2009_12_2:fig} shows the flow charts of MHO\,2009, 2011, 
2012 and part of MHO\,2001\,A and B, which move from west to east and from WSW 
to ENE, being indeed the prosecution of the flows of 
Figure~\ref{PM_MHO2001_2:fig}.  Figure~\ref{PM_MHO2000_13:fig} shows P.M.s in 
MHO\,2000 and 2013.  Our multi-epoch images allow us to get a better 
measurement of the P.M. for MHO\,2000\,A, which appears to originate from a 
region around IRS\,9 and CrA-24.  On the other hand, MHO\,2000\,B is moving 
eastward and might not be related to this flow, but rather to MHO\,2013, which 
appears as an expanding shell moving towards ENE, in this case IRS\,6 is 
likely the driving source.

MHO\,2014 (see Figure~\ref{PM_MHO2014:fig}) is roughly moving towards NNE, and 
it is likely an expanding bow-shock, which originates from the western 
outskirts of the Coronet.  Unfortunately, the lack of a second-epoch image for 
knots D and E does not allow us to completely reconstruct its motion.

\subsection{Notes on individual MHOs - South and west regions \label{SW_MHOs:sec}}

\subsubsection{Morphology}

\begin{figure}
\includegraphics[width=16.0 cm]{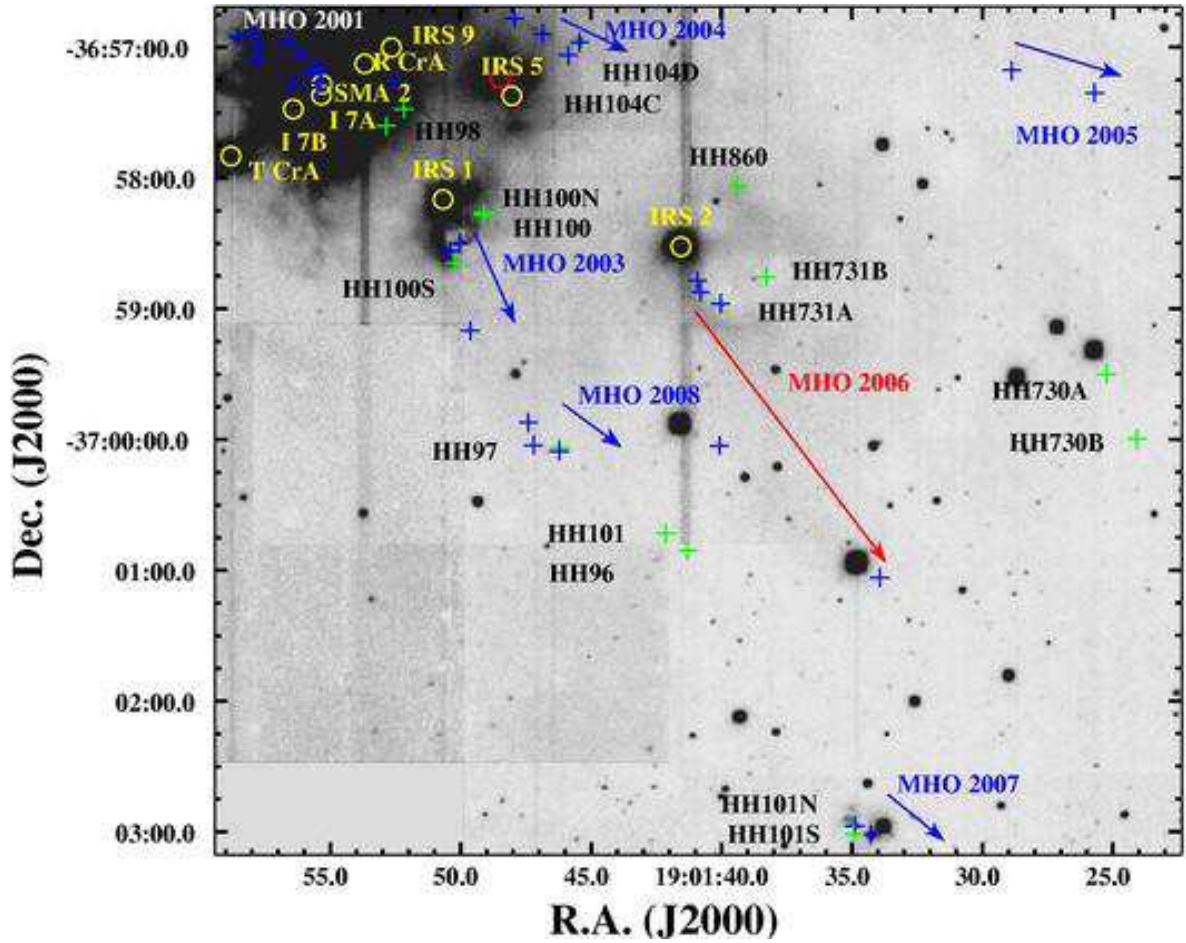}
\caption{\textit{SofI} H$_2$ image of the south-western flows detected outside 
the Coronet.  MHO 2001 knots inside the Coronet are labeled with blue crosses. 
\label{SW_flows:fig}}
\end{figure}

\begin{figure}
\includegraphics[width=10.0 cm]{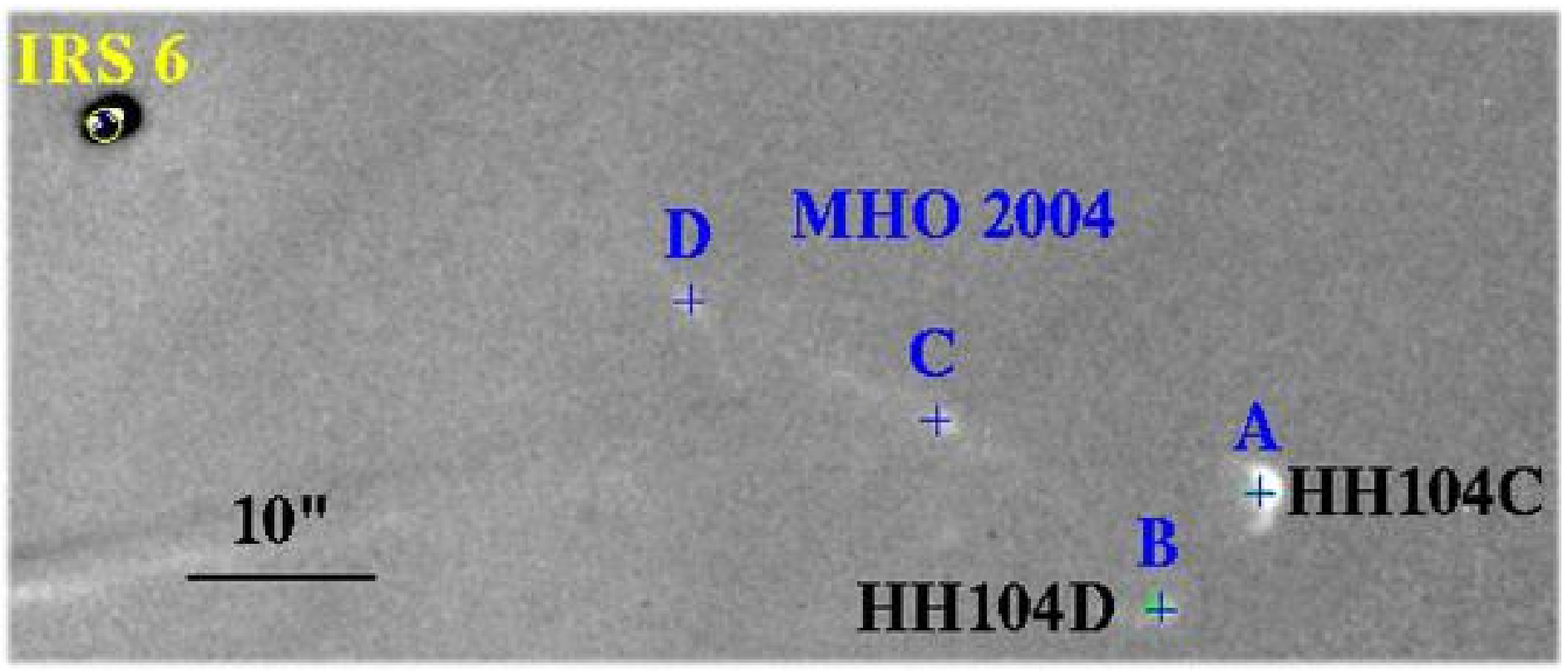}
\caption{\textit{ISAAC} H$_2$ continuum-subtracted image of MHO\,2004 (see 
Figure~\ref{SW_flows:fig}).  Labels and crosses indicate names and positions 
of detected knots and sub-structures along the flows.\label{MHO2004:fig}}
\end{figure}

\begin{figure}
\includegraphics[width=10.0 cm]{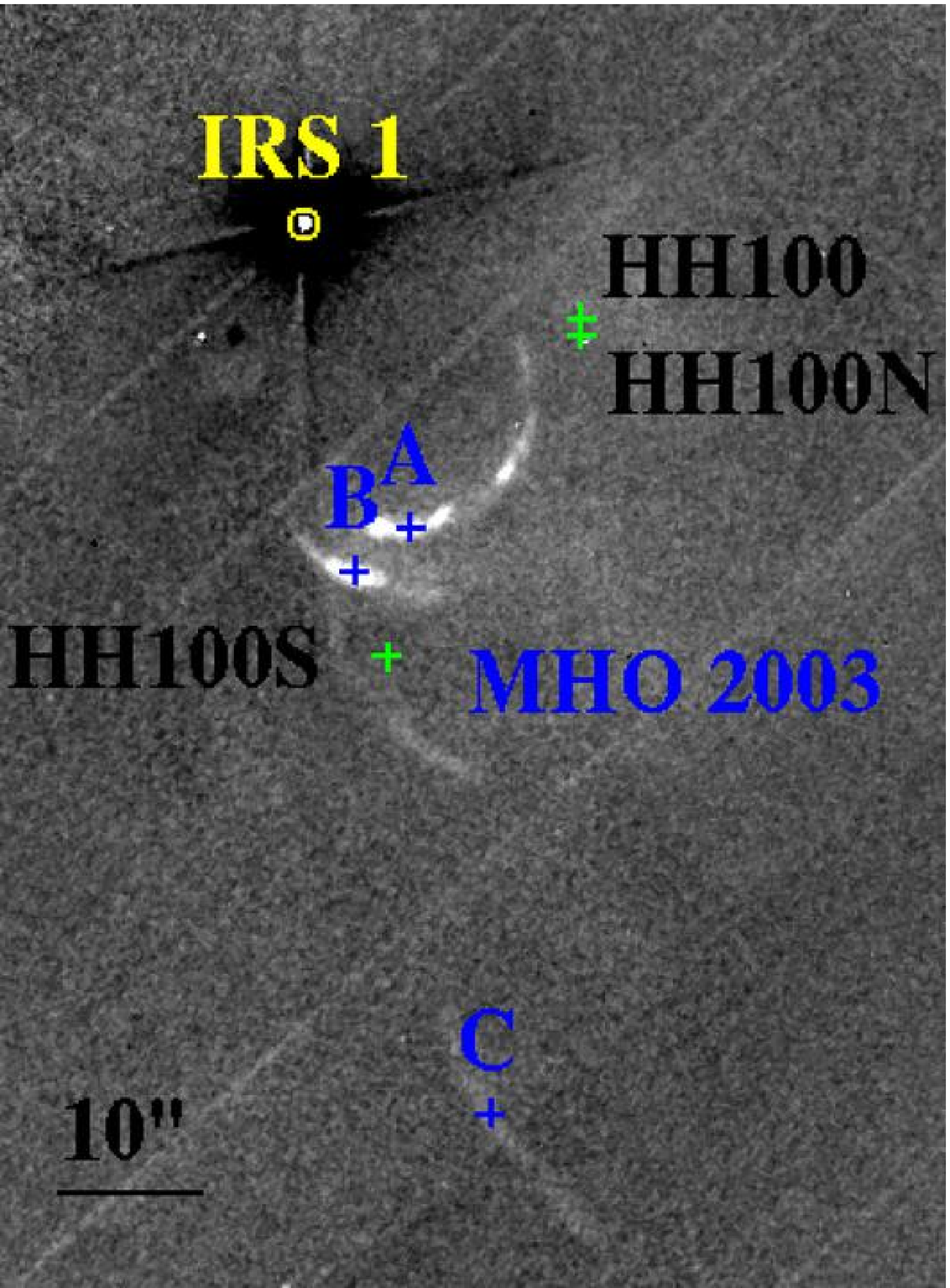}
\caption{\textit{ISAAC} H$_2$ continuum-subtracted image of MHO\,2003 (see 
Figure~\ref{SW_flows:fig}).  Labels and crosses indicate names and positions 
of detected knots and sub-structures along the flows.\label{MHO2003:fig}}
\end{figure}

\begin{figure}
\includegraphics[width=10.0 cm]{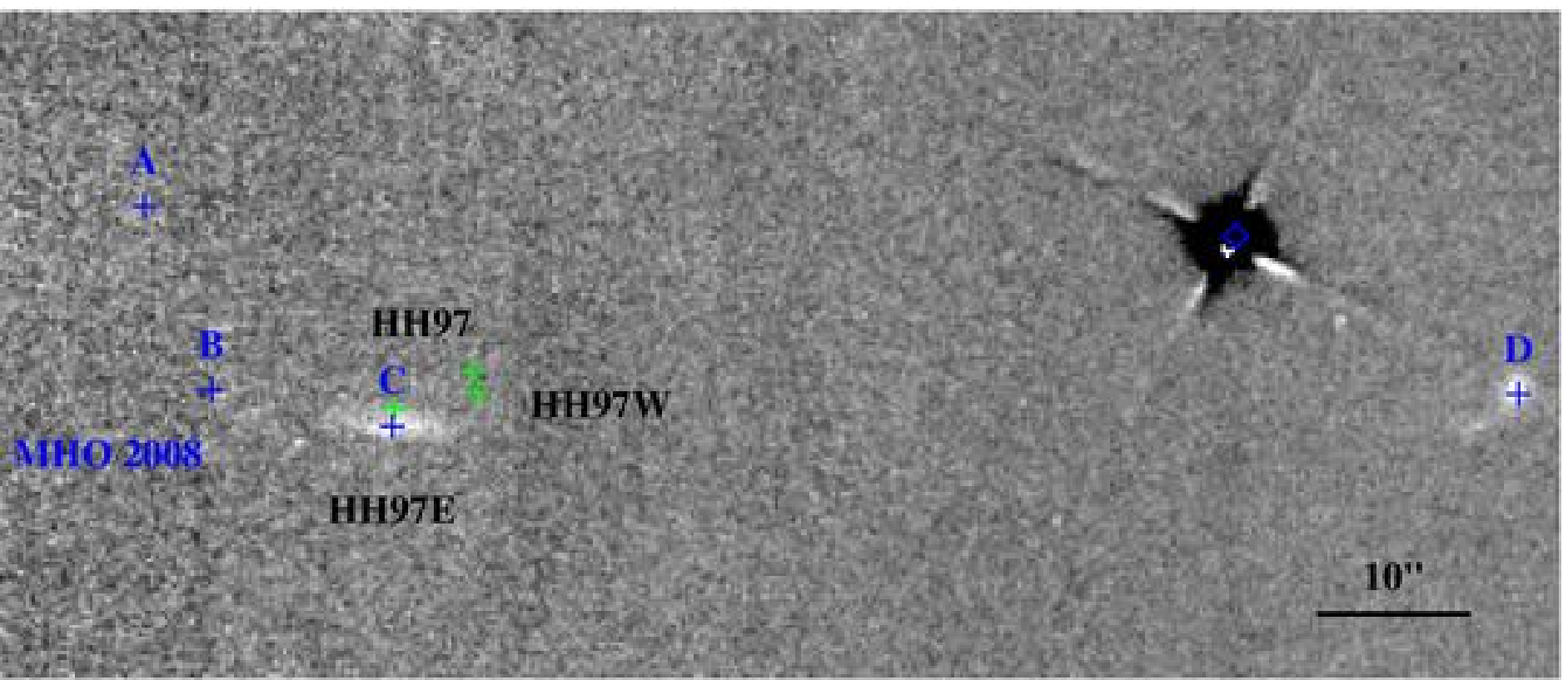}
\caption{\textit{ISAAC} H$_2$ continuum-subtracted image of MHO\,2008 (see 
Figure~\ref{SW_flows:fig}).  Labels and crosses indicate names and positions 
of detected knots and sub-structures along the flows.\label{MHO2008:fig}}
\end{figure}

\begin{figure}
\includegraphics[width=10.0 cm]{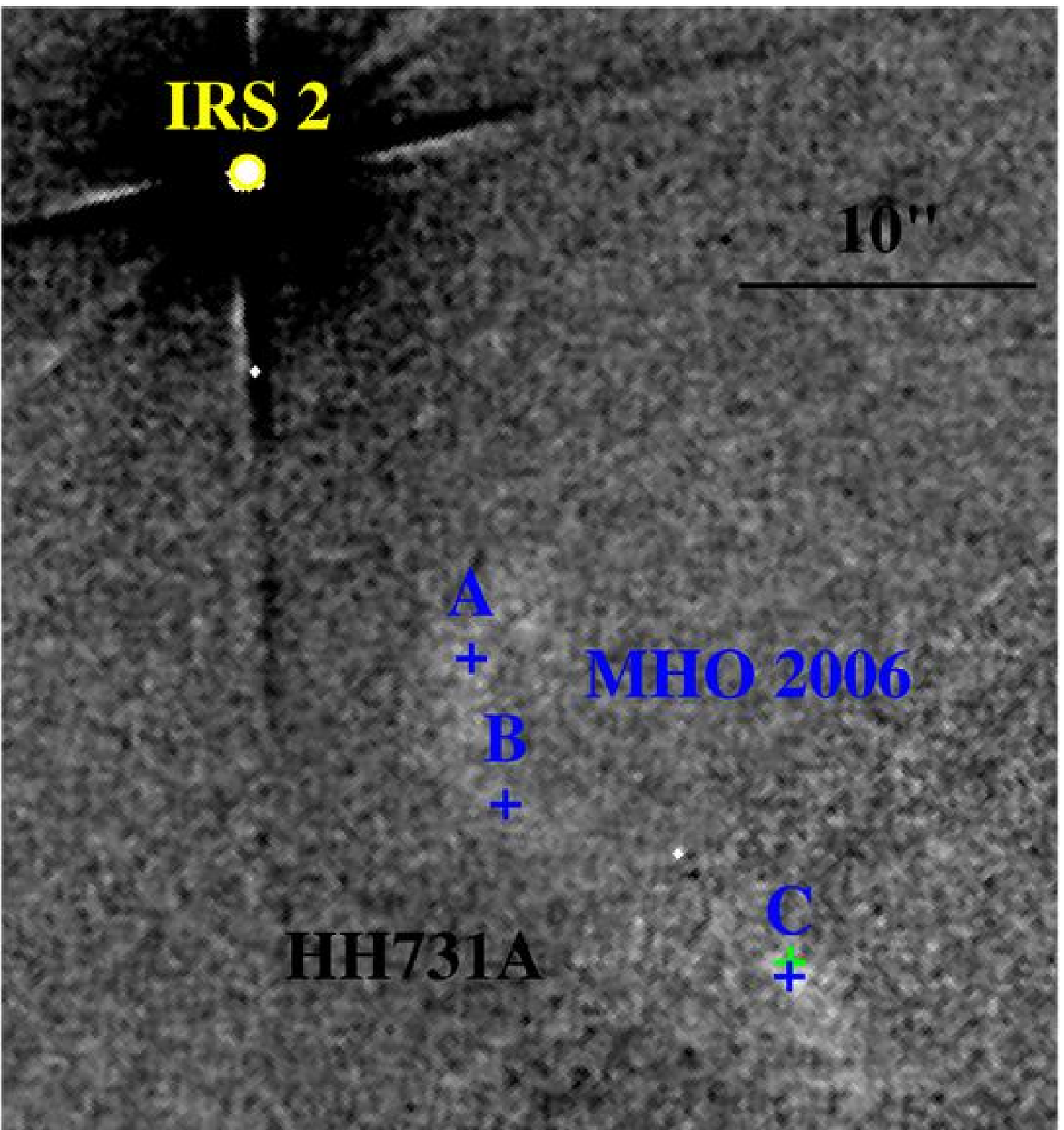}
\caption{\textit{ISAAC} H$_2$ continuum-subtracted image of MHO\,2006 (see 
Figure~\ref{SW_flows:fig}).  Labels and crosses indicate names and positions 
of detected knots and sub-structures along the flows.\label{MHO2006:fig}}
\end{figure}

\begin{figure}
\includegraphics[width=10.0 cm]{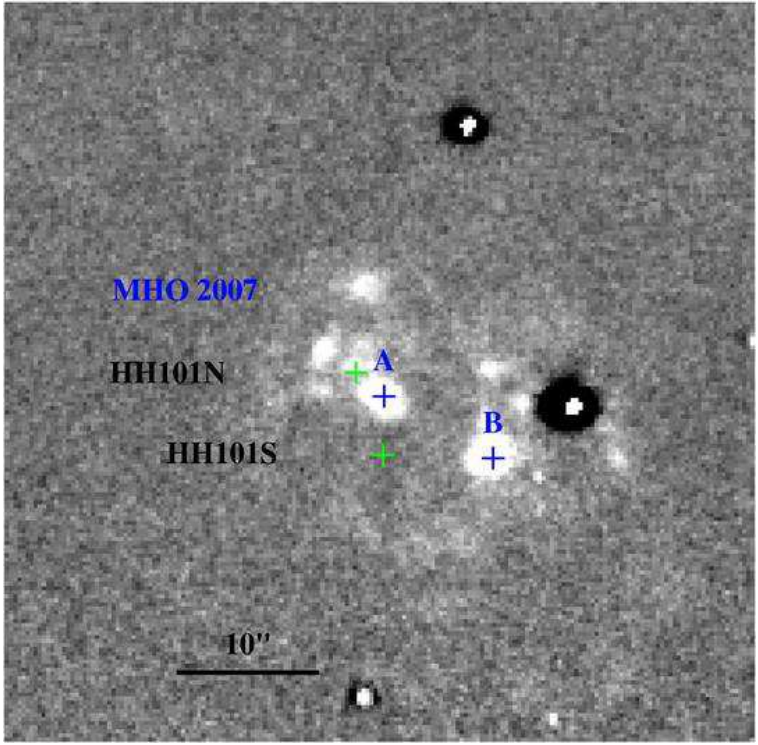}
\caption{\textit{ISAAC} H$_2$ continuum-subtracted image of MHO\,2007, the 
molecular counterpart of HH\,101 (see Figure~\ref{SW_flows:fig}).  Labels and 
crosses indicate names and positions of detected knots and sub-structures 
along the flows. \label{MHO2007:fig}}
\end{figure}

\begin{figure}
\includegraphics[width=18.0 cm]{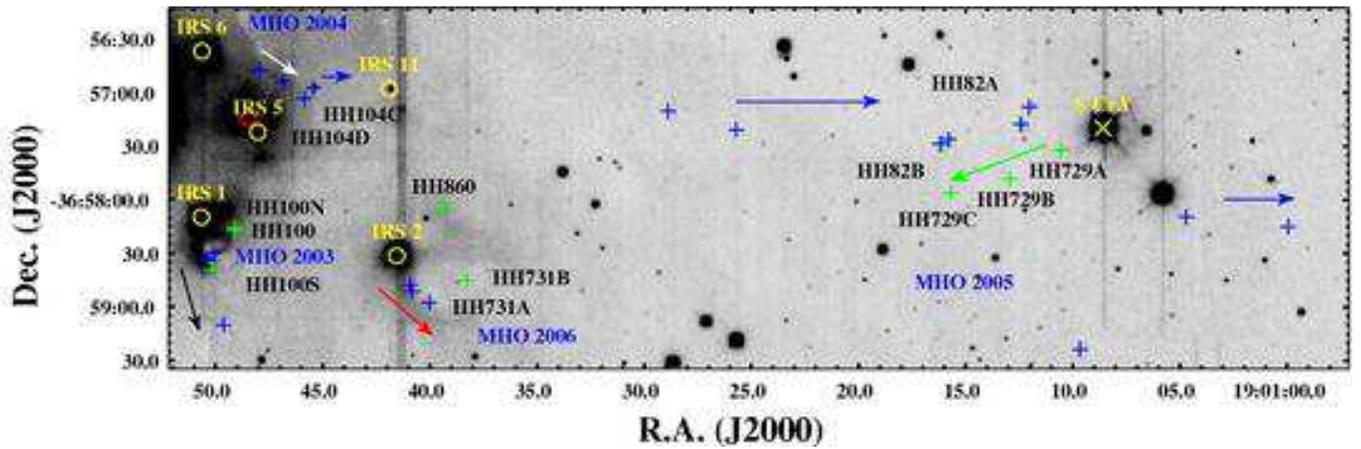}
\caption{\textit{SofI} H$_2$ image of the western flows detected outside 
the Coronet.\label{W_flows:fig}}
\end{figure}

\begin{figure}
\includegraphics[width=18.0 cm]{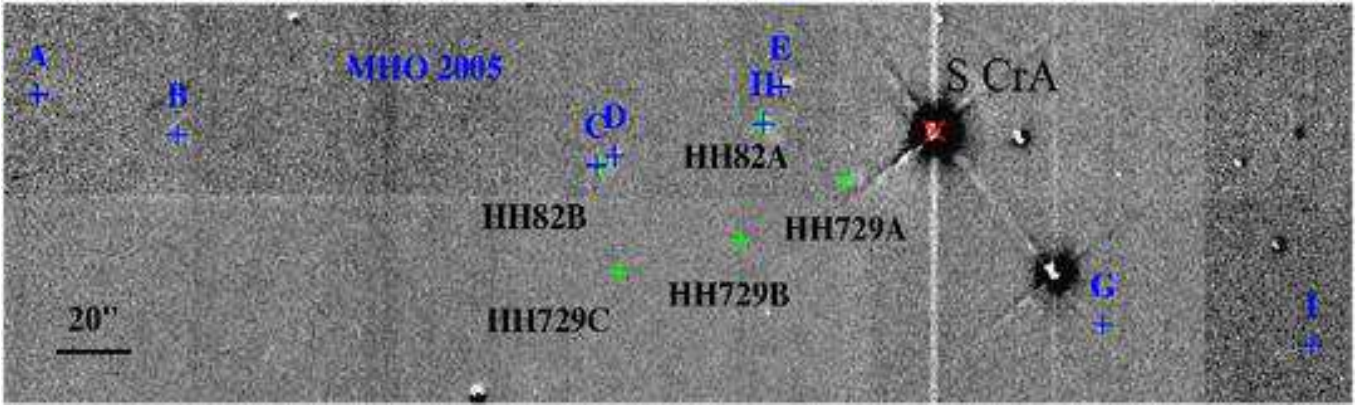}
\caption{\textit{SofI} H$_2$ continuum-subtracted image of MHO\,2005 (see 
Figure~\ref{W_flows:fig}). 
\label{MHO2005:fig}}
\end{figure}

\begin{figure}
\includegraphics[width=10.0 cm]{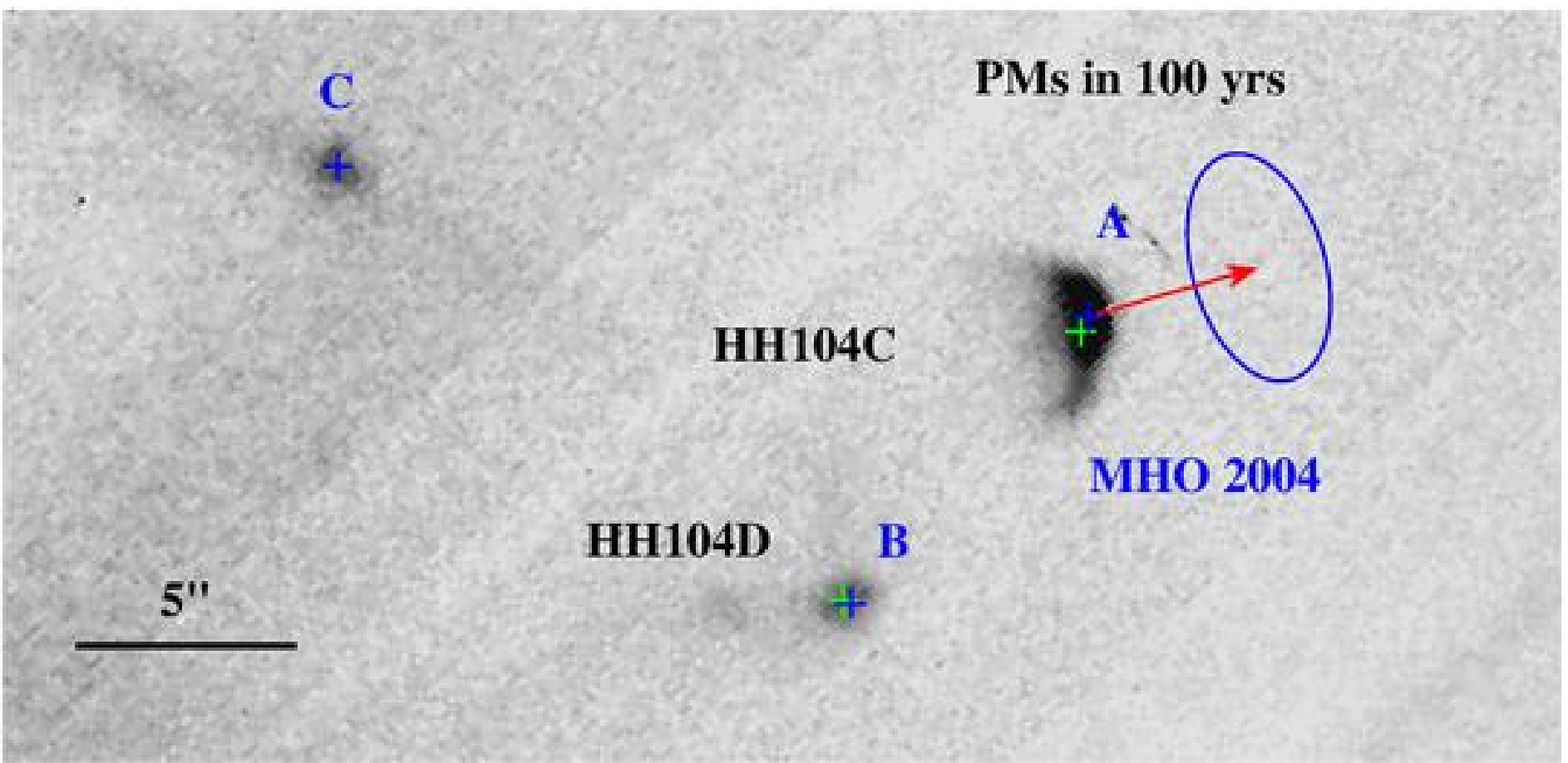}
\caption{H$_2$ flow chart of MHO\,2004. Proper motions in 100\,yrs and their error bars are indicated by arrows and ellipses, respectively. \label{PM_MHO2004:fig}}
\end{figure}

Figure~\ref{SW_flows:fig} shows a \textit{SofI} H$_2$ image of flows located 
to the south-west of the Coronet.  There are four MHO flows, roughly NE-SW 
oriented, escaping from the Coronet (MHO\,2003, 2004, 2006--2008).  MHO\,2003, 
2004 \citep{wilking97,caratti06,davis09b} and the newly detected MHO\,2006, 
2007 and 2008 are the molecular counterparts of HH\,100, HH\,104, HH\,97, 
HH\,101, and HH\,731 \citep[see e.\,g.,][]{ssg74,hartigan87,wang}, 
respectively.

Figure~\ref{MHO2004:fig} shows an \textit{ISAAC} H$_2$ continuum-subtracted 
image of MHO\,2004 (i.\,e. HH\,104\,C and D).  The molecular object, 
positioned about 20$\arcsec$ SW of IRS\,6, appears as a faint curved jet with 
three bright condensations (namely B--D), NE-SW oriented, and a brighter 
bow-shock (A or HH\,104\,C) placed $\sim$8$\arcsec$ NW off-axis from the jet 
and east-west oriented.  The bow-shock should not be associated with the main 
jet, which is likely driven by IRS\,6.

MHO\,2003 (see Figure~\ref{SW_flows:fig}) is positioned about 20$\arcsec$ SW 
of IRS\,1 (or HH\,100\,IR), which is likely the driving source.  A close up 
view of the flow (see Figure~\ref{MHO2003:fig}) with \textit{ISAAC} shows two 
bright asymmetric arcs (A and B) close to the source, and two fainter trails 
about 10$\arcsec$ and 40$\arcsec$ farther to the SW.  The H$_2$ emission  
overlaps well with the CO\,3-2 blue-shifted outflow ejected by 
IRS\,1~\citep[][]{vankempen}.  The flow has a P.A.$\sim$203$\degr$.

MHO\,2008 (A--C), located about 1$\arcmin$ to the SW and well aligned with the 
previous flow (see Figure~\ref{SW_flows:fig}), could be part of MHO\,2003.  On 
the other hand, MHO\,2008\,D is a bow-shock east-west oriented and should not 
be part of the flow (see Figure~\ref{MHO2008:fig}), but it is likely driven by 
the Class II source CrA\,41.

MHO\,2006 (A--C) is a trail of faint precessing knots (P.A.$\sim$206$\degr$), 
located about 15$\arcsec$ to the SW of IRS\,2 (see Figure~\ref{MHO2006:fig}), 
which is the driving source. MHO\,2006\,D (see Figure~\ref{SW_flows:fig}) is a 
faint patch and could be part of the flow.

About 6$\arcmin$ south-westward of the Coronet, we detect MHO\,2007, the 
molecular counterpart of HH\,101 (see Figure~\ref{SW_flows:fig}).  The MHO 
extends for roughly 40$\arcsec$, NNE-SSW elongated, with two faint arcs in 
front, likely a bow-shock structure and several bright scattered knots behind 
(A and B are particularly bright).  The MHO is shown in 
Figure~\ref{MHO2007:fig}, where an \textit{ISAAC} H$_2$ continuum-subtracted 
image is shown.

The last newly detected flow, MHO\,2005 (knots A--I), is located 
$\sim$4$\arcmin$ to the west of the Coronet and extends for about 6$\arcmin$ 
around S\,CrA (see Figure~\ref{W_flows:fig}), roughly showing a broad 
S-shape.  Knots C and E, the brightest knots of the flow, along with D, are 
the molecular counterpart of HH\,82\,B and A, respectively.  Knot F is 
positioned about 1$\arcmin$ south and might not be part of the flow.  The 
HH\,729 knots, which should be driven by S\,CrA 
\citep[see e.\,g.,][]{ssg74,wang}, are not detected in our H$_2$ images.

\subsubsection{P.M.s}

Figure~\ref{PM_MHO2004:fig} shows the MHO\,2004\,A P.M., confirming that the 
bow-shock is not related to the main jet, but is moving WNW, possibly 
associated with the same flow as MHO\,2001\,Q.

\begin{figure}
\includegraphics[width=10.0 cm]{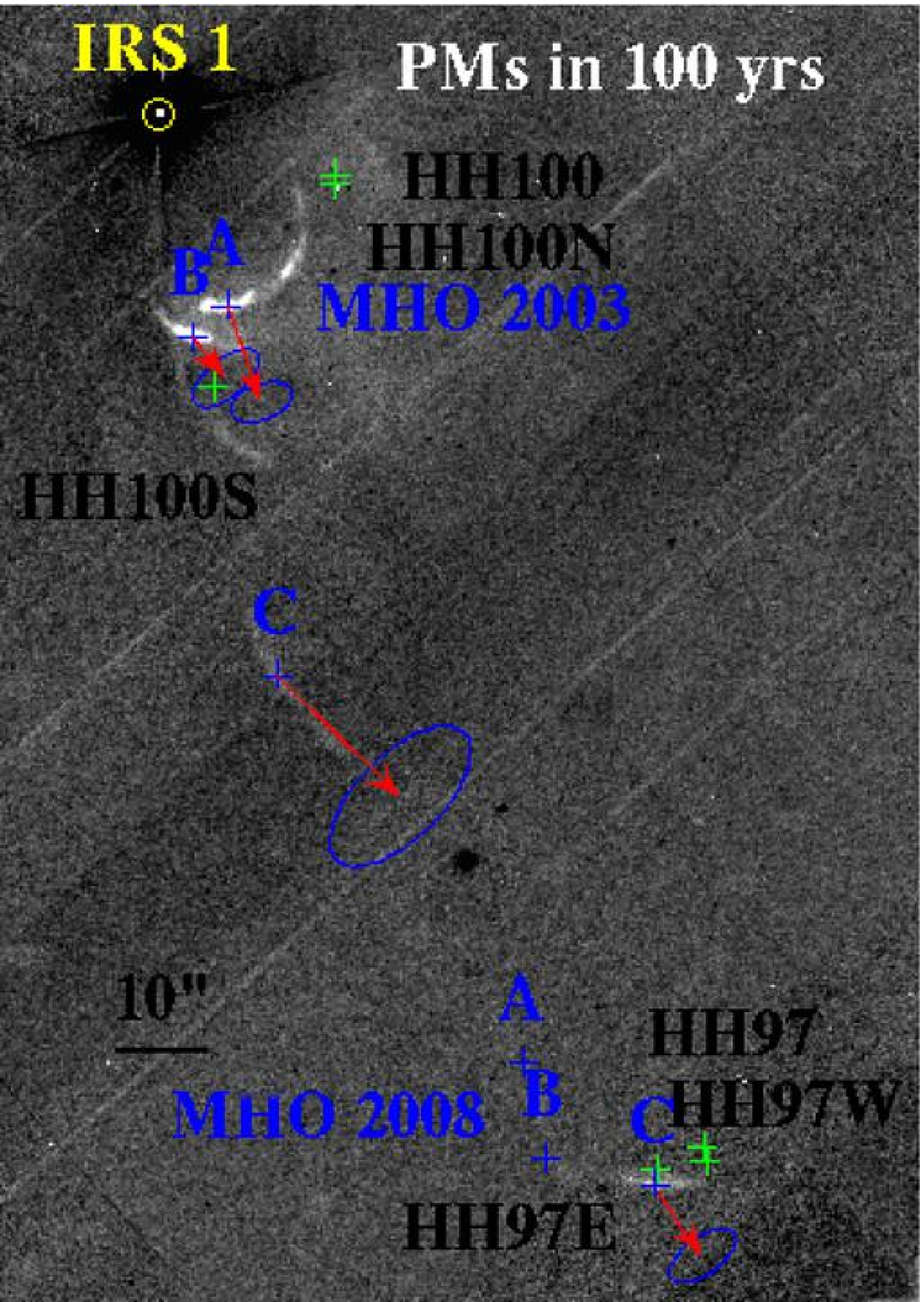}
\caption{H$_2$ flow chart of MHO\,2003 and part of MHO\,2008.  Proper motions in 100\,yrs and their error bars are indicated by arrows and ellipses, respectively. \label{PM_MHO2003_8:fig}}
\end{figure}

\begin{figure}
\includegraphics[width=10.0 cm]{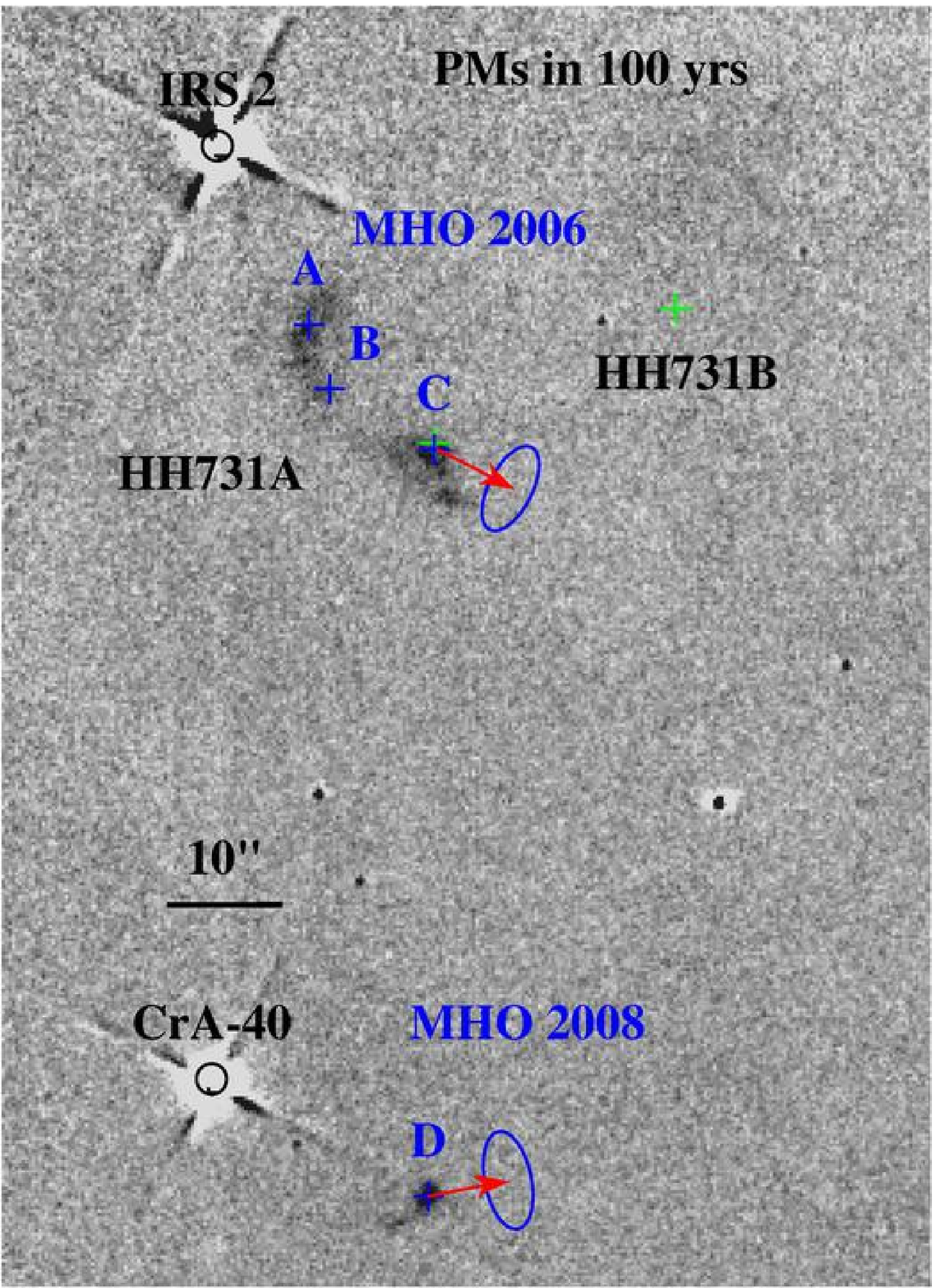}
\caption{H$_2$ flow chart of MHO\,2006 and part of MHO\,2008. Proper motions in 100\,yrs and their error bars are indicated by arrows and ellipses, respectively. \label{PM_MHO2006_8:fig}}
\end{figure}

\begin{figure}
\includegraphics[width=10.0 cm]{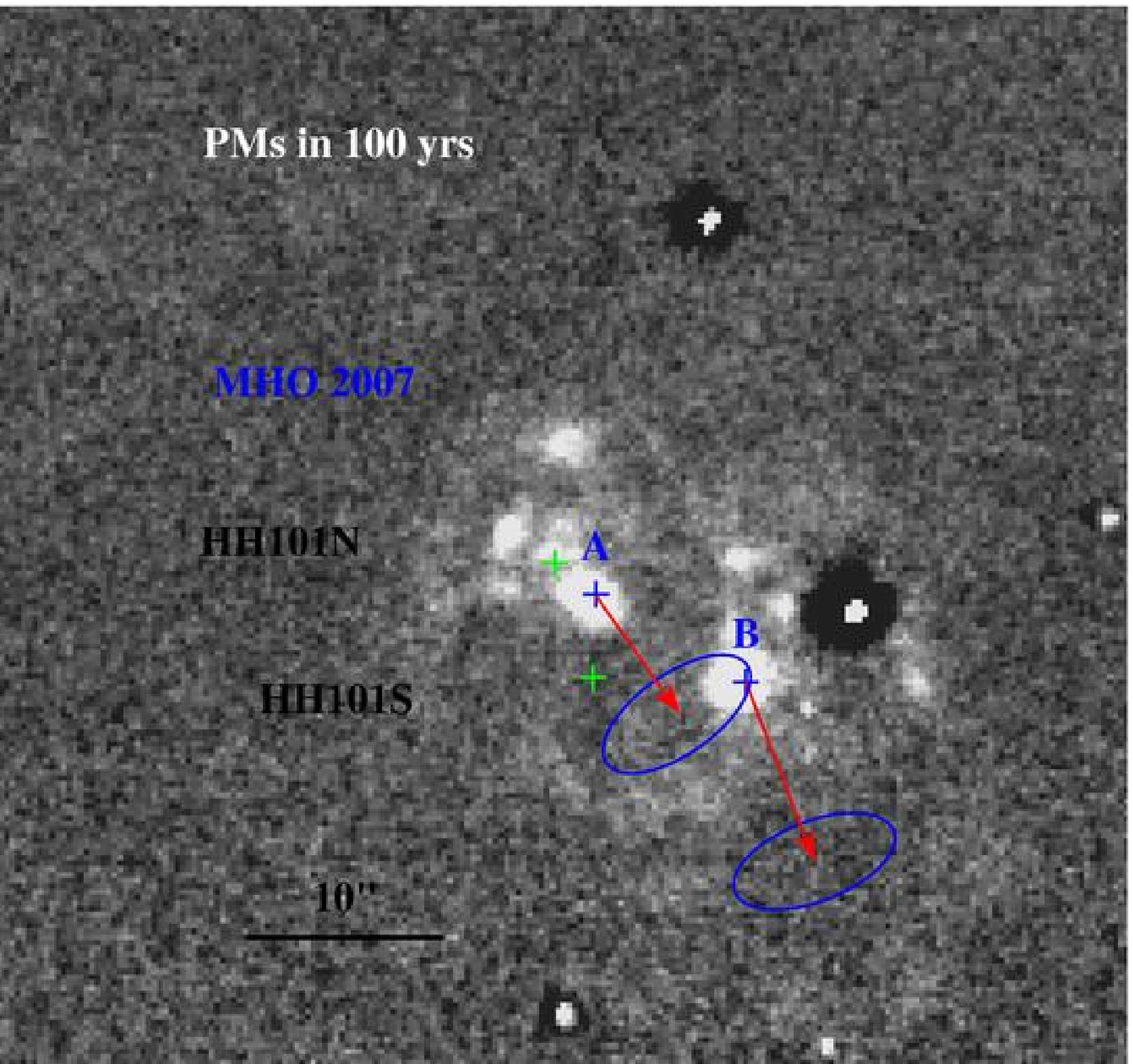}
\caption{H$_2$ flow chart of MHO\,2007. Proper motions in 100\,yrs and their error bars are indicated by arrows and ellipses, respectively. \label{PM_MHO2007:fig}}
\end{figure}

Finally, our kinematical analysis reveals three or four different flows in the 
CrA south-western region, as shown in Figures~\ref{PM_MHO2003_8:fig}, 
\ref{PM_MHO2006_8:fig}, and \ref{PM_MHO2007:fig}.  MHO\,2003 and knots A--C 
of MHO\,2008 are likely part of the same flow (Figure~\ref{PM_MHO2003_8:fig}) 
moving towards the SW.  On the other hand, knot D stands alone moving towards 
the SSW (Figure~\ref{PM_MHO2006_8:fig}), likely driven by the nearby YSO 
candidate CrA-41, and MH0\,2006 (same figure) represents another flow.  
MHO\,2007 (Figure~\ref{PM_MHO2007:fig}) is roughly aligned with the 
MHO\,2003--6 flow, with similar tangential velocities.  However, it is not 
clear whether or not it could be part of it, because of the different 
orientation of the leading bow-shock, which rather points to the SSW.

\subsection{Notes on individual \textit{Spitzer} flows \label{Spitzer_flows:sec}}

\begin{figure}
\includegraphics[width=10.0 cm]{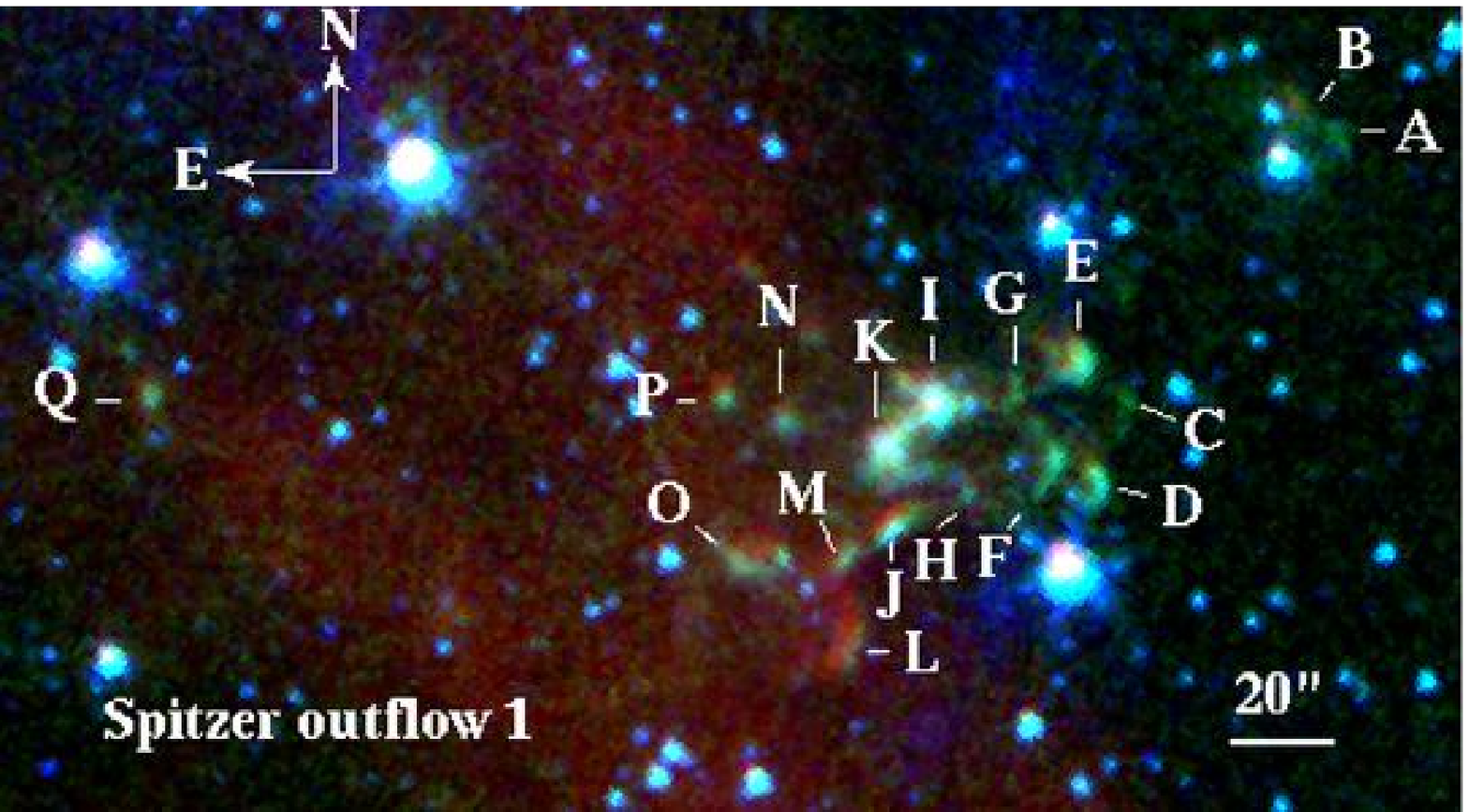}
\caption{\textit{Spitzer} three-color image of \textit{Spitzer} outflow\,1, 
made from the IRAC 3.6\,$\mu$m (blue), 4.5\,$\mu$m (green), and 8\,$\mu$m 
(red) images.  Labels indicate the position of the newly detected knots 
(EGOs).\label{spitzer1:fig}}
\end{figure}

The first outflow detected in our \textit{Spitzer} maps, named 
\textit{Spitzer} outflow\,1, is located on the western edge of the CrA star 
forming region, about 17$\arcmin$ from the Coronet.  Figure~\ref{spitzer1:fig} 
shows the morphology of the outflow, which is composed of several knots and 
bow-shocks, roughly east-west oriented.  These knots likely originate from a 
precessing jet, as indicated by their scattered placement.  No obvious 
counterpart on the eastern edge of CrA is observed. 

\begin{figure}
\includegraphics[width=10.0 cm]{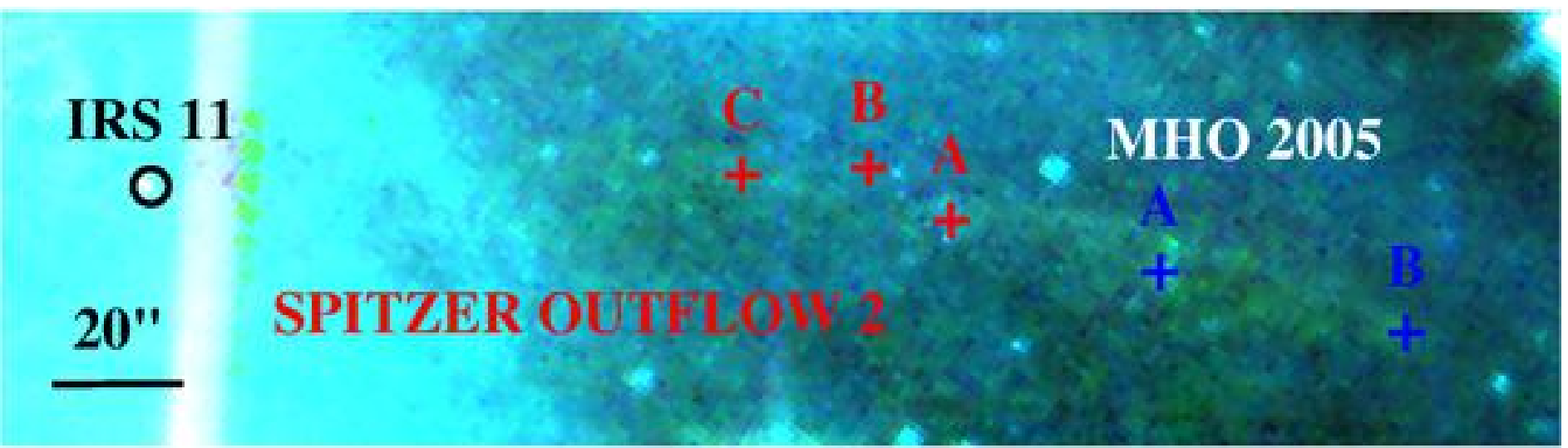}
\caption{\textit{Spitzer} three-color image of \textit{Spitzer} outflow\,2, 
made from the IRAC 3.6\,$\mu$m (blue), 4.5\,$\mu$m (green), and 8\,$\mu$m 
(red) images.  Labels indicate the position of the newly detected knots 
(EGOs) and MHO\,2005\,A and B.\label{spitzer2:fig}}
\end{figure}

\begin{figure}
\includegraphics[width=16.0 cm]{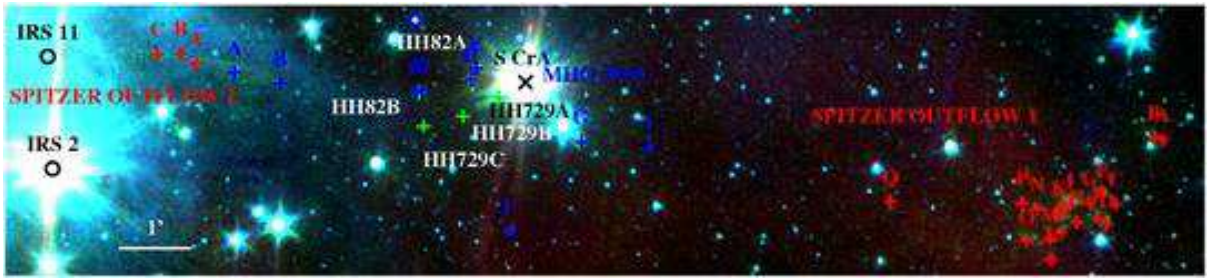}
\caption{\textit{Spitzer} three-color image of the region around 
\textit{Spitzer} outflow\,1, 2 and MHO\,2005.\label{westflow:fig}}
\end{figure}

\textit{Spitzer} outflow\,2, located $\sim$ 1$\arcmin$ west of IRS\,11 (see 
Figure~\ref{spitzer2:fig}), is a curved jet which includes MHO\,2005 A and B, 
but is not detected in our H$_2$ \textit{SofI} map, because it is slightly 
below the detection limit.  It is worth noting that \textit{Spitzer} 
outflow\,1 and 2, along with MHO\,2005, could be part of a larger east-west 
precessing flow (see Figure~\ref{westflow:fig}), possibly driven by a YSO 
inside the Coronet.  On the other hand, as seen in Figure~\ref{westflow:fig}, 
the bulk of \textit{Spitzer} outflow\,1 is well aligned with IRS\,2, and thus 
it is not clear which is the driving source.

\begin{figure}
\includegraphics[width=14.0 cm]{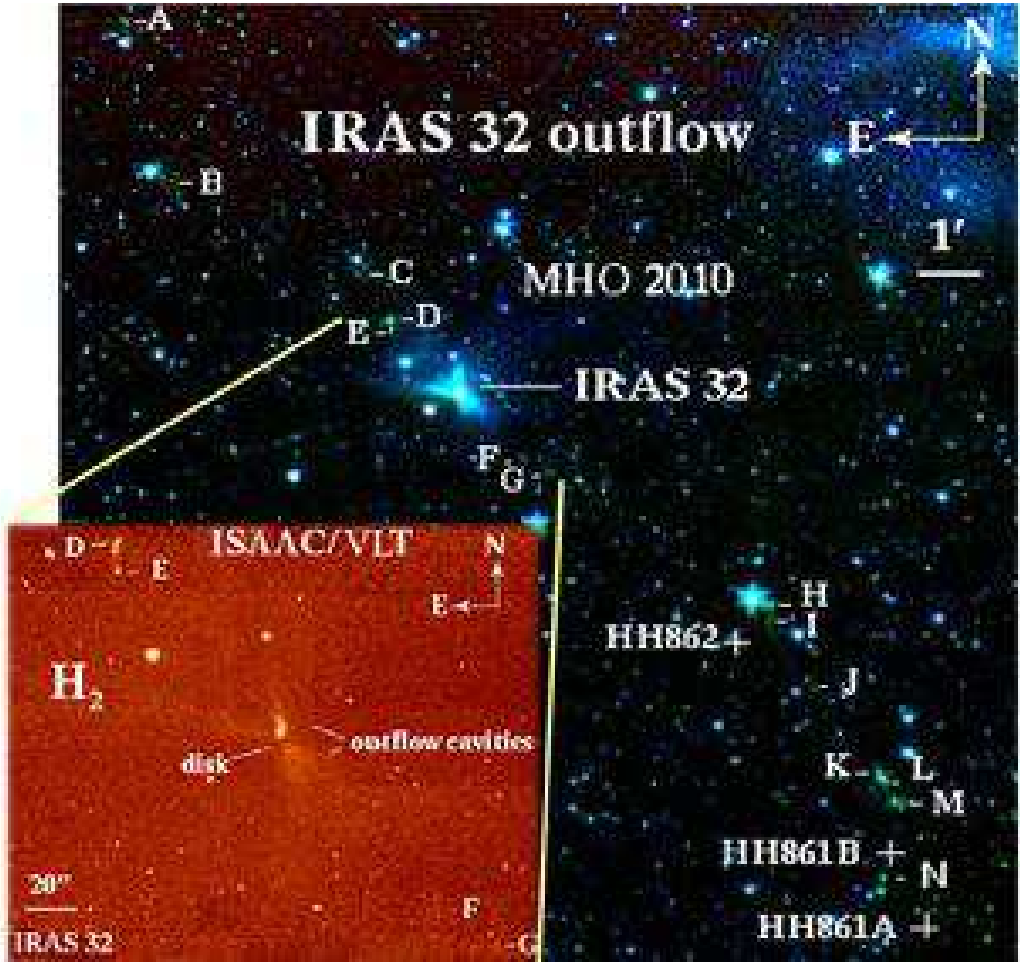}
\caption{\textit{Spitzer} three-color image of the IRAS 32 (IRAS\,18595-3712) 
outflow, made from the IRAC 3.6\,$\mu$m (blue), 4.5\,$\mu$m (green), and 
8\,$\mu$m (red) images.  Labels indicate the position of the newly detected 
knots (EGOs), MHO\,2010 and previously known HH objects.  \textit{Lower left 
panel} displays a close up view of the inner region of the outflow at 
2.12\,$\mu$m, showing the inner knots, the outflow cavities, and a 
possible disk.\label{irasoutflow:fig}}
\end{figure}

The IRAS\,32 outflow, the third \textit{Spitzer} flow discovered 
(P.A.$\sim$39$\degr$), originates from IRAS\,32, and has a clear precessing 
morphology, as seen in Figure~\ref{irasoutflow:fig}.  The outflow extends for 
$\sim$20.3$\arcmin$, i.\,e. it has a projected length of about 0.77\,pc.  
Knots from A to E are part of the north-eastern lobe, while knots from F to N, 
along with HH\,861\,A, B, and HH\,862 are located in the south-western lobe.  
The shock-excited nature of these knots is revealed by our H$_2$ image taken 
with \textit{ISAAC/VLT} (lower left panel of Figure~\ref{irasoutflow:fig}), 
centered around the YSO.  Due to the small FoV, we could only identify the 
closest knots (i.\,e. D-G), which thus we also labeled as MHO\,2010.  Our 
H$_2$ and K$_s$ images also show the two outflow cavities, illuminated by 
scattered light from the YSO.  In between the two cavities, a dark band is 
well visible, possibly indicating the presence of a disk.




\clearpage

%% the file, uncomment the \plotone, \plottwo, and
%% If you need a layout that cannot be achieved with \plotone or
%% \plottwo, you can invoke the graphicx package directly with the
%% \includegraphics command or use \plotfiddle. For more information,
%% please see the tutorial on "Using Electronic Art with AASTeX" in the
%% documentation section at the AASTeX Web site,
%% http://www.journals.uchicago.edu/AAS/AASTeX.
%%
%% The examples below also include sample markup for submission of
%% supplemental electronic materials. As always, be sure to check
%% the instructions to authors for the journal you are submitting to
%% for specific submissions guidelines as they vary from

%% This example uses \plotone to include an EPS file scaled to
%% 80% of its natural size with \epsscale. Its caption
%% has been written to indicate that additional figure parts will be
%% available in the electronic journal.

\clearpage

\clearpage

\clearpage

\clearpage

\clearpage

\clearpage

%% If you are not including electonic art with your submission, you may
%% mark up your captions using the \figcaption command. See the
%% No more than seven \figcaption commands are allowed per page,
%% so if you have more than seven captions, insert a \clearpage

%% Tables should be submitted one per page, so put a \clearpage before

%% Two options are available to the author for producing tables:  the
%% deluxetable environment provided by the AASTeX package or the LaTeX
%% table environment.  Use of deluxetable is preferred.
%%

%% Three table samples follow, two marked up in the deluxetable environment,
%% one marked up as a LaTeX table.

%% In this first example, note that the \tabletypesize{}
%% command has been used to reduce the font size of the table.
%% We also use the \rotate command to rotate the table to
%% landscape orientation since it is very wide even at the
%% Note also that the \label command needs to be placed

%% This table also includes a table comment indicating that the full
%% version will be available in machine-readable format in the electronic

\clearpage

\begin{deluxetable}{crrr}
\tabletypesize{\scriptsize}
\tablecaption{{\it Spitzer} Observations of CrA}
\tablewidth{0pt}
\tablehead{
\colhead{Instrument}& \colhead{AOR ID} & \colhead{PID} & \colhead{Observation Date}
}
\startdata
IRAC & 0003650816 &     6 & 2004-04-20 \\
     & 0017672960 & 30784 & 2006-09-25 \\
     & 0017673472 & 30784 & 2006-09-25 \\
     & 0027041280 & 30574 & 2008-05-10 \\
MIPS & 0003664640 &     6 & 2004-04-11 \\
     & 0017673216 & 30784 & 2007-05-30 \\
     & 0017673728 & 30784 & 2007-05-29 \\
     & 0027042816 & 30574 & 2008-10-23 
\enddata
\label{tab:spitzerobs}
\end{deluxetable}

\begin{deluxetable}{lr}
\tabletypesize{\scriptsize}
\tablecaption{Detection Statistics (S/N$\ge$3) in CrA}
\tablewidth{0pt}
\tablehead{
\colhead{Detected with/in...}& \colhead{Number of Detections}
}
\startdata
...IRAC Band 1 & 73254\\
...IRAC Band 2 & 50096\\
...IRAC Band 3 & 11023\\
...IRAC Band 4 & 6381\\
...MIPS Band 1 & 1919\\
...MIPS Band 2 & 65\\
...MIPS Band 1 and Band 2 & 54\\
...2MASS (total) & 8974\\
...all 4 IRAC bands, but not 2MASS & 350 \\
...2MASS alone & 4 \\
...IRAC, but not 2MASS & 51745 \\
...MIPS Band 1 and 2MASS K$_S$ & 557\\
\enddata
\label{tab:spitzerdetect}
\end{deluxetable}

\clearpage

\begin{deluxetable}{ccccccccc}
\tabletypesize{\scriptsize}
\rotate
\tablecaption{Journal of Observations - Ancillary Near-Infrared Data}
\tablewidth{0pt}
\tablehead{
\colhead{Date of Obs.}& \colhead{Telescope/} & \colhead{Filter} & \colhead{Resolution} & \colhead{Seeing} & \colhead{Exp. Time} & \colhead{FoV} & \colhead{Centered on/}\\
\colhead{(d,m,y)} & \colhead{Instrument} & \colhead{Band} & \colhead{($\arcsec$~pixel$^{-1}$)} & \colhead{($\arcsec$)} & \colhead{(s)} & \colhead{($\arcmin\times\arcmin$)} & \colhead{Target}
}
\startdata
2007 08 23     &    ESO-NTT/SofI     &   H$_2$, K$_s$   &   0.29   & 1.3   & 1440, 290  &   10$\times$10   &  HH 734   \\
2007 08 23     &    ESO-NTT/SofI     &   H$_2$, K$_s$   &   0.29   & 1.3   & 1440, 290  &   10$\times$10   &  HH 730    \\
2007 08 23     &    ESO-NTT/SofI     &   H$_2$, K$_s$   &   0.29   & 1.3   & 1440, 290  &   10$\times$10   &  HH 101    \\ 
2007 06 26     &    ESO-NTT/SofI     &   H$_2$, K$_s$   &   0.29   & 1.1   & 1440, 290  & 10$\times$10     &  HH 99         \\
2005 06 17     &    ESO-VLT/ISAAC    &   H$_2$, K$_s$   &   0.15   & 0.8   & 500, 45    &  5$\times$2.5    &  IRAS\,32   \\
2005 06 16     &    ESO-VLT/ISAAC    &   H$_2$          &   0.15   & 0.8   & 4800       &   5$\times$7     &  R CrA   \\
2003 07 12     &    ESO-VLT/ISAAC    &   H$_2$, NB 2.09 &   0.15   & 0.9   & 1440, 290  &   6.3$\times$6.3 &  HH 101 \\
2000 07 19     &    ESO-VLT/ISAAC    &   H$_2$          &   0.15   & 0.9   & 1200       &   6.8$\times$6   &  HH 99   \\
1999 06 06     &    ESO-NTT/SofI     &   H$_2$, K$_s$   &   0.29   & 0.8   & 600, 100   & 5.2$\times$5.2   &  R CrA\tablenotemark{a}\\
\enddata
\tablenotetext{a}{Data already published in \citet{caratti06}}.
\label{obs_ancillary:tab}
\end{deluxetable}

\clearpage

\begin{deluxetable}{lccccccccrl}
\tabletypesize{\tiny}
\rotate
\tablecaption{Sources Classified as YSO Candidates in CrA based on {\it Spitzer} IRAC and MIPS}
\tablewidth{0pt}
\tablehead{
\colhead{ID} & \colhead{RA} & \colhead{Dec} & \colhead{3.6 $\mu$m} &
\colhead{4.5 $\mu$m} & \colhead{5.8 $\mu$m} & \colhead{8.0 $\mu$m} & 
\colhead{24 $\mu$m} & \colhead{70 $\mu$m} & \colhead{$\alpha$} & 
\colhead{Other Names\tablenotemark{a}}\\
 & \colhead{(J2000)} & \colhead{(J2000)} & \colhead{[mJy]} & \colhead{[mJy]}
& \colhead{[mJy]} & \colhead{[mJy]} & \colhead{[mJy]} & \colhead{[mJy]} & \colhead{(Class)} &
}
\startdata
CrA-1 & 18 56 39.76 & $-$37 07 20.8 & 47.4$\pm$2.4  & 38.8$\pm$2.0  & 37.7$\pm$1.8  & 42.6$\pm$2.1  & 35.6$\pm$3.3  &      \nodata & $-$1.17 (II)  & \\
CrA-2 & 18 57 52.63 & $-$37 14 40.1 & 2.04$\pm$0.12 & 8.90$\pm$0.44 & 35.8$\pm$1.8  & 93.6$\pm$4.6  & 171.$\pm$16.  & 1690$\pm$196 & 1.73 (G)   & Leda Galaxy 90315\\
CrA-3 & 18 59 43.92 & $-$37 04 0.11 & 1.69$\pm$0.09 & 2.22$\pm$0.11 & 2.93$\pm$0.20 & 3.58$\pm$0.19 & 6.97$\pm$0.66 &      \nodata & $-$0.18 (F)   & ISO-CrA 55$^1$\\
CrA-4 & 18 59 50.95 & $-$37 06 31.6 & 4.69$\pm$0.25 & 3.93$\pm$0.20 & 3.16$\pm$0.17 & 2.79$\pm$0.19 & 1.93$\pm$0.22 &      \nodata & $-$1.53 (II)  & DENIS-J185950.9-370632\\
CrA-5 & 19 00 15.55 & $-$36 57 57.7 & 1.07$\pm$0.06 & 2.84$\pm$0.15 & 5.79$\pm$0.30 & 8.04$\pm$0.40 & 13.3$\pm$1.2  &      \nodata & 0.51 (I)   & ISO-CrA 76$^1$\\
CrA-6 & 19 00 29.07 & $-$36 56 03.8 & 51.6$\pm$2.7  & 36.0$\pm$1.8  & 26.1$\pm$1.3  & 18.5$\pm$0.9  & 36.0$\pm$3.3  & 52.0$\pm$6.4 & $-$1.57 (II)  & CrAPMS 8$^2$; GP g2$^3$\\
CrA-7 & 19 00 38.94 & $-$36 58 14.7 & 40.8$\pm$2.2  & 33.8$\pm$1.7  & 25.8$\pm$1.3  & 17.8$\pm$0.9  & 4.49$\pm$0.43 &      \nodata & $-$2.08 (III) & ISO-CrA 93$^1$\\
CrA-8 & 19 00 45.31 & $-$37 11 48.2 & 7.73$\pm$0.40 & 6.66$\pm$0.33 & 5.73$\pm$0.29 & 6.85$\pm$0.34 & 14.6$\pm$1.4  &      \nodata & $-$0.90 (II)  & CrA-444$^4$\\
CrA-9 & 19 00 58.05 & $-$36 45 05.0 & 56.3$\pm$3.3  & 43.8$\pm$2.3  & 31.7$\pm$1.5  & 25.1$\pm$1.2  & 179.$\pm$17.  & 154.$\pm$20.\tablenotemark{b} & $-$1.04 (II) & \\
CrA-10& 19 00 59.75 & $-$36 47 11.2 & 4.05$\pm$0.22 & 3.95$\pm$0.20 & 3.46$\pm$0.18 & 3.87$\pm$0.20 & 4.36$\pm$0.41 & 35.3$\pm$6.0 & $-$1.03 (II)  & CrA-432$^4$\\
CrA-11& 19 01 03.26 & $-$37 03 39.4 & 473.$\pm$29.  & 335.$\pm$18.  & 243.$\pm$12.  & 174.$\pm$9.   & 225.$\pm$21.  &      \nodata & $-$1.75 (III) & HD 176269; HR 7169$^5$ \\
CrA-12& 19 01 16.29 & $-$36 56 28.3 & 8.69$\pm$0.45 & 7.81$\pm$0.38 & 6.77$\pm$0.33 & 6.08$\pm$0.29 & 7.46$\pm$0.70 &      \nodata & $-$1.23 (II)  & V 667 CrA; CrA-4110$^4$ \\% ISO-CrA 123$^1$ \\
CrA-13& 19 01 18.95 & $-$36 58 28.4 & 36.5$\pm$1.9  & 34.8$\pm$1.8  & 36.2$\pm$1.7  & 35.0$\pm$1.7  & 52.8$\pm$4.9  & 54.6$\pm$7.2 & $-$0.92 (II)  & CrA-466$^4$; ISO-CrA 127$^1$\\
CrA-14& 19 01 29.01 & $-$37 01 48.4 & 19.5$\pm$1.0  & 13.8$\pm$0.7  & 9.62$\pm$0.48 & 6.61$\pm$0.32 & 2.93$\pm$0.29 &      \nodata & $-$2.12 (III) & HH 101 IRS 1$^6$; G-94$^7$ \\%; ISO-CrA 137$^1$\\
CrA-15& 19 01 32.31 & $-$36 58 03.0 & 21.2$\pm$1.1  & 18.9$\pm$0.9  & 16.4$\pm$0.8  & 13.8$\pm$0.7  & 16.2$\pm$1.5  &      \nodata & $-$1.19 (II)  & IRS 14$^8$; TS 2.9$^8$; G-87$^7$ \\%; ISO-CrA-139$^1$ \\
CrA-16& 19 01 33.85 & $-$36 57 44.9 & 68.3$\pm$5.1  & 70.9$\pm$3.7  & 66.6$\pm$3.2  & 71.9$\pm$3.6  & 131.$\pm$12.  & 173.$\pm$30. & $-$0.65 (II)  & IRS 13$^8$; TS 2.8$^8$; G-85$^7$\\
CrA-17& 19 01 36.26 & $-$36 58 03.0 & 0.816$\pm$0.047 & 0.678$\pm$0.041 & 0.513$\pm$0.128 & 0.340$\pm$0.060 & 5.73$\pm$0.92\tablenotemark{c} & \nodata & $-$ & removed from list\\
CrA-18& 19 01 40.41 & $-$36 51 42.3 & 49.0$\pm$2.7  & 42.0$\pm$2.1  & 41.3$\pm$2.1  & 46.7$\pm$2.8  & 118.$\pm$14.  &      \nodata & $-$0.70 (II)  & G-65$^7$\\
CrA-19& 19 01 48.03 & $-$36 57 22.2 & 389.$\pm$24.  & 798.$\pm$50.  & 1270.$\pm$76. & 1810.$\pm$107.& 4500.$\pm$570.\tablenotemark{c} &      \nodata & 0.78 (I)   & IRS 5$^8$; TS 2.4$^8$; MMS 12$^9$; SMM 4$^{10}$ \\
CrA-20& 19 01 48.46 & $-$36 57 14.7 & 8.56$\pm$0.51 & 26.2$\pm$1.3  & 52.8$\pm$3.5  & 80.9$\pm$4.2  & 1110.$\pm$155.\tablenotemark{c} &      \nodata & 1.41 (I)   & IRS 5N$^{11}$ \\
CrA-21& 19 01 51.12 & $-$36 54 12.4 & 30.3$\pm$1.5  & 32.6$\pm$1.7  & 37.2$\pm$1.8  & 45.2$\pm$2.2  & 39.9$\pm$3.7  &      \nodata & $-$0.68 (II)  & IRS 8$^8$; TS 2.2$^8$ \\
CrA-22& 19 01 51.86 & $-$37 10 44.7 & 3.75$\pm$0.19 & 3.32$\pm$0.17 & 2.79$\pm$0.15 & 2.27$\pm$0.13 & 1.79$\pm$0.21 &      \nodata & $-$1.47 (II)  & \\
CrA-23& 19 01 53.75 & $-$37 00 33.9 & 3.16$\pm$0.17 & 2.96$\pm$0.15 & 2.74$\pm$0.17 & 3.01$\pm$0.16 & 3.09$\pm$0.34 & 113.$\pm$40.\tablenotemark{d} & $-$1.09 (II) & Star A$^{12}$\\
CrA-24& 19 01 55.60 & $-$36 56 51.1 & 11.0$\pm$1.6  & 30.3$\pm$3.2  & 54.0$\pm$5.2  & 93.1$\pm$13.5\tablenotemark{c} & 297.$\pm$50.\tablenotemark{c}  &      \nodata & 0.60 (I)   &  \\ %J190155.61-365651.1 [FP07]\\
CrA-25& 19 01 57.27 & $-$36 52 05.0 & 0.588$\pm$0.179\tablenotemark{c} & 1.30$\pm$0.38\tablenotemark{c} & 1.43$\pm$0.21 & 1.06$\pm$0.32\tablenotemark{c} & 3.22$\pm$0.47\tablenotemark{c} &    \nodata & HH & removed from list\\
CrA-26& 19 02 06.80 & $-$36 58 41.0 & 2.29$\pm$0.12 & 2.22$\pm$0.11 & 2.35$\pm$0.14 & 2.48$\pm$0.14 & 3.11$\pm$0.64\tablenotemark{c} &      \nodata & $-$0.84 (II)  & \\ %WMB 185844.2-370304\\
CrA-27& 19 02 10.44 & $-$36 53 44.5 & 6.94$\pm$0.36 & 4.55$\pm$0.23 & 3.55$\pm$0.20 & 2.12$\pm$0.11 & 0.962$\pm$0.177 &    \nodata & $-$2.24 (III) & \\ %ISO-CrA 142$^1$\\
CrA-28& 19 02 12.00 & $-$37 03 09.4 & 6.15$\pm$0.31 & 5.04$\pm$0.25 & 4.26$\pm$0.22 & 4.62$\pm$0.24 & 4.63$\pm$0.45 &      \nodata & $-$1.27 (II)  & ISO-CrA 143$^1$; G-14$^7$\\
CrA-29& 19 02 14.63 & $-$37 00 32.9 & 9.21$\pm$0.49 & 11.4$\pm$0.57 & 12.2$\pm$0.59 & 11.8$\pm$0.6  & 14.2$\pm$1.3  &      \nodata & $-$0.64 (II)  & ISO-CrA 145$^1$\\
CrA-30& 19 02 27.07 & $-$36 58 13.1 & 300.$\pm$17.  & 242.$\pm$14.  & 256.$\pm$13.  & 248.$\pm$13.  & 150.$\pm$14.  &      \nodata & $-$1.37 (II)  & H$\alpha$ 14$^{13}$; HBC 680$^{14}$ \\
CrA-31& 19 02 33.07 & $-$36 58 21.2 & 356.$\pm$19.  & 335.$\pm$20.  & 347.$\pm$17.  & 339.$\pm$17.  & 238.$\pm$22.  & 167.$\pm$18.\tablenotemark{d} & $-$1.09 (II) & H$\alpha$ 3$^{13}$; ISO-CrA 159$^1$ \\ %; KS-15E\\
CrA-32& 19 02 56.82 & $-$37 07 19.4 & 0.314$\pm$0.018 & 0.374$\pm$0.021 & 0.351$\pm$0.039 & 0.234$\pm$0.038 & 5.21$\pm$0.92\tablenotemark{c} & \nodata & $-$ & removed from list\\
CrA-33& 19 03 01.03 & $-$37 07 53.4 & 7.04$\pm$0.36 & 10.2$\pm$0.5  & 15.3$\pm$0.8  & 16.9$\pm$0.8  & 32.9$\pm$3.1  &      \nodata & 0.06 (F)   & IRAS 32d$^{15}$ \\ %; ISO-CrA 185$^1$\\
CrA-34& 19 03 09.16 & $-$36 57 22.0 & 149.$\pm$8.   & 121.$\pm$6.   & 92.2$\pm$4.4  & 68.2$\pm$3.4  & 20.4$\pm$1.9  &      \nodata & $-$2.02 (III) & ISO-CrA 198$^1$\\
CrA-35& 19 03 11.84 & $-$37 09 02.1 & 19.6$\pm$1.0  & 14.9$\pm$0.7  & 11.7$\pm$0.6  & 10.4$\pm$0.5  & 11.7$\pm$1.1  &      \nodata & $-$1.46 (II)  & ISO-CrA 201$^1$\\
CrA-36& 19 03 24.29 & $-$37 15 07.7 & 44.9$\pm$2.5  & 44.9$\pm$2.3  & 52.1$\pm$2.7  & 61.3$\pm$3.0  & 132.$\pm$12.  & 96.1$\pm$9.7 & $-$0.39 (II)  & \\
CrA-37& 19 03 55.24 & $-$37 09 35.9 & 1.43$\pm$0.07 & 1.95$\pm$0.10 & 2.63$\pm$0.14 & 3.84$\pm$0.20 & 20.8$\pm$1.9  & 78.1$\pm$8.3\tablenotemark{d} & 0.36 (I) & \\
CrA-38& 18 58 56.40 & $-$37 07 37.9 & 519.$\pm$28.  & 298.$\pm$17.  & 264.$\pm$13.  & 176.$\pm$9.   & 51.3$\pm$4.7  &      \nodata & $-$2.21 (III) & ISO-CrA 13$^1$\\
CrA-39& 19 00 23.47 & $-$37 12 24.2 & 3.81$\pm$0.33 & 2.21$\pm$0.18 & 1.65$\pm$0.15 & 1.96$\pm$0.27 & 6.92$\pm$0.72 & 943.$\pm$97.\tablenotemark{d} & $-$0.95 (G) & 6dFGSgJ190023.5-37\\
CrA-40& 19 01 25.75 & $-$36 59 19.1 & 259.$\pm$18.  & 223.$\pm$12.  & 216.$\pm$11.  & 170.$\pm$8.   & 103.$\pm$10.  &      \nodata & $-$1.38 (II)  & VSSt 18$^{16}$ \\ %; ISO-CrA 134$^1$\\
CrA-41& 19 01 41.62 & $-$36 59 52.7 & 298.$\pm$17.  & 295.$\pm$16.  & 293.$\pm$18.  & 324.$\pm$16.  & 293.$\pm$27.  &      \nodata & $-$1.04 (II)  & H$\alpha$ 2$^{13}$; HBC 677$^{14}$\\
CrA-42& 19 01 50.48 & $-$36 56 38.4 & 106.$\pm$7.   & 143.$\pm$9.   & 179.$\pm$10.  & 196.$\pm$12.  & 348.$\pm$35.  &      \nodata & $-$0.39 (II)  & IRS 6$^8$\\
CrA-43& 19 01 58.54 & $-$36 57 08.5 & 22.5$\pm$1.3  & 72.1$\pm$3.8  & 136.$\pm$8.   & 195.$\pm$11.  & 807.$\pm$78.  &      \nodata & 0.88 (I)   & SMM2$^{10}$; CHLT 15$^{17}$\\
CrA-44& 19 02 58.67 & $-$37 07 35.9 & 6.30$\pm$0.71 & 17.2$\pm$1.2  & 17.3$\pm$1.1  & 12.9$\pm$0.7  & 2040.$\pm$193.& 28500.$\pm$2870 & 1.66 (I) & IRAS 32c$^{15}$\\
CrA-45& 19 03 16.09 & $-$37 14 08.2 & 223.$\pm$12.  & 224.$\pm$12.  & 256.$\pm$13.  & 278.$\pm$14.  & 676.$\pm$63.  & 680.$\pm$69. & $-$0.42 (II)  & VSSt 10$^{16}$; IRAS 34$^{15}$\\
\enddata
\tablenotetext{a}{References are as follows: $^1$\citet{olof99}; $^2$\citet{walter97}; $^3$\citet{gp75}; $^4$\citet{lopezmarti05}; $^5$\citet{knacke73}; $^6$\citet{rw83}; $^7$\citet{s-a08}; $^8$\citet{ts84}; $^9$\citet{chini03}; $^{10}$\citet{nutter05}; $^{11}$\citet{fp07}; $^{12}$\citet{graham93}; $^{13}$\citet{mr81}; $^{14}$\citet{hbc88}; $^{15}$\citet{wilking92}; $^{16}$\citet{vrba76}; $^{17}$\citet{chlt08}.}
\tablenotetext{b}{The flux for CrA-9 was computed using aperture photometry from the 70 $\mu$m image; the catalog entry for this source shows it as a non-detection.  See Section \ref{sec:cra9}}
\tablenotetext{c}{The flux for this source at this wavelength has been ``band-filled'' (described in \S \ref{sec:ysoselect}).}
\tablenotetext{d}{This 70 $\mu$m flux is listed in the catalog with a ``P'' flag, meaning that there were multiple catalog detections near this 70 $\mu$m source.  The catalog source chosen to match the 70 $\mu$m detection gave the most ``consistent'' SED, and was also the one closest to the 70 $\mu$m detection \citep[see the c2d Final Delivery Document,][]{evans07}.}
\label{tab:ysos_wfluxes}
\end{deluxetable}

\clearpage

\begin{deluxetable}{lcccccccccl}
\tabletypesize{\tiny}
\rotate
\tablecaption{Sources Classified as YSO Candidates in CrA based on 2MASS and MIPS 24 $\mu$m}
\tablewidth{0pt}
\tablehead{
\colhead{ID} & \colhead{RA\tablenotemark{a}} & \colhead{Dec\tablenotemark{a}} & \colhead{3.6 $\mu$m} &
\colhead{4.5 $\mu$m} & \colhead{5.8 $\mu$m} & \colhead{8.0 $\mu$m} & 
\colhead{24 $\mu$m} & \colhead{70 $\mu$m} & \colhead{$\alpha$} & 
\colhead{Other Names\tablenotemark{b}}\\
 & \colhead{(J2000)} & \colhead{(J2000)} & \colhead{[mJy]} & \colhead{[mJy]}
& \colhead{[mJy]} & \colhead{[mJy]} & \colhead{[mJy]} & \colhead{[mJy]} & 
\colhead{(Class)} &
}
\startdata
CrA-46 & 18 55 56.32 & $-$37 00 07.1 & \nodata & \nodata & \nodata & \nodata & 15.6$\pm$1.5  & \nodata & $-$0.11 (F) & \\
CrA-47 & 18 56 40.28 & $-$36 55 20.8 & \nodata & \nodata & \nodata & \nodata & 4.46$\pm$0.43 & \nodata & $-$1.04 (II) & \\
CrA-48 & 18 57 07.86 & $-$36 54 04.4 & \nodata & \nodata & \nodata & \nodata & 2.53$\pm$0.26 & \nodata & $-$1.21 (II) & \\
CrA-49 & 18 57 22.48 & $-$37 34 42.5 & \nodata & \nodata & \nodata & \nodata & 3.00$\pm$0.30 & \nodata & $-$0.96 (II) & \\
CrA-50 & 18 57 52.22 & $-$37 14 45.8 & 0.883$\pm$0.053 & 0.567$\pm$0.031 & 0.403$\pm$0.060 & 0.286$\pm$0.098\tablenotemark{c} & 9.87$\pm$1.87 & \nodata & $-$0.66 (G) & Leda Galaxy 90315\\% (extended with CrA-2) \\
CrA-51 & 18 59 34.26 & $-$37 21 41.5 & 1.26$\pm$0.11  & \nodata &  0.848$\pm$0.083  & \nodata & 5.69$\pm$0.55 & \nodata & $-$0.49 (II) & \\
CrA-52 & 19 00 01.58 & $-$36 37 06.2 & \nodata & \nodata & \nodata & \nodata & 124.0$\pm$11.5 & \nodata &  $-$0.95 (II) & CrAPMS 7$^1$; HBC 674$^2$; H$\alpha$ 6$^3$\\
CrA-53 & 19 01 11.49 & $-$36 45 33.8 & 5.99$\pm$0.31 & \nodata & 3.49$\pm$0.19 & \nodata & 4.35$\pm$0.42 & \nodata & $-$1.31 (II) & LM10-1$^4$ \\ 
CrA-54 & 19 01 39.15 & $-$36 53 29.4 & 93.4$\pm$4.9 & 59.4$\pm$3.0 & 67.4$\pm$23.9\tablenotemark{c} & 131.0$\pm$48.4\tablenotemark{c} & 602.$\pm$101. & \nodata & $-$1.01 (II) & HD 176386B \\
CrA-55 & 19 01 55.23 & $-$37 23 41.0 & \nodata & \nodata & \nodata & \nodata & 1420.$\pm$134. & 1540.$\pm$161. & $-$0.51 (II) & DG CrA; HBC 289$^2$; IRAS 24$^5$;\\
& & & & & & & & & & IRAS 18585-3728\\
CrA-56 & 19 02 16.66 & $-$36 45 49.4 & 15.2$\pm$0.8  & \nodata & 7.3$\pm$0.4 & \nodata & 10.4$\pm$1.0 & \nodata & $-$1.39 (II) & ISO-CrA 146$^6$; CrA-4109$^7$; \\
& & & & & & & & & & DENIS-P J190216.6-364549\\
CrA-57 & 19 03 25.48 & $-$36 55 05.3 & 12.4$\pm$0.6 & \nodata & 5.74$\pm$0.29 & \nodata & 5.43$\pm$0.52 & \nodata & $-$1.60 (II) & ISO-CrA 217$^6$\\
\enddata
\tablenotetext{a}{Coordinates from our {\it Spitzer} catalog.}
\tablenotetext{b}{References are as follows: $^1$\citet{walter97}; $^2$\citet{hbc88}; $^3$\citet{mr81}; $^4$\citet{lopezmarti10}; $^5$\citet{wilking92}; $^6$\citet{olof99}; $^7$\citet{lopezmarti05}.}
\tablenotetext{c}{This flux has a quality of detection of ``D'' meaning it has a 2 $\leq$ S/N $<$ 3.  Fluxes that do not have this note all have a quality of detection of ``A'' (S/N $\geq$7), ``B'' (S/N $\geq$5), or ``C'' (S/N $\geq$ 3).}
\label{tab:mips_ysos}
\end{deluxetable}

\clearpage

\begin{deluxetable}{llccccccccll}
\tabletypesize{\tiny}
\rotate
%\begin{landscape}
\tablecaption{Class III YSO Candidates from {\it Spitzer} + ROSAT}
\tablewidth{0pt}
\tablehead{
\colhead{{\it Spitzer} ID} & \colhead{ROSAT ID\tablenotemark{a}} & \colhead{Offset\tablenotemark{b}} & \colhead{RA\tablenotemark{c}} & \colhead{Dec\tablenotemark{c}} & \colhead{3.6 $\mu$m} &
\colhead{4.5 $\mu$m} & \colhead{5.8 $\mu$m} & \colhead{8.0 $\mu$m} & 
\colhead{24 $\mu$m} & \colhead{$\alpha$} & \colhead{Other Names\tablenotemark{d}}\\
 & \colhead{RX \#}  & \colhead{$^{\prime\prime}$} & \colhead{(J2000)} & \colhead{(J2000)} & \colhead{[mJy]} & \colhead{[mJy]}
& \colhead{[mJy]} & \colhead{[mJy]} & \colhead{[mJy]}  &  &
}
\startdata
CrA-58 & J1855.3-3730 & 6.26 & 18 55 18.91 & $-$37 29 56.7 & 1350.$\pm$77. & 804.$\pm$50. & 610.$\pm$34. & 338.$\pm$18. & 34.9$\pm$3.2 & $-$2.86 & HD 175073; LTT 7499\\
CrA-59\tablenotemark{e} & J1857.5-3732 & 2.54 & 18 57 34.19 & $-$37 32 33.3 & \nodata & \nodata & \nodata & \nodata & 0.826$\pm$0.144 & $-$2.76 & \\ 
CrA-60 & J1857.7-3719 & 13.1 & 18 57 45.15 & $-$37 19 37.9 & 32.0$\pm$1.6 & 22.7$\pm$1.1 & 15.7$\pm$0.8 & 9.00$\pm$0.43 & 0.882$\pm$0.150 & $-$2.68 & [P98c] R01c$^1$ \\
CrA-61 & J185802.3-365348 & 6.29 & 18 58 01.82 & $-$36 53 45.4 & \nodata & \nodata & \nodata & \nodata & 1.87$\pm$0.21 & $-$2.82 & CrAPMS 5$^2$\\ 
CrA-62\tablenotemark{f} & J1858.7-3706  & 6.24 & 18 58 43.33 & $-$37 06 27.3 & 5980.$\pm$690.\tablenotemark{g} & 2370.$\pm$286. & 3120.$\pm$158. & 1960.$\pm$111. & 228.$\pm$21. & $-$2.67 & $\epsilon$ CrA; HR 7152; R04$^1$ \\
CrA-63 & J185856.2-364032\tablenotemark{h} & 4.35 & 18 58 56.51 & $-$36 40 29.7 & \nodata & \nodata & \nodata & \nodata & 0.976$\pm$0.207 & $-$2.67 & [P98c] R17c$^1$ \\
CrA-64 & J1859.2-3711  & 1.04 & 18 59 14.82 & $-$37 11 32.6 & 21.6$\pm$1.3 & 14.6$\pm$0.9 & 9.15$\pm$0.52 & 5.90$\pm$0.32 & 0.979$\pm$0.138 & $-$2.74 & R02$^1$; CrAPMS 6$^2$; H$\alpha$ 11$^3$ \\
CrA-65 & J190039.0-364805 & 8.17 & 19 00 39.30 & $-$36 48 12.3 & 27.2$\pm$1.4 & 18.0$\pm$0.9 & 12.4$\pm$0.6 & 7.45$\pm$0.37 & 0.898$\pm$0.146 & $-$2.64 & R14$^1$; CrAPMS 9$^2$ \\
CrA-66 & J190109.4-364751 & 4.00 & 19 01 09.70 & $-$36 47 52.9 & 118.$\pm$7.  & \nodata & 57.1$\pm$2.8 & \nodata & 3.58$\pm$0.34 & $-$2.79 & VSS VIII-27$^4$\\
CrA-67 & J1901.4-3704 & 8.60 & 19 01 25.63 & $-$37 04 53.7 & 11.5$\pm$0.6 & 8.56$\pm$0.42 & 5.94$\pm$0.29 & 3.70$\pm$0.18 & 0.371$\pm$0.163\tablenotemark{i,j} & $-$2.40 & ISO-CrA 133$^5$\\
CrA-68 & J190126.8-365911 & 4.88 & 19 01 27.15 & $-$36 59 08.5 & 37.1$\pm$2.0 & 26.2$\pm$1.3 & 18.4$\pm$0.9 & 11.0$\pm$0.5 & 1.12$\pm$0.18\tablenotemark{i} & $-$2.62 & R07a$^1$; ISO-CrA 135$^5$ \\
CrA-69 & J190128.6-365933 & 1.82 & 19 01 28.71 & $-$36 59 31.8 & 95.1$\pm$5.5 & 66.1$\pm$3.4 & 47.7$\pm$2.3 & 28.9$\pm$1.4 & 2.86$\pm$0.29 & $-$2.66 &  ISO-CrA 136$^5$; CXO 12$^6$ \\
CrA-70 & J190134.6-370058 & 3.27 & 19 01 34.84 & $-$37 00 56.5 & 258.$\pm$15. & 152.$\pm$8. & 111.$\pm$6. & 67.5$\pm$3.4 & 7.87$\pm$0.74 & $-$2.74 & CrAPMS 1$^2$; VSSt 13$^7$ \\
CrA-71 & J190139.0-370207 & 3.88 & 19 01 39.32 & $-$37 02 07.4 & 25.9$\pm$1.4 & 16.8$\pm$0.9 & 11.5$\pm$0.6 & 6.65$\pm$0.33 & 0.718$\pm$0.193\tablenotemark{i} & $-$2.64 &[GP75] R CrA o$^8$ \\
CrA-72 & J190140.3-364429 & 4.20 & 19 01 40.55 & $-$36 44 32.0 & 91.9$\pm$5.0 & \nodata & 40.9$\pm$2.1 & \nodata & 2.92$\pm$0.30 & $-$2.72 & VSS VIII-26$^4$\\
CrA-73 & J190200.1-370227 & 4.93 & 19 02 00.11 & $-$37 02 22.1 & 13.7$\pm$0.7 & 9.21$\pm$0.46 & 6.38$\pm$0.34 & 3.87$\pm$0.22 & 0.491$\pm$0.250\tablenotemark{i,k} & $-$2.53 & H$\alpha$ 13$^3$\\
CrA-74 & J190201.7-370746 & 4.02 & 19 02 01.96 & $-$37 07 43.5 & 129.$\pm$7.  & 76.6$\pm$4.3  & 56.2$\pm$2.7  & 31.6$\pm$1.6  & 3.88$\pm$0.38 & $-$2.79 & V702 CrA; CrAPMS 2$^2$\\
CrA-75 & J190222.0-365541 & 1.41 & 19 02 22.12 & $-$36 55 40.9 & 65.4$\pm$3.4 & 43.2$\pm$2.2  & 30.5$\pm$1.5  & 19.0$\pm$0.9  & 2.22$\pm$0.25 & $-$2.66 & VSSt 24$^7$; HBC 679$^9$ \\
CrA-76 & J190249.7-364618 & 5.79 & 19 02 49.80 & $-$36 46 23.7 & 37.0$\pm$2.0 & \nodata       & 15.7$\pm$0.8  & \nodata       & 1.07$\pm$0.17 & $-$2.69 & ISO-CrA 174$^5$ \\
CrA-77 & J190408.2-370249 & 3.20 & 19 04 07.95 & $-$37 02 47.8 & 36.7$\pm$2.0 & \nodata       & 16.7$\pm$0.8  & \nodata       & 1.04$\pm$0.15 & $-$2.76 & VSS VIII-21$^4$\\
\enddata
\tablenotetext{a}{From Patten et al. (private communication).  The shorter coordinate names are sources from the PSPC (Position Sensitive Proportional Counter) catalog (they are less precise than the longer coordinate names). All the ROSAT IDs have a prefix of ``RX''}
\tablenotetext{b}{Offset is calculated as the distance between our {\it Spitzer} position and the ROSAT position from Patten et al. (priv. comm).  Note that the ROSAT positions are computed from centroiding using the PROS package in IRAF.}
\tablenotetext{c}{Coordinates from our {\it Spitzer} catalog.}
\tablenotetext{d}{References are as follows: $^1$\citet{patten98}; $^2$\citet{walter97}; $^3$\citet{mr81}; $^4$\citet{vss76}; $^5$\citet{olof99}; $^6$\citet[][CXO is the listing number from Table 2 their paper]{fp07}; $^7$\citet{vrba76}; $^8$\citet{gp75}; $^9$\citet{hbc88}}
\tablenotetext{e}{In Table 13 of \citet{fern08}, CrA-59 (RX J1857.5-3732) is referred to as a binary, where the eastern component has a spectral type of M5 and the western component has a spectral type of M6.  Only one source is detected in our {\it Spitzer} observations.  Note that this source does not have any IRAC fluxes because it is off the field in the IRAC images (the MIPS field extends a little further).}
\tablenotetext{f}{\citet{patten98} says this is not a member because it is an eclipsing binary and if you work out the numbers, it is likely foreground.  This could be true because these *are* very bright fluxes.]}
\tablenotetext{g}{This flux is flagged ``K'' in the catalog meaning that it is complex, i.e. there are two or more detections within 2$^{\prime\prime}$ in the same band in the same epoch.  The quality is likely still quite good, however.}
\tablenotetext{h}{There is also a PSPC coordinate name for this source, RX J1858.9-3640, and although it is about 42'' away, it is the same source due to imprecise PSPC coordinates.}
\tablenotetext{i}{The flux for this source at this wavelength has been ``band-filled'' (described in \S \ref{sec:ysoselect}).}
\tablenotetext{j}{This flux has a quality of detection of ``D'' meaning it has a 2 $\leq$ S/N $<$ 3.  Fluxes that do not have this note all have a quality of detection of ``A'' (S/N $\geq$7), ``B'' (S/N $\geq$5), or ``C'' (S/N $\geq$ 3).}
\tablenotetext{k}{This flux is flagged ``U'' in the catalog indicating that it is an upper limit because it has been band-filled and has a S/N $<$ 5.}
\label{tab:classIII}
%\end{landscape}
\end{deluxetable}

\clearpage

\begin{deluxetable}{lcccccccccl}
\tabletypesize{\tiny}
\rotate
\tablecaption{{\it Spitzer} Matches to Known YSOs and Candidate YSOs in CrA}
\tablewidth{0pt}
\tablehead{
\colhead{Name\tablenotemark{a}}  & \colhead{RA\tablenotemark{b}} & \colhead{Dec\tablenotemark{b}} & \colhead{3.6 $\mu$m} &
\colhead{4.5 $\mu$m} & \colhead{5.8 $\mu$m} & \colhead{8.0 $\mu$m} & 
\colhead{24 $\mu$m}  & \colhead{$\alpha$} & \colhead{c2d Catalog} & \colhead{Other Names\tablenotemark{c}}\\
 & \colhead{(J2000)} & \colhead{(J2000)} & \colhead{[mJy]} & \colhead{[mJy]}
& \colhead{[mJy]} & \colhead{[mJy]} & \colhead{[mJy]} & (Class) & \colhead{Designation} &
}
\startdata
R CrA & 19 01 53.67 & $-$36 57 08.0 & $\cdots$\tablenotemark{d,e} & $\cdots$\tablenotemark{d,e} & 10500.$\pm$1980 & $\cdots$\tablenotemark{d,e} & 76.1$\pm$644.0\tablenotemark{d,e} & $-$2.55 (III)\tablenotemark{f} & star & TS 2.10$^1$; VSSt 1$^2$\\ 
S CrA & 19 01 08.62 & $-$36 57 20.3 & 2580.$\pm$187. & 3040.$\pm$218. & 3510.$\pm$224. & 2640.$\pm$234. & $<$195\tablenotemark{d,e} & $-$0.80 (II) & star+dust(IR1) & VSSt 3$^2$; SMM 7$^3$ \\ 
T CrA & 19 01 58.78 & $-$36 57 49.9 & 1930.$\pm$131. & 2110.$\pm$127. & 2360.$\pm$130. & 3520.$\pm$202. & \nodata & $-$0.35 (II) & rising & TS 4.3$^1$, VSSt 2$^2$\\
VV CrA & 19 03 06.80 & $-$37 12 49.1 & 1570.$\pm$203. & 2900.$\pm$479. & 12200.$\pm$915. & 6770.$\pm$678. & $<$5800\tablenotemark{d,e} & 0.36 (I) & rising & VSSt 4$^2$; SMM 9$^3$ \\
TY CrA & 19 01 40.81 & $-$36 52 33.8 & 844.$\pm$82. & 631.$\pm$59. & 1060.$\pm$114. & 2900.$\pm$372. & \nodata & $-$0.76 (II) & PAH-em & VSSt 5$^2$\\
HD 176270 & 19 01 04.30 & $-$37 03 41.7 & 684.$\pm$42. & 468.$\pm$25. & 359.$\pm$18. & 190.$\pm$9. & 21.7$\pm$2.1 & $-$2.81 (III) & star & HR 7170$^4$\\
IRS 1\tablenotemark{g} & 19 01 50.68 & $-$36 58 09.7 & 2960.$\pm$352. & 4520.$\pm$526. & 14000.$\pm$1150. & $<$150\tablenotemark{d,e} & 22300.$\pm$4600.\tablenotemark{d} & 0.92 (I) & rising & VSSt 15$^2$; TS 2.6$^1$\\ 
IRS 2 & 19 01 41.56 & $-$36 58 31.2 & 1770.$\pm$235. & 2330.$\pm$239. & 6580.$\pm$375. & 7940.$\pm$506. & $<$1450\tablenotemark{d,e} & 0.75 (I) & rising & TS 13.1$^1$; SMM 5$^3$ \\
IRS 7A (=7W) & 19 01 55.32 & $-$36 57 21.9 & 50.9$\pm$4.5 & 239.$\pm$24. & 546.$\pm$43. & 1050.$\pm$78. & $<$2790\tablenotemark{d,e} & 2.64 (I) & rising & \\
IRS 7B (=7E) & 19 01 56.40 & $-$36 57 28.1 & 36.7$\pm$2.6 & 257.$\pm$18. & 697.$\pm$38. & 957.$\pm$51. & \nodata & 2.78 (I) & rising & \\
IRS 10 & 19 02 04.09 & $-$36 57 01.2 & 5.28$\pm$0.31 & 4.49$\pm$0.24 & 3.70$\pm$0.37 & 2.66$\pm$0.16 & 1.27$\pm$0.83\tablenotemark{d,e} & $-$1.48 (II) & star & TS 4.2$^1$; [HHD2008]A 569$^5$ \\
IRS 11 & 19 01 41.89 & $-$36 56 58.4 & 3.77$\pm$0.20 & 3.02$\pm$0.15 & 2.32$\pm$0.17 & 1.47$\pm$0.13 & $<$1.9\tablenotemark{d,e} & $-$1.77 (III) & star & TS 2.5$^1$; [HHD2008]A 367$^5$ \\
IRS 15 & 19 01 47.89 & $-$36 59 30.1 & 7.76$\pm$0.39 & 5.29$\pm$0.27 & 4.66$\pm$0.24 & 2.65$\pm$0.14 & 0.0966$\pm$0.5590\tablenotemark{d,e} & $-$2.01 (III) & star & TS 13.4$^1$; GP n$^6$ \\
VSSt 8 & 19 01 37.27 & $-$36 49 21.0 & 206.$\pm$11. & 131.$\pm$7. & 100.$\pm$5. & 62.3$\pm$3.2 & 6.45$\pm$0.61 & $-$2.67 (III) & star & TS 1.8$^1$; GP q$^6$  \\
H$\alpha$ 10 & 18 58 51.86 & $-$37 19 22.7 & 7.80$\pm$0.39 & 4.70$\pm$0.23 & 3.18$\pm$0.17 & 2.04$\pm$0.11 & 0.132$\pm$0.136\tablenotemark{d,e} & $-$2.68 (III) & star & HBC 673$^7$\\
H$\alpha$ 15 & 19 04 17.25 & $-$36 59 03.2 & 20.5$\pm$1.1 & \nodata & 13.4$\pm$0.7 & \nodata & 17.3$\pm$1.6 & $-$1.24 (II) & star+dust(IR3) & \\
ISO-CrA 46 & 18 59 31.79 & $-$36 50 26.3 & 39.1$\pm$2.0 & 24.6$\pm$1.2 & 17.7$\pm$0.9 & 10.5$\pm$0.5 & 1.11$\pm$0.17 & $-$2.70 (III) & star & \\
ISO-CrA 53 & 18 59 40.99 & $-$36 56 01.4 & 18.9$\pm$0.9 & 11.5$\pm$0.6 & 7.85$\pm$0.39 & 4.79$\pm$0.24 & 0.399$\pm$0.154\tablenotemark{d,h} & $-$2.73 (III) & star & [P98c] R13a$^8$\\
ISO-CrA 140 & 19 02 10.03 & $-$36 56 49.0 & 1.32$\pm$0.08 & 1.40$\pm$0.09 & 1.35$\pm$0.31 & 0.662$\pm$0.097\tablenotemark{d} & 0.375$\pm$0.290\tablenotemark{d,e} & $-$1.66 (III) & star & 185847.7-370111$^9$\\
ISO-CrA 177 & 19 02 54.64 & $-$36 46 19.2 & 14.5$\pm$0.7 & \nodata & 11.9$\pm$0.6 & \nodata & 11.2$\pm$1.0 & $-$1.19 (II) & star+dust(IR3) & CrA-4107$^{10}$\\
CXO 23 & 19 01 43.07 & $-$36 50 18.9 & 0.549$\pm$0.032 & 0.370$\pm$0.023 & 0.279$\pm$0.054 & 0.121$\pm$0.104\tablenotemark{d,e} & 0.178$\pm$0.230\tablenotemark{d,e} & $-$2.18 (III) & star & \\
CXO 34 & 19 01 55.77 & $-$36 57 27.5 & 11.1$\pm$1.1 & 41.3$\pm$2.7 & 71.5$\pm$4.0 & 66.5$\pm$5.0 & \nodata & 0.87 (I) & cup-down & \\
CXO 37 & 19 01 57.40 & $-$37 03 11.8 & 3.00$\pm$0.16 & 2.31$\pm$0.12 & 1.67$\pm$0.10 & 1.03$\pm$0.07 & 0.150$\pm$0.206\tablenotemark{d,e} & $-$2.10 (III) & star & 185834.7-370735$^9$\\
CXO 38 & 19 01 58.32 & $-$37 00 26.7 & 3.25$\pm$0.16 & 2.68$\pm$0.15 & 2.19$\pm$0.14 & 1.30$\pm$0.09 & 0.045$\pm$0.258\tablenotemark{d,e} & $-$1.71 (III) & star & G-32$^{11}$; 185835.7-370450$^9$\\
CXO 42 & 19 02 01.96 & $-$36 54 00.0 & 0.292$\pm$0.026 & 0.258$\pm$0.024 & 0.217$\pm$0.140\tablenotemark{d,e} & 1.04$\pm$0.10 & 3.31$\pm$0.35 & 0.38 (I) & red & 185839.6-365823$^9$ \\
CrA-205 & 19 01 11.72 & $-$37 22 21.5 & \nodata & 2.89$\pm$0.16 & \nodata & 1.06$\pm$0.09 & 0.163$\pm$0.153\tablenotemark{d,e} & $-$2.49 (III) & star & \\
CrA-452  & 19 00 44.55 & $-$37 02 11.0 & 24.4$\pm$1.2 & 16.5$\pm$0.8 & 11.7$\pm$0.6 & 7.01$\pm$0.35 & 0.951$\pm$0.236\tablenotemark{d} & $-$2.52 (III) & star & ISO-CrA 98$^{12}$ \\
CrA-453  & 19 01 04.62 & $-$37 01 29.4 & 5.74$\pm$0.29 & 4.31$\pm$0.22 & 3.03$\pm$0.15 & 1.82$\pm$0.10 & 0.00174$\pm$0.13000\tablenotemark{d,e} & $-$2.36 (III) & star & CXO 1$^{13}$\\
CrA-468  & 19 01 49.35 & $-$37 00 28.6 & 8.30$\pm$0.42 & 5.94$\pm$0.29 & 3.91$\pm$0.20 & 2.32$\pm$0.12 & 0.453$\pm$0.256\tablenotemark{d,e} & $-$2.52 (III) & star & CXO 26$^{13}$; LS-RCrA 2$^{14}$ \\
CrA-4108 & 19 02 09.67 & $-$36 46 34.5 & 7.85$\pm$0.42 & \nodata & 3.88$\pm$0.20 & \nodata & 0.351$\pm$0.173\tablenotemark{d,e} & $-$2.40 (III) & star & ISO-CrA 141$^{12}$ \\
CrA-4111 & 19 01 20.85 & $-$37 03 02.9 & 4.16$\pm$0.22 & 3.08$\pm$0.15 & 2.16$\pm$0.12 & 1.35$\pm$0.09 & 0.568$\pm$0.186 & $-$2.29 (III) & star\_no\_MP1 & CXO 7$^{13}$ \\
Haas 4 & 19 01 40.67 & $-$36 56 04.9 & 0.978$\pm$0.068 & 1.41$\pm$0.09 & 1.38$\pm$0.12 & 1.77$\pm$0.29 & 0.458$\pm$0.338\tablenotemark{d,e} & $-$0.28 (F) & cup-down & \\
Haas 7 & 19 01 43.67 & $-$36 57 17.2 & 0.477$\pm$0.032 & 0.426$\pm$0.030 & 0.284$\pm$0.071 & 0.127$\pm$0.095\tablenotemark{d,e} & $<$35\tablenotemark{d,e} & $-$1.70 (III) & star & \\
Haas 17 & 19 01 51.74 & $-$36 55 14.2 & 0.724$\pm$0.066 & 0.579$\pm$0.038 & 0.484$\pm$0.081 & 0.236$\pm$0.125\tablenotemark{d,e} & $<$8.3\tablenotemark{d,e} & $-$1.48 (II) & star & \\
Haas 33 & 19 02 00.85 & $-$36 58 50.6 & 2.02$\pm$0.11 & 1.36$\pm$0.07 & 1.11$\pm$0.09 & 0.619$\pm$0.069 & $<$5\tablenotemark{d,e} & $-$2.17 (III) & star & 185838.4-370312$^9$\\
Haas 34 & 19 02 01.41 & $-$36 58 51.7 & 0.685$\pm$0.042 & 0.445$\pm$0.031 & 0.407$\pm$0.065 & 0.220$\pm$0.059 & $<$4\tablenotemark{d,e} & $-$2.02 (III) & star & 185838.9-370314$^9$\\ 
LS-RCrA 1\tablenotemark{i} & 19 01 33.56 & $-$37 00 30.3 & 1.25$\pm$0.07 & 1.20$\pm$0.06 & 1.33$\pm$0.09 & 2.01$\pm$0.12 & 12.2$\pm$1.2  & $-$0.08 (F) & Galc\_star+dust(IR2) & G09 CrA-18$^{15}$ \\
G09 CrA-9 & 19 01 58.32 & $-$37 01 06.0 & 0.302$\pm$0.019 & 0.453$\pm$0.028 & 0.469$\pm$0.055 & 0.336$\pm$0.051 & 0.0422$\pm$0.2480\tablenotemark{d,e} & $-$0.55 (II) & cup-down & \\
G09 CrA-11 & 19 01 01.11 & $-$36 58 34.5 & 1.08$\pm$0.06   & 0.777$\pm$0.042 & 0.544$\pm$0.044 & 0.162$\pm$0.061\tablenotemark{h} & 0.0151$\pm$0.1610\tablenotemark{d,e} & $-$2.38 (III) & star & \\
GP p\tablenotemark{j}   &  19 01 38.91 & $-$36 53 26.4 & 583.$\pm$33. & 366.$\pm$23. & 272.$\pm$50.\tablenotemark{d} & 296.$\pm$74.\tablenotemark{d} & \nodata & $-$2.63 (III) & star & HD 176386\\
GP wa\tablenotemark{k} & 19 02 22.43 & $-$36 55 38.6 & 14.2$\pm$1.1\tablenotemark{l} & 10.3$\pm$0.5 & 6.78$\pm$0.35 & 4.14$\pm$0.23 & 0.316$\pm$0.208\tablenotemark{d,e} & $-$2.44 (III) & star & CrAPMS 3B$^{16}$ \\
LM10-2 & 19 01 39.11 & $-$37 00 17.3 & 6.24$\pm$0.32 & 4.70$\pm$0.24 & 3.33$\pm$0.19 & 2.07$\pm$0.12 & 0.317$\pm$0.224\tablenotemark{d,e} & $-$2.19 (III) & star & 185816.6-370439$^9$\\
LM10-3 & 19 01 51.80 & $-$37 10 47.9 & 14.8$\pm$1.0 & 10.1$\pm$0.5 & 6.77$\pm$0.34 & 3.99$\pm$0.21 & 0.088$\pm$0.208\tablenotemark{d,e} & $-$2.52 (III) & star & \\
LM10-4 & 19 02 07.68 & $-$37 11 55.9 & 4.35$\pm$0.23 & 3.14$\pm$0.16 & 2.40$\pm$0.13 & 1.33$\pm$0.08 & 0.302$\pm$0.169\tablenotemark{d,e} & $-$2.30 (III) & star & \\
LM10-5 & 19 02 39.40 & $-$36 53 11.2 & 6.11$\pm$0.31 & 4.23$\pm$0.21 & 3.02$\pm$0.19 & 1.74$\pm$0.10 & 0.229$\pm$0.153\tablenotemark{d,e} & $-$2.44 (III) & star & ISO-CrA 163$^{12}$\\
\enddata
\tablenotetext{a}{The name listed in this column can be used to search for the source in SIMBAD, except in the following cases: IRS \citep{ts84}, H$\alpha$ \citep{mr81}, CXO \citep[][listing number from their Table 2]{fp07}, CrA- \citep{lopezmarti05}; Haas \citep[][listing number from their Table 3]{haas08}; G09 CrA- \citep[][listing number from their Table 4]{rguter09}; GP \citep{gp75}; LM10- \citep{lopezmarti10}.  There are 3 known YSOs that do not appear in this list because they do not have {\it Spitzer} counterparts.  These sources are (SIMBAD coordinates in J2000):  IRS 9 \citep[RA: 19 01 52.63, Dec: $-$36 57 00.2;][]{ts84}, SMA\,2 (RS\,9) \citep[RA: 19 01 55.28, Dec: $-$36 57 16.6;][]{brown87}, and Anon 2 \citep[RA: 19 01 06.92, Dec: $-$36 58 07.3;][]{knacke73}.}
\tablenotetext{b}{Coordinates from our {\it Spitzer} catalog.}
\tablenotetext{c}{References are as follows: $^1$\citet{ts84}; $^2$\citet{vrba76}; $^3$\citet{nutter05}; $^4$\citet{knacke73}; $^5$\citet{haas08}; $^6$\citet{gp75}; $^7$\citet{hbc88}; $^8$\citet{patten98}; $^9$\citet{wilking97}; $^{10}$\citet{lopezmarti05}; $^{11}$\citet{s-a08}; $^{12}$\citet{olof99}; $^{13}$\citet{fp07}; $^{14}$\citet{fc01}; $^{15}$\citet{rguter09}; $^{16}$\citet{walter97}.}
\tablenotetext{d}{The flux for this source at this wavelength has been ``band-filled'' (described in \S \ref{sec:ysoselect}).}
\tablenotetext{e}{This flux is flagged ``U'' in the catalog indicating that it is an upper limit because it has been band-filled and has a S/N $<$ 5.}
\tablenotetext{f}{Due to band-filling and saturation, the spectral index for R\,CrA has been derived from the K$_s$ and 5.8 $\mu$m bands alone.  Its value and resulting classification are likely incorrect.  It should be noted that R\,CrA has long been known to be a PMS star with an accretion disk \citep[e.g.][]{knacke73}, and an SED slope of a Class~II \citep{hillenbrand92}.}
\tablenotetext{g}{IRS 1 is the only source in this table that has a 70 $\mu$m flux in the catalog; it is an excellent ``A'' quality flux with a value of: 33700.$\pm$4700.}
\tablenotetext{h}{This flux has a quality of detection of ``D'' meaning it has a 2 $\leq$ S/N $<$ 3.  Fluxes that do not have this note all have a quality of detection of ``A'' (S/N $\geq$7), ``B'' (S/N $\geq$5), or ``C'' (S/N $\geq$ 3).}
\tablenotetext{i}{Discussed in \citet{fc01}.}
\tablenotetext{j}{This source is a double with CrA-54 (HD 176386B).}
\tablenotetext{k}{This source name comes from \citet{wilking97}; it is located to the northeast of source GP w (called GP wb in the same paper, where GP is from \citet{gp75}) and was also seen by \citet{walter97} who call the two sources CrAPMS 3 and 3B.  CrAPMS 3 is also a YSO candidate, and can be found in Table \ref{tab:classIII} as CrA-75.  It seems that in \citet{mw09} there was some confusion when referring to GP wa - in that paper, it actually corresponds to our CrA-75.}
\tablenotetext{l}{This flux is flagged ``K'' in the catalog meaning that it is complex, i.e. there are two or more detections within 2$^{\prime\prime}$ in the same band in the same epoch.  The quality is likely still quite good, however.}
\label{tab:knownysos_spitzercounterpart}
\end{deluxetable}

\clearpage

\begin{deluxetable}{lccc}
\tablecaption{Clustering Analysis Comparison}
\tablewidth{0pt}
\tablehead{ & \colhead{G09\tablenotemark{a}} & \colhead{All YSO candidates} & \colhead{Class I/II YSOs} \\
& \colhead{(35 sources)} & \colhead{(116 sources)} & \colhead{(62 sources)}
}
\startdata
L$_{crit}$ & 0.116 pc & 0.164 pc & 0.152 pc \\
N$_{cores}$ (\# members) & 1 (24)& 1 (68) & 1 (34)\\
R$_{hull} (core)$ & 0.200 pc & 0.385 pc & 0.237 pc\\
R$_{circ} (core) $ & 0.317 pc & 0.590 pc & 0.326 pc\\
Aspect Ratio\tablenotemark{b} & 2.51 & 2.36 & 1.89\\
Density\tablenotemark{c} & 190 pc$^{-2}$ & 150 pc$^{-2}$ & 190 pc$^{-2}$\\
\enddata
\tablenotetext{a}{Values from \citet{rguter09}.  Note that they have been modified for a distance of 130 pc (a distance of 170 pc was used in their paper).}
\tablenotetext{b}{Aspect Ratio = (R$_{circ}$/R$_{hull}$)$^2$}
\tablenotetext{c}{Density = Number of members in each core / ($\pi \times$ R$_{hull}^2$).}
\label{tab:clustering_analysis}
\end{deluxetable}
\clearpage

\begin{deluxetable}{llcccc}
\tablewidth{0pt}
\tablecaption{Submillimeter Array Results \label{tab-sma}}
\tablehead{
\colhead{Source} & \colhead{Other Name} & \colhead{RA
(J2000)\tablenotemark{a}} &
\colhead{Dec (J2000)\tablenotemark{a}} & 
\colhead{$F_{\rm 1.3mm}$\tablenotemark{b,c}} & 
\colhead{$M_{\rm disk}$\tablenotemark{d}} \\
& & \colhead{~$h$~~$m$~~$s$}      &
\colhead{\phs\phs\arcdeg~~~\arcmin~~~\arcsec} & \colhead{[mJy]} &
\colhead{[M$_{\sun}$]}
}
\startdata
CrA-20 & IRS 5N & 19 01 48.47 & $-$36 57 14.9 & 95 (7) & 0.023 \\
& R CrA & 19 01 53.65 & $-$36 57 08.6 & 36 (25) & 0.009 \\
& SMA\,2 & 19 01 55.29 & $-$36 57 17.0 & 269 (20) & 0.063 \\
& IRS 7B & 19 01 56.40 & $-$36 57 28.3 & 320 (19) & 0.076 \\
CrA-43 & WMB 55 & 19 01 58.52 & $-$36 57 09.3 & 135 (27) & 0.032 \\
CrA-44 & IRAS 32 & 19 02 58.68 & $-$37 07 36.7 & 103 (8) & 0.024 
\enddata
\tablenotetext{a}{RA, Dec values from point source fits to the
continuum image data.}
\tablenotetext{b}{Point source fit to continuum images using baselines
longer than 20 k$\lambda$.}
\tablenotetext{c}{The uncertainties are reported in braces.  These are
the uncertainties returned by the fitting program IMFIT in MIRIAD.}
\tablenotetext{d}{Disk mass assuming a dust temperature of 30 K.}
\end{deluxetable}

\clearpage

\begin{deluxetable}{lcclcc}
\tabletypesize{\scriptsize}
\tablecaption{Known HH objects in CrA}
\tablewidth{0pt}
\tablehead{
\colhead{Knot}   & \colhead{R.A.}    & \colhead{Dec.}     & \colhead{Knot}   & \colhead{R.A.}    & \colhead{Dec.} \\
\colhead{Name}   & \colhead{(J2000)} & \colhead{(J2000)}  & \colhead{Name}   & \colhead{(J2000)} & \colhead{(J2000)} \\
}
\startdata
HH82A	 &    19 01 12.480 & -36 57 17.50   &  HH729C	&    19 01 15.679 & -36 57 56.41 \\
HH82B	 &    19 01 16.090 & -36 57 29.30   &  HH730A	&    19 01 25.26 & -36 59 30.0   \\
HH96	 &    19 01 41.309 & -37 00 51.08   &  HH730B	&    19 01 24.079 & -36 59 59.78 \\
HH97	 &    19 01 45.780 & -37 00 01.51   &  HH730C	&    19 01 15.720 & -37 00 29.99 \\
HH97E	 &    19 01 46.200 & -37 00 03.89   &  HH731A	&    19 01 40.030 & -36 58 57.29 \\
HH97W	 &    19 01 45.749 & -37 00 02.92   &  HH731B	&    19 01 38.270 & -36 58 45.59 \\
HH98	 &    19 01 52.829 & -36 57 36.22   &  HH732A	&    19 01 57.199 & -36 52 07.21 \\
HH98B	 &    19 01 52.150 & -36 57 28.40   &  HH732B	&    19 01 56.410 & -36 51 38.41 \\
HH99	 &    19 02 05.304 & -36 54 39.96   &  HH732C	&    19 01 57.730 & -36 52 17.00 \\
HH99A	 &    19 02 05.410 & -36 54 39.60   &  HH733	&    19 02 09.470 & -36 54 46.19 \\
HH99B	 &    19 02 05.609 & -36 54 34.99   &  HH734A	&    19 02 11.011 & -36 58 28.60 \\
HH99C	 &    19 02 00.910 & -36 55 17.51   &  HH734B	&    19 02 11.729 & -36 58 22.69 \\
HH99E	 &    19 01 59.040 & -36 57 18.00   &  HH735A	&    19 02 12.559 & -36 56 37.79 \\
HH100	 &    19 01 49.104 & -36 58 15.96   &  HH735B	&    19 02 12.761 & -36 56 42.11 \\
HH100N	 &    19 01 49.104 & -36 58 17.04   &  HH736	&    19 02 14.410 & -36 55 22.19 \\
HH100S	 &    19 01 50.208 & -36 58 39.00   &  HH860	&    19 01 39.401 & -36 58 03.79 \\
HH101	 &    19 01 42.000 & -37 00 43.20   &  HH861A	&    19 02 21.569 & -37 16 54.98 \\
HH101N	 &    19 01 35.090 & -37 02 55.39   &  HH861B	&    19 02 24.300 & -37 15 38.02 \\
HH101S	 &    19 01 34.930 & -37 03 01.19   &  HH862	&    19 02 36.581 & -37 11 57.19 \\
HH104	 &    19 01 58.992 & -36 57 15.84   &  HH863A	&    19 03 09.010 & -37 23 55.72 \\
HH104D	 &    19 01 45.890 & -36 57 03.71   &  HH863B	&    19 03 10.819 & -37 19 12.29 \\
HH104C	 &    19 01 45.439 & -36 56 57.59   &  HH863C	&    19 03 18.600 & -37 23 31.20 \\
HH729A	 &    19 01 10.560 & -36 57 32.40   &  HH863D	&    19 03 26.909 & -37 21 54.40 \\
HH729B	 &    19 01 12.941 & -36 57 48.20   &  HH863E	&    19 03 31.699 & -37 20 32.60
\enddata
\label{HHs:tab}
\end{deluxetable}

\begin{deluxetable}{lccl}
\tabletypesize{\scriptsize}
\tablecaption{Identified knots in the CrA star-forming region}
\tablewidth{0pt}
\tablehead{
\colhead{Knot}   & \colhead{R.A.}    & \colhead{Dec.}     & \colhead{Comments} \\
\colhead{Name}   & \colhead{(J2000)} & \colhead{(J2000)}  & \colhead{} \\
}
\startdata
\multicolumn{4}{l}{\textit{Detected in IRAC maps - EGOs }}\\
\hline\\[-5pt]
\textit{Spitzer} outflow 1 A & 19 00 24.139 & -36 58 07.20 & Part of a larger flow with outflow 2 and MHO\,2005?	\\ 
\textit{Spitzer} outflow 1 B & 19 00 24.744 & -36 58 05.19 & Part of a larger flow with outflow 2 and MHO\,2005?	\\  
\textit{Spitzer} outflow 1 C & 19 00 27.379 & -36 59 00.99 & Part of a larger flow with outflow 2 and MHO\,2005?	\\ 
\textit{Spitzer} outflow 1 D & 19 00 28.033 & -36 59 18.27  & Part of a larger flow with outflow 2 and MHO\,2005?   \\
\textit{Spitzer} outflow 1 E & 19 00 28.304 & -36 58 53.48 & Part of a larger flow with outflow 2 and MHO\,2005?	\\  
\textit{Spitzer} outflow 1 F & 19 00 28.602 & -36 59 13.01 & Part of a larger flow with outflow 2 and MHO\,2005?	\\  
\textit{Spitzer} outflow 1 G & 19 00 29.218 & -36 59 00.20 & Part of a larger flow with outflow 2 and MHO\,2005?	\\  
\textit{Spitzer} outflow 1 H & 19 00 30.132 & -36 59 16.10 & Part of a larger flow with outflow 2 and MHO\,2005?	\\  
\textit{Spitzer} outflow 1 I & 19 00 30.648 & -36 59 01.46 & Part of a larger flow with outflow 2 and MHO\,2005?	\\ 
\textit{Spitzer} outflow 1 J & 19 00 31.198 & -36 59 25.28 & Part of a larger flow with outflow 2 and MHO\,2005?	\\
\textit{Spitzer} outflow 1 K & 19 00 31.308 & -36 59 08.80 & Part of a larger flow with outflow 2 and MHO\,2005?	\\  
\textit{Spitzer} outflow 1 L & 19 00 31.850 & -36 59 47.26 & Part of a larger flow with outflow 2 and MHO\,2005?	\\  
\textit{Spitzer} outflow 1 M & 19 00 31.962 & -36 59 30.18 & Part of a larger flow with outflow 2 and MHO\,2005?	\\  
\textit{Spitzer} outflow 1 N & 19 00 32.892 & -36 59 05.16 & Part of a larger flow with outflow 2 and MHO\,2005?	\\  
\textit{Spitzer} outflow 1 O & 19 00 33.592 & -36 59 32.03 & Part of a larger flow with outflow 2 and MHO\,2005?	\\  
\textit{Spitzer} outflow 1 P & 19 00 33.913 & -36 59 00.30 & Part of a larger flow with outflow 2 and MHO\,2005?	\\  
\textit{Spitzer} outflow 1 Q & 19 00 43.098 & -36 58 59.21 & Part of a larger flow with outflow 2 and MHO\,2005?	\\  
\textit{Spitzer} outflow 2 A & 19 01 31.574 & -36 57 03.69 &  Jet-like morphology, part of MHO\,2005 \\ 
\textit{Spitzer} outflow 2 B & 19 01 32.635 & -36 56 55.58 &  Jet-like morphology, part of MHO\,2005 \\  
\textit{Spitzer} outflow 2 C & 19 01 34.272 & -36 56 56.71 &  Jet-like morphology, part of MHO\,2005  \\  
IRAS\,32 A    & 19 03 25.207 & -37 01 07.39 &  Part of IRAS\,32 outflow - NE, blue lobe \\ 
IRAS\,32 B    & 19 03 22.159 & -37 03 53.27 &  Part of IRAS\,32 outflow - NE, blue lobe\\  
IRAS\,32 C    & 19 03 06.643 & -37 05 33.85  & Part of IRAS\,32 outflow - NE, blue lobe \\ 
IRAS\,32 H    & 19 02 34.231 & -37 11 19.14  & Part of IRAS\,32 outflow - SW, red lobe \\ 
IRAS\,32 I    & 19 02 34.128 & -37 11 39.90  & Part of IRAS\,32 outflow - SW, red lobe \\ 
IRAS\,32 J    & 19 02 30.290 & -37 12 46.61  & Part of IRAS\,32 outflow - SW, red lobe \\ 
IRAS\,32 K    & 19 02 25.219 & -37 14 14.13  &   Part of IRAS\,32 outflow - SW, red lobe \\ 
IRAS\,32 L    & 19 02 24.948 & -37 14 20.23  &   Part of IRAS\,32 outflow - SW, red lobe \\ 
IRAS\,32 M    & 19 02 23.923 & -37 14 42.25  &    Part of IRAS\,32 outflow - SW, red lobe \\ 
IRAS\,32 N    & 19 02 24.840 & -37 16 05.23  &    Part of IRAS\,32 outflow - SW, red lobe\\ 
\hline\\[-5pt]
\multicolumn{4}{l}{\textit{Detected in H$_2$ and IRAC maps}}\\
\hline\\[-5pt]
IRAS\,32 D   &  19 03 04.142 & -37 06 21.01 &  MHO\,2010 - Part of IRAS\,32 outflow - NE, blue lobe \\
IRAS\,32 E   &  19 03 04.092 & -37 06 30.78 &  MHO\,2010 - Part of IRAS\,32 outflow - NE, blue lobe \\ 
IRAS\,32 F   &  19 02 53.472 & -37 08 47.56 &  MHO\,2010 - Part of IRAS\,32 outflow - SW, red lobe \\ 
IRAS\,32 G   &  19 02 51.478 & -37 09 01.72 &  MHO\,2010 - Part of IRAS\,32 outflow - SW, red lobe\\ 
MHO\,2000 A         &  19 02 05.561 & -36 54 37.60 &  HH 99\,B 	- Driven by IRS\,9 or CrA\,24.     \\ 
MHO\,2000 B         &  19 02 04.668 & -36 54 56.40 &  HH 99\,A	- Same flow of A? Possibly driven by IRS\,6.     \\  
MHO\,2001 A         &  19 02 01.085 & -36 56 37.47 & 	Driven by CXO 34?	      \\ 
MHO\,2001 B         &  19 02 00.523 & -36 56 40.25 & 	Driven by CXO 34?	     \\
MHO\,2001 C         &  19 01 59.873 & -36 56 32.08 & 	Driven by YSO in the SMA\,2 region    \\  
MHO\,2001 D	   &  19 01 59.568 & -36 56 50.98 & 	Driven by CXO 34?	      \\  
MHO\,2001 E	   &  19 01 58.512 & -36 56 55.32 & 	Driven by CXO 34?	      \\  
MHO\,2001 F	   &  19 01 57.900 & -36 56 53.74 & 	Driven by YSO in the SMA\,2 region	      \\  
MHO\,2001 G	   &  19 01 57.828 & -36 57 05.84 &     Driven by CrA-43 (SMM\,2)  \\ 
MHO\,2001 H	   &  19 01 57.660 & -36 56 59.79 & 	Driven by CrA-43 (SMM\,2)?	      \\
MHO\,2001 I	   &  19 01 56.626 & -36 56 57.79 & 	Driven by YSO in the SMA\,2 region   \\  
MHO\,2001 J         &  19 01 56.412 & -36 57 17.09 & 	Driven by SMA\,2?    \\  
MHO\,2001 K        &  19 01 56.110 & -36 57 02.84 & 	Driven by YSO in the SMA\,2 region	      \\ 
MHO\,2001 L        &  19 01 55.792 & -36 57 11.62 & 	Driven by YSO in the SMA\,2 region	      \\ 
MHO\,2001 M         &  19 01 55.475 & -36 57 18.80 & 	Driven by IRS\,7A?  \\  
MHO\,2001 N         &  19 01 55.355 & -36 57 16.21 & 	Driven by SMA\,2?     \\  
MHO\,2001 O	   &  19 01 55.330 & -36 57 12.20 & 	Driven by YSO in the SMA\,2 region	      \\  
MHO\,2001 P	   &  19 01 55.474 & -36 57 08.74 & 	Driven by YSO in the SMA\,2 region	      \\  
MHO\,2001 Q         &  19 01 52.520 & -36 57 16.12 & 	Driven by SMA\,2?     \\  
MHO\,2002 A         &  19 01 59.462 & -36 57 17.63 & 	Driven by SMA\,2?     \\ 
MHO\,2002 B         &  19 01 59.222 & -36 57 14.46 & 	Driven by CrA-43 (SMM\,2)	 \\  
MHO\,2002 C         &  19 01 58.622 & -36 57 19.08 & 	Driven by SMA\,2 or IRS\,7B?	      \\  
MHO\,2003 A         &  19 01 50.081 & -36 58 30.23 & 	Driven by IRS\,1	      \\ 
MHO\,2003 B         &  19 01 50.394 & -36 58 33.69 & 	Driven by IRS\,1	      \\  
MHO\,2003 C         &  19 01 49.620 & -36 59 10.20 & 	Driven by IRS\,1	       \\ 
MHO\,2004 A         &  19 01 45.432 & -36 56 57.74 &    HH 104C - bow-shock $\sim$east-west oriented - Part of a larger flow with MHO\,2005?\\ 
MHO\,2004 B         &  19 01 45.864 & -36 57 03.79 &    HH 104D - part of a curved jet ne-sw oriented. Driven by IRS\,6 \\ 
MHO\,2004 C         &  19 01 46.849 & -36 56 53.99 &	Part of a curved jet ne-sw oriented. Driven by IRS\,6	       \\ 
MHO\,2004 D         &  19 01 47.929 & -36 56 47.64 &	Part of a curved jet ne-sw oriented. Driven by IRS\,6	       \\ 
MHO\,2005 A	   &  19 01 28.879 & -36 57 10.66 &	Part of a larger east-west flow?     \\ 
MHO\,2005 B	   &  19 01 25.702 & -36 57 21.11 &	Part of a larger east-west flow? 	       \\ 
MHO\,2005 C	   &  19 01 16.202 & -36 57 28.73 &     HH 82B - Part of a larger east-west flow?       \\ 
MHO\,2005 D	   &  19 01 15.790 & -36 57 26.35 &	Part of a larger east-west flow?		  \\ 
MHO\,2005 E	   &  19 01 12.038 & -36 57 08.52 &	Part of a larger east-west flow?		  \\  
MHO\,2005 F	   &  19 01 09.670 & -36 59 23.75 &	Part of a larger east-west flow?		  \\ 
MHO\,2005 G	   &  19 01 04.706 & -36 58 09.74 &	Part of a larger east-west flow?		 \\
MHO\,2005 H	   &  19 01 12.426 & -36 57 17.09 &	HH 82A - Part of a larger east-west flow?	  \\  
MHO\,2005 I	   &  19 00 59.969 & -36 58 15.00 &	Part of a larger east-west flow?		  \\  
MHO\,2006 A         &  19 01 40.923 & -36 58 47.18 &	Driven by IRS\,2		  \\  
MHO\,2006 B         &  19 01 40.827 & -36 58 52.08 &	Driven by IRS\,2		  \\  
MHO\,2006 C         &  19 01 40.033 & -36 58 57.85 &	Driven by IRS\,2		  \\ 
MHO\,2006 D         &  19 01 33.924 & -37 01 03.54 & 	Driven by IRS\,2		  \\
MHO\,2007 A         &  19 01 34.937 & -37 02 55.89 & HH 101N - Driven by IRS\,5?		  \\  
MHO\,2007 B         &  19 01 34.282 & -37 03 01.41 & HH 101S - Driven by IRS\,5?	  \\  
MHO\,2008 A         &  19 01 47.393 & -36 59 52.30 & Part of a larger outflow incl. MH0\,2003?			  \\  
MHO\,2008 B         &  19 01 47.198 & -37 00 02.71 & Part of a larger outflow incl. MH0\,2003?			  \\  
MHO\,2008 C         &  19 01 46.214 & -37 00 05.50 & HH 97E - Part of a larger outflow incl. MH0\,2003?	  \\  
MHO\,2008 D         &  19 01 40.080 & -37 00 02.97 & Bow-shock ENE-WSW oriented, likely driven by CrA-41. \\  
MHO\,2009 A         &  19 02 12.343 & -36 57 10.80 & Driven by SMA 2?			  \\  
MHO\,2009 B         &  19 02 11.357 & -36 57 09.12 & Driven by SMA 2?			  \\ 
MHO\,2009 C         &  19 02 11.031 & -36 57 09.67 & Driven by SMA 2?			  \\  
MHO\,2009 D         &  19 02 10.786 & -36 57 12.80 & Driven by SMA 2?		\\  
MHO\,2009 E         &  19 02 09.240 & -36 57 13.97 & Driven by SMA 2?			  \\ 
MHO\,2009 F         &  19 02 06.463 & -36 57 23.72 & 	Driven by SMA 2?		  \\  
MHO\,2009 G         &  19 02 04.601 & -36 57 10.32 &	Driven by SMA 2?		   \\ 
MHO\,2011 A	   &  19 02 21.247 & -36 55 53.97 &  Part of a large west-east precessing outflow			   \\ 
MHO\,2011 B	   &  19 02 19.037 & -36 56 11.80 &  Part of a large west-east precessing outflow		   \\ 
MHO\,2011 C         &  19 02 17.796 & -36 56 24.24 & Part of a large west-east precessing outflow	   \\ 
MHO\,2011 D	   &  19 02 12.622 & -36 56 43.94 &  HH 735B - Part of a large west-east precessing outflow		   \\ 
MHO\,2011 E	   &  19 02 11.002 & -36 56 49.38 &	Part of a large west-east precessing outflow		   \\ 
MHO\,2011 F	   &  19 02 10.284 & -36 56 52.56 &	Part of a large west-east precessing outflow		   \\ 
MHO\,2012 A         &  19 02 13.980 & -36 56 05.99 & Driven by IRS\,7B?			   \\ 
MHO\,2012 B         &  19 02 07.356 & -36 56 36.37 & Driven by IRS\,7B?			    \\ 
MHO\,2012 C         &  19 02 06.360 & -36 56 41.61 & Driven by IRS\,7B?			    \\  
MHO\,2013 A         &  19 02 09.447 & -36 54 41.50 &	Driven by CXO 34?		     \\ 
MHO\,2013 B         &  19 02 09.454 & -36 54 49.02 &   HH 733 - Driven by CXO 34?	     \\ 
MHO\,2013 C         &  19 02 09.221 & -36 54 52.45 &	Driven by CXO 34?		     \\ 
MHO\,2014 A	   &  19 02 00.641 & -36 52 27.15 &	Driven by IRS\,5?		     \\ 
MHO\,2014 B	   &  19 02 00.281 & -36 52 20.94 &	Driven by IRS\,5?		     \\ 
MHO\,2014 C	   &  19 01 59.801 & -36 52 22.11 &   HH 732C - Driven by IRS\,5?	     \\ 
MHO\,2014 D	   &  19 01 57.662 & -36 52 13.50 &	Driven by IRS\,5?		     \\ 
MHO\,2014 E	   &  19 01 57.158 & -36 52 04.57 &   HH 732A - Driven by IRS\,5?	     \\ 
MHO\,2014 F	   &  19 01 56.916 & -36 52 17.29 & 	Driven by IRS\,5?
\enddata
\label{identified_knots:tab}
\end{deluxetable}

\begin{deluxetable}{lccc}
\tablecaption{Measured H$_2$ proper motions in the CrA knots}
\tablewidth{0pt}
\tablehead{
\colhead{Knot} & \colhead{P.M.} &  \colhead{$v_{tan}$} & \colhead{P.A.}\\
\colhead{Name} & \colhead{($\arcsec$\,yr$^{-1}$)} & \colhead{(km\,s$^{-1}$)} & \colhead{($\degr$)}
}								  
\startdata
MHO\,2000 A &  0.079$\pm$0.009 & 49$\pm$6   & 43$\pm$6	  \\  
MHO\,2000 B &  0.024$\pm$0.007 & 15$\pm$4   & 91$\pm$16   \\ 
MHO\,2001 A &  0.065$\pm$0.007 & 40$\pm$5   & 81$\pm$5  \\
MHO\,2001 B &  0.065$\pm$0.007 & 40$\pm$5   & 81$\pm$5   \\  
MHO\,2001 C &  0.040$\pm$0.010 & 24$\pm$6   & 43$\pm$15	  \\  
MHO\,2001 D &  0.074$\pm$0.010 & 46$\pm$7   & 54$\pm$8	  \\  
MHO\,2001 E &  0.110$\pm$0.010 & 70$\pm$7   & 41$\pm$5	  \\  
MHO\,2001 F &  0.090$\pm$0.015 & 55$\pm$10  & 45$\pm$10	  \\ 
MHO\,2001 G &  0.039$\pm$0.007 & 24$\pm$4   & 328$\pm$10	  \\
MHO\,2001 J &  0.175$\pm$0.006 & 108$\pm$4  & 79$\pm$2	      \\  
MHO\,2001 K &  0.067$\pm$0.009 & 41$\pm$6   & 45$\pm$8	\\  
MHO\,2001 M &  0.084$\pm$0.018 & 52$\pm$11  & 97$\pm$8	\\  
MHO\,2001 N &  0.023$\pm$0.009\tablenotemark{a} & 14$\pm$6\tablenotemark{a}   & 62$\pm$23  	\\  
MHO\,2001 O &  0.044$\pm$0.012 & 27$\pm$7   & 59$\pm$14	  \\  
MHO\,2001 P &  0.110$\pm$0.010 & 70$\pm$7   & 40$\pm$5	      \\ 
MHO\,2001 Q &  0.058$\pm$0.008 & 36$\pm$5   & 289$\pm$8  	\\  
MHO\,2002 A &  0.083$\pm$0.014 & 52$\pm$9   & 81$\pm$9	  \\ 
MHO\,2002 B &  0.056$\pm$0.008 & 35$\pm$5   & 133$\pm$9	\\  
MHO\,2002 C &  0.040$\pm$0.008 & 25$\pm$5   & 65$\pm$12	\\  
MHO\,2003 A &  0.110$\pm$0.013 & 68$\pm$8   & 199$\pm$5	\\ 
MHO\,2003 B &  0.056$\pm$0.013 & 34$\pm$8   & 220$\pm$13   \\  
MHO\,2003 C &  0.186$\pm$0.028 & 115$\pm$17 & 225$\pm$9	   \\ 
MHO\,2004 A &  0.038$\pm$0.011 & 24$\pm$7   & 286$\pm$11    \\ 
MHO\,2006 C &  0.075$\pm$0.012 & 46$\pm$8   & 243$\pm$9     \\ 
MHO\,2007 A &  0.073$\pm$0.012 & 45$\pm$8   & 215$\pm$10     \\ 
MHO\,2007 B &  0.097$\pm$0.012 & 60$\pm$8   & 201$\pm$7     \\ 
MHO\,2008 C &  0.085$\pm$0.012 & 53$\pm$7   & 215$\pm$8     \\ 
MHO\,2008 D &  0.071$\pm$0.012 & 44$\pm$8   & 261$\pm$10  	 \\ 
MHO\,2009 D &  0.147$\pm$0.045 & 91$\pm$28  & 63$\pm$15     \\ 
MHO\,2009 E &  0.137$\pm$0.018 & 85$\pm$11  & 66$\pm$7	    \\ 
MHO\,2009 F &  0.127$\pm$0.019 & 79$\pm$12  & 80$\pm$8	    \\  
MHO\,2009 G &  0.147$\pm$0.036 & 91$\pm$22  & 63$\pm$12	    \\ 
MHO\,2011 D &  0.068$\pm$0.018 & 42$\pm$11  & 68$\pm$15     \\
MHO\,2011 E &  0.256$\pm$0.019 & 158$\pm$12 & 84$\pm$4	    \\  
MHO\,2011 F &  0.080$\pm$0.019 & 49$\pm$12  & 76$\pm$13	    \\  
MHO\,2012 B &  0.134$\pm$0.018 & 83$\pm$11  & 68$\pm$8	    \\  
MHO\,2012 C &  0.140$\pm$0.018 & 87$\pm$11  & 62$\pm$7	    \\  
MHO\,2013 A &  0.086$\pm$0.017 & 53$\pm$11  & 45$\pm$11     \\ 
MHO\,2013 B &  0.041$\pm$0.017\tablenotemark{a} & 26$\pm$11\tablenotemark{a}  & 101$\pm$24     \\
MHO\,2013 C &  0.091$\pm$0.017 & 56$\pm$11  & 63$\pm$11	       \\  
MHO\,2014 A &  0.065$\pm$0.017 & 40$\pm$11  & 32$\pm$15	    	   \\  
MHO\,2014 B &  0.133$\pm$0.017 & 82$\pm$11  & 34$\pm$7	       \\  
MHO\,2014 C &  0.071$\pm$0.017 & 44$\pm$11  & 31$\pm$14	 \\  
MHO\,2014 F &  0.037$\pm$0.017\tablenotemark{a} & 23$\pm$11\tablenotemark{a}  & 347$\pm$27\\
\enddata
\tablenotetext{a}{Below 3$\sigma$}
\label{PMs:tab}
\end{deluxetable}

\clearpage

\begin{deluxetable}{llc}
\tabletypesize{\scriptsize}
\tablecaption{Candidate driving sources and possible flows}
\tablewidth{0pt}
\tablehead{ \colhead{YSO} &  \colhead{Flow\tablenotemark{a}} &  \colhead{P.A.($\degr$) - blue lobe} }					  
\startdata
IRS\,1 &  MHO\,2003, MH0\,2008\,A--C (b)  & 203	  \\  
IRS\,2 &  MHO\,2006 (r?)  & 26	  \\  
IRS\,5N (?) &  MHO\,2007 (b); MH0\,2014 (r) & 25	  \\  
IRS\,6 &  MHO\,2000\,B (?), HH\,99\,C (?) (r); MH0\,2004\,B--D (b?)  & 240	  \\  
IRS\,9 or CrA-24 &  MHO\,2000\,A (r)  & 229 or 223  \\  
IRS\,7A (?) &  MHO\,2001\,F (?)  & 55	  \\ 
IRS\,7B &  MHO\,2012, HH\,736 (?), MHO\,2002\,C (r?); HH\,731\,B (?) (r?) & 248	  \\  
CXO\,34 (?) &  MHO\,2013 (?), MHO\,2001\,A, B, E (?), F (?) (b?)  & 45	  \\  
SMA\,2 region &   MHO\,2013 (?), MHO\,2001\,O, P, K, I (?) (b?); MHO\,2001\,Q (r?) & $\sim$48    \\ 
SMA\,2 &  MHO\,2009, MHO\,2001\,J, A (b?); MHO\,2001\,Q (r?)  & 90	  \\  
IRS7\,A or SMA\,2 &  MHO\,2011 (b); MHO\,2004\,D; Sptz. outflow 1, 2, MHO\,2005 (r) & $\sim$90 \\
CrA-43 (SMM\,2) & MHO\,2002\,B, HH\,734 (?) (b?); HH\,MHO\,2001\,G, H(?) (r?)     &  125   \\
IRAS\,32  & IRAS\,32 outflow\,A--E (b); F--N, HH\,861 (r)  & 39 \\
CrA-41 &  MH0\,2008\,D (b?) & 242  \\ 
CrA-40 (?) & HH\,730\,A--C (?) & 201 (?) \\
S CrA  (?) & HH\,762  (b?)   &   115     \\
VV CrA (?) & HH\,862  (b?)    &    274\\
\enddata
\tablenotetext{a}{b = blue lobe, r = red lobe} 
\label{YSOs_flows:tab}
\end{deluxetable}

%% Tables may also be prepared as separate files. See the accompanying
%% sample file table.tex for an example of an external table file.
%% To include an external file in your main document, use the \input
%% command. Uncomment the line below to include table.tex in this
%% sample file. (Note that you will need to comment out the \documentclass,
%% \begin{document}, and \end{document} commands from table.tex if you want
%% to include it in this document.)


%% The following command ends your manuscript. LaTeX will ignore any text

\end{document}